\documentclass[12pt]{article}
\pdfoutput=1

\usepackage{epsfig}
\usepackage{psfrag}
\usepackage{latexsym}
\usepackage{indentfirst}
\usepackage{fancyhdr}
\usepackage{dsfont}
\usepackage{amsmath}
\usepackage{amssymb}
\usepackage{amsfonts}
\usepackage{mathrsfs}

\usepackage{amsthm}
\usepackage{pifont}
\usepackage{dsfont}
\usepackage{multirow}
\usepackage{array}
\usepackage{chngpage}
\usepackage{longtable}
\usepackage{slashbox}
\usepackage{cite}
\usepackage{bbold}
\usepackage{color}
\usepackage{colordvi}
\usepackage{fancybox}
\usepackage[footnotesize]{caption2}
\usepackage{graphicx}
\usepackage[center,footnotesize,hang]{subfigure}
\usepackage{bbm}
\usepackage[colorlinks, linkcolor=red, anchorcolor=black, citecolor=green]{hyperref}
\usepackage{multirow}
\usepackage{array}
\usepackage{textcomp}  
\usepackage{ulem}
\usepackage{enumitem}
\usepackage{mathrsfs}  
\newcommand{\PreserveBackslash}[1]{\let\temp=\\#1\let\\=\temp}
\newcolumntype{C}[1]{>{\PreserveBackslash\centering}p{#1}}
\newcolumntype{R}[1]{>{\PreserveBackslash\raggedleft}p{#1}}
\newcolumntype{L}[1]{>{\PreserveBackslash\raggedright}p{#1}}
\allowdisplaybreaks  

\makeatletter
\@addtoreset{equation}{section}
\makeatother

\begin{document}

\title{\hfill ~\\[-30mm] \hfill\mbox{\small USTC-ICTS-16-11}\\[10mm]
        \textbf{{\Large Alternative Schemes of Predicting Lepton Mixing Parameters from Discrete Flavor and CP Symmetry}  }}

\date{}

\author{\\[1mm]Jun-Nan Lu\footnote{Email: {\tt hitman@mail.ustc.edu.cn}}~,~Gui-Jun Ding\footnote{Email: {\tt dinggj@ustc.edu.cn}}\\ \\
\it{\small Interdisciplinary Center for Theoretical Study and  Department of Modern Physics, }\\
\it{\small University of Science and
    Technology of China, Hefei, Anhui 230026, China}\\[4mm] }
\maketitle

\begin{abstract}

We suggest two alternative schemes to predict lepton mixing angles as well as $CP$ violating phases from a discrete flavor symmetry group combined with $CP$ symmetry. In the first scenario, the flavor and $CP$ symmetry is broken  to the residual groups of the structure $Z_2\times CP$ in the neutrino and charged lepton sectors. The resulting lepton mixing matrix depends on two free parameters $\theta_{\nu}$ and $\theta_{l}$. This type of breaking pattern is extended to the quark sector. In the second scheme, an abelian subgroup contained in the flavor group is preserved by the charged lepton mass matrix and the neutrino mass matrix is invariant under a single remnant $CP$ transformation, all lepton mixing parameter are determined in terms of three free parameters $\theta_{1, 2, 3}$. We derive the most general criterion to determine whether two distinct residual symmetries lead to the same mixing pattern if the redefinition of the free parameters $\theta_{\nu, l}$ and $\theta_{1, 2, 3}$ is taken into account. We have studied the lepton mixing patterns arising from the flavor group $S_4$ and $CP$ symmetry which are subsequently broken to all of the possible residual symmetries discussed in this work.

\end{abstract}
\thispagestyle{empty}
\vfill

\newpage
\setcounter{page}{1}

\section{Introduction}

The neutrino oscillation experiments have made great progress in the last twenty years~\cite{Kajita:2016cak,McDonald:2016ixn,nobel_NPB}. It has been firmly established that neutrinos must be massive particles and different flavor eigenstates are mixed. The three lepton mixing angles $\theta_{12}$, $\theta_{13}$ and $\theta_{23}$ as well as two mass squared difference $\Delta m^2_{21}$ and $\Delta m^2_{31}$ have been precisely measured~\cite{Capozzi:2013csa,Forero:2014bxa,Gonzalez-Garcia:2014bfa,Capozzi:2016rtj}. However, we still don't know the neutrino mass ordering ($\Delta m^2_{31}>0$ or $\Delta m^2_{31}<0$) and the signal of $CP$ violation in the lepton sector has not been observed. The preliminary T2K data favor a maximal Dirac $CP$ violation phase $\delta_{CP}\simeq-\pi/2$~\cite{Abe:2015awa}, and the latest global fits of neutrino mixing parameters show a weak evidence for a negative Dirac phase $-\pi<\delta_{CP}<0$~\cite{Forero:2014bxa,Gonzalez-Garcia:2014bfa,Capozzi:2016rtj}. The primary objectives of near future neutrino experiments are to determine the ordering of the neutrino masses and to measure the value of $\delta_{CP}$.

On the theoretical side, the origin of neutrino mass and lepton flavor mixing is still unknown although there has been lots of theoretical studies. Motivated by the observation that the simple tri-bimaximal mixing possibly originates from a $A_4$ flavor group, non-abelian discrete flavor symmetry has been extensively exploited to explain the observed lepton mixing angles. Many other symmetries such as $S_4$, $A_5$, $\Delta(3n^2)$ and $\Delta(6n^2)$ etc have been considered over the years. Please see Refs.~\cite{Altarelli:2010gt,Ishimori:2010au,King:2013eh,King:2014nza,King:2015aea} for review on discrete flavor symmetry and its application in model
building. A significant progress in recent years is the precise measurement of the reactor mixing angle $\theta_{13}$~\cite{Abe:2011sj,Adamson:2011qu,Abe:2011fz,An:2012eh,Ahn:2012nd}. The discovery of a somewhat large value of $\theta_{13}$ rules out the tri-bimaximal mixing patterns and many flavor models which predicted small or zero $\theta_{13}$. Many approaches have been pursued to explain such a largish $\theta_{13}$. Within the paradigm of the discrete flavor symmetry, model-independent scan of the lepton sector reveals that only large flavor symmetry groups (e.g. $(Z_{18}\times Z_6)\rtimes S_3$ with the group id [648, 259]) can produce mixing patterns compatible with experimental data and the Dirac $CP$ phase is generally trivial if the lepton mixing matrix is fully fixed by the symmetry alone~\cite{Holthausen:2012wt,King:2013vna,Fonseca:2014koa,Talbert:2014bda,Yao:2015dwa}.

In order to accommodate a non-zero $\theta_{13}$ and a nontrivial Dirac $CP$ phase simultaneously, it is interesting to combine flavor symmetry with $CP$ symmetry. This approach can generate a rich structure of mixing patterns which are in good agreement with the experimental data, and it allows us to predict all the mixing angles and $CP$ phases in terms of a small number of input parameters~\cite{Feruglio:2012cw,Chen:2015siy,Chen:2016ica}. From the bottom-up point of view, the generic neutrino and charged lepton mass matrices have both residual $CP$ symmetry and residual flavor symmetry, and the residual flavor symmetry can be generated from the residual $CP$ transformations~\cite{Chen:2014wxa,Everett:2015oka,Chen:2015nha}. Hence it is natural to assume that the residual flavor and $CP$ symmetry arise from a large flavor and $CP$ symmetry group at high energy scale. In this approach, the $CP$ symmetry nontrivially acts on the flavor space such that the so called consistency condition has to be fulfilled in order for the theory to be consistent~\cite{Feruglio:2012cw,Grimus:1995zi,Holthausen:2012dk,Chen:2014tpa}. There has been intense theoretical activity on flavor symmetry in combination with $CP$ symmetry. Many flavor symmetry groups and their predictions for lepton mixing parameters have been studied such as $A_4$~\cite{Ding:2013bpa,Nishi:2016jqg,He:2015gba,Ma:2015pma,Li:2016nap}, $S_4$~\cite{Feruglio:2012cw,Mohapatra:2012tb,Ding:2013hpa,Feruglio:2013hia,Luhn:2013vna,Li:2013jya,Li:2014eia}, $A_5$~\cite{Li:2015jxa,DiIura:2015kfa,Ballett:2015wia,Turner:2015uta}, $\Delta(27)$~\cite{Branco:2015hea,Branco:2015gna}, $\Delta(48)$~\cite{Ding:2013nsa,Ding:2014hva} and $\Delta(96)$~\cite{Ding:2014ssa} as well as $\Delta(3n^2)$~\cite{Hagedorn:2014wha,Ding:2015rwa}, $\Delta(6n^{2})$~\cite{Hagedorn:2014wha,King:2014rwa,Ding:2014ora} and $D^{(1)}_{9n, 3n}$~\cite{Li:2016ppt} group series for a generic integer $n$. Recently a comprehensive scan of leptonic mixing parameters which can be obtained from finite discrete groups of order less than 2000 and $CP$ symmetry has been performed~\cite{Yao:2016zev}. Moreover, the phenomenological implications of flavor and $CP$ symmetry in neutrinoless double decay~\cite{Li:2016nap,Ding:2013hpa,Li:2014eia,Li:2015jxa,Ding:2014ora,Li:2016ppt,Yao:2016zev,Hagedorn:2016lva} and leptogenesis~\cite{Yao:2016zev,Chen:2016ptr,Hagedorn:2016lva} have been investigated. It is remarkable that the residual $CP$ symmetry provides a bridge between flavored leptogenesis and low energy leptonic $CP$ violation.

It is usually assumed that the residual flavor symmetry in the charged lepton is an abelian subgroup which can distinguish among the three generations, and the residual symmetry in the neutrino sector is a direct product of $Z_2$ and $CP$. As a consequence, the lepton mixing matrix turns out to depend on a single real parameter $\theta$ and all mixing parameters are strongly correlated with each other. In the present work, we shall discuss the other possible approaches to predict lepton mixing parameters from flavor and $CP$ symmetry, and two scenarios would be considered. In the first one, the neutrino and charged lepton mass matrices are invariant under two distinct $Z_2\times CP$ subgroups. Consequently all mixing parameters including mixing angles and $CP$ phases are predicted in terms of two real parameters $\theta_{l}$ and $\theta_{\nu}$. In the second scenario, the postulated flavor symmetry is broken to a residual abelian subgroup with three or more elements in the charged lepton sector while a single residual $CP$ transformation is preserved by the neutrino mass matrix, the PMNS mixing matrix would depend on three real parameters $\theta_{1, 2, 3}$. As an example, we present a detailed analysis for the $S_4$ flavor symmetry group and $CP$ symmetry. All possible independent combinations of remnant symmetries and the predictions for lepton mixing parameters are studied analytically and numerically.

The paper is organized as follows. In section~\ref{sec:lepton_Z2xCP} we study the symmetry breaking pattern in which a flavor symmetry combined with a $CP$ symmetry is broken to $Z_2\times CP$ in both the neutrino and charged lepton sectors. The resulting consequence for the prediction of the lepton mixing matrix is discussed, and the technical steps in the derivation are explained. We derive the conditions under which two distinct residual symmetries give rise to the same mixing pattern. Moreover we analyze the independent mixing patterns which can be obtained from the popular flavor group $S_4$ and CP in this scheme. In section~\ref{sec:quark_Z2xCP} our approach is extended to the quark sector. In section~\ref{sec:single_res_CP} we explore another proposal in which the charged lepton and neutrino mass matrices are invariant under the action of a residual abelian subgroup and a single CP transformation respectively. Finally section~\ref{sec:conclusion} concludes this paper. Moreover, Appendix~\ref{sec:S4_group_app} contains the necessary group theory of $S_4$ as well as its abelian subgroups. Appendix~\ref{sec:equivalence_quark_n0_n1_app} gives the conditions under which two distinct residual symmetries of the structure $Z_2\times CP$ in both the up and down quark sectors lead to the same CKM mixing matrix in the case that the fixed element is neither 0 nor 1.

\section{\label{sec:lepton_Z2xCP}Lepton flavor mixing from residual symmetry $Z_2\times CP$ in both charged lepton and neutrino sectors }

In the widely studied direct and semidirect approaches~\cite{King:2013eh,King:2014nza,King:2015aea}, it is assumed that the neutrino mass matrix $m_{\nu}$ possesses residual symmetry $Z_2\times Z_2$ and $Z_2\times CP$ respectively, and the charged lepton mass matrix is invariant under an abelian subgroup contained in the flavor group. In this section, we shall be concerned with the scenario that the remnant symmetry preserved by both the neutrino and charged lepton mass matrices is of the structure $Z_2\times CP$. The three generations of left-handed leptons are assigned to a faithful irreducible triplet $\mathbf{3}$ of the flavor symmetry group.

\subsection{\label{subsec:framework_Z2xCP}General form of the PMNS matrix}

We shall denote the residual $Z_2$ flavor symmetry of the charged lepton sector as $Z^{g_l}_{2}\equiv\{1, g_{l}\}$ with $g^2_{l}=1$, and the remnant $CP$ transformation is $X_{l}$. In order for the theory to be consistent, the following consistency condition has to be fulfilled
\begin{equation}
\label{eq:res_cons_cha}X_{l}\rho^{*}_{\mathbf{3}}(g_{l})X^{-1}_{l}=\rho_{\mathbf{3}}(g_{l})\,,
\end{equation}
where $\rho_{\mathbf{3}}(g_{l})$ denote the representation matrix of the element $g_{l}$ in the three dimensional representation $\mathbf{3}$. The charged lepton mass matrix $m^{\dagger}_{l}m_{l}$ is invariant under the action of the residual symmetry $Z^{g_l}_2\times X_{l}$, and it fulfills
\begin{subequations}
\begin{eqnarray}
\label{eq:cons_charged_lepton_z2xCP_CP} X^{\dagger}_{l}m^{\dagger}_{l}m_{l}X_{l}&=&(m^{\dagger}_{l}m_{l})^{*}\,,\\
\label{eq:cons_charged_lepton_z2xCP_fla}\rho^{\dagger}_{\mathbf{3}}(g_{l})m^{\dagger}_{l}m_{l}\rho_{\mathbf{3}}(g_{l})&=&m^{\dagger}_{l}m_{l}\,.
\end{eqnarray}
\end{subequations}
The unitary transformation $U_{l}$ which diagonalizes the hermitian matrix $m^{\dagger}_{l}m_{l}$ with $U^{\dagger}_{l}m^{\dagger}_{l}m_{l}U_{l}=\text{diag}(m^2_{e}, m^2_{\mu}, m^2_{\tau})$ are strongly constrained by the postulated residual symmetry. In the following, we shall show how to determine $U_{l}$ from $\rho_{\mathbf{3}}(g_{l})$ and $X_{l}$.
As the order of $g_{l}$ is 2, the eigenvalues of $\rho_{\mathbf{3}}(g_{l})$ are $(1,-1,-1)$ or $(-1,1,1)$, we take the first case as an example without loss of generality. Assuming $\Sigma_{l1}$ is a diagonalization matrix of $\rho(g_{l})$ and it satisfies
\begin{equation}
\label{eq:z2lep1}
\Sigma_{l1}^{\dagger}\rho_{\mathbf{3}}(g_{l})\Sigma_{l1}=\text{diag}(1,-1,-1)\equiv \hat{\rho}_{\mathbf{3}}(g_{l}),
\end{equation}
then we have
\begin{equation}
\label{eq:z2lep2}
\rho_{\mathbf{3}}(g_{l})=\Sigma_{l1}\hat{\rho}_{\mathbf{3}}(g_{l})\Sigma_{l1}^{\dagger}
\end{equation}
The residual $CP$ has to be consistent with the residual flavor symmetry, therefore the following consistency condition should be fulfilled~\cite{Li:2014eia,Li:2016ppt}
\begin{equation}
\label{eq:z2lep3}
X_{l}\rho^{*}_{\mathbf{3}}(g_{l})X_{l}^{\dagger}=\rho_{\mathbf{3}}(g_{l}^{-1}).
\end{equation}
Inserting Eq.~\eqref{eq:z2lep2} into the above equation and considering $g^2_{l}=1$, we get
\begin{equation}
\label{eq:z2lep4}
X_{l}\Sigma_{l1}^{*}\hat{\rho}_{\mathbf{3}}(g_{l})\Sigma_{l1}^{T}X_{l}^{\dagger}=\Sigma_{l1}\hat{\rho}_{\mathbf{3}}(g_{l})\Sigma_{l1}^{\dagger},
\end{equation}
which leads to
\begin{equation}
\label{eq:z2lep5}
(\Sigma_{l1}^{\dagger}X_{l}\Sigma_{l1}^{*})\hat{\rho}_{\mathbf{3}}(g_{l})(\Sigma_{l1}^{T}X_{l}^{\dagger}\Sigma_{l1})=\hat{\rho}_{\mathbf{3}}(g_{l})\,,
\end{equation}
which means
\begin{equation}
\label{eq:z2lep6}
(\Sigma_{l1}^{\dagger}X_{l}\Sigma_{l1}^{*})\hat{\rho}_{\mathbf{3}}(g_{l})=\hat{\rho}_{\mathbf{3}}(g_{l})(\Sigma_{l1}^{\dagger}X_{l}\Sigma_{l1}^{*})\,.
\end{equation}
Therefore $\Sigma_{l1}^{\dagger}X_{l}\Sigma_{l1}^{*}$ is a block diagonal and symmetric matrix and its most general form is given by
\begin{equation}
\label{eq:z2lep7}
\Sigma_{l1}^{\dagger}X_{l}\Sigma_{l1}^{*}=\left(
\begin{array}{cc}
e^{i\xi_{1}} & 0  \\
0 & u^{l}_{2\times2}
\end{array}
\right)
\end{equation}
where $\xi_{1}$ is an arbitrary real number and $u^{l}_{2\times2}$ is a two-dimensional symmetric unitary matrix. We denote the Takagi factorization of $u^{l}_{2\times2}$ as $\sigma^{l}_{2\times2}$ fulfilling $u^{l}_{2\times2}=\sigma^{l}_{2\times2}\sigma^{lT}_{2\times2}$, where $\sigma^{l}_{2\times2}$ is a two-dimensional unitary matrix. As a result, the matrix $\Sigma_{l1}^{\dagger}X_{l}\Sigma^{*}_{l1}$ can be written into
\begin{equation}
\label{eq:z2lep8}
\Sigma_{l1}^{\dagger}X_{l}\Sigma^{*}_{l1}=\left(
\begin{array}{cc}
e^{i\xi_{1}/2} & 0  \\
0 & \sigma^{l}_{2\times2}
\end{array}
\right)\left(
\begin{array}{cc}
e^{i\xi_{1}/2} & 0  \\
0 & \sigma^{lT}_{2\times2}
\end{array}
\right)\,.
\end{equation}
Then we can obtain the Takagi factorization of $X_{l}$ as
\begin{equation}
\label{eq:z2lep9}
X_{l}=[\Sigma_{l1}\left(
\begin{array}{cc}
e^{i\xi_{1}/2} & 0  \\
0 & \sigma^{l}_{2\times2}
\end{array}
\right)][\Sigma_{l1}\left(
\begin{array}{cc}
e^{i\xi_{1}/2} & 0  \\
0 & \sigma^{l}_{2\times2}
\end{array}
\right)]^{T}\equiv\Sigma_{l}\Sigma_{l}^{T}
\end{equation}
with
\begin{equation}
\Sigma_{l}=\Sigma_{l1}\left(
\begin{array}{cc}
e^{i\xi_{1}/2} & 0  \\
0 & \sigma^{l}_{2\times2}
\end{array}
\right)\,.
\end{equation}
It is straightforward to check that the remnant flavor transformation $\rho_{\mathbf{3}}(g_{l})$ is diagonalized by $\Sigma_{l}$,
\begin{equation}
\label{eq:z2lep10}
\Sigma_{l}^{\dagger}\rho_{\mathbf{3}}(g_{l})\Sigma_{l}=\text{diag}(1,-1,-1).
\end{equation}
From Eq.~\eqref{eq:cons_charged_lepton_z2xCP_CP} we can obtain that the constraint on the unitary transformation $U_{l}$ from the residual $CP$ transformation $X_{l}$ is
\begin{equation}
U_{l}^{\dagger}X_{l}U_{l}^{*}=\text{diag}(e^{i\beta_{e}},e^{i\beta_{\mu}},e^{i\beta_{\tau}})\equiv Q^2_{l}\,,
\end{equation}
where $\beta_{e, \mu, \tau}$ are arbitrary real parameters. Thus we have
\begin{equation}
U_{l}^{\dagger}\Sigma_{l}\Sigma^{T}_{l}U_{l}^{*}=Q^2_{l}\,,
\end{equation}
which leads to
\begin{equation}
\label{eq:z2lep11}
(\Sigma_{l}^{T}U_{l}^{*}Q_{l}^{-1})^{T}(\Sigma_{l}^{T}U_{l}^{*}Q_{l}^{-1})=1\,.
\end{equation}
Hence the combination $\Sigma_{l}^{T}U_{l}^{*}Q_{l}^{-1}$ is an orthogonal matrix, and it is also a unitary matrix. Therefore $\Sigma_{l}^{T}U_{l}^{*}Q_{l}^{-1}$ is a real orthogonal matrix denoted by $O_{3\times3}$. Then the unitary transformation $U_{l}$ takes the following form
\begin{equation}
\label{eq:z2lep14}
U_{l}=\Sigma_{l}O_{3\times3}Q_{l}^{-1}\,.
\end{equation}
Furthermore, Eq.~\eqref{eq:cons_charged_lepton_z2xCP_fla} implies that $U_{l}$ is also subject to the constraint of the residual flavor symmetry as follows,
\begin{equation}
\label{eq:z2lep15}
U_{l}^{\dagger}\rho_{\mathbf{3}}(g_{l})U_{l}=P_{l}\,\text{diag}(1,-1,-1)P^{T}_{l}
\end{equation}
where $P_{l}$ is a generic permutation matrix, and it can take six possible forms $1$, $P_{12}$, $P_{13}$, $P_{23}$, $P_{23}P_{12}$ and $P_{23}P_{13}$ with
\begin{equation}
  \label{eq:2.1}
  P_{12}=\left( \begin{array}{ccc}
0&1&0\\
1&0&0\\
0&0&1 \end{array} \right)
,~~
  P_{13}=\left( \begin{array}{ccc}
0&0&1\\
0&1&0\\
1&0&0 \end{array} \right)
,~~
  P_{23}=\left( \begin{array}{ccc}
1&0&0\\
0&0&1\\
0&1&0 \end{array} \right)\,.
\end{equation}
Plugging the expression of $U_{l}$ in Eq.~\eqref{eq:z2lep14} into Eq.~\eqref{eq:z2lep15}, we obtain
\begin{equation}
\label{eq:z2lep16}P^{T}_{l}Q_{l}O_{3\times3}^{\dagger}\Sigma^{\dagger}_{l}\rho_{\mathbf{3}}(g_{l})\Sigma_{l} O_{3\times3}Q_{l}^{-1}P_{l}=\text{diag}(1,-1,-1)\,.
\end{equation}
Using Eq.\eqref{eq:z2lep10} we have
\begin{equation}
\label{eq:z2lep17}\left[O_{3\times3}Q_{l}^{-1}P_{l}\right]^{\dagger}\text{diag}(1,-1,-1)\left[O_{3\times3}Q_{l}^{-1}P_{l}\right]=\text{diag}(1,-1,-1)
\end{equation}
Therefore the combination $O_{3\times3}Q_{l}^{-1}P_{l}$ is a block diagonal unitary matrix, and it can be parameterized as
\begin{equation}
\label{eq:z2lep18}
O_{3\times3}Q_{l}^{-1}P_{l}=\left( \begin{array}{cc}
e^{i\xi_{2}} & 0\\
0& v^{l}_{2\times2} \end{array} \right)\,,
\end{equation}
where $\xi_{2}$ is a real number and $v^{l}_{2\times2}$ is a two-dimensional unitary matrix. Thus we have
\begin{equation}
\label{eq:z2lep19}
(O_{3\times3}Q_{l}^{-1}P_{l})^{T}(O_{3\times3}Q_{l}^{-1}P_{l})=P^{T}_{l}Q_{l}^{-2}P_{l}=\left( \begin{array}{cc}
e^{2i\xi_{2}} & 0 \\
0 & v_{2\times2}^{lT}v^{l}_{2\times2} \end{array} \right)\,,
\end{equation}
which implies
\begin{equation}
\left[\left(\begin{array}{cc}
e^{i\xi_{2}}  &  0 \\
0   &   v^{l}_{2\times2}
\end{array}\right)
P^{T}_{l}Q_{l}P_{l}\right]^{T}\left[\left(\begin{array}{cc}
e^{i\xi_{2}}  &  0 \\
0   &   v^{l}_{2\times2}
\end{array}\right)
P^{T}_{l}Q_{l}P_{l}\right]=1\,.
\end{equation}
Hence $\left(\begin{array}{cc}
e^{i\xi_{2}}  &  0 \\
0   &   v^{l}_{2\times2}
\end{array}\right)
P^{T}_{l}Q_{l}P_{l}$ is a block diagonal real orthogonal matrix, and it takes the form
\begin{equation}
\left(\begin{array}{cc}
e^{i\xi_{2}}  &  0 \\
0   &   v^{l}_{2\times2}
\end{array}\right)
P^{T}_{l}Q_{l}P_{l}=S^{T}_{23}(\theta_{l})\,,
\end{equation}
where $S_{23}(\theta_{l})$ is a rotation matrix with
\begin{equation}
\label{eq:z2lep27}
S_{23}(\theta_{l})\equiv\left( \begin{array}{ccc}
1 & 0 & 0 \\
0 & \cos\theta_{l} & \sin\theta_{l} \\
0 & -\sin\theta_{l} & \cos\theta_{l} \end{array} \right)\,.
\end{equation}
As a consequence, the unitary transformation $U_{l}$ is fixed by the residual symmetry $Z_{2}\times CP$ to be
\begin{equation}
\label{eq:z2lep29}
U_{l}=\Sigma_{l}S^{T}_{23}(\theta_{l})P^{T}_{l}Q^{-1}_{l}\,.
\end{equation}
Similarly the residual flavor symmetry of the neutrino mass matrix is denoted as $Z^{g_\nu}_2\equiv\{1, g_{\nu}\}$ with $g_{\nu}^{2}=1$, the residual $CP$ transformation is $X_{\nu}$, and $CP$ should commute with $Z^{g_{\nu}}_2$ as well
\begin{equation}
\label{eq:res_cons_neutrino}X_{\nu}\rho^{*}_{\mathbf{3}}(g_{\nu})X^{-1}_{\nu}=\rho_{\mathbf{3}}(g_{\nu})\,.
\end{equation}
The invariance of the neutrino mass matrix under the residual symmetry $Z^{g_\nu}_2\times X_{\nu}$ requires
\begin{equation}
\rho^T_{\mathbf{3}}(g_\nu)m_\nu\rho_{\mathbf{3}}(g_\nu)=m_\nu,\qquad X_{\nu}^Tm_\nu X_{\nu}=m_\nu^{\ast}
\end{equation}
Plugging $U_\nu^Tm_\nu U_\nu=\text{diag}(m_1, m_2, m_3)$ into this equation, we can derive the following constraints on the unitary transformation $U_{\nu}$,
\begin{subequations}
\begin{eqnarray}
\label{eq:Unu_Gnu}&&U_{\nu}^{\dagger}\rho_{\mathbf{3}}(g_\nu)U_\nu=\text{diag}(\pm 1,\pm 1,\pm 1)\,,\\
\label{eq:Unu_Xnu}&&U_\nu^{\dagger}X_{\nu}U_{\nu}^*=\text{diag}(\pm 1,\pm 1,\pm 1)\equiv Q^2_{\nu}\,,
\end{eqnarray}
\end{subequations}
where $Q_{\nu}$ is a diagonal and unitary matrix with non-vanishing entries equal to $\pm1$ and $\pm i$. Without loss of generality $Q_{\nu}$ can be parameterized as
\begin{equation}
\label{eq:Qnu_para}Q_{\nu}=\left(\begin{array}{ccc}
1   ~&~  0  ~&~  0 \\
0  ~&~  i^{k_1}  ~&~ 0 \\
0  ~&~  0  ~&~  i^{k_2}
\end{array}
\right)\,,
\end{equation}
with $k_{1,2}=0, 1, 2, 3$. Firstly we can diagonalize the residual flavor symmetry transformation $\rho_{\mathbf{3}}(g_{\nu})$ by a unitary transformation $\Sigma_{\nu1}$ as
\begin{equation}
\label{eq:z2lep30}
\Sigma_{\nu1}^{\dagger}\rho_{\mathbf{3}}(g_{\nu})\Sigma_{\nu1}=\hat{\rho}_{\mathbf{3}}(g_{\nu})=\text{diag}(1,-1,-1)
\end{equation}
The consistency condition of remnant symmetry is
\begin{equation}
  \label{eq:z2lep31}
  X_{\nu}\rho^{*}_{\mathbf{3}}(g_{\nu})X_{\nu}^{\dagger}=\rho_{\mathbf{3}}(g_{\nu}^{-1})=\rho_{\mathbf{3}}(g_{\nu})\,,
\end{equation}
which leads to
\begin{equation}
  \label{eq:z2lep32}
  X_{\nu}\Sigma_{\nu1}^{*}\hat{\rho}^{*}_{\mathbf{3}}(g_{\nu})\Sigma_{\nu1}^{T}X_{\nu}^{\dagger}=\Sigma_{\nu1}\hat{\rho}_{\mathbf{3}}(g_{\nu})\Sigma_{\nu1}^{\dagger}\,.
\end{equation}
Thus we have
\begin{equation}
  \label{eq:z2lep33}
  (\Sigma_{\nu1}^{\dagger}X_{\nu}\Sigma_{\nu1}^{*})\hat{\rho}^{*}_{\mathbf{3}}(g_{\nu})(\Sigma_{\nu1}^{\dagger}X_{\nu}\Sigma_{\nu1}^{*})^{\dagger}=\hat{\rho}_{\mathbf{3}}(g_{\nu})\,.
\end{equation}
Hence $\Sigma_{\nu1}^{\dagger}X_{\nu}\Sigma_{\nu1}^{*}$ is a block diagonal matrix, and it is of the following form
\begin{equation}
\label{eq:z2lep34}
\Sigma_{\nu1}^{\dagger}X_{\nu}\Sigma_{\nu1}^{*}=\left( \begin{array}{cc}
e^{i\zeta_{1}} & 0\\
0& u^{\nu}_{2\times2} \\
\end{array} \right),
\end{equation}
where $\zeta_{1}$ is an arbitrary real number and $u^{\nu}_{2\times2}$ is a two-dimensional symmetric unitary matrix. $u^{\nu}_{2\times2}$ can be factorized into the form $u^{\nu}_{2\times2}=\sigma^{\nu}_{2\times2}\sigma_{2\times2}^{\nu T}$ with $\sigma^{\nu}_{2\times2}\sigma_{2\times2}^{\nu\dagger}=1$. Then we obtain
\begin{equation}
\Sigma_{\nu1}^{\dagger}X_{\nu}\Sigma_{\nu1}^{*}=\Sigma_{\nu2}\Sigma^{T}_{\nu2}\,,
\end{equation}
where
\begin{equation}
\label{eq:z2lep35}
  \Sigma_{\nu2}=\left( \begin{array}{cc}
e^{i\zeta_{1}/2} & 0 \\
0 & \sigma^{\nu}_{2\times2} \end{array} \right)\,.
\end{equation}
As a consequence, the Takagi factorization of the residual $CP$ transformation $X_{\nu}$ is given by
\begin{equation}
\label{eq:z2lep38}
X_{\nu}=\Sigma_{\nu}\Sigma_{\nu}^{T}.
\end{equation}
with $\Sigma_{\nu}=\Sigma_{\nu1}\Sigma_{\nu2}$. It is easy to check that the residual flavor transformation $\rho_{\mathbf{3}}(g_{\nu})$ is diagonalized by $\Sigma_{\nu}$ as well,
\begin{equation}
\label{eq:z2lep39}
\Sigma_{\nu}^{\dagger}\rho_{\mathbf{3}}(g_{\nu})\Sigma_{\nu}=
\Sigma_{\nu2}^{\dagger}\Sigma_{\nu1}^{\dagger}\rho_{\mathbf{3}}(g_{\nu})\Sigma_{\nu1}\Sigma_{\nu2}=\Sigma_{\nu2}^{\dagger}\text{diag}(1,-1,-1)\Sigma_{\nu2}=\text{diag}(1,-1,-1)\,.
\end{equation}
Now we proceed to discuss the constraint on $U_{\nu}$ from the remnant $CP$ transformation. Substituting the relation $X_{\nu}=\Sigma_{\nu}\Sigma_{\nu}^T$ into Eq.~\eqref{eq:Unu_Xnu}, we get
\begin{equation}
\left(Q_{\nu}U_\nu^{\dagger}\Sigma_{\nu}\right)\left(Q_{\nu}U_\nu^{\dagger}\Sigma_{\nu}\right)^{T}=1\,.
\end{equation}
This implies that $Q_{\nu}U_\nu^{\dagger}\Sigma_{\nu}$ is a real orthogonal matrix denoted as $O_{3\times3}$. Therefore the unitary transformation $U_{\nu}$ is of the form
\begin{equation}
\label{eq:UnuSigma}
U_{\nu}=\Sigma_{\nu} O_{3\times 3}^T Q_{\nu}\,.
\end{equation}
Subsequently we consider the constraint from the residual flavor symmetry given in Eq.~\eqref{eq:Unu_Gnu},
\begin{equation}
\label{eq:GnuUnu2_Pnu}
U_{\nu}^{\dagger}\rho_{\mathbf{3}}(g_\nu)U_\nu=P^T_\nu\text{diag}(1,-1,-1)P_\nu\,,
\end{equation}
where $P_{\nu}$ is a permutation matrix, since the neutrino masses are unconstrained in the present framework and the neutrino mass spectrum can be either normal hierarchy (NH) or inverted hierarchy (IH). Inserting Eq.~\eqref{eq:UnuSigma} into Eq.~\eqref{eq:GnuUnu2_Pnu}, one finds
\begin{equation}
\label{eq:z2lep47}
Q^{-1}_{\nu}O_{3\times3}\Sigma_{\nu}^{\dagger}\rho_{\mathbf{3}}(g_{\nu})\Sigma_{\nu}O^{T}_{3\times3}Q_{\nu}
=Q^{-1}_{\nu}O_{3\times3}\text{diag}(1,-1,-1)O^{T}_{3\times3}Q_{\nu}=P^{T}_{\nu}\text{diag}(1,-1,-1)P_{\nu}\,.
\end{equation}
which gives rise to
\begin{equation}
\text{diag}(1,-1,-1)\left(O^{T}_{3\times3}Q_{\nu}P^{T}_{\nu}\right)=\left(O^{T}_{3\times3}Q_{\nu}P^{T}_{\nu}\right)\text{diag}(1,-1,-1)\,.
\end{equation}
Therefore $O^{T}_{3\times3}Q_{\nu}P^{T}_{\nu}$ is a block-diagonal unitary matrix, and we can parameterize it as
\begin{equation}
\label{eq:z2lep48}
O^{T}_{3\times3}Q_{\nu}P^{T}_{\nu}=\left( \begin{array}{cc}
e^{i\zeta_{2}} & 0 \\
0 & v^{\nu}_{2\times2} \end{array} \right)\,,
\end{equation}
where $\zeta_{2}$ is real and $v^{\nu}_{2\times2}$ is a two-dimensional unitary matrix. Both sides of this equation multiply with their transpose, we obtain
\begin{equation}
\left(O^{T}_{3\times3}Q_{\nu}P^{T}_{\nu}\right)^{T}\left(O^{T}_{3\times3}Q_{\nu}P^{T}_{\nu}\right)=P_{\nu}Q^{2}_{\nu}P^{T}_{\nu}=
\left( \begin{array}{cc}
e^{2i\zeta_{2}} & 0 \\
0 & v^{\nu T}_{2\times2}v^{\nu}_{2\times2} \end{array} \right)\,,
\end{equation}
which implies
\begin{equation}
\left[\left( \begin{array}{cc}
e^{i\zeta_{2}} & 0 \\
0 & v^{\nu}_{2\times2} \end{array} \right)P_{\nu}Q^{-1}_{\nu}P^{T}_{\nu}\right]^{T}\left[\left( \begin{array}{cc}
e^{i\zeta_{2}} & 0 \\
0 & v^{\nu}_{2\times2} \end{array} \right)P_{\nu}Q^{-1}_{\nu}P^{T}_{\nu}\right]=1\,.
\end{equation}
Therefore $\left( \begin{array}{cc}
e^{i\zeta_{2}} & 0 \\
0 & v^{\nu}_{2\times2} \end{array} \right)P_{\nu}Q^{-1}_{\nu}P^{T}_{\nu}$ is a block diagonal real orthogonal matrix, and it is of the following form
\begin{equation}
\left( \begin{array}{cc}
e^{i\zeta_{2}} & 0 \\
0 & v^{\nu}_{2\times2} \end{array} \right)P_{\nu}Q^{-1}_{\nu}P^{T}_{\nu}=S_{23}(\theta_{\nu})\,,
\end{equation}
where $\theta_{\nu}$ is real. Consequently, the unitary transformation $U_{\nu}$ is fixed to be
\begin{equation}
\label{eq:z2lep56}
U_{\nu}=\Sigma_{\nu}S_{23}(\theta_{\nu})P_{\nu}Q_{\nu}\,.
\end{equation}
The lepton mixing matrix $U_{PMNS}$ is a result of the mismatch between $U_{l}$ and $U_{\nu}$. Hence we find $U_{PMNS}$ is of the form
\begin{equation}
\label{eq:z2lep57}
U_{PMNS}=U_{l}^{\dagger}U_{\nu}=Q_{l}P_{l}S_{23}(\theta_{l})\Sigma_{l}^{\dagger}
\Sigma_{\nu}S_{23}(\theta_{\nu})P_{\nu}Q_{\nu}\,,
\end{equation}
where the phase matrix $Q_{l}$ can be absorbed by redefinition of the charged lepton fields. We see that the lepton mixing matrix depends on two free continuous parameters $\theta_{l}$ and $\theta_{\nu}$, and one entry of the PMNS matrix is fixed to be some constant value by the postulated residual symmetry. Notice that $S_{23}(\theta+\pi)=S_{23}(\theta)\text{diag}(1,-1,-1)=\text{diag}(1,-1,-1)S_{23}(\theta)$ where the diagonal matrix can be absorbed into the matrices $Q_{l}$ and $Q_{\nu}$, consequently the fundamental interval of the parameters $\theta_{l}$ and $\theta_{\nu}$ are $\left[0, \pi\right)$.

If two pairs of residual subgroups $\{Z^{g'_{l}}_2\times X'_{l}, Z^{g'_{\nu}}_2\times X'_{\nu}\}$ and $\{Z^{g_{l}}_2\times X_{l}, Z^{g_{\nu}}_2\times X_{\nu}\}$ are related by a similarity transformation
\begin{eqnarray}
\nonumber
hg_{l}h^{-1}&=&g_{l}',\qquad \rho_{\mathbf{3}}(h)X_{l}\rho_{\mathbf{3}}(h)^{T}=X'_{l},\\
\label{eq:z2lep58} hg_{\nu}h^{-1}&=&g_{\nu}',\qquad \rho_{\mathbf{3}}(h)X_{\nu}\rho_{\mathbf{3}}(h)^{T}=X'_{\nu}
\end{eqnarray}
with $h\in S_{4}$, then the unitary transformations of the changed lepton and neutrino fields are related by
\begin{equation}
\label{eq:Utrans_similar}U'_{l}=\rho_{\mathbf{3}}(h)U_{l},\qquad U'_{\nu}=\rho_{\mathbf{3}}(h)U_{\nu}\,.
\end{equation}
Therefore the same result for the PMNS matrix would be obtained.

\subsection{\label{subsec:criterion_Z2xCP}The criterion for the equivalence of two mixing patterns}

In some cases, two distinct residual symmetries lead to the same mixing pattern, if a possible shift in the continuous free parameters $\theta_{l}$ and $\theta_{\nu}$ is taken into account. Then we shall call these two mixing patterns are equivalent. In this section, we shall derive the criterion to determine whether two resulting mixing patterns are equivalent or not. In our approach, the lepton mixing matrices derived from two generic residual symmetries take the form
\begin{align}
U_{PMNS}&=Q_{l}P_{l}S_{23}(\theta_{l})\Sigma_{l}^{\dagger}
\Sigma_{\nu}S_{23}(\theta_{\nu})P_{\nu}Q_{\nu},\\
U'_{PMNS}&=Q'_{l}P'_{l}S_{23}(\theta'_{l})\Sigma'^{\dagger}_{l}\Sigma'_{\nu}S_{23}(\theta'_{\nu})P'_{\nu}Q'_{\nu}\,.
\end{align}
Obviously the fixed element has to be equal if the two mixing patterns are equivalent, and without loss of generality we assume it is the (11) entry of the PMNS matrix. As a result, the permutation matrices $P_{l}$, $P_{\nu}$, $P'_{l}$ and $P'_{\nu}$ can only be $1$ and $P_{23}$. Because the following identities
\begin{equation}
\label{eq:z2lep61}
P_{23}S_{23}(\theta_{l})=\text{diag}(1, -1, 1)S_{23}(\theta_{l}-\pi/2),\quad S_{23}(\theta_{\nu})P_{23}=S_{23}(\theta_{\nu}+\pi/2)\text{diag}(1, -1, 1)
\end{equation}
are satisfied, and the diagonal matrix can be absorbed into the matrices $Q_{l}$ and $Q_{\nu}$, we could choose $P_{l}=P_{\nu}=P'_{l}=P'_{\nu}=1$. For any given values of $\theta_l$, $\theta_{\nu}$ and the matrices $Q_{l}$, $P_{l}$, $Q_{\nu}$, $P_{\nu}$, if the corresponding solutions of $\theta'_l$, $\theta'_{\nu}$  as well as $Q'_{l}$, $P'_{l}$, $Q'_{\nu}$, $P'_{\nu}$ can be found such that the equality $U_{PMNS}=U'_{PMNS}$ is fulfilled, these two mixing patterns would be equivalent, i.e.,
\begin{equation}
\label{eq:z2lep64}
Q_{l}S_{23}(\theta_{l})US_{23}(\theta_{\nu})Q_{\nu}=Q'_{l}S_{23}(\theta'_{l})U'S_{23}(\theta'_{\nu})Q_{\nu}'\,,
\end{equation}
where $U\equiv \Sigma^{\dagger}_{l}\Sigma_{\nu}$ and $U'\equiv \Sigma'^{\dagger}_{l}\Sigma'_{\nu}$. Then we have
\begin{equation}
\label{eq:z2lep67}
Q_{L}S_{23}(\theta_{l})US_{23}(\theta_{\nu})Q_{N}=S_{23}(\theta'_{l})U'S_{23}(\theta'_{\nu})\,,
\end{equation}
where $Q_{L}=Q'^{\dagger}_{l}Q_{l}$ is a generic diagonal phase matrix, and $Q_{N}=Q_{\nu}Q_{\nu}'^{\dagger}$ is also diagonal with entries $\pm1$ and $\pm i$. The matrices on both sides of Eq.~\eqref{eq:z2lep67} multiplying with their transpose leads to
\begin{equation}
\label{eq:z2lep68}
Q_{L}S_{23}(\theta_{l})US_{23}(\theta_{\nu})Q^2_{N}S^{T}_{23}(\theta_{\nu})U^{T}S_{23}^T(\theta_{l})Q_{L}=S_{23}(\theta'_{l})U'U'^{T}S^{T}_{23}(\theta'_{l})\,.
\end{equation}
Subsequently taking trace, we obtain
\begin{equation}
\label{eq:z2lep69}
\mathrm{Tr}\left[S^{T}_{23}(\theta_l)Q_{L}^{2}S_{23}(\theta_{l})US_{23}(\theta_{\nu})Q_{N}^{2}S_{23}^{T}(\theta_{\nu})U^{T}\right]=\mathrm{Tr}\left[U'U'^{T}\right]\,.
\end{equation}
Since the right-handed side of this equality is a constant and it doesn't depend on $\theta_{l}$ and $\theta_{\nu}$, the phase matrices $Q_{L}$ and $Q_{N}$ should be of the form
\begin{equation}
\label{eq:z2lep71}
Q_{L}=\left( \begin{array}{ccc}
e^{i \delta_{1}} & 0 & 0\\
0 & e^{i\delta_{2}} & 0 \\
0 & 0 & k_1e^{i \delta_{2}} \end{array} \right),
~~Q_{N}=\left( \begin{array}{ccc}
\eta_{1} & 0 & 0\\
0 & \eta_{2} & 0 \\
0 & 0 & k_2\eta_{2} \end{array} \right)\,,
\end{equation}
where $k_{1,2}=\pm1$, $\delta_{1,2}$ are real parameters, and $\eta_{1,2}$ are $\pm1$ and $\pm i$ with $e^{i\delta_{1}}\eta_1=1$. Thus from Eq.~\eqref{eq:z2lep67} we can derive
\begin{equation}
\label{eq:z2lep81}
Q_{L}UQ_{N}=S_{23}(\theta''_{l})U'S_{23}(\theta''_{\nu})\,,
\end{equation}
with
\begin{equation}
\theta''_{l}=\theta'_{l}-k_{1}\theta_{l},~~\theta''_{\nu}=\theta'_{\nu}-k_{2}\theta_{\nu}.
\end{equation}
Once the residual symmetries are specified, the unitary matrices $U$ and $U'$ can be determined by following the procedures listed in section~\ref{subsec:framework_Z2xCP}. Generically $U$ and $U'$ can be written as
\begin{equation}
\label{eq:z2lep82}
U=\left( \begin{array}{ccc}
a_{1} &a_{2} &a_{3}\\
a_{4} &a_{5} &a_{6}\\
a_{7} &a_{8} &a_{9} \end{array}
\right),~~
U'=\left( \begin{array}{ccc}
b_{1} &b_{2} &b_{3}\\
b_{4} &b_{5} &b_{6}\\
b_{7} &b_{8} &b_{9} \end{array}
\right)\,.
\end{equation}
A necessary condition for the equivalence of $U_{PMNS}$ and $U'_{PMNS}$ is $a_1=b_1$ which can not be 0 or 1 in order to be compatible with experimental data. Firstly let's consider a special case with
\begin{equation}
\label{eq:z2lep83}
Q_{L}=\left( \begin{array}{ccc}
1 & 0 & 0\\
0 & e^{i\delta} & 0 \\
0 & 0 & e^{i \delta} \end{array} \right),
~~Q_{N}=\left( \begin{array}{ccc}
1~ & 0 & ~0\\
0~ & 1 & ~0 \\
0~ & 0 & ~1 \end{array} \right)\,.
\end{equation}
Solving the equation Eq.~\eqref{eq:z2lep81} for the variables $\theta''_{l}$, $\theta''_{\nu}$ and $\delta$, we can obtain the condition for the existence of solution.

\begin{itemize}[labelindent=-0.6em, leftmargin=1.0em]
\item{$b_{2}^{2}+b_{3}^{2}\neq 0,~~b_{4}^{2}+b_{7}^{2}\neq 0$}

In this case, the solutions for $\theta''_{l}$, $\theta''_{\nu}$ and $\delta$ are given by
\begin{eqnarray}
\nonumber
&&\cos\theta''_{l}=\frac{a_{4}b_{4}+a_{7}b_{7}}{b_{4}^2+b_{7}^2}e^{i\delta},~~\sin\theta''_{l}=\frac{a_{4}b_{7}-a_{7}b_{4}}{b_{4}^2+b_{7}^2}e^{i\delta},~~e^{-2i\delta}=\frac{a_{4}^2+a_{7}^2}{b_{4}^2+b_{7}^2}\,,\\
\label{eq:z2lep84_lepton}
&&\cos\theta''_{\nu}=\frac{a_{2}b_{2}+a_{3}b_{3}}{b_{2}^2+b_{3}^2},~~~\sin\theta''_{\nu}=\frac{a_{3}b_{2}-a_{2}b_{3}}{b_{2}^2+b_{3}^2}\,.
\end{eqnarray}
Since $\theta''_{l}$, $\theta''_{\nu}$ and $\delta$ are real parameters, $a_i$ and $b_i$ should be subject to the following constraints
\begin{eqnarray}
\nonumber&&(a_{4}b_{4}+a_{7}b_{7})(a^{*}_4b^{*}_7-a^{*}_7b^{*}_4)\in\mathbb{R},\quad \left|a^2_4+a^2_7\right|=\left|b^2_4+b^2_7\right|\,,\\
&&(a_{2}b_{2}+a_{3}b_{3})(a^{*}_2b^{*}_3-a^{*}_3b^{*}_2)\in\mathbb{R},\quad a^2+a^2_3=b^2_2+b^2_3\,.
\end{eqnarray}
Inserting Eq.~\eqref{eq:z2lep84_lepton} into Eq.~\eqref{eq:z2lep81}, we find that the equivalence of these two mixing patterns requires
\begin{eqnarray}
\nonumber
a_{5}=\frac{(xb_{5}+yb_{6})z+(xb_{8}+yb_{9})w}{(b_{2}^{2}+b_{3}^{2})(b_{4}^{2}+b_{7}^{2})},~~a_{6}=\frac{(xb_{6}-yb_{5})z+(xb_{9}-yb_{8})w}{(b_{2}^{2}+b_{3}^{2})(b_{4}^{2}+b_{7}^{2})}\,,\\
\label{eq:z2lep86_lepton}
a_{8}=\frac{(xb_{8}+yb_{9})z-(xb_{5}+yb_{6})w}{(b_{2}^{2}+a_{3}^{2})(b_{4}^{2}+b_{7}^{2})},~~a_{9}=\frac{(xb_{9}-yb_{8})z-(xb_{6}-yb_{5})w}{(b_{2}^{2}+b_{3}^{2})(b_{4}^{2}+b_{7}^{2})}\,,
\end{eqnarray}
with
\begin{equation}
\label{eq:z2lep87_lepton}
x=a_{2}b_{2}+a_{3}b_{3},~~y=a_{2}b_{3}-a_{3}b_{2},~~z=a_{4}b_{4}+a_{7}b_{7},~~w=a_{4}b_{7}-a_{7}b_{4}.
\end{equation}
\item{$b_{2}^2+b_{3}^2=0,~~b_{4}^2+b_{7}^2\neq 0$}

This case requires
\begin{equation}
b_{3}=is_{1}b_{2},\quad a_{3}=is_{1}a_{2},\quad \text{with}\quad s_{1}=\pm 1\,.
\end{equation}
The parameters $\theta''_{l}$, $\theta''_{\nu}$ and $\delta$ are determined to be
\begin{eqnarray}
\nonumber
&&\cos\theta''_{l}=\frac{a_{4}b_{4}+a_{7}b_{7}}{b_{4}^2+b_{7}^2}e^{i\delta},~~\sin\theta''_{l}=\frac{a_{4}b_{7}-a_{7}b_{4}}{b_{4}^2+b_{7}^2}e^{i\delta},~~e^{-2i\delta}=\frac{a_{4}^2+a_{7}^2}{b_{4}^2+b_{7}^2}\,,\\
\label{eq:z2lep89}
&&\cos\theta''_{\nu}=\Re(a_{2}/b_{2}),~~~\sin\theta''_{\nu}=-s_{1}\Im(a_{2}/b_{2})\,,
\end{eqnarray}
with the constraints
\begin{equation}
(a_{4}b_{4}+a_{7}b_{7})(a^{*}_{4}b^{*}_{7}-a^{*}_{7}b^{*}_{4})\in\mathbb{R},\quad \left|a_{4}^{2}+a_{7}^{2}\right|=\left|b_{4}^{2}+b_{7}^{2}\right|\,.
\end{equation}
These two PMNS matrices would be equivalent if and only if the following conditions are fulfilled:
\begin{eqnarray}
\nonumber
&&a_{5}(b^2_4+b^2_7)=s_{1}(zb_{6}+wb_{9})\Im(a_{2}/b_2)+(zb_{5}+wb_{8})\Re(a_{2}/b_2),\\
\nonumber
&&a_{6}(b^2_4+b^2_7)=-s_{1}(zb_{5}+wb_{8})\Im(a_{2}/b_2)+(zb_{6}+wb_{9})\Re(a_{2}/b_2),\\
\nonumber
&&a_{8}(b^2_4+b^2_7)=s_{1}(zb_{9}-wb_{6})\Im(a_{2}/b_2)+(zb_{8}-wb_{5})\Re(a_{2}/b_2),\\
\label{eq:z2lep91}
&&a_{9}(b^2_4+b^2_7)=-s_{1}(zb_{8}-wb_{5})\Im(a_{2}/b_2)+(zb_{9}-wb_{6})\Re(a_{2}/b_2)\,.
\end{eqnarray}

\item{$b_{2}^2+b_{3}^2\neq 0,~~b_{4}^2+b_{7}^2=0$}

From $b_{4}^2+b_{7}^2=0$, we obtain $b_{7}=is_{2}b_{4}$ with $s_2=\pm1$. Moreover, the equality $a_{7}=is_{2}a_{4}$ should be satisfied otherwise $U_{PMNS}$ and $U'_{PMNS}$ are two different mixing patterns. The condition of equivalence in Eq.~\eqref{eq:z2lep81} gives rise to
\begin{equation}
t_{i}T_{j}-t_{j}T_{i}=0,\quad t_i/T_i\in\mathbb{R},\quad\text{with}\quad i,j=5, 6, 8, 9\,,
\end{equation}
where
\begin{eqnarray}
\nonumber &&t_{5}=za_{5}-wb_{5}-vb_{6},\qquad t_{6}=za_{6}-wb_{6}+vb_{5},\\
\nonumber&&t_{8}=-za_{8}+wb_{8}+vb_{9},\qquad t_{9}=-za_{9}+wb_{9}-vb_{8},\\
\nonumber &&T_{5}=-iza_{5}+s_{2}(wb_{8}+vb_{9}),\quad T_{6}=-iza_{6}+s_{2}(wb_{9}-vb_{8}),\\
\label{eq:z2lep98}
&&T_{8}=iza_{8}+s_{2}(wb_{5}+vb_{6}),\quad T_{9}=iza_{9}+s_{2}(wb_{6}-vb_{5})\,,
\end{eqnarray}
and
\begin{equation}
\label{eq:z2lep99}
z\equiv b_{4}(b_{2}^2+b_{3}^2),~~w\equiv a_{4}(a_{2}b_{2}+a_{3}b_{3}),~~v\equiv a_{4}(a_{2}b_{3}-a_{3}b_{2})\,.
\end{equation}
The values of the rotation angles $\theta''_{l}$ and $\theta''_{\nu}$ are
\begin{eqnarray}
\nonumber&&\cos\theta''_{l}=\Re(a_{4}e^{i\delta}/b_{4}),~~\sin\theta''_{l}=s_{2}\Im(a_{4}e^{i\delta}/b_{4}),\\
\label{eq:z2lep94}
&&\cos\theta''_{\nu}=\frac{a_{2}b_{2}+a_{3}b_{3}}{b_{2}^2+b_{3}^2},~~\sin\theta''_{\nu}=\frac{a_{3}b_{2}-a_{2}b_{3}}{b_{2}^2+b_{3}^2}\,,
\end{eqnarray}
with the constraints
\begin{equation}
(a_{2}b_{2}+a_{3}b_{3})(a^{*}_{2}b^{*}_{3}-a^{*}_{3}b^{*}_{2})\in\mathbb{R},\quad a_{2}^2+a_{3}^2=b_{2}^2+b_{3}^2\,.
\end{equation}
The phase $\delta$ is determined by
\begin{equation}
\frac{\Im(a_{4}e^{i\delta}/b_{4})}{\Re(a_{4}e^{i\delta}/b_{4})}=\frac{t_i}{T_i}\,.
\end{equation}
\item{$b_{2}^2+b_{3}^2=0,b_{4}^2+b_{7}^2=0$}

In the same fashion as previous cases, we find
\begin{eqnarray}
\nonumber&&b_{3}=is_{3}b_{2},\quad a_{3}=is_{3}a_{2},\quad s_{3}=\pm 1,\\
\label{eq:z2lep100}&&b_{7}=is_{4}b_{4},\quad a_{7}=is_{4}a_{4},\quad s_{4}=\pm 1\,.
\end{eqnarray}
The condition of equivalence in Eq.~\eqref{eq:z2lep81} would be fulfilled if
\begin{equation}
\label{eq:z2lep102}t'_{i}T'_{j}-t'_{j}T'_{i}=0,\quad t'_i/T'_i\in\mathbb{R},\quad\text{with}\quad i,j=5, 6, 8, 9\,,
\end{equation}
where
\begin{eqnarray}
\nonumber&&t'_{5}=a_{5}b_{4}-a_{4}[s_{3}b_{6}\Im(a_{2}/b_{2})+b_{5}\Re(a_{2}/b_{2})],\\
\nonumber&&t'_{6}=a_{6}b_{4}+a_{4}[s_{3}b_{5}\Im(a_{2}/b_{2})-b_{6}\Re(a_{2}/b_{2})],\\
\nonumber&&t'_{8}=a_{8}b_{4}-a_{4}[s_{3}b_{9}\Im(a_{2}/b_{2})+b_{8}\Re(a_{2}/b_{2})],\\
\nonumber&&t'_{9}=a_{9}b_{4}+a_{4}[s_{3}b_{8}\Im(a_{2}/b_{2})-b_{9}\Re(a_{2}/b_{2})]\,,\\
\nonumber&&T'_{5}=-ia_{5}b_{4}+s_{4}a_{4}[s_{3}b_{9}\Im(a_{2}/b_{2})+b_{8}\Re(a_{2}/b_{2})],\\
\nonumber&&T'_{6}=-ia_{6}b_{4}-s_{4}a_{4}[s_{3}b_{8}\Im(a_{2}/b_{2})-b_{9}\Re(a_{2}/b_{2})],\\
\nonumber&&T'_{8}=-ia_{8}b_{4}-s_{4}a_{4}[s_{3}b_{6}\Im(a_{2}/b_{2})+b_{5}\Re(a_{2}/b_{2})],\\
\label{eq:z2lep104}&&T'_{9}=-ia_{9}b_{4}+s_{4}a_{4}[s_{3}b_{5}\Im(a_{2}/b_{2})-b_{6}\Re(a_{2}/b_{2})]\,.
\end{eqnarray}
The solutions for $\theta''_{l}$, $\theta''_{\nu}$ and $\delta$ are
\begin{eqnarray}
\nonumber&&\cos\theta''_{l}=\Re(a_{4}e^{i\delta}/b_{4}),~~\sin\theta''_{l}=s_{4}\Im(a_{4}e^{i\delta}/b_{4}),~~\frac{\Im(a_{4}e^{i\delta}/b_{4})}{\Re(a_{4}e^{i\delta}/b_{4})}=\frac{t'_i}{T'_i},\\
\label{eq:z2lep101}
&&\cos\theta''_{\nu}=\Re(a_{2}/b_{2}),~~\sin\theta''_{\nu}=-s_{3}\Im(a_{2}/b_{2})\,.
\end{eqnarray}
\end{itemize}
For the most general values of the diagonal matrices $Q_{L}$ and $Q_{N}$
\begin{equation}
Q_{L}=\left( \begin{array}{ccc}
\eta^{-1}_{1} & 0 & 0\\
0 & e^{i\delta} & 0 \\
0 & 0 & k_1e^{i\delta} \end{array} \right),
~~Q_{N}=\left( \begin{array}{ccc}
\eta_{1} & 0 & 0\\
0 & \eta_{2} & 0 \\
0 & 0 & k_2\eta_{2} \end{array} \right)\,,
\end{equation}
the condition for the equivalence of two generic mixing patterns can be obtained from the above results by making the following substitutions
\begin{eqnarray}
\nonumber
&&a_{1}\rightarrow a_{1},~~a_{2}\rightarrow \eta_{1}^{-1}\eta_{2}a_{2},~~a_{3}\rightarrow k_{2}\eta_{1}^{-1}\eta_{2}a_{3},\\
\nonumber
&&a_{4}\rightarrow \eta_{1}a_{4},~~a_{5}\rightarrow \eta_{2}a_{5},~~a_{6}\rightarrow k_{2}\eta_{2}a_{6},\\
\label{eq:z2lep105}
&&a_{7}\rightarrow k_{1}\eta_{1}a_{7},~~a_{8}\rightarrow k_{1}\eta_{2}a_{8},~~a_{9}\rightarrow k_{1}k_{2}\eta_{2}a_{9}\,.
\end{eqnarray}

\subsection{Possible mixing patterns from $S_4$ and CP and numerical results}
We shall perform a comprehensive study of the lepton mixing patterns arising from the breaking of $S_4$ and $CP$ symmetry into two distinct residual groups of the structure $Z_2\times CP$ in the charged lepton and neutrino sectors. The basic properties of the $S_4$ group and its representation are collected in appendix~\ref{sec:S4_group_app}. It turns out that the most general $CP$ transformation compatible with $S_4$ is of the same form as the flavor symmetry transformation in  our chosen basis~\cite{Ding:2013hpa,Li:2013jya}. Each of the nine different $Z_2$ symmetries in Eq.~\eqref{eq:Z2-subgroups} together with the compatible $CP$ transformation can be residual symmetry of the neutrino and charged lepton mass matrices.

By applying the similarity transformation and the equivalence criterion derived in section~\ref{subsec:criterion_Z2xCP}, we find that it is sufficient to only consider a number of independent cases which lead to different results for mixing angles and $CP$ phases. All possible permutations of the rows and columns of the mixing matrix would be considered. We exclude all patterns that can not describe the experimental data on lepton mixing angles at the $3\sigma$ level for certain values of the free parameters $\theta_{l}$
and $\theta_{\nu}$. As a result, we find totally eighteen phenomenologically viable cases. The residual flavor symmetry of the neutrino and charged lepton sectors can be chosen to be $Z^{ST^2SU}_2$, $Z^{TU}_2$ or $Z^{S}_2$, the corresponding residual $CP$ transformation $X_{r}$ and the Takagi factorization matrix $\Sigma$ are summarized in table~\ref{tab:res_symm_lepton}. As shown in section~\ref{subsec:framework_Z2xCP}, the Takagi factorization $\Sigma$ satisfies
\begin{equation}
X_{r}=\Sigma\Sigma^{T},\qquad \Sigma^{\dagger}\rho_{\mathbf{3}}(g_{r})\Sigma=\text{diag}(1, -1, -1)\,,
\end{equation}
where $g_{r}$ is the generator of $G_{r}$. Notice that $\rho_{\mathbf{3}}(g_{r})X_{r}$ is also a residual $CP$ symmetry of the neutrino sector, and it leads to the same constraint on the neutrino mass matrix as $X_{r}$. For each possible residual symmetry, the lepton mixing matrix can be straightforwardly obtained by using the master formula of Eq.~\eqref{eq:z2lep57}. If two cases possess the same residual symmetry, but differ in the choice of the row permutation with $P_{l}=P_{12}$ and $P_{l}=P_{13}$ respectively, then the resulting mixing matrices are effectively related through the exchange of the second and the third rows, because the following identity
\begin{equation}
P_{23}P_{12}S_{23}(\theta)=\text{diag}(-1, 1, 1)P_{13}S_{23}(\theta-\pi/2)
\end{equation}
is satisfied. Subsequently we can extract the lepton mixing parameters, and the results for the mixing angles $\sin^2\theta_{13}$, $\sin^2\theta_{12}$, $\sin^2\theta_{23}$ and the $CP$ invariants $J_{CP}$, $I_1$, $I_2$ are listed in table~\ref{tab:Uab} and table~\ref{tab:Uac} for all the viable cases. Here $J_{CP}$, $I_1$ and $I_2$ conventionally defined as
\begin{eqnarray}
\nonumber J_{CP}&=&\Im\left(U_{PMNS,11}U_{PMNS,33}U_{PMNS,13}^{\ast}U_{PMNS, 31}^{\ast}\right)\,,\\
\nonumber &=&\frac 1{8}\sin2\theta_{12}\sin2\theta_{13}\sin2\theta_{23}\cos\theta_{13}\sin\delta_{CP}\,,\\
I_{1}&=&\Im\left(U_{PMNS,12}^{2}U_{PMNS,11}^{*2}\right)=\sin^{2}\theta_{12}\cos^{2}\theta_{12}\cos^{4}\theta_{13}\sin\alpha_{21},\\
I_{2}&=&\Im\left(U_{PMNS,13}^{2}U_{PMNS,11}^{*2}\right)=\sin^{2}\theta_{13}\cos^{2}\theta_{13}\cos^{2}\theta_{12}\sin(\alpha_{31}-2\delta_{CP})\,,
\end{eqnarray}
where $\delta_{CP}$ is the Dirac $CP$ violating phase, $\alpha_{21}$ and $\alpha_{31}$ are the Majorana $CP$ phases in the standard parametrization~\cite{Agashe:2014kda}. One notices that the invariants $J_{CP}$, $I_1$ and $I_2$ are exactly vanishing such that the all the three $CP$ phases $\delta_{CP}$, $\alpha_{21}$ and $\alpha_{31}$ are trivial in some cases. Furthermore, we perform a conventional $\chi^2$ analysis that includes the three mixing angles, and the results for the mixing parameters and the best fit values $(\theta_l, \theta_{\nu})_{\mathrm{bf}}$ are displayed in table~\ref{tab:Uab_chi2_NH}, table~\ref{tab:Uab_chi2_IH}, table~\ref{tab:Uac_chi2_NH} and table~\ref{tab:Uac_chi2_IH}. For the residual flavor symmetry $(G_{l}, G_{\nu})=(Z_{2}^{ST^2SU}, Z_{2}^{TU})$, one element of the PMNS matrix is fixed to be $1/2$. From table~\ref{tab:Uab_chi2_NH} and table~\ref{tab:Uab_chi2_IH}, we can see that the $CP$ phases are predicted to be $\delta_{CP}\simeq1.569\pi$, $\alpha_{21}~(\mathrm{mod}~\pi)\simeq0.728\pi$ and  $\alpha_{31}~(\mathrm{mod}~\pi)\simeq0.808\pi$ in the case of
$(X_{l}, X_{\nu}, P_{l}, P_{\nu})=(T^2, T, P_{12}, P_{12})$, while all the three $CP$ phases are conserved for the remaining cases. In the same manner, for another residual flavor symmetry $(G_{l}, G_{\nu})=(Z_{2}^{ST^2SU}, Z_{2}^{S})$, the fixed element is $1/\sqrt{2}$, we find that all that both Dirac and Majorana phases are trivial except $(X_{l}, X_{\nu}, P_{l}, P_{\nu})=(T^2, SU, P_{12}, P_{13})$, $(T^2, SU, P_{12}, P_{13})$ which give rise to $\delta_{CP}\simeq0.458\pi$, $0.542\pi$, $1.458\pi$ or $1.542\pi$. Moreover, the atmospheric mixing angle $\theta_{23}$ is predicted to be non-maximal in all the cases studied. The latest results from T2K and NO$\nu$A show a weak evidence for a nearly maximal CP-violating phase $\delta_{CP}\sim3\pi/2$~\cite{T2K_delta_CP, NovA_delta_CP}, and hits of $\delta_{CP}\sim3\pi/2$ also show up in the global analysis of neutrino oscillation data~\cite{Capozzi:2013csa,Forero:2014bxa,Gonzalez-Garcia:2014bfa,Capozzi:2016rtj}. On the other hand, NO$\nu$A excludes maximal mixing at $2.5\sigma$ while the experimental data of T2K are consistent with maximal mixing~\cite{T2K_delta_CP, NovA_delta_CP}. Hence the above mixing patterns predicting $\delta_{CP}\simeq1.569\pi$, $1.458\pi$ and $1.542\pi$ are slightly favored over the remaining cases by the present experimental data.

\begin{figure}[hptb!]
\centering
\includegraphics[width=0.40\textwidth]{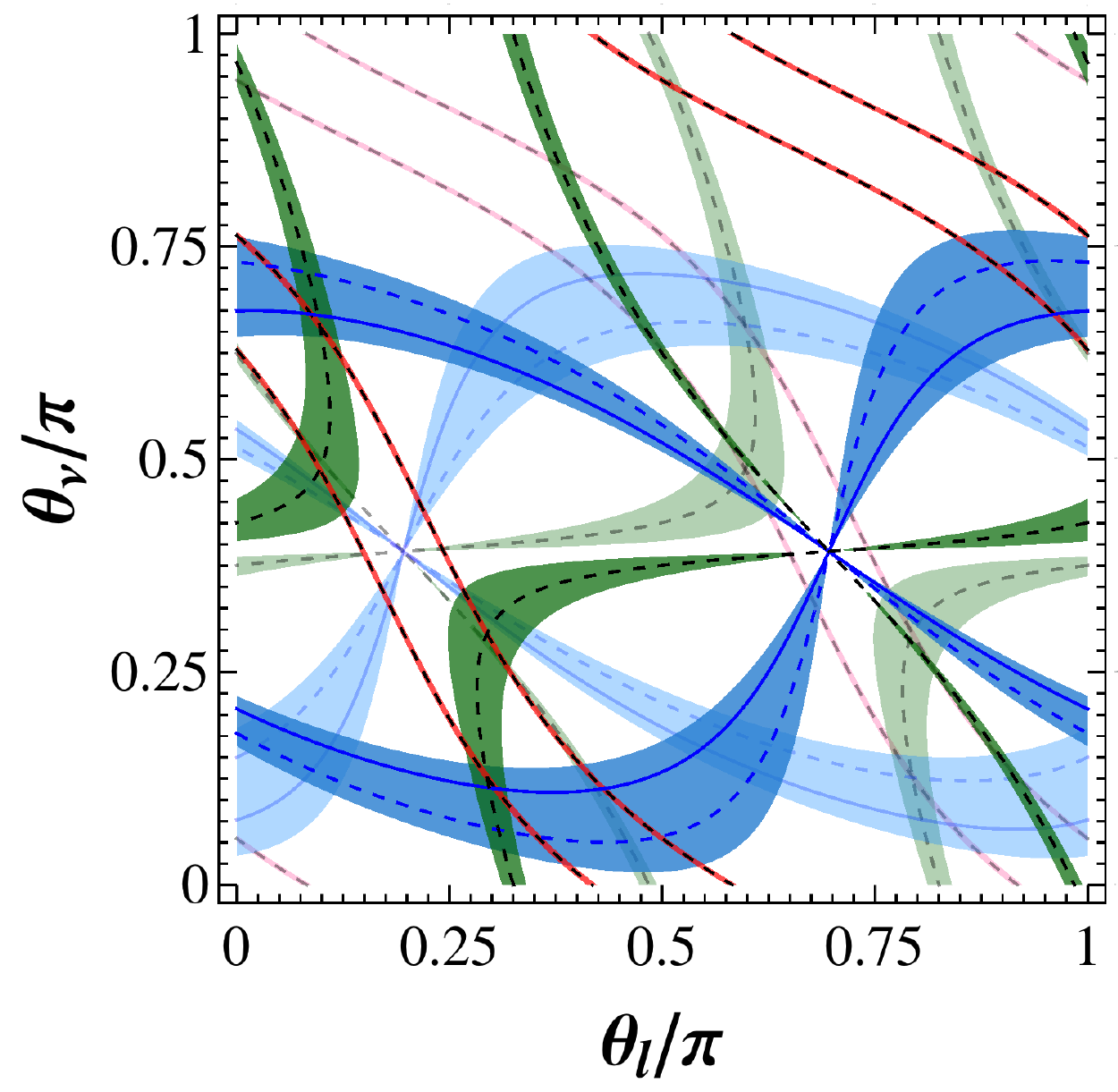}
\includegraphics[width=0.40\textwidth]{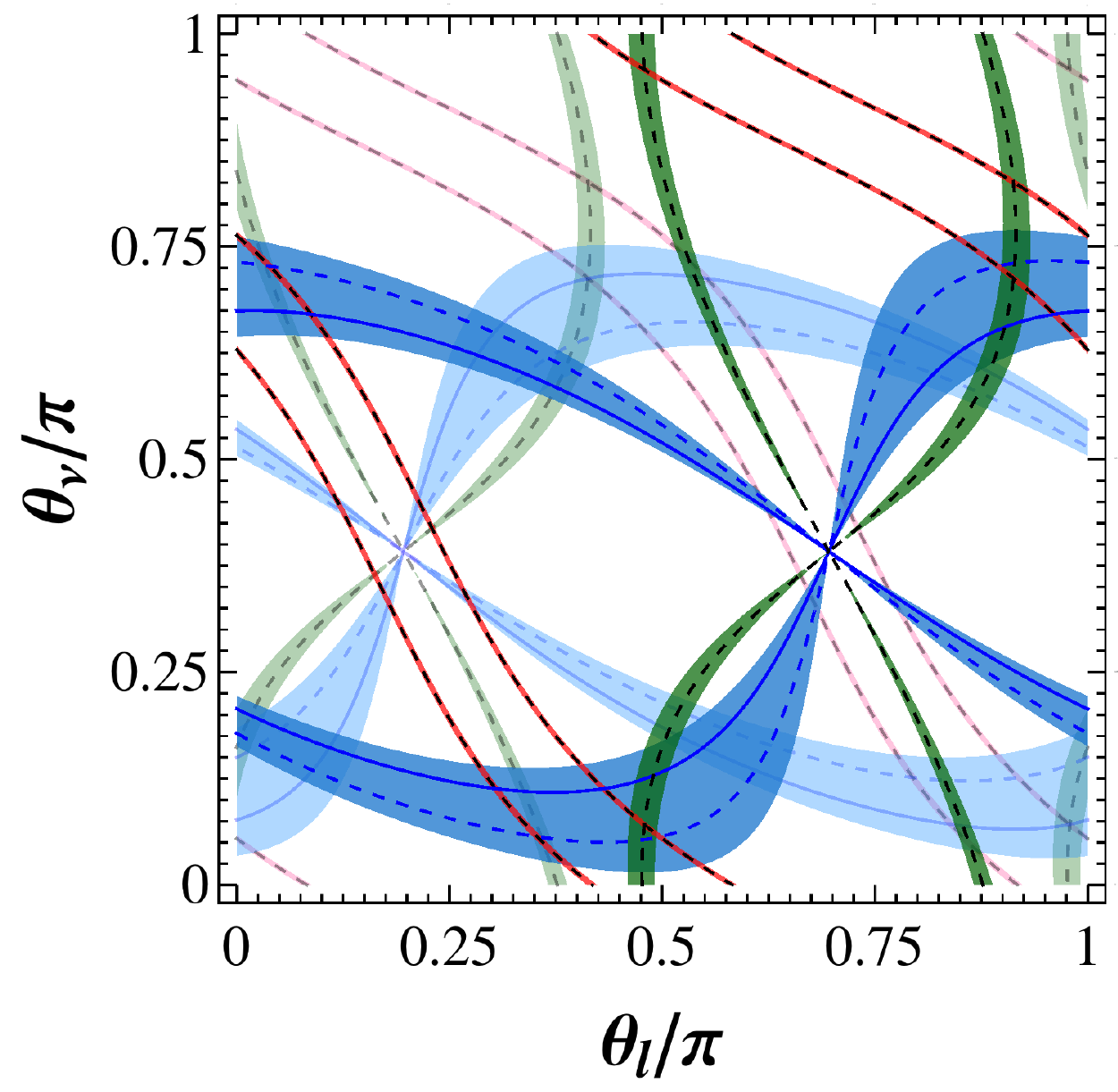}
\includegraphics[width=0.40\textwidth]{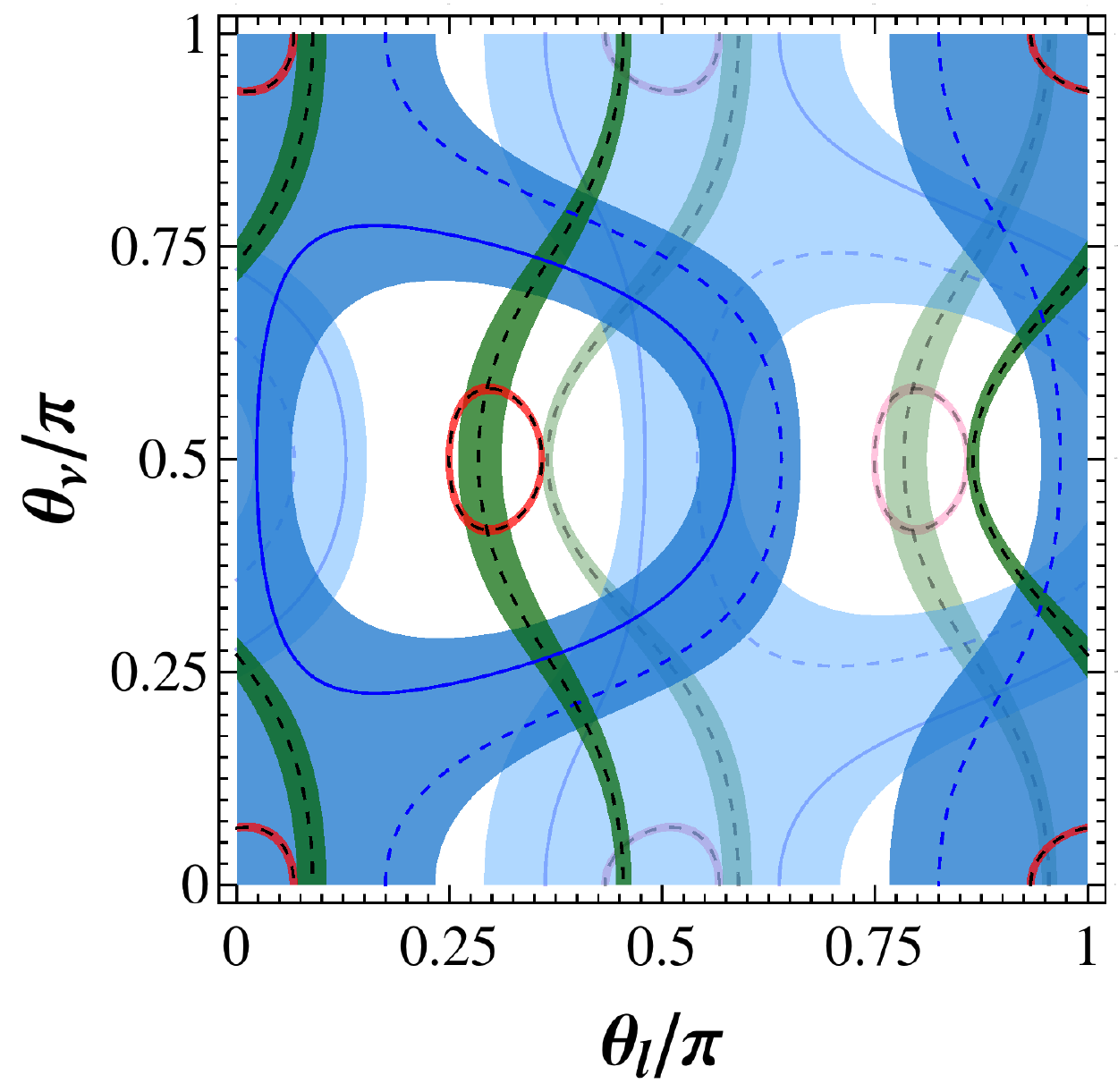}
\includegraphics[width=0.40\textwidth]{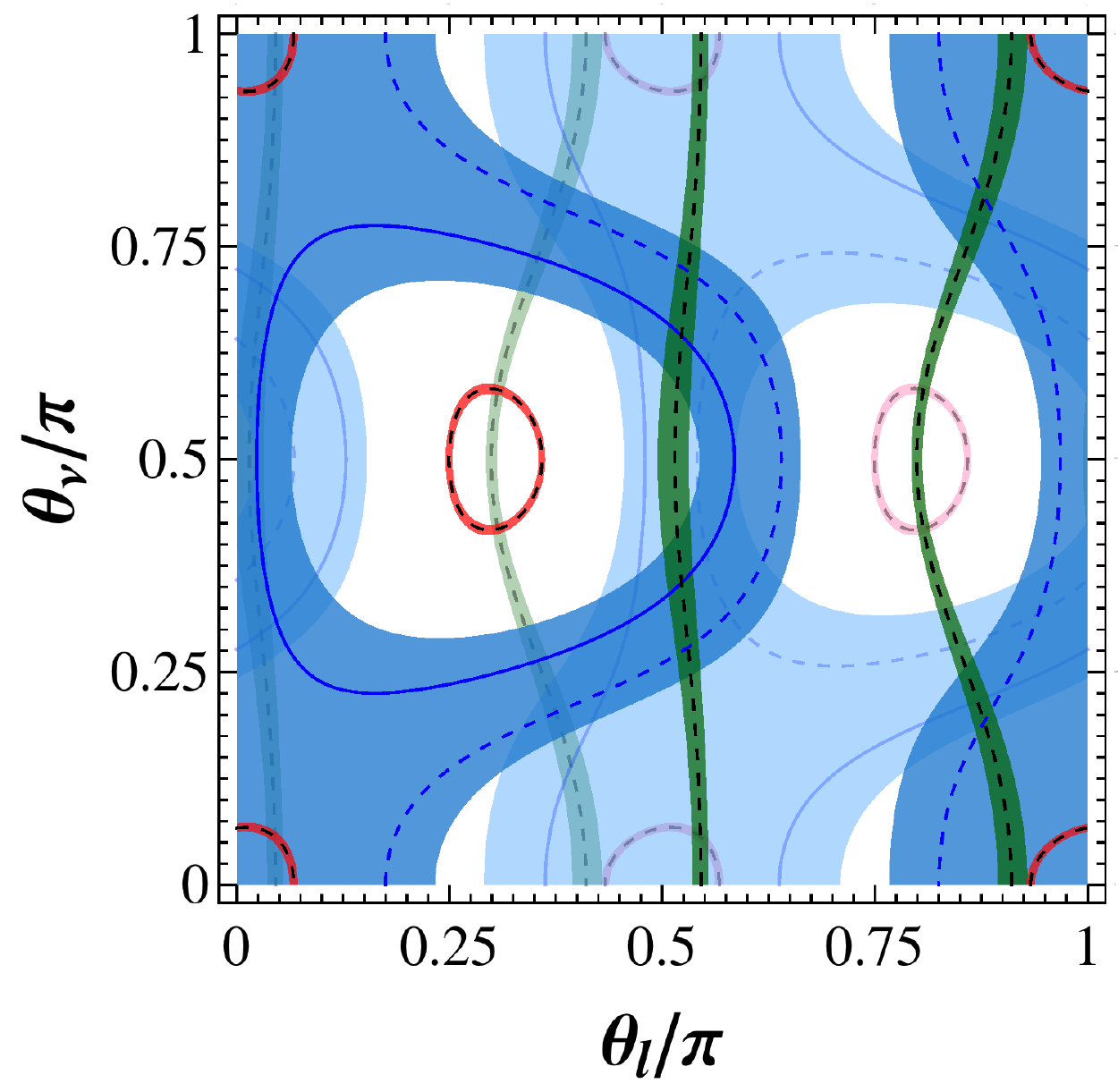}
\caption{\label{fig:Uab_contour}The contour plots of $\sin^2\theta_{ij}$ in the plane $\theta_{\nu}$ versus $\theta_{l}$. The red, blue and green areas denote the $3\sigma$ regions of $\sin^{2}\theta_{13},\sin^{2}\theta_{23}$ and $\sin^{2}\theta_{12}$ respectively. The dashed (or solid) lines indicate the best fit values of the mixing angles. Notice that the best fit value of $\sin^2\theta_{23}$ depends on the neutrino mass ordering, the solid and dashed lines are for NH and IH respectively. The residual flavor symmetry is $(G_{l}, G_{\nu})=(Z^{ST^2SU}_2, Z^{TU}_2)$ in this case. The first row corresponds to $(X_{l}, X_{\nu}, P_{l}, P_{\nu})=(U,T, P_{12},1)$ on the left panel and $(X_{l}, X_{\nu}, P_{l}, P_{\nu})=(U,T, P_{12},P_{12})$ on the right panel, and the last row is for $(X_{l}, X_{\nu}, P_{l}, P_{\nu})=(U, STS, P_{12}, 1)$, $(U, STS, P_{12}, P_{12})$. The foreground and background differ in the values of $P_{l}$ which are equal to $P_{12}$ and $P_{13}$ respectively.}
\end{figure}

\begin{figure}[hptb!]
\centering
\includegraphics[width=0.4\textwidth]{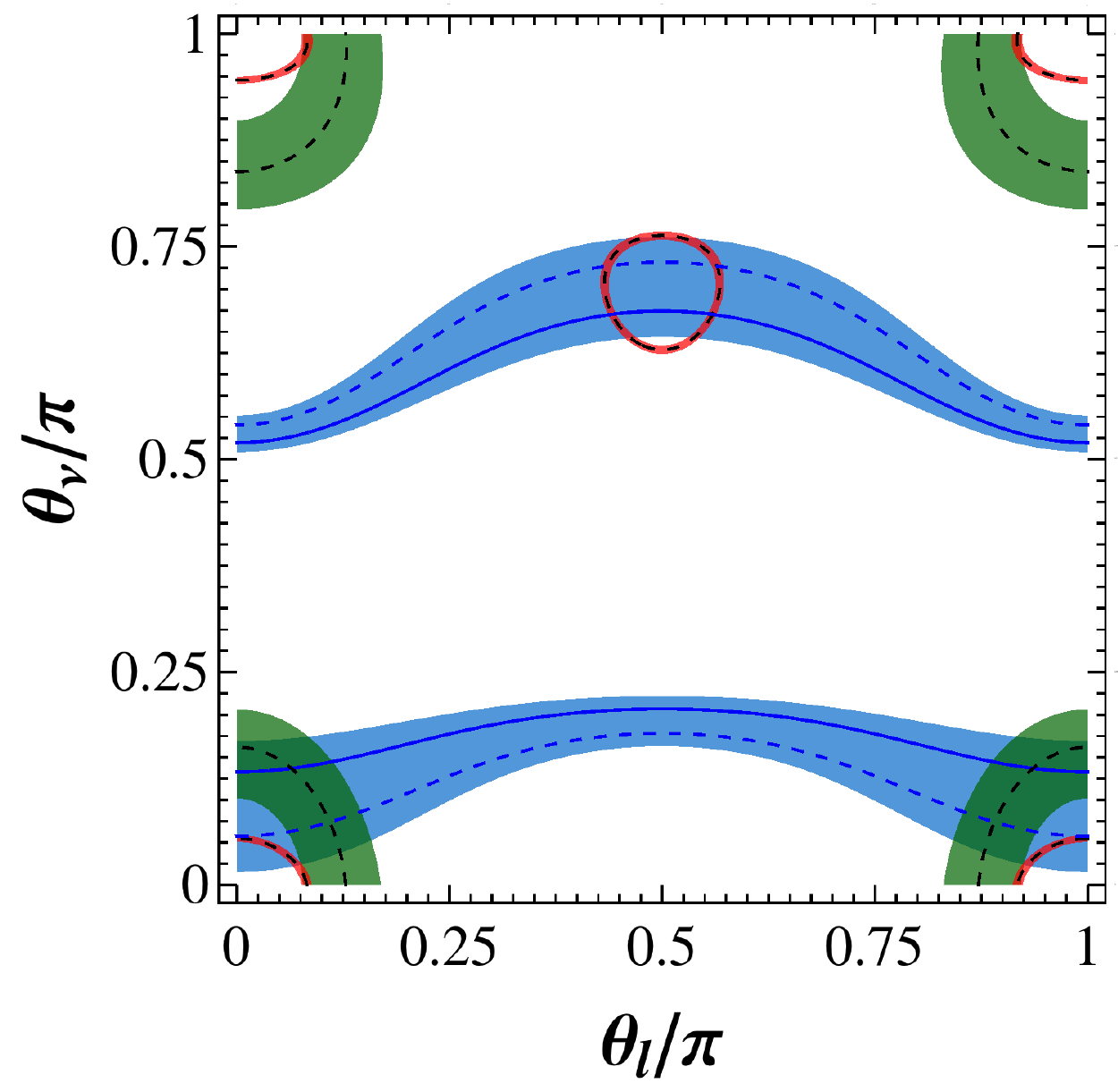}
\includegraphics[width=0.4\textwidth]{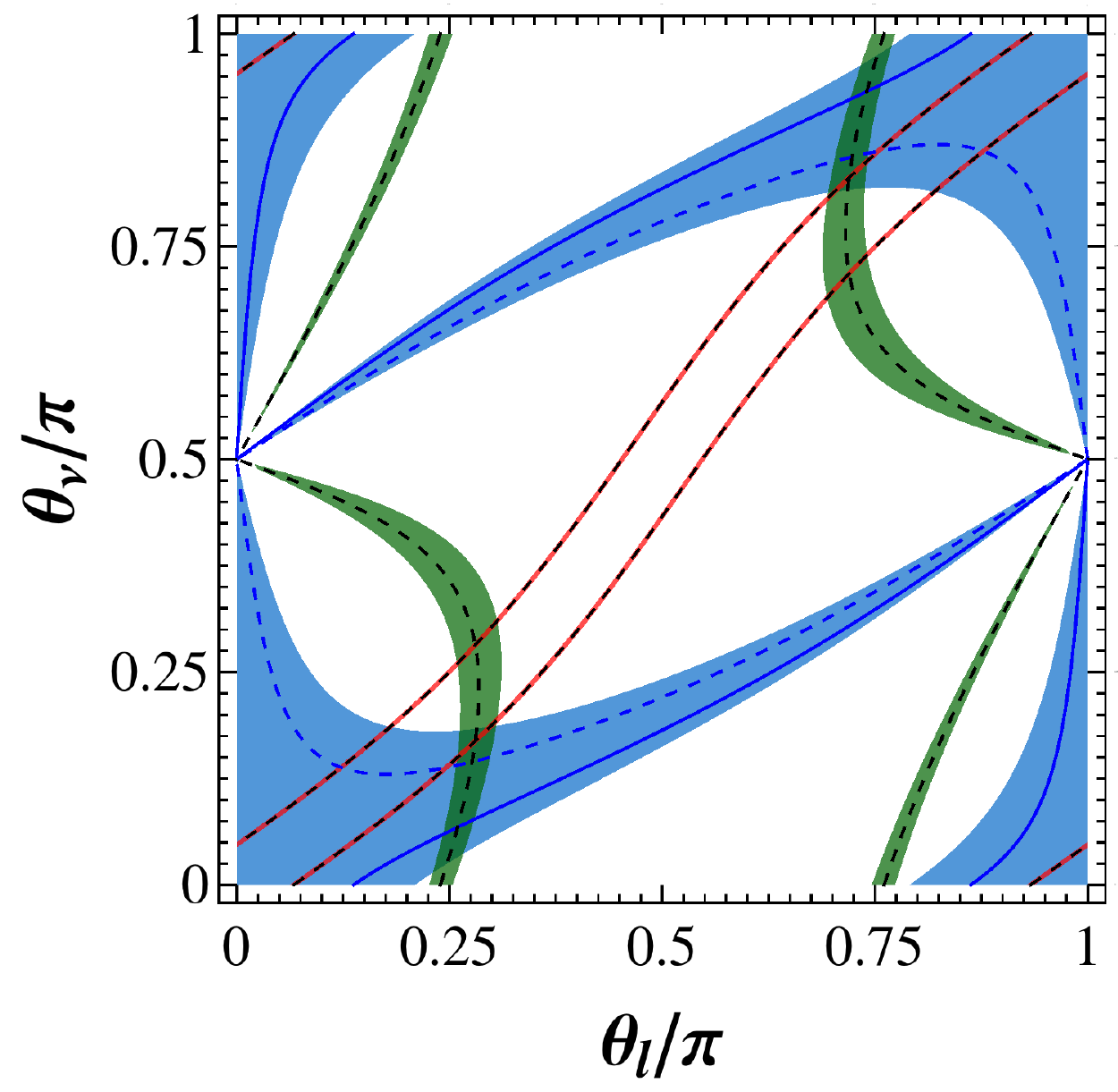}
\caption{\label{fig:Uab_Uac_contour}The contour plots of $\sin^2\theta_{ij}$ in the plane $\theta_{\nu}$ versus $\theta_{l}$. The red, blue and green areas denote the $3\sigma$ regions of $\sin^{2}\theta_{13},\sin^{2}\theta_{23}$ and $\sin^{2}\theta_{12}$ respectively. The dashed (or solid) lines indicate the best fit values of the mixing angles. Notice that the best fit value of $\sin^2\theta_{23}$ depends on the neutrino mass ordering, the solid and dashed lines are for NH and IH respectively. The left and right panels correspond to $(G_{l}, G_{\nu}, X_{l}, X_{\nu}, P_{l}, P_{\nu})=(Z^{ST^2SU}_2, Z^{TU}_2, T^2, T, P_{12}, P_{12})$ and $(Z^{ST^2SU}_2, Z^{S}_2, T^2, TST^2U, P_{13}, P_{12})$ respectively.}
\end{figure}

\begin{figure}[hptb!]
\centering
\includegraphics[width=0.4\textwidth]{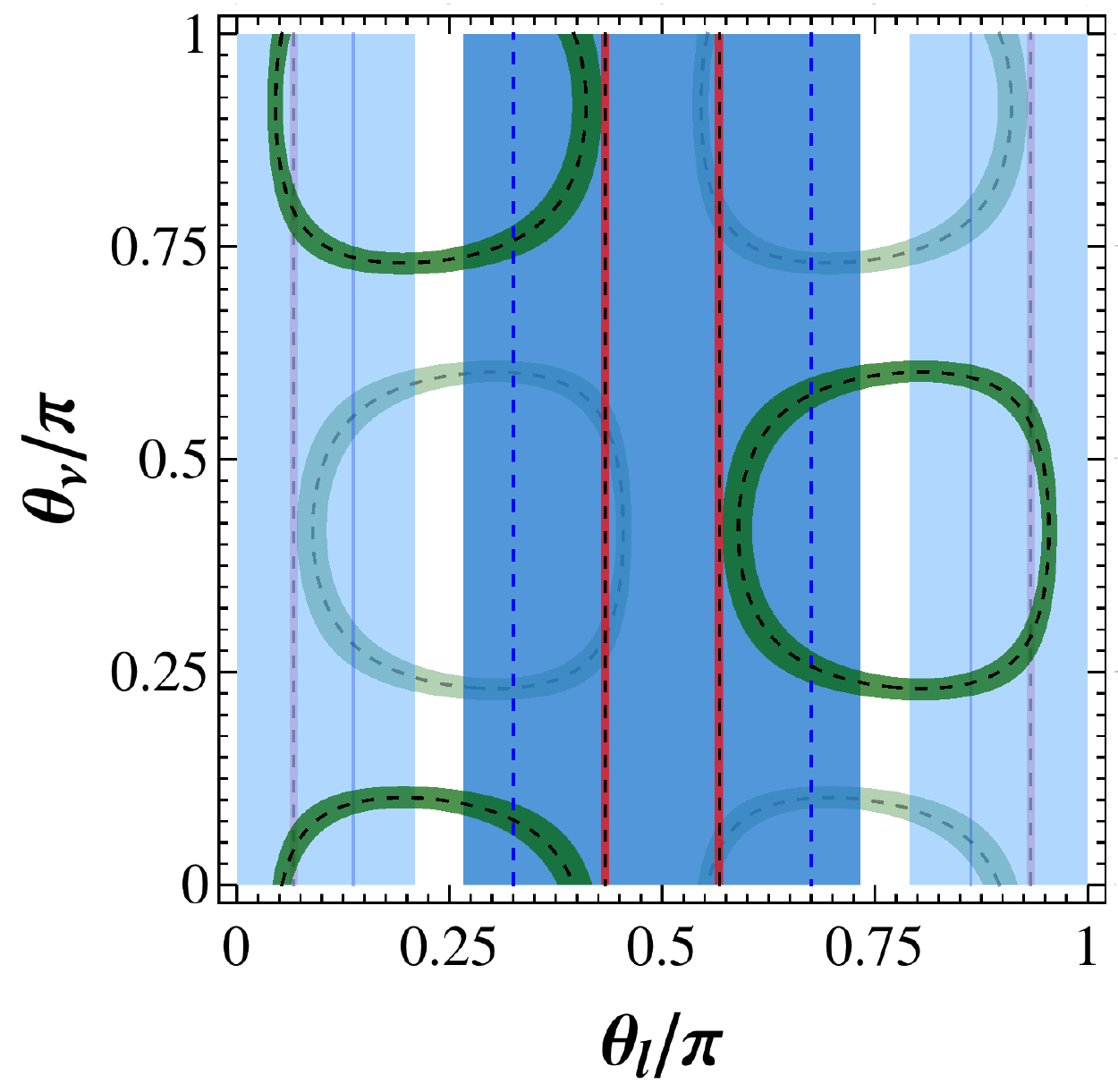}
\includegraphics[width=0.4\textwidth]{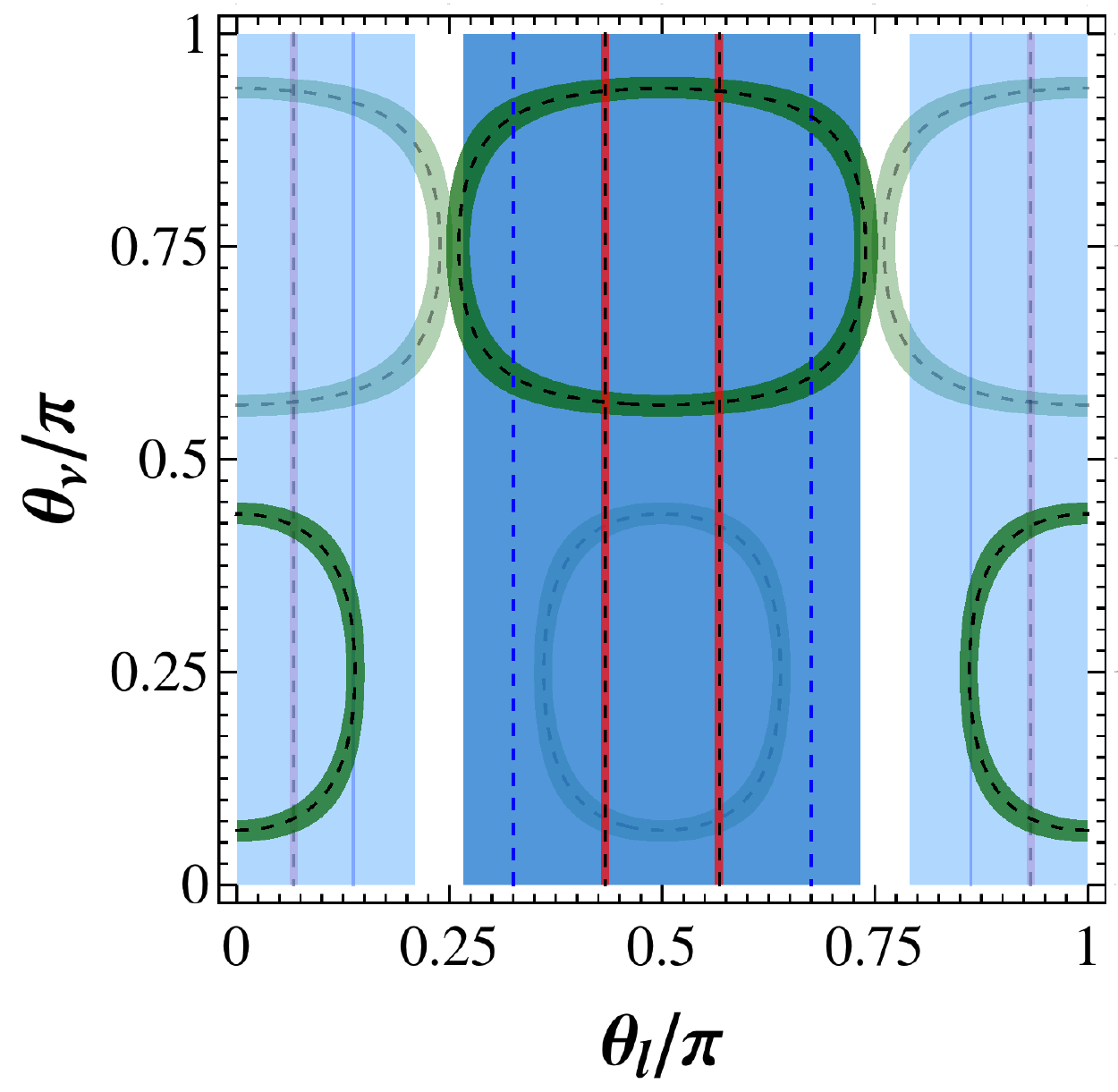}
\includegraphics[width=0.4\textwidth]{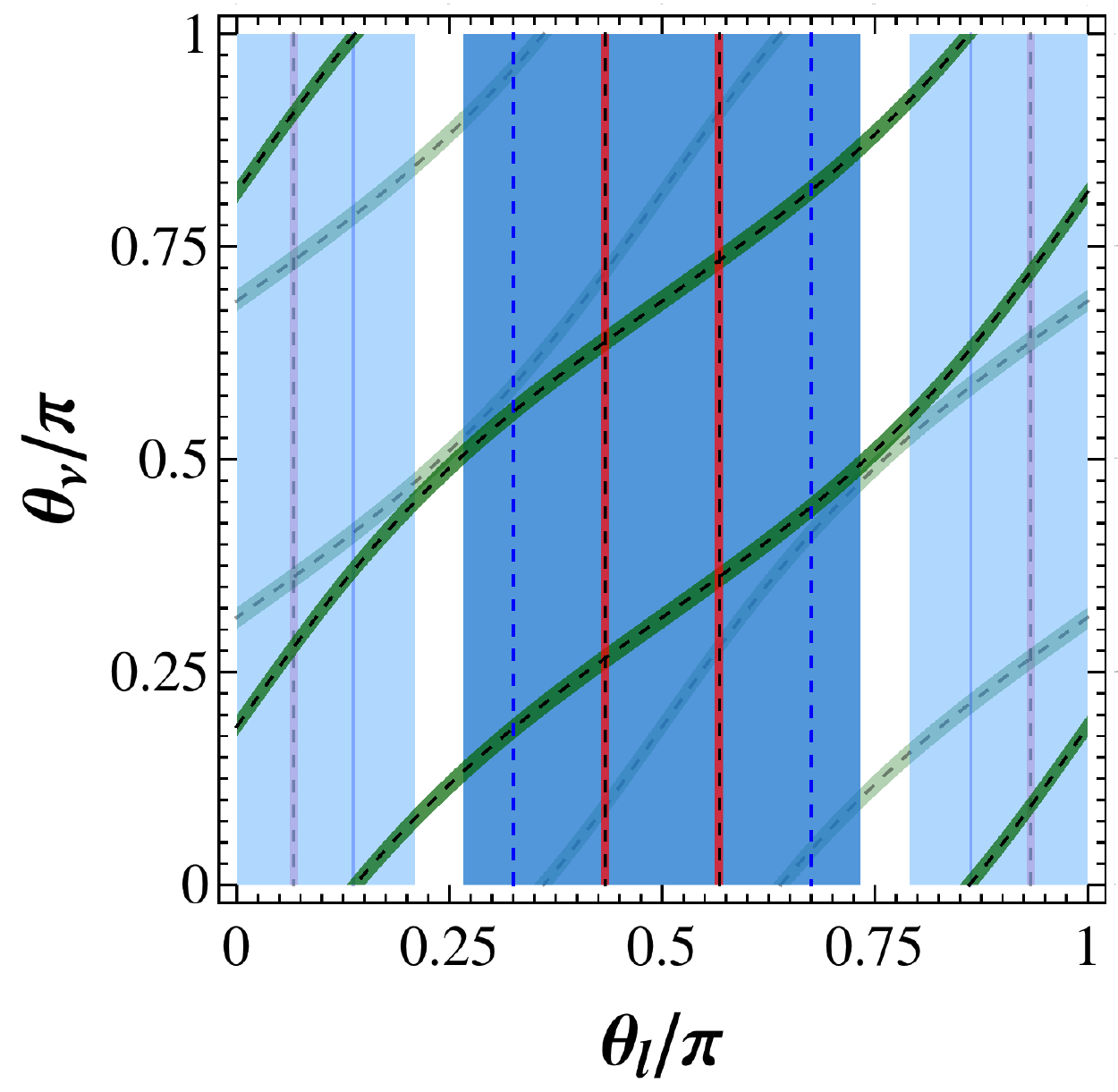}
\includegraphics[width=0.4\textwidth]{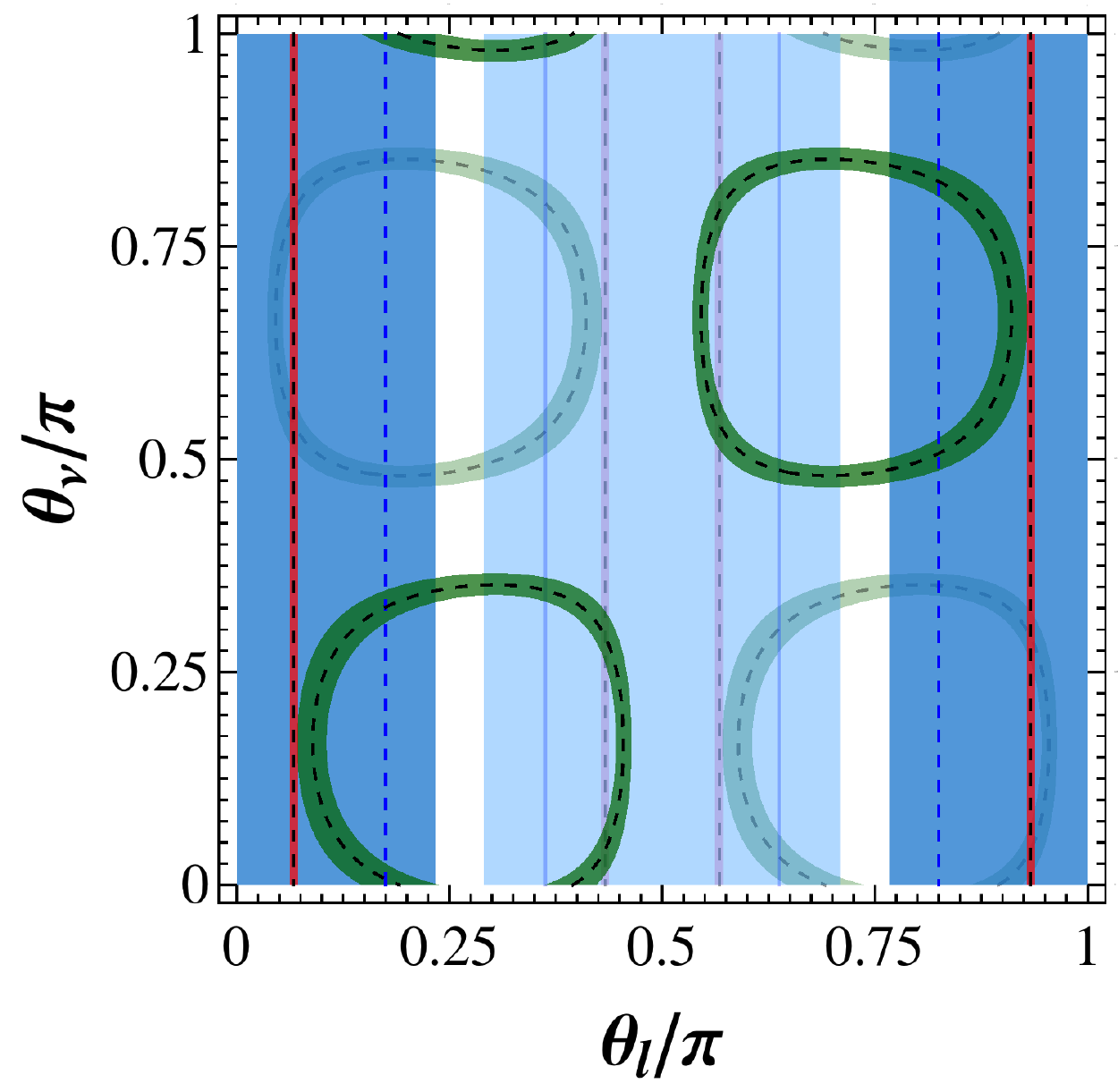}
\caption{\label{fig:Uac_contour}The contour plots of $\sin^2\theta_{ij}$ in the plane $\theta_{\nu}$ versus $\theta_{l}$. The red, blue and green areas denote the $3\sigma$ regions of $\sin^{2}\theta_{13},\sin^{2}\theta_{23}$ and $\sin^{2}\theta_{12}$ respectively. The dashed (or solid) lines indicate the best fit values of the mixing angles. Notice that the best fit value of $\sin^2\theta_{23}$ depends on the neutrino mass ordering, the solid and dashed lines are for NH and IH respectively. The residual flavor symmetry is $(G_{l}, G_{\nu})=(Z^{ST^2SU}_2, Z^{S}_2)$ in this case. The first row corresponds to $(X_{l}, X_{\nu}, P_{l}, P_{\nu})=(T^2, 1, P_{12}, P_{13})$ on the left panel and $(X_{l}, X_{\nu}, P_{l}, P_{\nu})=(T^2, SU, P_{12}, P_{13})$ on the right panel, and the last row is for $(X_{l}, X_{\nu}, P_{l}, P_{\nu})=(T^2, TST^2U, P_{13}, P_{13})$, $(U, 1, P_{12}, P_{13})$. The foreground and background differ in the values of $P_{l}$ which are equal to $P_{12}$ and $P_{13}$ respectively.}
\end{figure}

\begin{figure}[hptb!]
\centering
\includegraphics[width=0.4\textwidth]{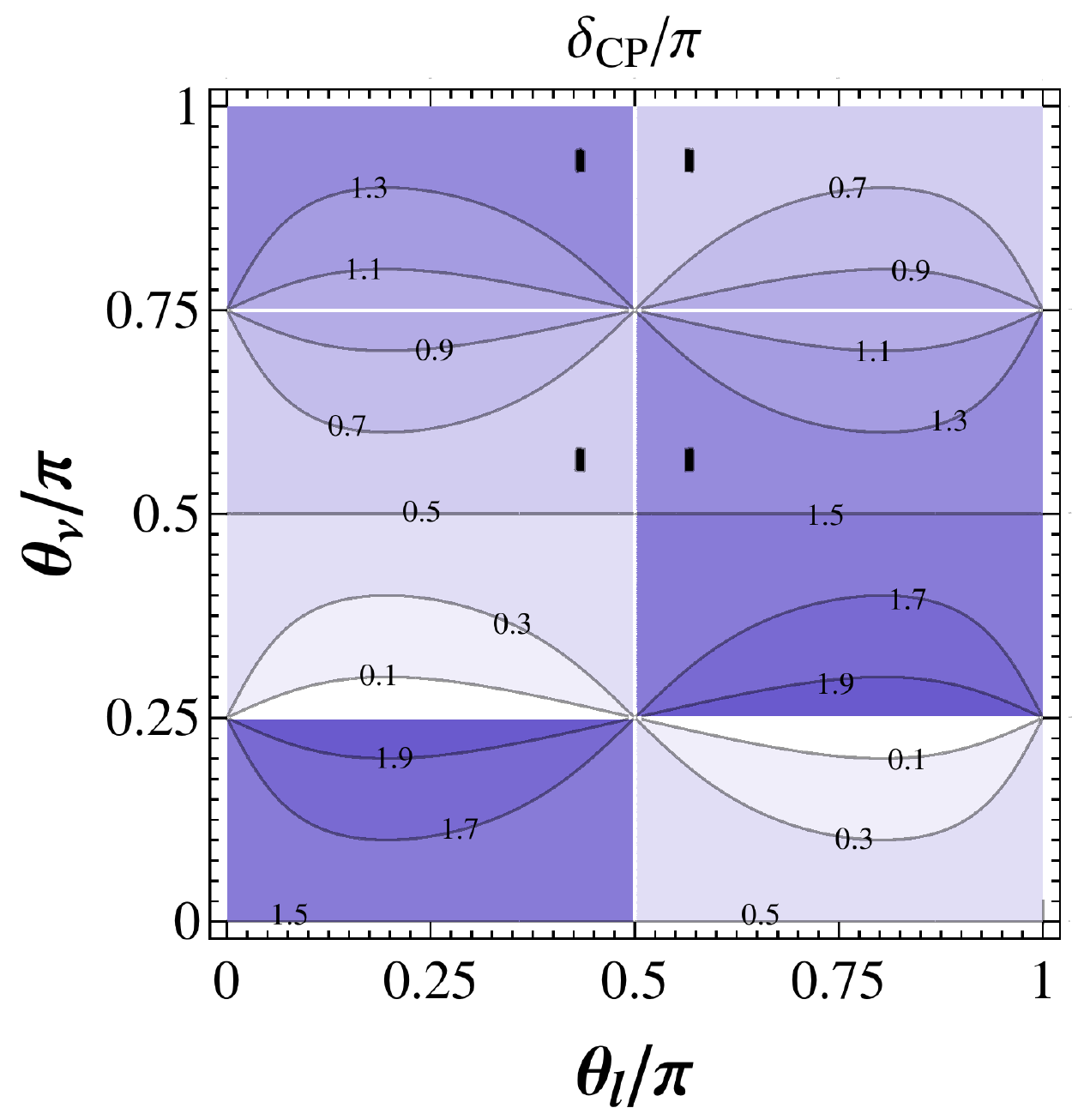}
\includegraphics[width=0.4\textwidth]{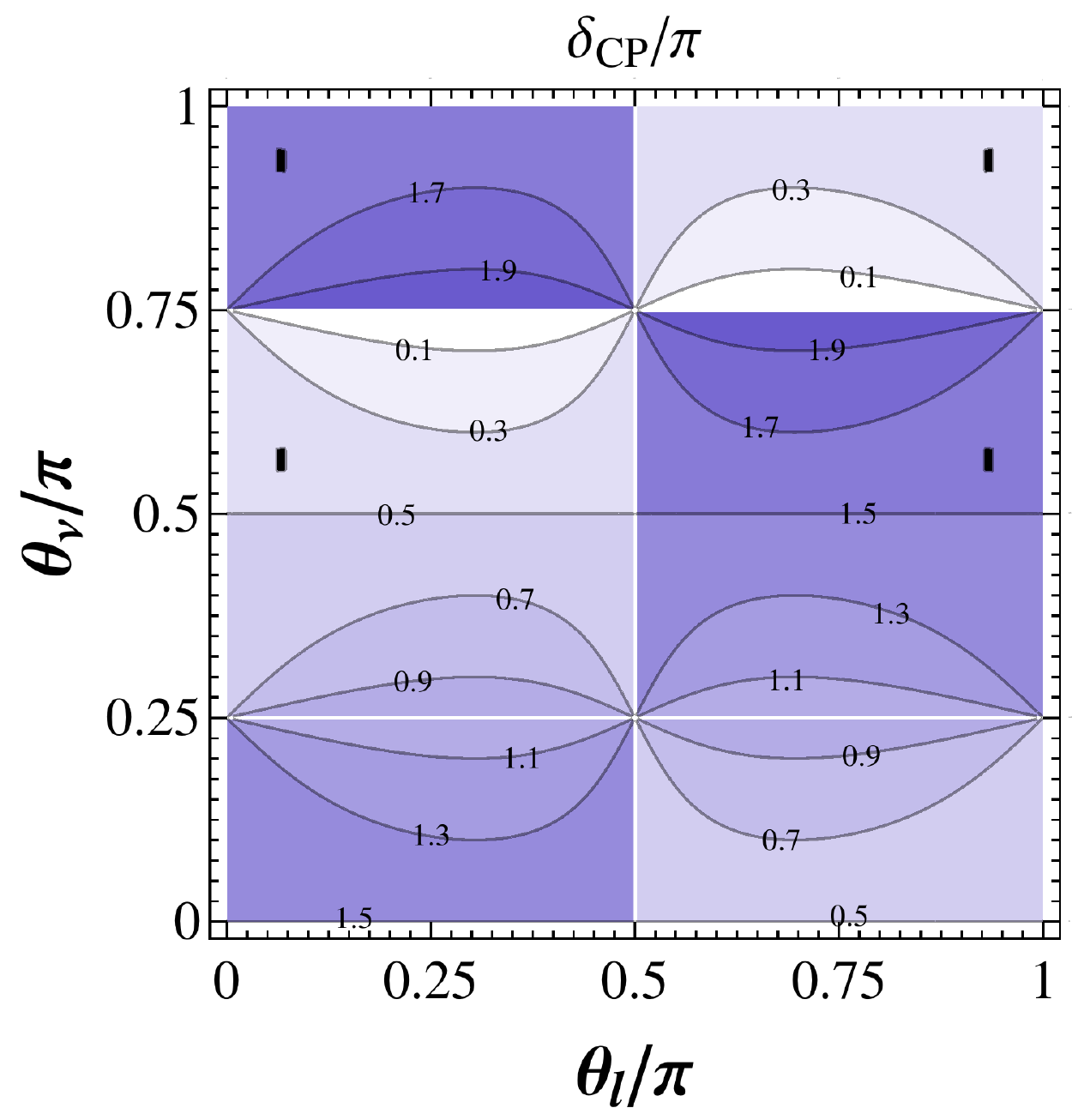}
\includegraphics[width=0.4\textwidth]{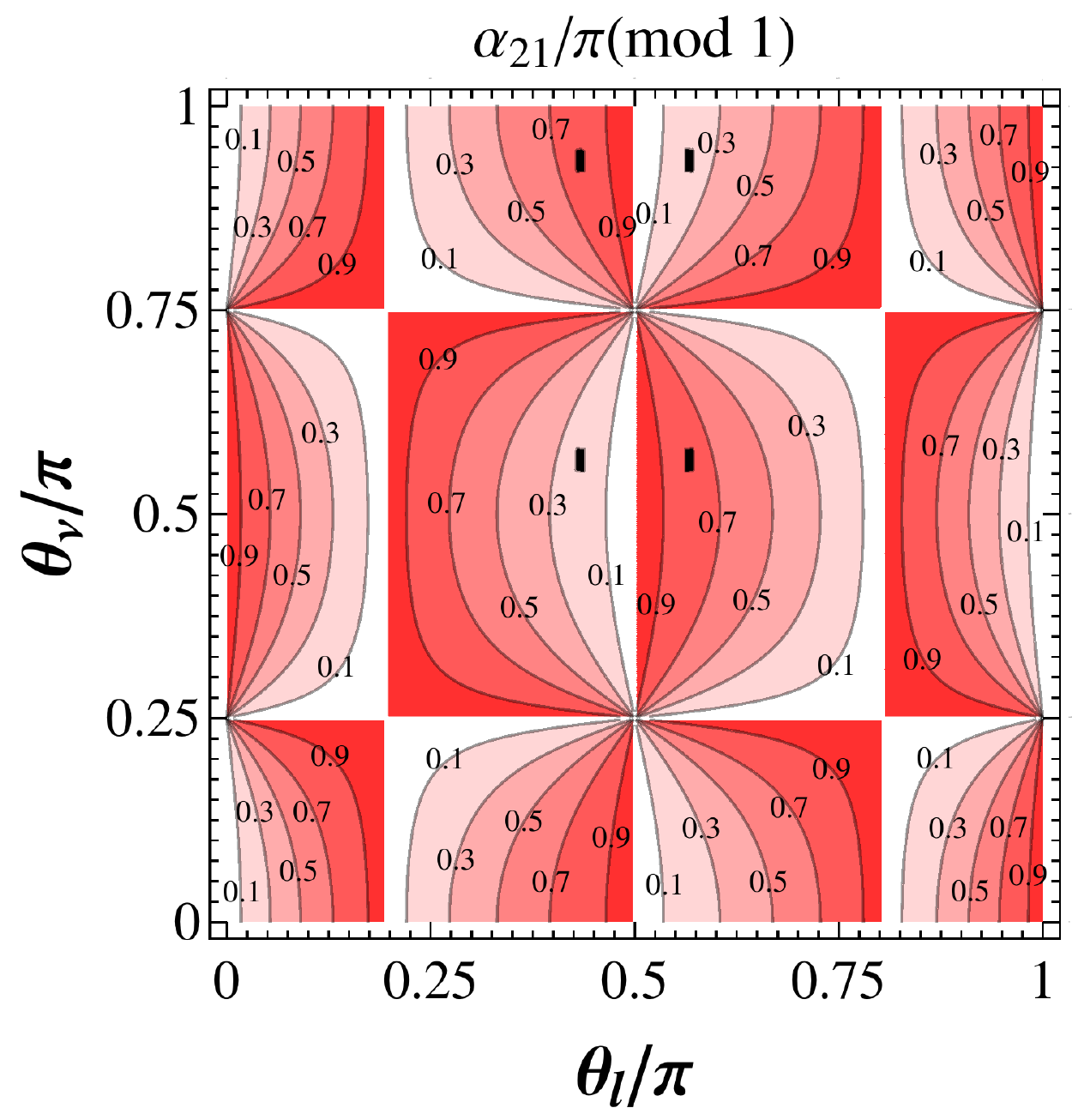}
\includegraphics[width=0.4\textwidth]{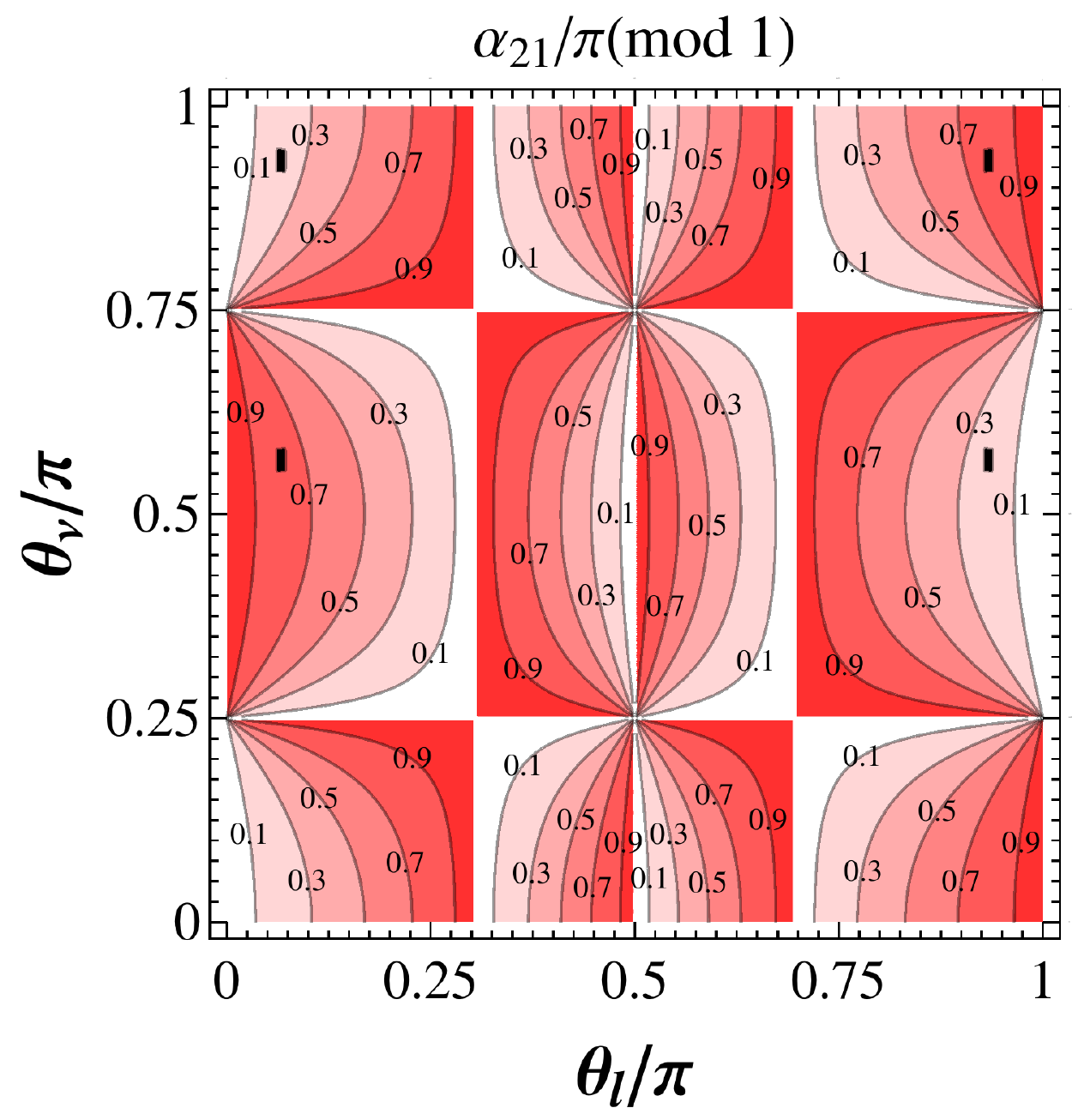}
\includegraphics[width=0.4\textwidth]{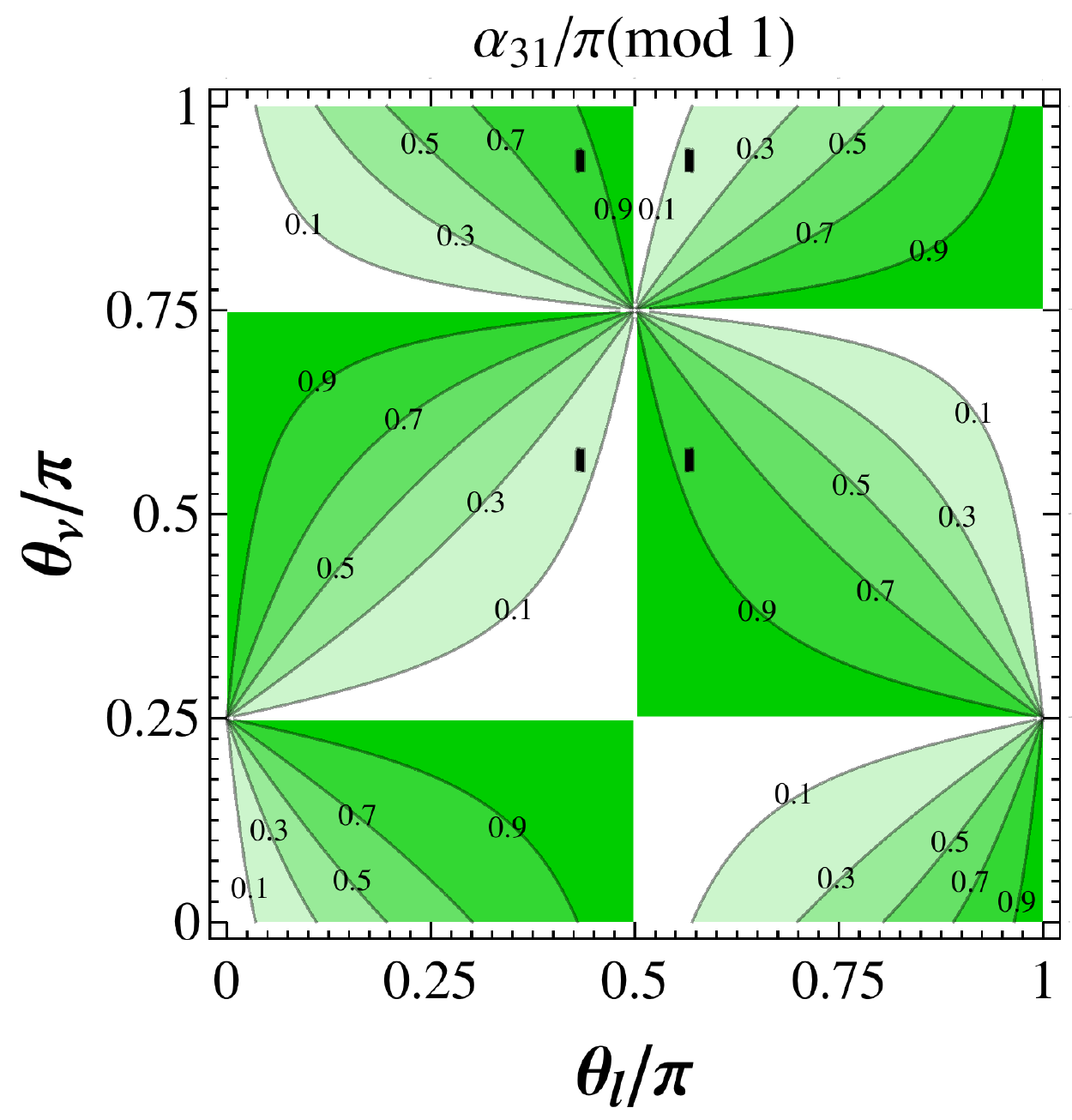}
\includegraphics[width=0.4\textwidth]{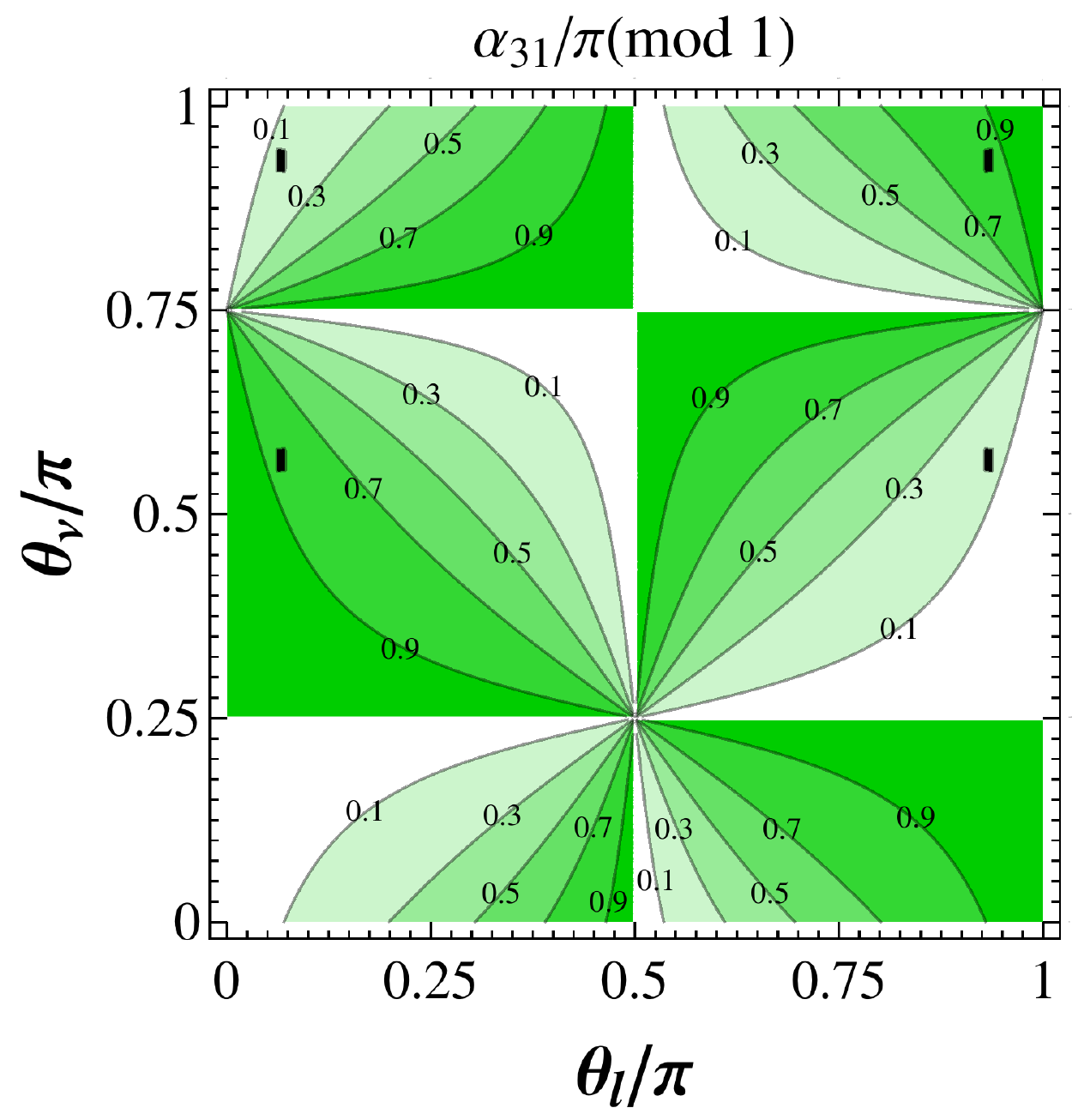}
\caption{\label{fig:contour_CP_phases} The contour plots of the $CP$ violation phases $\delta_{CP}$, $\alpha_{21}$ and $\alpha_{31}$ in the plane $\theta_{\nu}$ versus $\theta_{l}$. The black areas denote the regions in which the lepton mixing angles are compatible with experimental data at $3\sigma$ level. The residual symmetry is $(G_{l}, G_{\nu}, X_{l}, X_{\nu})=(Z^{ST^2SU}_2, Z^{S}_2, T^2, SU)$. The figures on the right-handed and left-handed sides correspond to the row and column permutations $(P_{l}, P_{\nu})=(P_{12}, P_{13})$ and $(P_{l}, P_{\nu})=(P_{13}, P_{13})$ respectively.  }
\end{figure}

The numerical results listed in table~\ref{tab:Uab_chi2_NH}, table~\ref{tab:Uab_chi2_IH}, table~\ref{tab:Uac_chi2_NH} and table~\ref{tab:Uac_chi2_IH} can be easily seen by plotting the contour regions of the mixing angle $\sin^2\theta_{ij}$ in the plane $\theta_{\nu}$ versus $\theta_{l}$, as shown in figure~\ref{fig:Uab_contour}, figure~\ref{fig:Uab_Uac_contour} and figure~\ref{fig:Uac_contour}. The most stringent constraint arises from the reactor neutrino mixing angle $\theta_{13}$ which has been measured quite precisely~\cite{Capozzi:2013csa,Forero:2014bxa,Gonzalez-Garcia:2014bfa,Capozzi:2016rtj}. One sees that the three lepton mixing angles $\theta_{12}$, $\theta_{13}$ and $\theta_{23}$ can be simultaneously compatible with the experimental data at $3\sigma$ level only in a rather narrow region of $\theta_{l}-\theta_{\nu}$ plane. Hence the mixing angles and $CP$ phases should be able to only vary a bit around the numerical values found in table~\ref{tab:Uab_chi2_NH}, table~\ref{tab:Uab_chi2_IH}, table~\ref{tab:Uac_chi2_NH} and table~\ref{tab:Uac_chi2_IH}, and consequently the present approach is very predictive. As an example, in figure~\ref{fig:contour_CP_phases} we display the predictions for the $CP$ phases $\delta_{CP}$, $\alpha_{21}$ and $\alpha_{31}$ in the plane $\theta_{\nu}$ versus $\theta_{l}$ for the residual symmetry $(G_{l}, G_{\nu}, X_{l}, X_{\nu})=(Z^{ST^2SU}_2, Z^{S}_2, T^2, SU)$ with $(P_{l}, P_{\nu})=(P_{12}, P_{13})$, $(P_{13}, P_{13})$, where the small black areas represent the regions in which the experimental data on lepton mixing can be accommodated.

Carefully examining all the numerical results, we see that the predictions for the reactor mixing angel $\theta_{13}$ are almost the same while the values of $\theta_{12}$, $\theta_{23}$ and $\delta_{CP}$ are considerably different in distinct cases. The current oscillation experiments T2K and NO$\nu$A are able to exclude certain ranges of  $\theta_{23}$ and $\delta_{CP}$ around the maximal values, if running in both the neutrino and the antineutrino modes is completed. The forthcoming reactor neutrino oscillation experiments such as JUNO~\cite{An:2015jdp} and RENO~\cite{Kim:2014rfa} expect to make very precise measurement of the solar mixing angle $\theta_{12}$, and the error of $\sin^2\theta_{12}$ can be reduced to about $0.3\%$~\cite{An:2015jdp}. The planned long baseline experiments such as DUNE~\cite{Acciarri:2016crz} and Hyper-K~\cite{Kearns:2013lea,Abe:2014oxa} could significantly improve the precision on $\theta_{23}$ and $\delta_{CP}$. Hence future neutrino facilities have the potential to discriminate among the above possible cases, or rule them out completely.

\begin{figure}[hptb!]
\centering
 \begin{tabular}{ >{\centering\arraybackslash} m{6.5cm} >{\centering\arraybackslash} m{6.5cm} }
\centering
\includegraphics[width=0.4\textwidth]{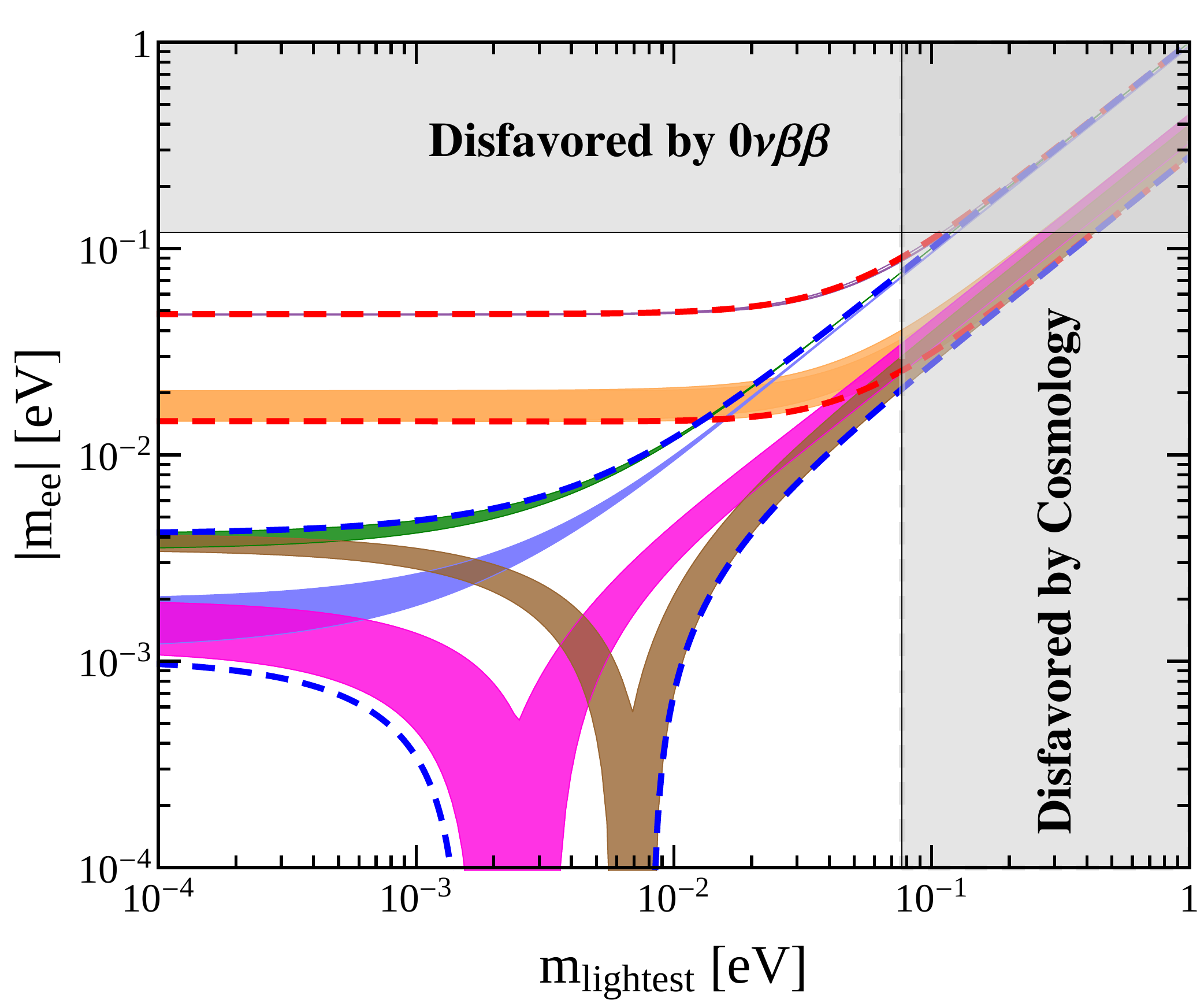}&
\includegraphics[width=0.4\textwidth]{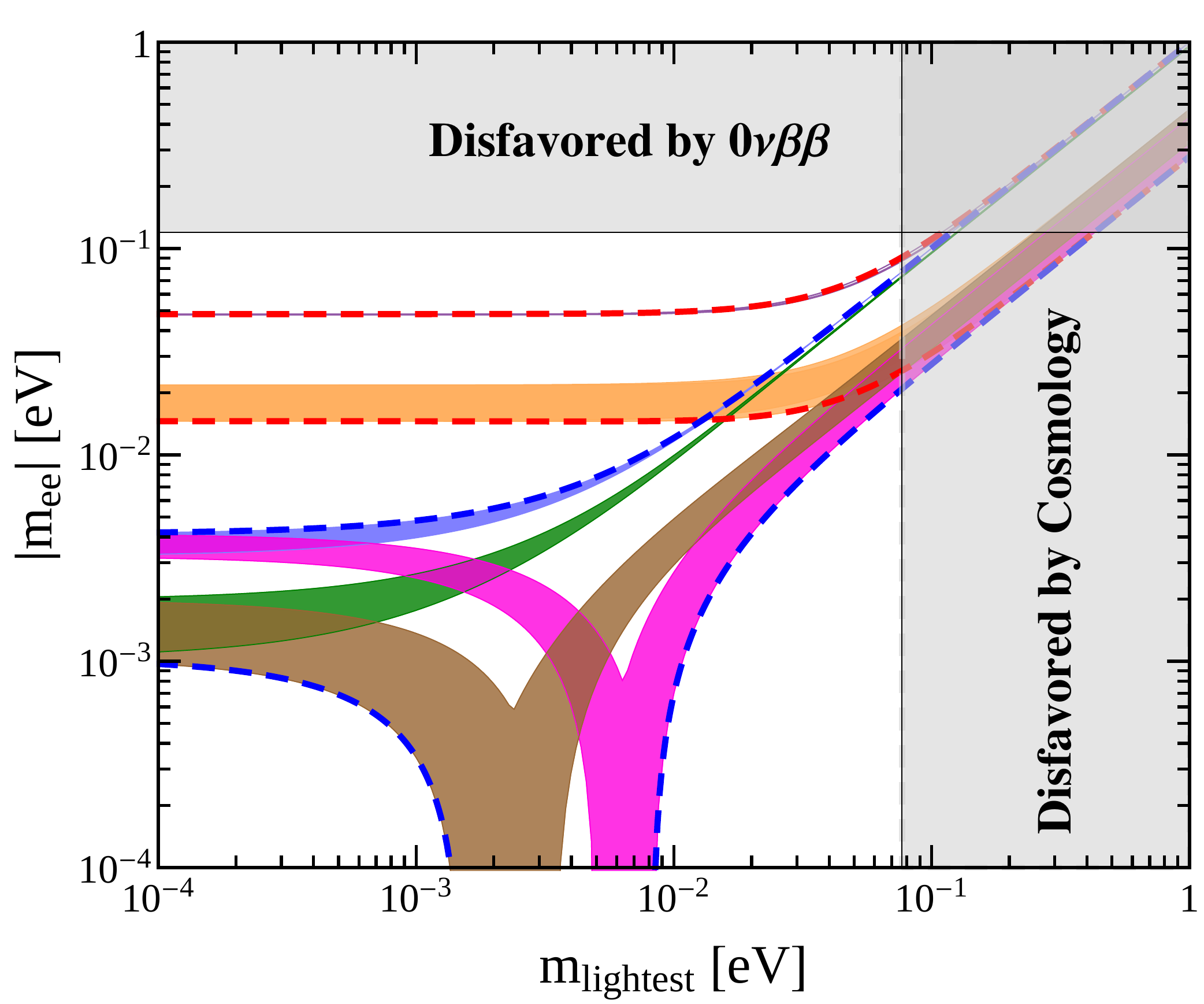}\\
\includegraphics[width=0.4\textwidth]{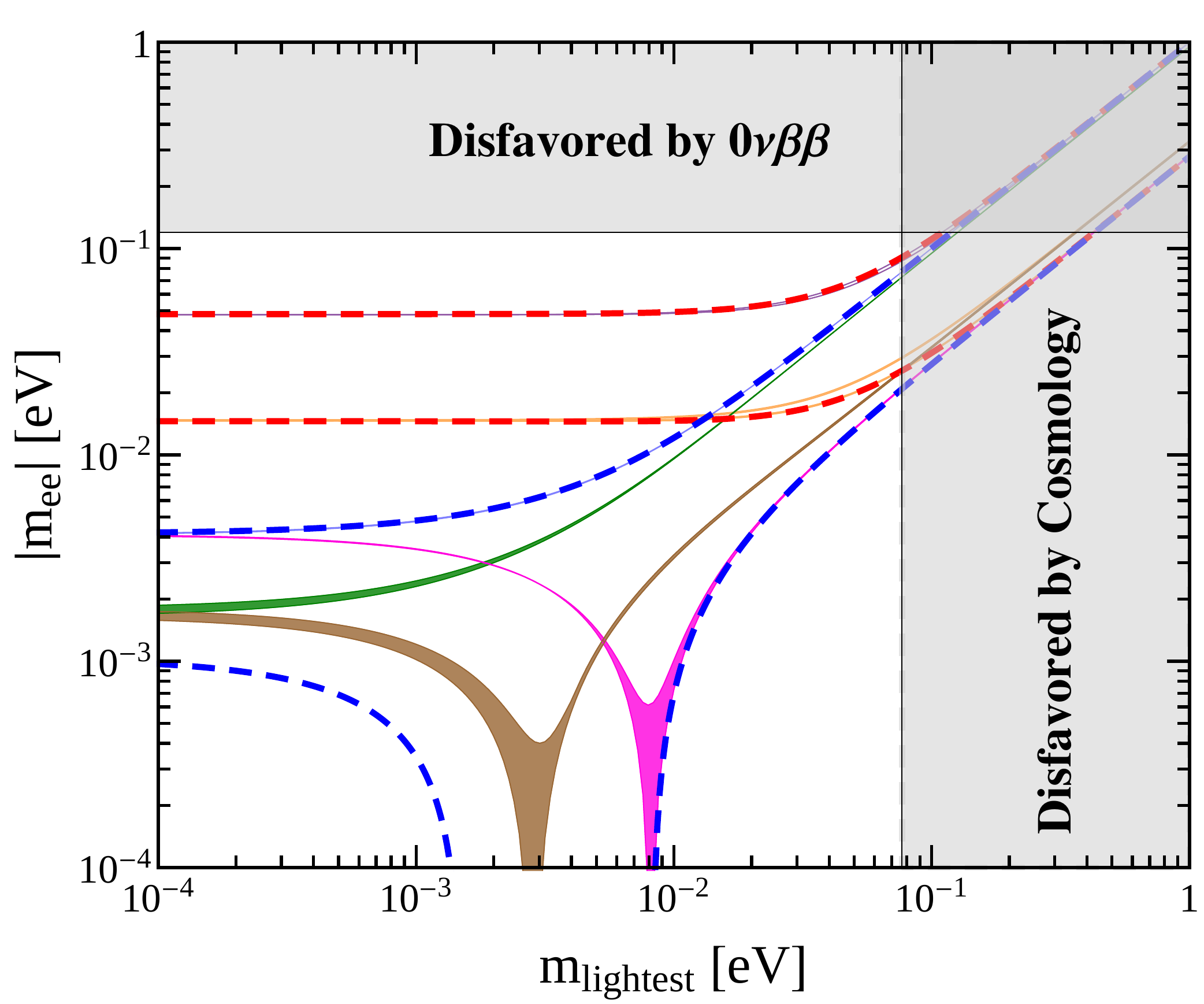}&
\includegraphics[width=0.4\textwidth]{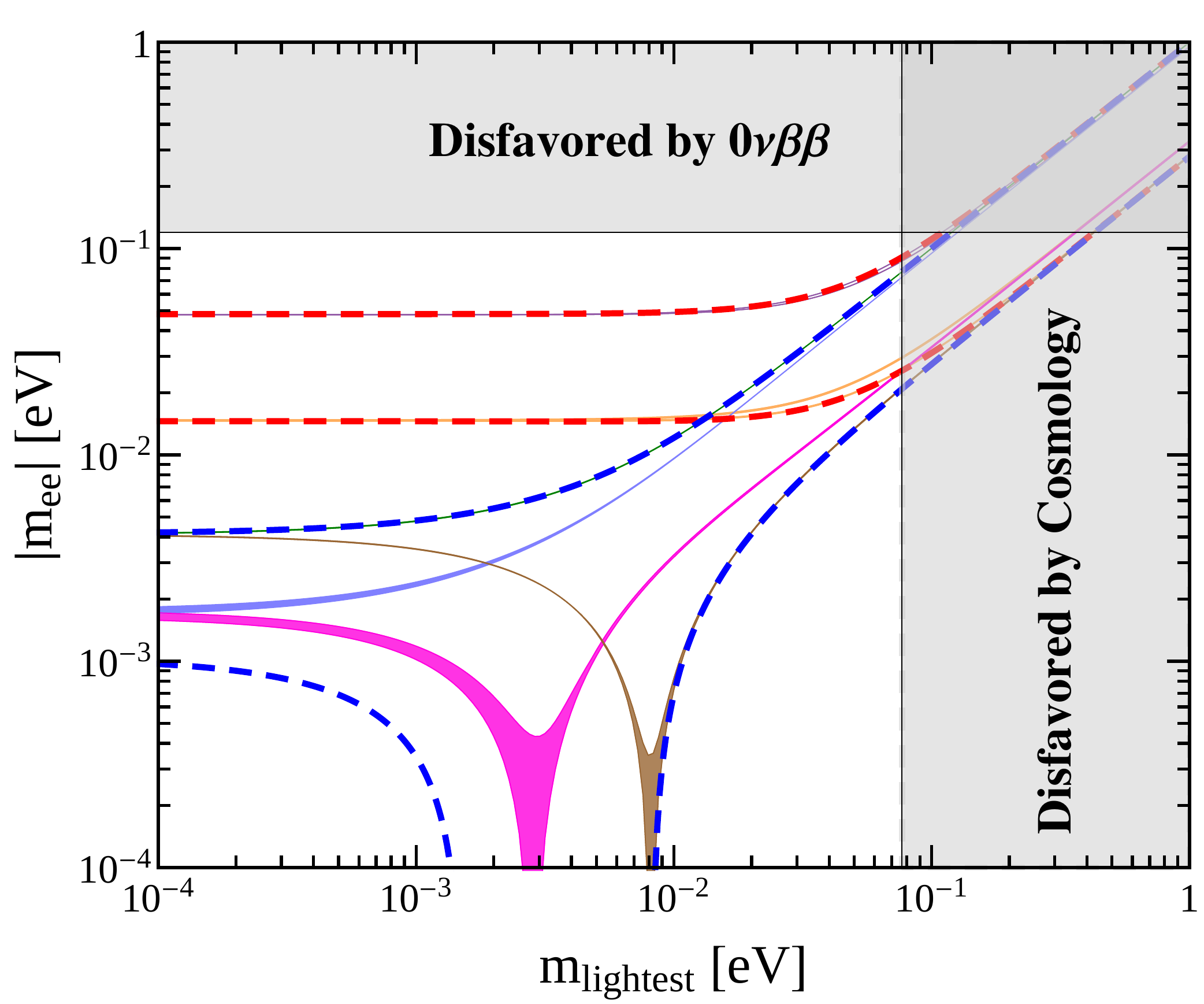}\\
\includegraphics[width=0.4\textwidth]{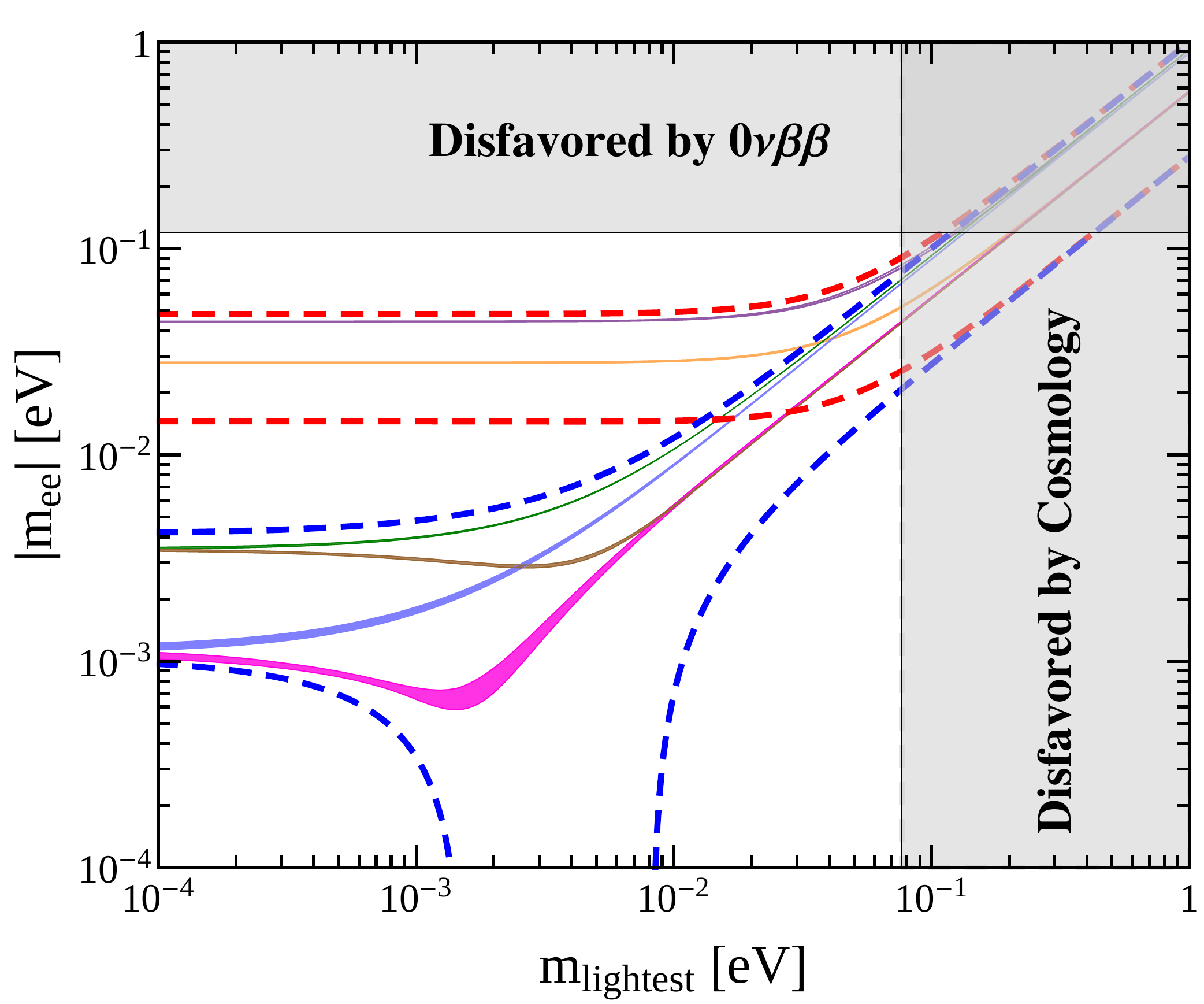}&~~
\includegraphics[width=0.3\textwidth]{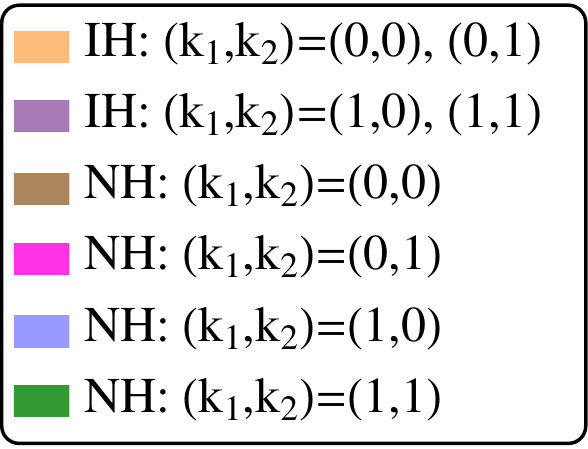}
\end{tabular}
\caption{\label{fig:mee_Uab}The predictions for the possible values of the effective Majorana mass $|m_{ee}|$ as a function of the lightest neutrino mass. The red (blue) dashed lines indicate the most general allowed regions for IH (NH) neutrino mass spectrum obtained by varying the mixing parameters over the $3\sigma$ ranges~\cite{Gonzalez-Garcia:2014bfa}. The residual flavor symmetry is $(G_{l}, G_{\nu})=(Z^{ST^2SU}_2, Z^{TU}_2)$ in this case. The first row corresponds to $(X_{l}, X_{\nu}, P_{l}, P_{\nu})=(U,T, P_{12},1)$ on the left and $(X_{l}, X_{\nu}, P_{l}, P_{\nu})=(U,T, P_{12}, P_{12})$ on the right, the middle row is for $(X_{l}, X_{\nu}, P_{l}, P_{\nu})=(U, STS, P_{12}, 1)$, $(U, STS, P_{12}, P_{12})$, and the last row for $(X_{l}, X_{\nu}, P_{l}, P_{\nu})=(T^2, T, P_{12}, P_{12})$.  The present most stringent upper limits $|m_{ee}|<0.120$ eV from EXO-200~\cite{Auger:2012ar, Albert:2014awa} and KamLAND-ZEN~\cite{Gando:2012zm} is shown by horizontal grey band. The vertical grey exclusion band is the current limit on the lightest neutrino masses from the cosmological data $\sum m_i<0.230$ eV at $95\%$ confidence level obtained by the Planck collaboration~\cite{Ade:2013zuv}.}
\end{figure}

\begin{figure}[hptb!]
\centering
 \begin{tabular}{ >{\centering\arraybackslash} m{6.5cm} >{\centering\arraybackslash} m{6.5cm} }
\includegraphics[width=0.4\textwidth]{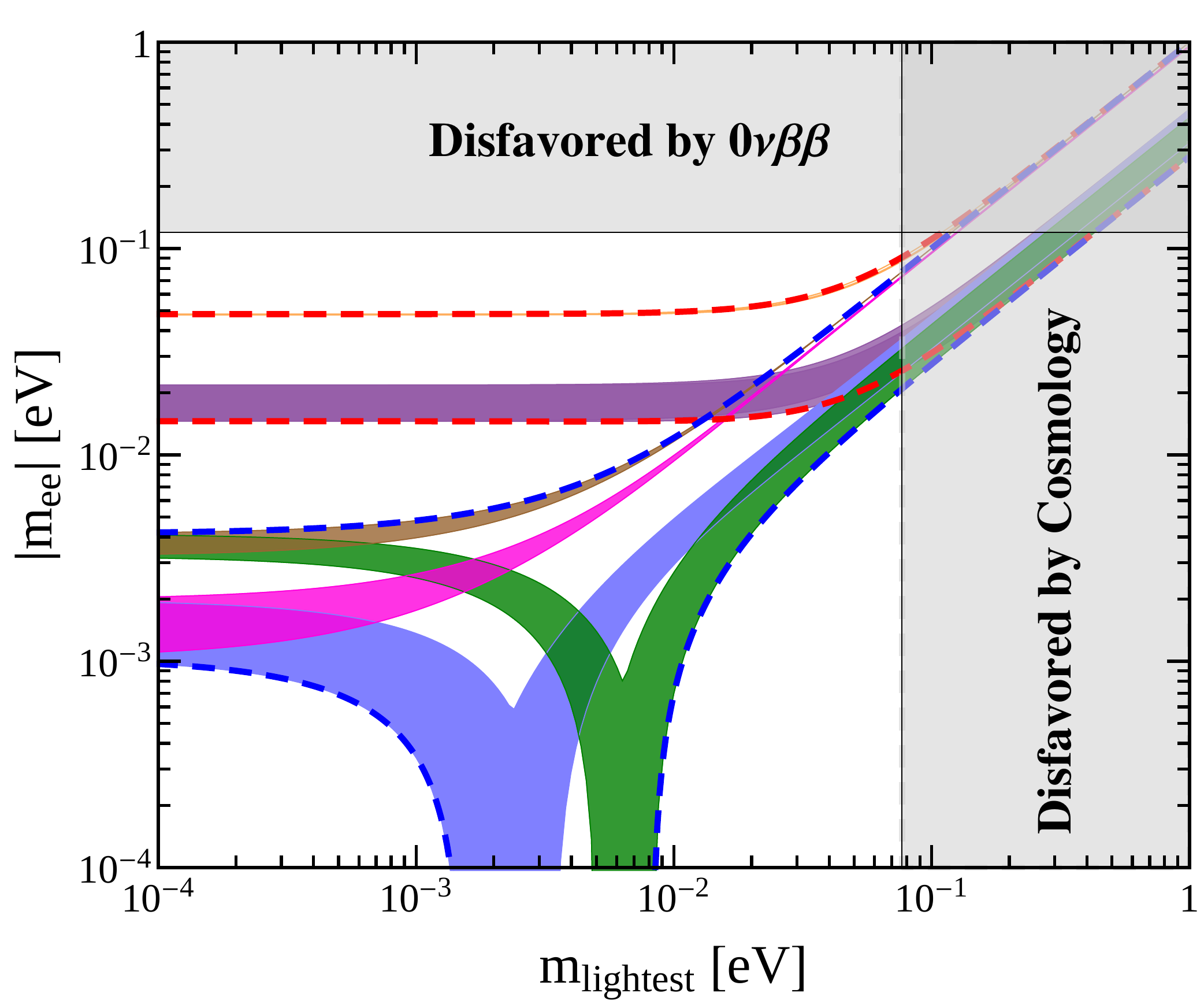}&
\includegraphics[width=0.4\textwidth]{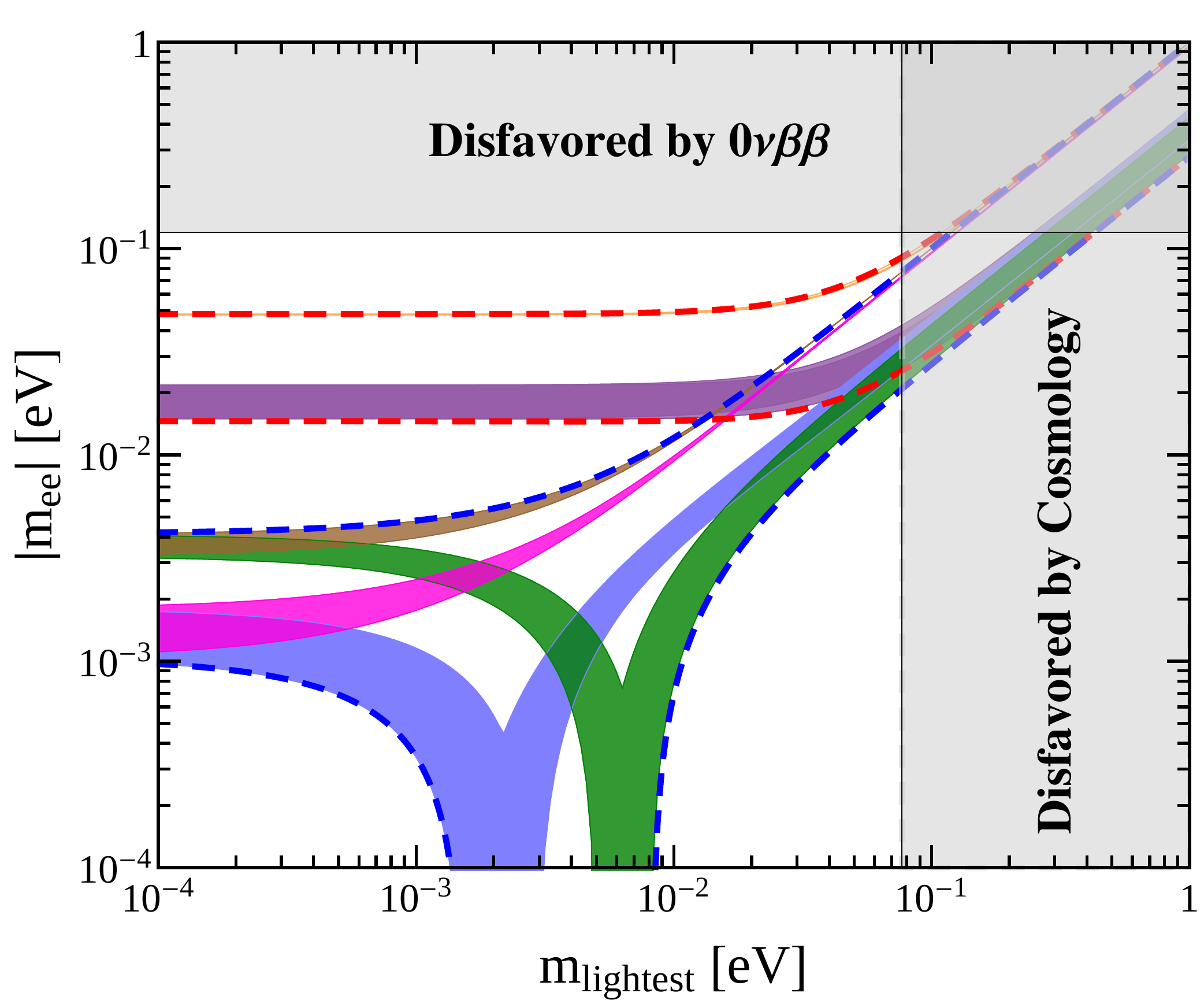}\\
\includegraphics[width=0.4\textwidth]{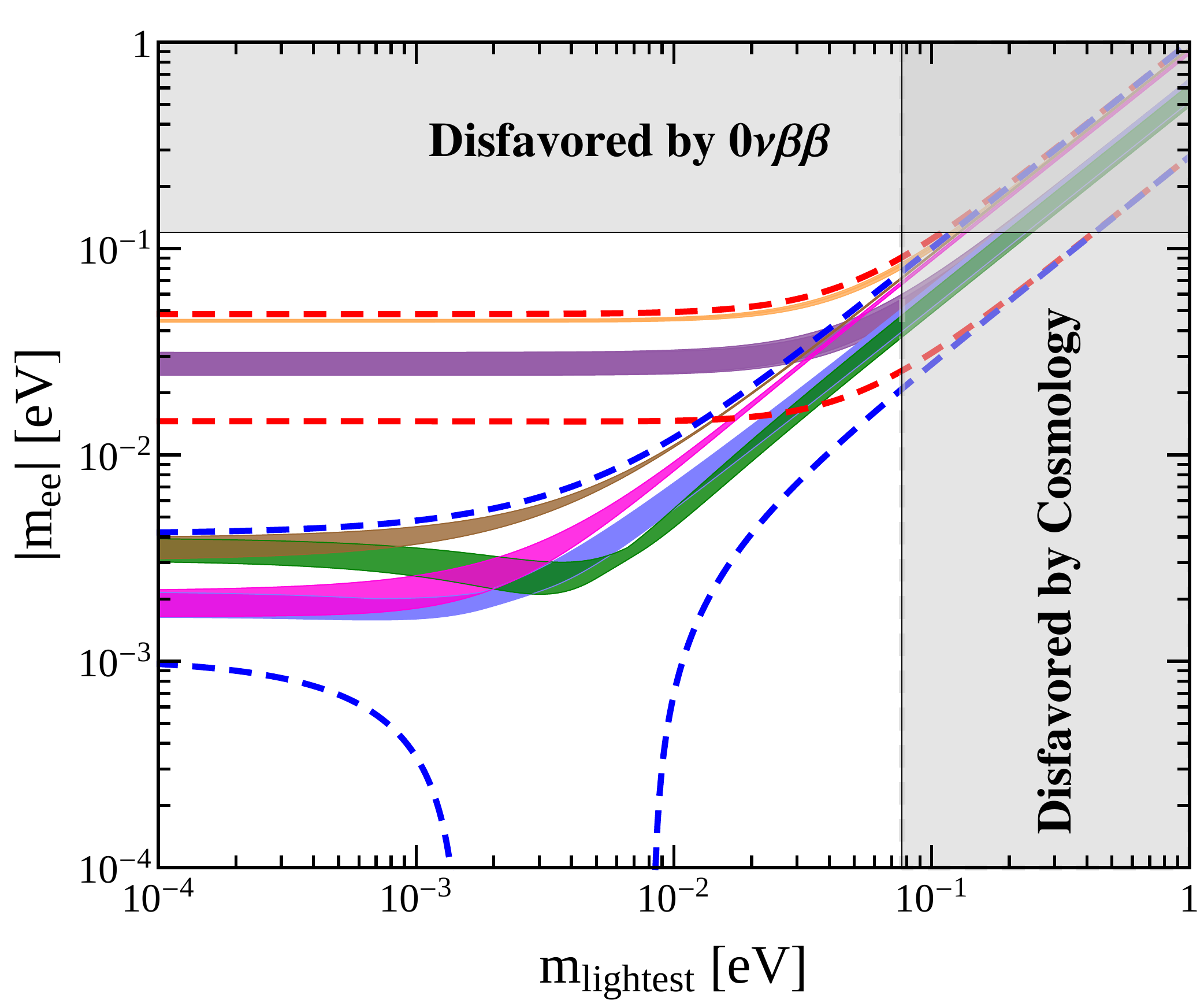}&~~
\includegraphics[width=0.30\textwidth]{note_v4.pdf}
\end{tabular}
\caption{\label{fig:mee_Uac}The predictions for the effective Majorana mass $|m_{ee}|$, where we use the same conventions as in figure~\ref{fig:mee_Uab}. The residual flavor symmetry is $(G_{l}, G_{\nu})=(Z^{ST^2SU}_2, Z^{S}_2)$ in this case. The top left panel corresponds to $(X_{l}, X_{\nu}, P_{l}, P_{\nu})=(T^2, TST^2U, P_{12}, P_{13})$, the top right panel is for $(X_{l}, X_{\nu}, P_{l}, P_{\nu})=(T^2, TST^2U, P_{13}, P_{12})$, and the last one for $(X_{l}, X_{\nu}, P_{l}, P_{\nu})=(T^2, SU, P_{12}, P_{13})$. }
\end{figure}

\begin{figure}[hptb!]
\centering
\includegraphics[width=0.4\textwidth]{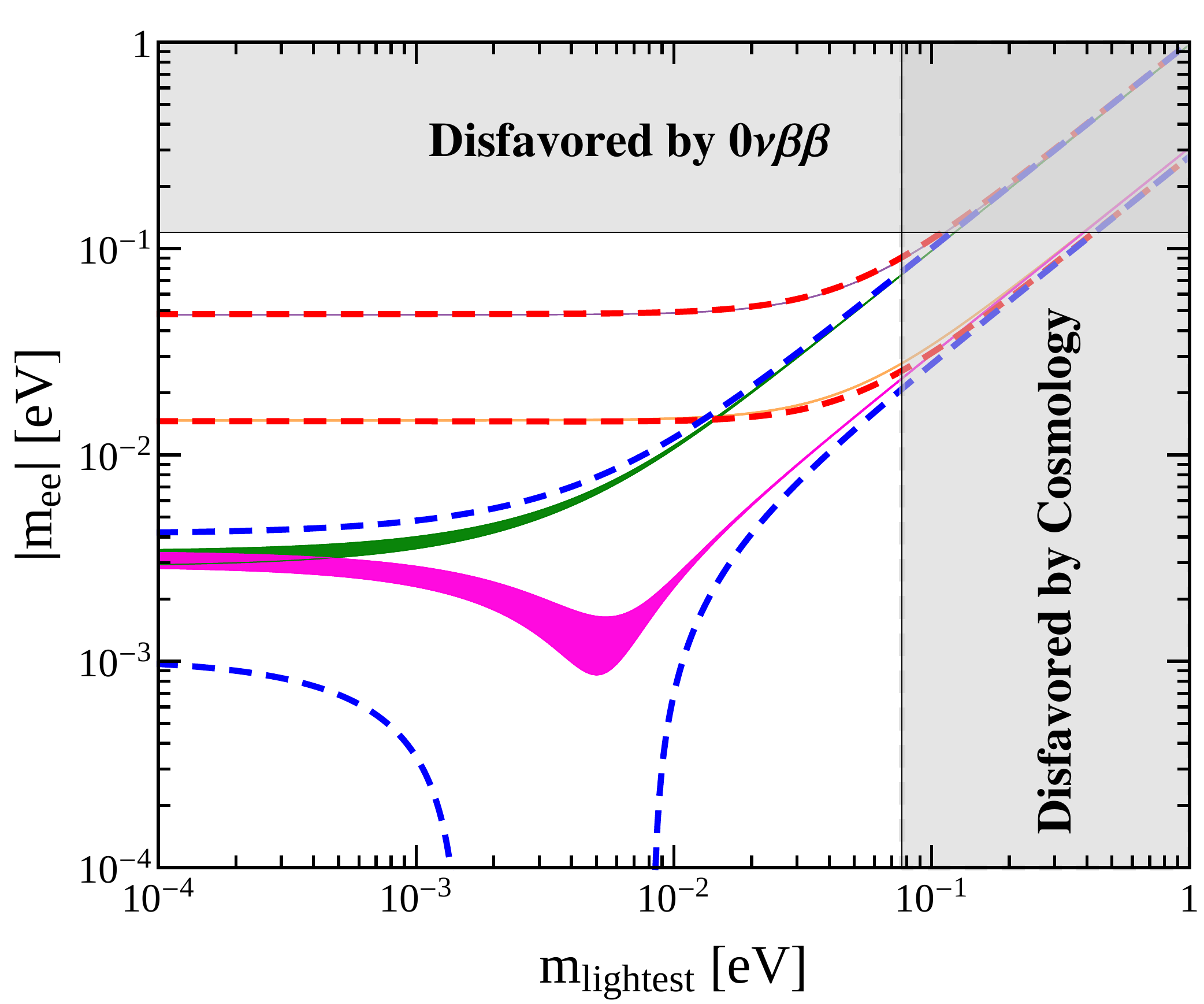}
\includegraphics[width=0.4\textwidth]{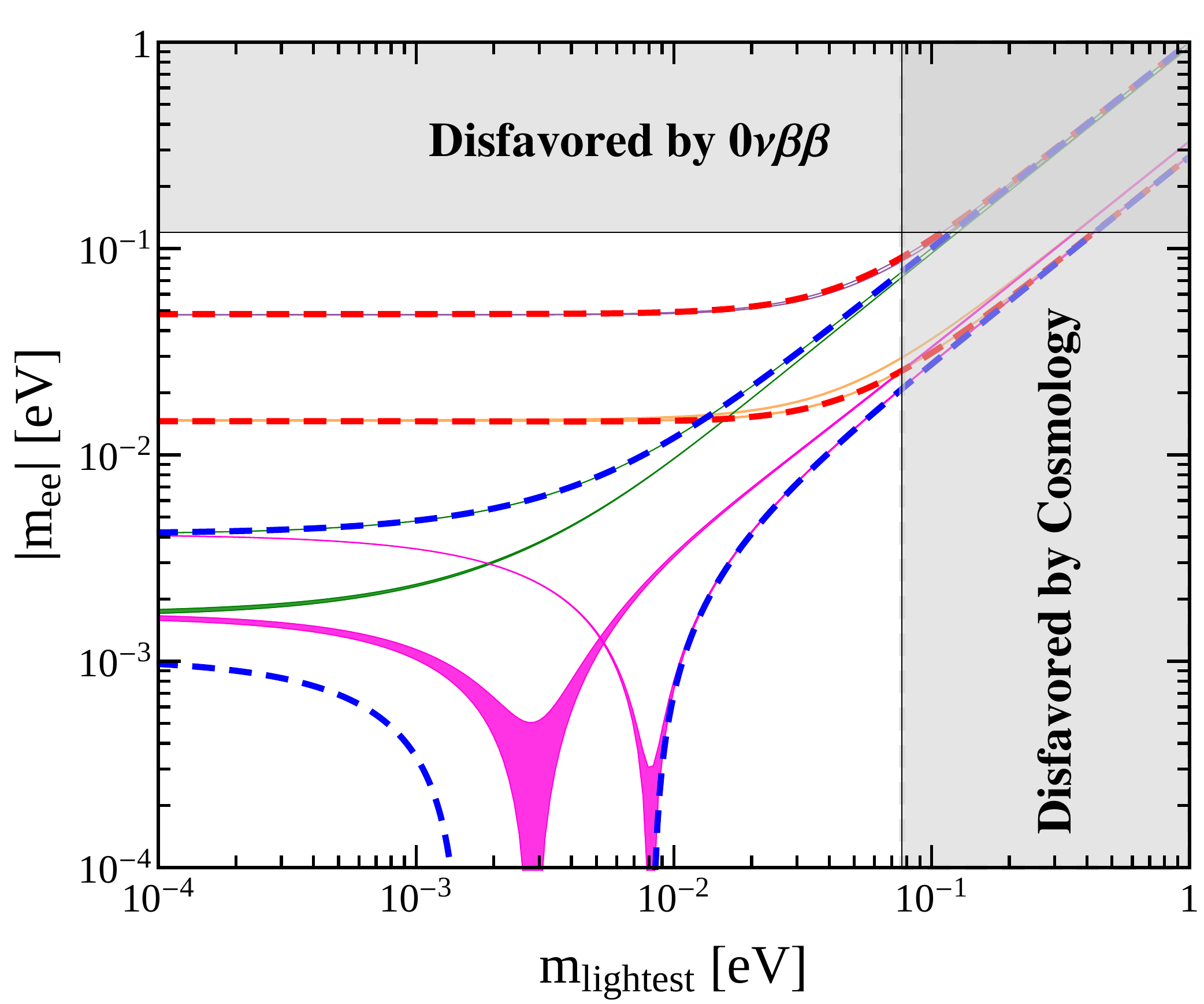}
\includegraphics[width=0.75\textwidth]{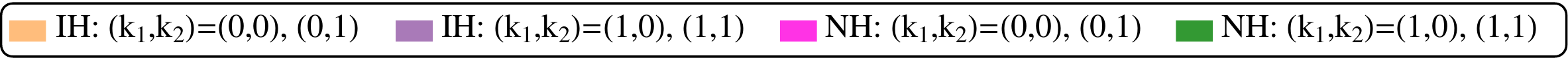}
\caption{ \label{fig:mee_Uac_2}The predictions for the effective Majorana mass $|m_{ee}|$, where we use the same conventions as in figure~\ref{fig:mee_Uab}. The residual flavor symmetry is $(G_{l}, G_{\nu})=(Z^{ST^2SU}_2, Z^{S}_2)$ in this case. The panels on the right-handed and left-handed sides correspond to $(X_{l}, X_{\nu}, P_{l}, P_{\nu})=(T^2, 1, P_{12}, P_{13})$ and $(X_{l}, X_{\nu}, P_{l}, P_{\nu})=(U, 1, P_{12}, P_{13})$ respectively. Notice that $|m_{ee}|$ is invariant under the transformations $\theta_{l}\rightarrow\pi-\theta_{l}$, $\theta_{\nu}\rightarrow\theta_{\nu}+\pi/2$ and $k_2\rightarrow k_2+1$, hence the effective mass is independent of $k_2$ in this case.}
\end{figure}

The neutrinoless double $(0\nu\beta\beta)$ decay is a lepton number violating process. It is an important probe of the Majorana nature of neutrinos, and it can provide us with precious information on
the neutrino mass scale and ordering. Searching for $0\nu\beta\beta$ decay has a long history. There are many new sensitive $0\nu\beta\beta$ experiments which are in various stages of planning and construction. The $0\nu\beta\beta$ decay rate is proportional to the effective Majorana mass $|m_{ee}|$ which is expressed in terms of neutrino masses and lepton mixing parameters as~\cite{Agashe:2014kda},
\begin{eqnarray}
\nonumber |m_{ee}|&=&|m_1U^2_{PMNS,11}+m_2U^2_{PMNS,12}+m_3U^2_{PMNS,13}|\\
 &=&\left|m_1\cos^2\theta_{12}\cos^2\theta_{13}+m_2\sin^2\theta_{12}\cos^2\theta_{13}e^{i\alpha_{21}}+m_3\sin^2\theta_{13}e^{i(\alpha_{31}-2\delta_{CP})}\right|\,,
\end{eqnarray}
where $m_{1,2,3}$ are light neutrino masses. For each admissible case, the allowed regions of the effective Majorana mass $|m_{ee}|$ as a function of the lightest neutrino mass are shown in figure~\ref{fig:mee_Uab}, figure~\ref{fig:mee_Uac} and figure~\ref{fig:mee_Uac_2}. Both parameters $\theta_{1}$ and $\theta_2$ freely vary between $0$ and $\pi$, and the three lepton mixing angles are required to lie in their current $3\sigma$ ranges~\cite{Gonzalez-Garcia:2014bfa}. Notice that $|m_{ee}|$ does not depend on $\theta_{23}$. Hence if two cases have the same residual symmetry but differ in the permutation matrices with $P_{l}=P_{12}$ and $P_{l}=P_{13}$ respectively, the same predictions for $|m_{ee}|$ would be obtained. For the case of IH neutrino mass spectrum, the effective Majorana mass is almost independent of the value of $k_2$. The reason is because the term in $|m_{ee}|$ proportional to $m_3$ is suppressed by both $\sin^2\theta_{13}$ and the small value of $m_3$ itself. Moreover, we see that $|m_{ee}|$ is predicted to be around the upper boundary 0.048~eV, lower boundary 0.015~eV or close to 0.028~eV for IH. Although these predictions are beyond the reach of the facilities in running, the next generation elaborate $0\nu\beta\beta$ decay experiments are capable of covering the full IH region, such that the present predictions could be tested in near future. For the case of NH mass spectrum, cancellation between different terms in $|m_{ee}|$ could occur for certain values of the lightest neutrino mass, consequently the effective mass can be smaller than $10^{-4}$~eV. However, the range of $m_{\text{lightest}}$ in which $|m_{ee}|$ can be quite small is significantly reduced with respect to the generic case. We can even find a non-trivial lower bound on $|m_{ee}|$ in some cases, see e.g. figure~\ref{fig:mee_Uab} for the remnant symmetry $(G_{l}, G_{\nu}, X_{l}, X_{\nu})=(Z^{ST^2SU}_2, Z^{TU}_2, T^2, T)$ with $P_{l}=P_{\nu}=P_{12}$.

\begin{table}[hptb!]
\centering
\footnotesize
\begin{tabular}{|c|c|c|}
\hline\hline
$G_{r}$ & $X_{r}$ & $\Sigma$\\
\hline
 &   &     \\ [-0.12in]
\multirow{4}{*}{$Z_{2}^{ST^{2}SU}$} & $T^{2}(TST^{2}U)$  & $\frac{1}{\sqrt{6}}\left(
\begin{array}{ccc}
 2 & 0 & -\sqrt{2} \\
 e^{\frac{i \pi }{3}} & -\sqrt{3} e^{\frac{i \pi }{3}} & \sqrt{2} e^{\frac{i \pi }{3}}
   \\
 e^{-\frac{i \pi }{3}} & \sqrt{3} e^{-\frac{i \pi }{3}} & \sqrt{2} e^{-\frac{i \pi }{3}}
   \\
\end{array}
\right)$\\
 &  &  \\[-0.12in] \cline{2-3}
 &    &     \\ [-0.12in]
 & $U(ST^{2}S)$ & $ \frac{1}{\sqrt{6}}\left(
\begin{array}{ccc}
 2 i & \sqrt{2} i & 0 \\
 -e^{-\frac{i \pi }{6}} & \sqrt{2} e^{-\frac{i \pi }{6}} & -\sqrt{3} e^{\frac{i \pi
   }{3}} \\
 e^{\frac{i \pi }{6}} & -\sqrt{2} e^{\frac{i \pi }{6}} & \sqrt{3} e^{-\frac{i \pi }{3}}
   \\
\end{array}
\right)$ \\
 &  &  \\[-0.12in] \hline
  &  &  \\[-0.12in]
\multirow{4}{*}{$Z_{2}^{TU}$} & $T(U)$ & $\frac{1}{\sqrt{2}}\left(
\begin{array}{ccc}
 0 & 0 & \sqrt{2} \\
 -e^{-\frac{i\pi }{3}} & e^{-\frac{i\pi}{3}} & 0 \\
 e^{\frac{i \pi }{3}} & e^{\frac{i \pi }{3}} & 0 \\
\end{array}
\right)$ \\
 &  &  \\[-0.12in] \cline{2-3}
  &  &  \\[-0.12in]
 & $STS(T^{2}STU)$ & $\frac{1}{\sqrt{6}}\left(
\begin{array}{ccc}
 0 & 2 i & \sqrt{2} \\
 \sqrt{3} e^{\frac{i \pi }{6}} & e^{\frac{i \pi }{6}} & -\sqrt{2} e^{-\frac{i \pi }{3}}
   \\
 \sqrt{3} e^{-\frac{i \pi }{6}} & -e^{-\frac{i\pi }{6}} & -\sqrt{2} e^{\frac{i \pi
   }{3}} \\
\end{array}
\right) $ \\
 &  &  \\[-0.12in] \hline
  &  &  \\[-0.12in]
\multirow{8}{*}{$Z_{2}^{S}$} & $1(S)$ & $\frac{1}{\sqrt{6}}\left(
\begin{array}{ccc}
 \sqrt{2} & -1 & -\sqrt{3} \\
 \sqrt{2} & 2 & 0 \\
 \sqrt{2} & -1 & \sqrt{3} \\
\end{array}
\right)$ \\
 &  &  \\[-0.12in] \cline{2-3}
  &  &  \\[-0.12in]
 & $SU(U)$ & $\frac{1}{\sqrt{6}}\left(
\begin{array}{ccc}
 \sqrt{2}\,i  & 0 & -2 \\
 \sqrt{2}\,i & -\sqrt{3}\,i  & 1 \\
 \sqrt{2}\,i  & \sqrt{3}\,i & 1 \\
\end{array}
\right)$ \\
 &  &  \\[-0.12in] \cline{2-3}
  &  &  \\[-0.12in]
 & $TST^{2}U(T^{2}STU)$ & $\frac{1}{\sqrt{3}}\left(
\begin{array}{ccc}
 1 & i & 1 \\
 1 & e^{-\frac{i \pi }{6}} & -e^{-\frac{i \pi }{3}} \\
 1 & -e^{\frac{i \pi }{6}} & -e^{\frac{i \pi }{3}} \\
\end{array}
\right) $ \\
\hline\hline
\end{tabular}
\caption{\label{tab:res_symm_lepton} The residual flavor symmetries $G_{r}=Z^{ST^2SU}_2$, $Z^{TU}_2$, $Z^{S}_2$, the corresponding residual $CP$ transformations $X_{r}$ consistent with $G_{r}$ and the Takagi factorization matrix $\Sigma$. Let's denote the generator of $G_{r}$ as $g_{r}$, then $\rho_{\mathbf{3}}(g_{r})X_{r}$ is also a residual $CP$ symmetry, and it is given in the parenthesis. For simplicity of notation, we do not distinguish between the abstract elements of the $S_4$ group and their representation matrices in $\mathbf{3}$.}
\end{table}

\begin{table}[hptb!]
\centering
\footnotesize
\begin{tabular}{|c|c|c|}
\hline\hline

\multicolumn{3}{|c|}{$(G_{l}, G_{\nu})=(Z^{ST^2SU}_2, Z^{TU}_2)$}\\
\hline\hline

$(X_{l}, X_{\nu}, P_{l}, P_{\nu})$ & $(U,T, P_{12}, 1)$ & $(U,STS, P_{12},1)$ \\\hline

$\sin^2\theta_{13}$ & $\frac{\left(\cos \theta _l \left(\sqrt{2} \sin \theta _{\nu}+2 \cos \theta _{\nu}\right)-3 \sin \theta _l \sin \theta _{\nu}\right){}^2}{12}$ & $\frac{\left(2 \sin ^2\theta _l \cos ^2\theta _{\nu}+\sin ^2\theta _{\nu} \left(\sin \theta _l-\sqrt{2} \cos \theta _l\right){}^2\right)}{4}$ \\
\hline
$\sin^2\theta_{12}$ & $1-\frac{6 \left(2 \sqrt{2} \sin 2 \theta _l+\cos 2 \theta _l+3\right)}{\Delta_{1}+7 \cos 2 \theta _{\nu}+33} $ & $1-\frac{2 \left(2 \sqrt{2} \sin 2 \theta _l+\cos 2 \theta _l+3\right)}{4 \sqrt{2} \sin 2 \theta _l \sin ^2\theta _{\nu}+\cos 2 \theta _{\nu}+\cos 2 \theta _l \left(3 \cos 2 \theta _{\nu}+1\right)+11} $ \\
\hline
$\sin^2\theta_{23} $ &$\frac{4 \left(\sin \theta _{\nu}-2 \sqrt{2} \cos \theta _{\nu}\right){}^2}{\Delta_{1}+7 \cos 2 \theta _{\nu}+33} $&$\frac{2 \left(\cos 2 \theta _{\nu}+3\right)}{4 \sqrt{2} \sin 2 \theta _l \sin ^2\theta _{\nu}+\cos 2 \theta _{\nu}+\cos 2 \theta _l \left(3 \cos 2 \theta _{\nu}+1\right)+11} $\\
\hline
$J_{CP}$ &$0$&$\frac{\sin 2 \theta _{\nu} \left(\sin 2 \theta _l-2 \sqrt{2} \cos 2 \theta _l\right)}{32}$\\
\hline
$I_{1}$ &$0 $&$\frac{\left(\sin \theta _l-3 \sin 3 \theta _l\right) \sin 2 \theta _{\nu} \left(\sqrt{2} \sin \theta _l+2 \cos \theta _l\right)}{64} $\\
\hline
$I_{2}$ &$0 $&$-\frac{\left(\sin \theta _l-3 \sin 3 \theta _l\right) \sin 2 \theta _{\nu} \left(\sqrt{2} \sin \theta _l+2 \cos \theta _l\right)}{64} $\\
\hline \hline

$(X_{l}, X_{\nu}, P_{l}, P_{\nu})$ & $(U,T, P_{12}, P_{12})$ & $(U,STS, P_{12}, P_{12})$ \\\hline

$\sin^2\theta_{13}$ & $\frac{\left(\cos \theta _l \left(\sqrt{2} \sin \theta _{\nu}+2 \cos \theta _{\nu}\right)-3 \sin \theta _l \sin \theta _{\nu}\right){}^2}{12}$ & $\frac{\left(2 \sin ^2\theta _l \cos ^2\theta _{\nu}+\sin ^2\theta _{\nu} \left(\sin \theta _l-\sqrt{2} \cos \theta _l\right){}^2\right)}{4} $  \\
\hline
$\sin^2\theta_{12}$ & $\frac{12 \left(\sin \theta _l+\sqrt{2} \cos \theta _l\right){}^2}{\Delta_{1}+7 \cos 2 \theta _{\nu}+33} $ & $\frac{4 \left(\sin \theta _l+\sqrt{2} \cos \theta _l\right){}^2}{4 \sqrt{2} \sin 2 \theta _l \sin ^2\theta _{\nu}+\cos 2 \theta _{\nu}+\cos 2 \theta _l \left(3 \cos 2 \theta _{\nu}+1\right)+11}  $  \\
\hline
$\sin^2\theta_{23}$ & $\frac{4 \left(\sin \theta _{\nu}-2 \sqrt{2} \cos \theta _{\nu}\right){}^2}{\Delta_{1}+7 \cos 2 \theta _{\nu}+33} $ & $\frac{2 \left(\cos 2 \theta _{\nu}+3\right)}{4 \sqrt{2} \sin 2 \theta _l \sin ^2\theta _{\nu}+\cos 2 \theta _{\nu}+\cos 2 \theta _l \left(3 \cos 2 \theta _{\nu}+1\right)+11} $  \\
\hline
$J_{CP}$ & $0 $ & $\frac{\sin 2 \theta _{\nu} \left(2 \sqrt{2} \cos 2 \theta _l-\sin 2 \theta _l\right)}{32}$  \\
\hline
$I_{1}$ & $0 $ & $-\frac{\left(\sin \theta _l-3 \sin 3 \theta _l\right) \sin 2 \theta _{\nu} \left(\sqrt{2} \sin \theta _l+2 \cos \theta _l\right)}{64} $  \\
\hline
$I_{2}$ & $0 $ & $\frac{\sin \theta _l \sin 2 \theta _{\nu} \left(\sqrt{2} \left(3 \sin \theta _l+7 \sin 3 \theta _l\right)-14 \cos \theta _l-2 \cos 3 \theta _l\right)}{64} $  \\
\hline \hline

$(X_{l}, X_{\nu}, P_{l}, P_{\nu})$ & $(U,T, P_{13}, 1)$ & $(U,STS, P_{13}, 1)$ \\\hline

$\sin^2\theta_{13}$ &$\frac{ \left(\sin \theta _l \left(\sqrt{2} \sin \theta _{\nu }+2 \cos \theta _{\nu }\right)+3 \sin \theta _{\nu } \cos \theta _l\right){}^2}{12}$ & $\frac{\left(2 \cos ^2\theta _l \cos ^2\theta _{\nu}+\sin ^2\theta _{\nu} \left(\cos \theta _l+\sqrt{2} \sin \theta _l\right){}^2\right)}{4}  $  \\
\hline
$\sin^2\theta_{12}$ & $ 1+\frac{6 \left(2 \sqrt{2} \sin 2 \theta _l+\cos 2 \theta _l-3\right)}{\Delta_{2}+7 \cos 2 \theta _{\nu}+33}  $ & $1-\frac{2 \left(2 \sqrt{2} \sin 2 \theta _l+\cos 2 \theta _l-3\right)}{4 \sqrt{2} \sin 2 \theta _l \sin ^2\theta _{\nu}-\cos 2 \theta _{\nu}+\cos 2 \theta _l\left(3 \cos 2 \theta _{\nu}+1\right)-11} $  \\
\hline
$\sin^2\theta_{23}$ & $1-\frac{4 \left(\sin \theta _{\nu}-2 \sqrt{2} \cos \theta _{\nu}\right){}^2}{\Delta_{2}+7 \cos 2 \theta _{\nu}+33}$ & $1+\frac{2 \left(\cos 2 \theta _{\nu}+3\right)}{4 \sqrt{2} \sin 2 \theta _l \sin ^2\theta _{\nu}-\cos 2 \theta _{\nu}+\cos 2 \theta _l\left(3 \cos 2 \theta _{\nu}+1\right)-11}  $  \\
\hline
$J_{CP}$ & $0 $ & $\frac{ \sin 2 \theta _{\nu} \left(\sin 2 \theta _l-2 \sqrt{2} \cos 2 \theta _l\right)}{32} $  \\
\hline
$I_{1}$ & $0 $ & $\frac{\sin 2 \theta _{\nu} \left(\cos \theta _l+3 \cos 3 \theta _l\right) \left(\sqrt{2} \cos \theta _l-2 \sin \theta _l\right)}{64}  $  \\
\hline
$I_{2}$ & $0 $ & $-\frac{\sin 2 \theta _{\nu} \left(\cos \theta _l+3 \cos 3 \theta _l\right) \left(\sqrt{2} \cos \theta _l-2 \sin \theta _l\right)}{64} $  \\
\hline \hline

$(X_{l}, X_{\nu}, P_{l}, P_{\nu})$ & $(U,T, P_{13}, P_{12})$ & $(U,STS, P_{13}, P_{12})$ \\\hline

$\sin^2\theta_{13}$ &$\frac{ \left(\sin \theta _l \left(\sqrt{2} \sin \theta _{\nu }+2 \cos \theta _{\nu }\right)+3 \sin \theta _{\nu } \cos \theta _l\right){}^2}{12}$  & $\frac{ \left(2 \cos ^2\theta _l \cos ^2\theta _{\nu}+\sin ^2\theta _{\nu} \left(\cos \theta _l+\sqrt{2} \sin \theta _l\right){}^2\right)}{4}  $  \\
\hline
$\sin^2\theta_{12}$ & $\frac{12 \left(\cos \theta _l-\sqrt{2} \sin \theta _l\right){}^2}{\Delta_{2}+7 \cos 2 \theta _{\nu}+33} $ & $ -\frac{4 \left(\cos \theta _l-\sqrt{2} \sin \theta _l\right){}^2}{4 \sqrt{2} \sin 2 \theta _l \sin ^2\theta _{\nu}-\cos 2 \theta _{\nu}+\cos 2 \theta _l \left(3 \cos 2 \theta _{\nu}+1\right)-11} $  \\
\hline
$\sin^2\theta_{23}$ & $1-\frac{4 \left(\sin \theta _{\nu}-2 \sqrt{2} \cos \theta _{\nu}\right){}^2}{\Delta_{2}+7 \cos 2 \theta _{\nu}+33} $ & $1+\frac{2 \left(\cos 2 \theta _{\nu}+3\right)}{4 \sqrt{2} \sin 2 \theta _l \sin ^2\theta _{\nu}-\cos 2 \theta _{\nu}+\cos 2 \theta _l \left(3 \cos 2 \theta _{\nu}+1\right)-11}$  \\
\hline
$J_{CP}$ & $0 $ & $\frac{ \sin 2 \theta _{\nu} \left(2 \sqrt{2} \cos 2 \theta _l-\sin 2 \theta _l\right)}{32} $  \\
\hline
$I_{1}$ & $0 $ & $-\frac{\left(\cos \theta _l+3 \cos 3 \theta _l\right)\sin 2 \theta _{\nu} \left(\sqrt{2} \cos \theta _l-2 \sin \theta _l\right)}{64}  $  \\
\hline
$I_{2}$ & $0 $ & $\frac{\cos \theta _l\sin 2 \theta _{\nu} \left(\sqrt{2}(3 \cos \theta _l-7 \cos 3 \theta _l)+14 \sin \theta _l-2 \sin 3 \theta _l\right)}{64}$  \\
\hline\hline

$(X_{l}, X_{\nu}, P_{l}, P_{\nu})$ & \multicolumn{2}{c|}{$(T^2,T, P_{12}, P_{12})$ } \\\hline

$\sin^2\theta_{13}$   & \multicolumn{2}{c|}{ $\frac{ \left(9 \sin ^2\theta _{\nu} \cos ^2\theta _l+\sin ^2\theta _l \left(\sqrt{2} \sin \theta _{\nu}+2 \cos \theta _{\nu}\right){}^2\right)}{12}$ }\\
\hline
$\sin^2\theta_{12}$   & \multicolumn{2}{c|}{ $-\frac{6 \left(\cos 2 \theta _l-3\right)}{-8 \sqrt{2} \sin 2 \theta _{\nu} \sin ^2\theta _l+7 \cos 2 \theta _{\nu}+\cos 2 \theta _l \left(11 \cos 2 \theta _{\nu}-3\right)+33} $ }\\
\hline
$\sin^2\theta_{23}$   & \multicolumn{2}{c|}{ $ -\frac{4 \left(\sin \theta _{\nu}-2 \sqrt{2} \cos \theta _{\nu}\right){}^2}{8 \sqrt{2} \sin 2 \theta _{\nu} \sin ^2\theta _l+\cos 2 \theta _l \left(3-11 \cos 2 \theta _{\nu}\right)-7 \cos 2 \theta _{\nu}-33} $ }\\
\hline
$J_{CP}$   & \multicolumn{2}{c|}{ $-\frac{\sin 2 \theta _l \left(7 \sin 2 \theta _{\nu}+4 \sqrt{2} \cos 2 \theta _{\nu}\right)}{96}$ }\\
\hline
$I_{1}$   & \multicolumn{2}{c|}{ $\frac{\left(\sin 4 \theta _l \left(12 \sqrt{2}-13 \sin 2 \theta _{\nu}\right)+14 \sin 2 \theta _l \sin 2 \theta _{\nu}+8 \sqrt{2} \left(\sin 2 \theta _l+\sin 4 \theta _l\right) \cos 2 \theta _{\nu}\right)}{192}  $} \\
\hline
$I_{2}$   & \multicolumn{2}{c|}{ $\frac{ \left(\left(14 \sin 2 \theta _l+11 \sin 4 \theta _l\right) \sin 2 \theta _{\nu}+32 \sqrt{2} \sin ^3\theta _l \cos \theta _l \cos 2 \theta _{\nu}\right)}{192} $} \\
\hline\hline
\end{tabular}
\caption{\label{tab:Uab}Results of the mixing parameters for the independent and viable cases with $(G_{l}, G_{\nu})=(Z^{ST^2SU}_2, Z^{TU}_2)$. Note that the factors $(-1)^{k_1}$ and $(-1)^{k_2}$ are omitted in the expressions of $I_1$ and $I_2$ respectively, and they arise from the $CP$ parity matrix $Q_{\nu}=\text{diag}(1, i^{k_1}, i^{k_2})$. For notational simplicity, here we introduce
$\Delta_{1}=\cos 2 \theta_{l} \left(3-11 \cos 2 \theta _{\nu}\right)-8 \sqrt{2} \sin 2 \theta _{\nu} \cos ^2\theta_{l}+12 \sin 2 \theta_{l} \sin \theta _{\nu} \left(\sqrt{2} \sin \theta _{\nu}+2 \cos \theta _{\nu}\right)$ and $\Delta_{2}=\cos 2 \theta_{l} \left(11 \cos 2 \theta _{\nu}-3\right)-8 \sqrt{2} \sin 2 \theta _{\nu} \sin ^2\theta_{l}-12 \sin 2 \theta_{l} \sin \theta _{\nu} \left(\sqrt{2} \sin \theta _{\nu}+2 \cos \theta _{\nu}\right)$.}
\end{table}

\begin{table}[hptb!]
\footnotesize
\begin{center}
\begin{tabular}{|c|c|c|}
\hline\hline

\multicolumn{3}{|c|}{$(G_{l}, G_{\nu})=(Z^{ST^2SU}_2, Z^{S}_2)$}\\
\hline\hline

$(X_{l}, X_{\nu}, P_{l}, P_{\nu})$ & $(T^2, 1, P_{12}, P_{13})$ & $(T^2, SU, P_{12}, P_{13})$ \\\hline

$\sin^2\theta_{13}$ & $\frac{1+\cos2\theta_{l}}{4} $ & $\frac{1+\cos2\theta_{l}}{4}  $ \\
\hline
$\sin^2\theta_{12}$ & $\frac{1}{2}-\frac{\sqrt{2} \sin 2 \theta _l \cos \left(2 \theta _{\nu}+\frac{\pi }{6}\right)}{3-\cos 2 \theta _l} $ & $\frac{1}{2}+\frac{\sin 2 \theta _{\nu} \left(1-3 \cos 2 \theta _l\right)}{2(3-\cos 2 \theta _l)}  $ \\
\hline
$\sin^2\theta_{23}$ & $\frac{2}{3-\cos 2 \theta _l} $ & $\frac{2}{3-\cos 2 \theta _l}  $ \\
\hline
$J_{CP}$ & $-\frac{\sin 2 \theta _l \sin \left(2 \theta _{\nu}+\frac{\pi }{6}\right)}{8 \sqrt{2}}  $ & $ -\frac{\sin 2 \theta _l \cos 2 \theta _{\nu}}{8 \sqrt{2}} $ \\
\hline
$I_{1}$ & $\frac{\left(2 \sin 2 \theta _l-3 \sin 4 \theta _l\right) \sin \left(2 \theta _{\nu}+\frac{\pi }{6}\right)}{16 \sqrt{2}}  $ & $\frac{\left(3 \sin 4 \theta _l-2 \sin 2 \theta _l\right) \cos 2 \theta _{\nu}}{16 \sqrt{2}} $ \\
\hline
$I_{2}$ & $\frac{\cos ^2\theta _l \left(2 \sqrt{2} \sin 2 \theta _l-\cos(2 \theta _{\nu}+\frac{\pi }{6}) \left(\cos 2 \theta _l-3\right)\right)}{16} $ & $ -\frac{\sin \theta _l \cos ^3\theta _l \cos 2 \theta _{\nu}}{2 \sqrt{2}} $ \\ \hline\hline

$(X_{l}, X_{\nu}, P_{l}, P_{\nu})$ & $(T^2, TST^2U, P_{12}, P_{13})$ & $(U, 1, P_{12}, P_{13})$ \\\hline

$\sin^2\theta_{13}$ & $\frac{1+\cos2\theta_{l}}{4} $ & $\frac{1-\cos2\theta_{l}}{4} $ \\
\hline
$\sin^2\theta_{12}$ & $\frac{\left(\sqrt{2} \sin \theta _{\nu} \cos \theta _l-2 \sin \theta _l \cos \theta _{\nu}\right){}^2}{3-\cos 2 \theta _l} $ & $\frac{1}{2}-\frac{\sqrt{2} \sin 2 \theta _l \sin \left(2 \theta _{\nu}+\frac{\pi }{6}\right)}{3+\cos 2 \theta _l} $ \\
\hline
$\sin^2\theta_{23}$ & $\frac{2}{3-\cos 2 \theta _l} $ & $ \frac{2}{3+\cos 2 \theta _l} $ \\
\hline
$J_{CP}$ & $0 $ & $\frac{\sin 2 \theta _l \cos \left(2 \theta _{\nu}+\frac{\pi }{6}\right)}{8 \sqrt{2}} $ \\
\hline
$I_{1}$ & $0 $ & $-\frac{\left(2 \sin 2 \theta _l+3 \sin 4 \theta _l\right) \cos \left(2 \theta _{\nu}+\frac{\pi }{6}\right)}{16 \sqrt{2}} $ \\
\hline
$I_{2}$ & $0 $ & $-\frac{\sin ^2\theta _l \left(3 \cos 2 \theta _l+1\right) \cos \left(2 \theta _{\nu}+\frac{\pi }{6}\right)}{16}  $ \\
\hline\hline

$(X_{l}, X_{\nu}, P_{l}, P_{\nu})$ & $(T^2, 1, P_{13}, P_{13})$ & $(T^2, SU, P_{13}, P_{13})$ \\\hline

$\sin^2\theta_{13}$ & $\frac{1-\cos2\theta_{l}}{4} $ & $\frac{1-\cos2\theta_{l}}{4} $ \\
\hline
$\sin^2\theta_{12}$ & $\frac{1}{2}+\frac{\sqrt{2} \sin 2 \theta _l \cos \left(2 \theta _{\nu}+\frac{\pi }{6}\right)}{3+\cos 2 \theta _l} $ & $\frac{1}{2}+\frac{\sin 2 \theta _{\nu} \left(1+3 \cos 2 \theta _l\right)}{2(3+\cos 2 \theta _l)} $ \\
\hline
$\sin^2\theta_{23}$ & $1-\frac{2}{3+\cos 2 \theta _l} $ & $1-\frac{2}{3+\cos 2 \theta _l} $ \\
\hline
$J_{CP}$ & $-\frac{\sin 2 \theta _l \sin \left(2 \theta _{\nu}+\frac{\pi }{6}\right)}{8 \sqrt{2}} $ & $-\frac{\sin 2 \theta _l \cos 2 \theta _{\nu}}{8 \sqrt{2}} $ \\
\hline
$I_{1}$ & $-\frac{\left(2 \sin 2 \theta _l+3 \sin 4 \theta _l\right) \sin \left(2 \theta _{\nu}+\frac{\pi }{6}\right)}{16 \sqrt{2}} $ & $\frac{\left(3 \sin 4 \theta _l+2 \sin 2 \theta _l\right) \cos 2 \theta _{\nu}}{16 \sqrt{2}}$ \\
\hline
$I_{2}$ & $\frac{\sin ^2\theta _l \left(\left(\cos 2 \theta _l+3\right) \cos \left(2 \theta _{\nu}+\frac{\pi }{6}\right)-2 \sqrt{2} \sin 2 \theta _l\right)}{16} $ & $\frac{\cos \theta _l \sin ^3\theta _l \cos 2 \theta _{\nu}}{2 \sqrt{2}} $ \\
\hline\hline

$(X_{l}, X_{\nu}, P_{l}, P_{\nu})$ & $(T^2, TST^2U, P_{13}, P_{13})$ & $(U, 1, P_{13}, P_{13})$ \\\hline

$\sin^2\theta_{13}$ & $\frac{1-\cos2\theta_{l}}{4} $ & $\frac{1+\cos2\theta_{l}}{4}$ \\
\hline
$\sin^2\theta_{12}$ & $\frac{\left(\sqrt{2} \sin \theta _l \sin \theta _{\nu}+2 \cos \theta _l \cos \theta _{\nu}\right){}^2}{3+\cos 2 \theta _l}$ & $\frac{1}{2}+\frac{ \sqrt{2} \sin 2 \theta _l \sin \left(2 \theta _{\nu}+\frac{\pi }{6}\right)}{3-\cos 2 \theta _l}$ \\
\hline
$\sin^2\theta_{23}$ & $1-\frac{2}{3+\cos 2 \theta _l}$ & $1-\frac{2}{3-\cos 2 \theta _l}$ \\
\hline
$J_{CP}$ & $0 $ & $\frac{\sin 2 \theta _l \cos \left(2 \theta _{\nu}+\frac{\pi }{6}\right)}{8 \sqrt{2}}$ \\
\hline
$I_{1}$ & $0 $ & $\frac{\left(2 \sin 2 \theta _l-3 \sin 4 \theta _l\right) \cos \left(2 \theta _{\nu}+\frac{\pi }{6}\right)}{16 \sqrt{2}} $ \\
\hline
$I_{2}$ & $0 $ & $\frac{\cos ^2\theta _l \left(3 \cos 2 \theta _l-1\right) \cos \left(2 \theta _{\nu}+\frac{\pi }{6}\right)}{16}$ \\
\hline\hline

$(X_{l}, X_{\nu}, P_{l}, P_{\nu})$ & \multicolumn{2}{c|}{$(T^2, TST^2U, P_{13}, P_{12})$}   \\ \hline

$\sin^2\theta_{13}$ & \multicolumn{2}{c|}{$\frac{\left(\sqrt{2} \sin \theta _l \cos \theta _{\nu}-2 \sin \theta _{\nu} \cos \theta _l\right){}^2}{4} $}\\
\hline
$\sin^2\theta_{12}$ &\multicolumn{2}{c|}{$\frac{4 \sin ^2\theta _l}{2 \sqrt{2} \sin 2 \theta _l \sin 2 \theta _{\nu}+\cos 2 \theta _{\nu}+\cos 2 \theta _l \left(3 \cos 2 \theta _{\nu}-1\right)+5} $}\\
\hline
$\sin^2\theta_{23}$ &\multicolumn{2}{c|}{$1-\frac{4 \cos ^2\theta _{\nu}}{2 \sqrt{2} \sin 2 \theta _l \sin 2 \theta _{\nu}+\cos 2 \theta _{\nu}+\cos 2 \theta _l\left(3 \cos 2 \theta _{\nu}-1\right)+5}$}\\
\hline
$J_{CP}$ &\multicolumn{2}{c|}{$0$}\\
\hline
$I_{1}$ &\multicolumn{2}{c|}{$0$}\\
\hline
$I_{2}$ &\multicolumn{2}{c|}{$0$}\\
\hline\hline
\end{tabular}
\end{center}
\caption{\label{tab:Uac} Results of the mixing parameters for the independent and viable cases with $(G_{l}, G_{\nu})=(Z^{ST^2SU}_2, Z^{S}_2)$. Note that the factors $(-1)^{k_1}$ and $(-1)^{k_2}$ are omitted in the expressions of $I_1$ and $I_2$ respectively, and they arise from the $CP$ parity matrix $Q_{\nu}=\text{diag}(1, i^{k_1}, i^{k_2})$. }
\end{table}

\begin{table}[hptb!]
\footnotesize
\begin{center}
\begin{tabular}{|c|c|c|c|c|c|c|c|c|}
\hline\hline
\multirow{2}{*}{$(X_{l}, X_{\nu}, P_{l}, P_{\nu})$} & \multirow{2}{*}{$\chi^{2}_{\mathrm{min}}$}  & \multirow{2}{*}{$(\theta_{l},\theta _{\nu})_{\mathrm{bf}}/\pi$} & \multirow{2}{*}{$\sin^{2}\theta_{13}$} & \multirow{2}{*}{$\sin^{2}\theta_{12}$} & \multirow{2}{*}{$\sin^{2}\theta_{23}$} & \multirow{2}{*}{$\delta_{CP}/\pi$} & $\alpha_{21}/\pi$ & $\alpha_{31}/\pi$ \\
& & & & & & &(mod 1) &(mod 1)\\
\hline
\multirow{2}{*}{$(U,T, P_{12},1)$} & \multirow{2}{*}{0.6354} &(0.299,0.120) & \multirow{2}{*}{0.022} & \multirow{2}{*}{0.311} & \multirow{2}{*}{0.437}  & \multirow{2}{*}{0} & \multirow{2}{*}{0} & \multirow{2}{*}{0}\\
\cline{3-3}
 & & (0.093,0.664) & & & & & & \\
\hline
\multirow{2}{*}{$(U,T, P_{13},1)$} & \multirow{2}{*}{4.6454} & (0.803,0.114) & \multirow{2}{*}{0.022} & \multirow{2}{*}{0.317} & \multirow{2}{*}{0.551}  & \multirow{2}{*}{1} & \multirow{2}{*}{0} & \multirow{2}{*}{0}\\
\cline{3-3}
 & & (0.589,0.669) & & & & & & \\
\hline
\multirow{2}{*}{$(U,T, P_{12}, P_{12})$} & \multirow{2}{*}{3.3522} & (0.477,0.072)& \multirow{2}{*}{0.022} & \multirow{2}{*}{0.308} & \multirow{2}{*}{0.546}  & \multirow{2}{*}{0} & \multirow{2}{*}{0} & \multirow{2}{*}{0}\\
\cline{3-3}
 & & (0.915,0.711) & & & & & & \\
\hline
\multirow{2}{*}{$(U,T, P_{13}, P_{12})$} & \multirow{2}{*}{0.0010} & (0.979,0.071)& \multirow{2}{*}{0.022} & \multirow{2}{*}{0.304} &\multirow{2}{*}{0.451}  & \multirow{2}{*}{1} & \multirow{2}{*}{0} & \multirow{2}{*}{0}\\
\cline{3-3}
 & & (0.413,0.713) & & & & & & \\
\hline
\multirow{2}{*}{$(U,STS, P_{12},1)$} & \multirow{2}{*}{17.3268} & (0.071,0) & \multirow{2}{*}{0.024} & \multirow{2}{*}{0.344} & \multirow{2}{*}{0.512}  & \multirow{2}{*}{0} & \multirow{2}{*}{0} & \multirow{2}{*}{0}\\
\cline{3-3}
 & & (0.071,1) & & & & & & \\
\hline
\multirow{2}{*}{$(U,STS, P_{13}, 1)$} & \multirow{2}{*}{16.4425} & (0.571,0) & \multirow{2}{*}{0.024} & \multirow{2}{*}{0.344} & \multirow{2}{*}{0.488}  & \multirow{2}{*}{1} & \multirow{2}{*}{0} & \multirow{2}{*}{0}\\
\cline{3-3}
 & & (0.571,1) & & & & & & \\
\hline
\multirow{2}{*}{$(U,STS, P_{12}, P_{12})$} & \multirow{2}{*}{17.3286} & (0.929,0) & \multirow{2}{*}{0.024} & \multirow{2}{*}{0.344} & \multirow{2}{*}{0.512}  & \multirow{2}{*}{0} & \multirow{2}{*}{0} & \multirow{2}{*}{0}\\
\cline{3-3}
 & & (0.929,1) & & & & & & \\
\hline
\multirow{2}{*}{$(U,STS, P_{13}, P_{12})$} & \multirow{2}{*}{16.4425} & (0.429,0) & \multirow{2}{*}{0.024} & \multirow{2}{*}{0.344} & \multirow{2}{*}{0.488}  & \multirow{2}{*}{1} & \multirow{2}{*}{0} & \multirow{2}{*}{0}\\
\cline{3-3}
 & & (0.429,1) & & & & & & \\
\hline
\multirow{2}{*}{$(T^2, T, P_{12}, P_{12})$} & \multirow{2}{*}{25.7405} & (0.075,0.024) & \multirow{2}{*}{0.024} & \multirow{2}{*}{0.270} & \multirow{2}{*}{0.644}  & \multirow{2}{*}{1.569} & \multirow{2}{*}{0.728} & \multirow{2}{*}{0.808} \\
\cline{3-3}
 & & (0.925,0.024) & & & & & & \\
\hline\hline
\end{tabular}
\caption{\label{tab:Uab_chi2_NH}
The results of the $\chi^{2}$ analysis for the independent and viable cases with $(G_{l}, G_{\nu})=(Z_{2}^{ST^2SU}, Z_{2}^{TU})$ under the
assumption of NH neutrino mass spectrum. $\chi^2_{\mathrm{min}}$ is the global minimum of $\chi^2$ at the best fitting values $(\theta_{l}, \theta_{\nu})_{\mathrm{bf}}$ for $\theta_{l}$ and $\theta_{\nu}$. We give the values of the mixing angles $\sin^2\theta_{13}$, $\sin^2\theta_{12}$, $\sin^2\theta_{23}$ and the $CP$ violating phases $\delta_{CP}$, $\alpha_{21}$ and $\alpha_{31}$ for $(\theta_{l}, \theta_{\nu})=(\theta_{l}, \theta_{\nu})_{\mathrm{bf}}$.}
\end{center}
\end{table}

\begin{table}[hptb!]
\footnotesize
\begin{center}
\begin{tabular}{|c|c|c|c|c|c|c|c|c|}
\hline\hline
\multirow{2}{*}{$(X_{l}, X_{\nu}, P_{l}, P_{\nu})$} & \multirow{2}{*}{$\chi^{2}_{\mathrm{min}}$} & \multirow{2}{*}{$(\theta_{l},\theta _{\nu})_{\mathrm{bf}}/\pi$} & \multirow{2}{*}{$\sin^{2}\theta_{13}$} & \multirow{2}{*}{$\sin^{2}\theta_{12}$} & \multirow{2}{*}{$\sin^{2}\theta_{23}$}  & \multirow{2}{*}{$\delta_{CP}/\pi$} & $\alpha_{21}/\pi$ & $\alpha_{31}/\pi$ \\
& & & & & & &(mod 1) &(mod 1)\\
\hline
\multirow{2}{*}{$(U,T, P_{12},1)$} & \multirow{2}{*}{11.7471} & (0.311,0.140) & \multirow{2}{*}{0.023} & \multirow{2}{*}{0.328} & \multirow{2}{*}{0.474}  & \multirow{2}{*}{0} & \multirow{2}{*}{0} & \multirow{2}{*}{0}\\
\cline{3-3}
 & & (0.081,0.680) & & & & & & \\
\hline
\multirow{2}{*}{$(U,T, P_{13},1)$ } & \multirow{2}{*}{0.0011} & (0.794,0.126) & \multirow{2}{*}{0.022} & \multirow{2}{*}{0.304} & \multirow{2}{*}{0.580}  & \multirow{2}{*}{1} & \multirow{2}{*}{0} & \multirow{2}{*}{0}\\
\cline{3-3}
 & & (0.597,0.657) & & & & & & \\
\hline
\multirow{2}{*}{$(U,T, P_{12}, P_{12})$} & \multirow{2}{*}{0.6316} & (0.480,0.070)& \multirow{2}{*}{0.022} & \multirow{2}{*}{0.302} & \multirow{2}{*}{0.550}  & \multirow{2}{*}{0} & \multirow{2}{*}{0} & \multirow{2}{*}{0}\\
\cline{3-3}
 & & (0.912,0.714) & & & & & & \\
\hline
\multirow{2}{*}{$(U,T, P_{13}, P_{12})$} & \multirow{2}{*}{11.0992} & (0.974,0.075)& \multirow{2}{*}{0.022} & \multirow{2}{*}{0.315} &\multirow{2}{*}{0.460}  & \multirow{2}{*}{1} & \multirow{2}{*}{0} & \multirow{2}{*}{0}\\
\cline{3-3}
 & & (0.417,0.709) & & & & & & \\
\hline
\multirow{2}{*}{$(U, STS, P_{12}, 1)$} & \multirow{2}{*}{17.6652} & (0.071,0) & \multirow{2}{*}{0.024} & \multirow{2}{*}{0.344} & \multirow{2}{*}{0.512}  & \multirow{2}{*}{0} & \multirow{2}{*}{0} & \multirow{2}{*}{0}\\
\cline{3-3}
 & & (0.071,1) & & & & & & \\
\hline
\multirow{2}{*}{$(U, STS, P_{13}, 1)$} & \multirow{2}{*}{20.5458} & (0.571,0) & \multirow{2}{*}{0.024} & \multirow{2}{*}{0.344} & \multirow{2}{*}{0.488}  & \multirow{2}{*}{1} & \multirow{2}{*}{0} & \multirow{2}{*}{0}\\
\cline{3-3}
 & & (0.571,1) & & & & & & \\
\hline
\multirow{2}{*}{$(U, STS, P_{12}, P_{12})$} & \multirow{2}{*}{17.6652} & (0.929,0) & \multirow{2}{*}{0.024} & \multirow{2}{*}{0.344} & \multirow{2}{*}{0.512}  & \multirow{2}{*}{0} & \multirow{2}{*}{0} & \multirow{2}{*}{0}\\
\cline{3-3}
 & & (0.929,1) & & & & & & \\
\hline
\multirow{2}{*}{$(U, STS, P_{13}, P_{12})$} & \multirow{2}{*}{20.5458} & (0.429,0) & \multirow{2}{*}{0.024} & \multirow{2}{*}{0.344} & \multirow{2}{*}{0.488}  & \multirow{2}{*}{1} & \multirow{2}{*}{0} & \multirow{2}{*}{0}\\
\cline{3-3}
 & & (0.429,1) & & & & & & \\
\hline
\multirow{2}{*}{$(T^2, T, P_{12}, P_{12})$} & \multirow{2}{*}{17.8338} & (0.075,0.024) & \multirow{2}{*}{0.024} & \multirow{2}{*}{0.270} & \multirow{2}{*}{0.644}  & \multirow{2}{*}{1.569} & \multirow{2}{*}{0.728} & \multirow{2}{*}{0.808} \\
\cline{3-3}
 & & (0.925,0.024) & & & & & & \\
\hline\hline
\end{tabular}
\caption{\label{tab:Uab_chi2_IH}
The results of the $\chi^{2}$ analysis for the independent and viable cases with $(G_{l}, G_{\nu})=(Z_{2}^{ST^2SU}, Z_{2}^{TU})$ under the
assumption of IH neutrino mass spectrum. $\chi^2_{\mathrm{min}}$ is the global minimum of $\chi^2$ at the best fitting values $(\theta_{l}, \theta_{\nu})_{\mathrm{bf}}$ for $\theta_{l}$ and $\theta_{\nu}$. We give the values of the mixing angles $\sin^2\theta_{13}$, $\sin^2\theta_{12}$, $\sin^2\theta_{23}$ and the $CP$ violating phases $\delta_{CP}$, $\alpha_{21}$ and $\alpha_{31}$ for $(\theta_{l}, \theta_{\nu})=(\theta_{l}, \theta_{\nu})_{\mathrm{bf}}$.}
\end{center}
\end{table}

\begin{table}[hptb!]
\footnotesize
\begin{center}
\begin{tabular}{|c|c|c|c|c|c|c|c|c|}
\hline\hline
\multirow{2}{*}{$(X_{l}, X_{\nu}, P_{l}, P_{\nu})$} & \multirow{2}{*}{$\chi^{2}_{\mathrm{min}}$}  & \multirow{2}{*}{$(\theta_{l},\theta _{\nu})_{\mathrm{bf}}/\pi$} & \multirow{2}{*}{$\sin^{2}\theta_{13}$} & \multirow{2}{*}{$\sin^{2}\theta_{12}$} & \multirow{2}{*}{$\sin^{2}\theta_{23}$} & \multirow{2}{*}{$\delta_{CP}/\pi$} & $\alpha_{21}/\pi$ & $\alpha_{31}/\pi$ \\
& & & & & & &(mod 1) &(mod 1)\\
\hline
\multirow{2}{*}{$(T^2, 1, P_{12}, P_{13})$ } & \multirow{2}{*}{17.3286} & (0.571,0.417) & \multirow{2}{*}{0.024} & \multirow{2}{*}{0.344} & \multirow{2}{*}{0.512} & \multirow{2}{*}{0}  & \multirow{2}{*}{0} & \multirow{2}{*}{0.5} \\
\cline{3-3}
 & & (0.429,0.917) & & & & & & \\
\hline
\multirow{2}{*}{$(T^2, 1, P_{13}, P_{13})$} & \multirow{2}{*}{16.4425} &(0.071,0.417) & \multirow{2}{*}{0.024} & \multirow{2}{*}{0.344} & \multirow{2}{*}{0.488} & \multirow{2}{*}{1} & \multirow{2}{*}{0} & \multirow{2}{*}{0.5} \\
\cline{3-3}
 & & (0.929,0.917) & & & & & & \\
\hline
\multirow{4}{*}{$(T^2, SU, P_{12}, P_{13})$} & \multirow{4}{*}{1.2935} &(0.433,0.567) & \multirow{4}{*}{0.022} & \multirow{4}{*}{0.304} & \multirow{4}{*}{0.511} & \multirow{2}{*}{0.542} & \multirow{2}{*}{0.208} & \multirow{2}{*}{0.146} \\
\cline{3-3}
 & & (0.567,0.933) & & & & & & \\
\cline{3-3}
\cline{7-9}
 & & (0.567,0.567) & & & &\multirow{2}{*}{1.458} &\multirow{2}{*}{0.792} &\multirow{2}{*}{0.854} \\
\cline{3-3}
& & (0.433,0.933) & & & & & & \\
\hline
\multirow{4}{*}{$(T^2, SU, P_{13}, P_{13})$} & \multirow{4}{*}{0.5023} & (0.933,0.567) & \multirow{4}{*}{0.022} & \multirow{4}{*}{0.304} & \multirow{4}{*}{0.489} &\multirow{2}{*}{1.542} &\multirow{2}{*}{0.208} &\multirow{2}{*}{0.146} \\
\cline{3-3}
 & & (0.067,0.933) & & & & & & \\
\cline{3-3}
\cline{7-9}
 & & (0.067,0.567) & & & &\multirow{2}{*}{0.458} & \multirow{2}{*}{0.792} & \multirow{2}{*}{0.854} \\
\cline{3-3}
 & & (0.933,0.933) & & & & & & \\
\hline
\multirow{4}{*}{$(T^2, TST^2U, P_{12}, P_{13})$} & \multirow{4}{*}{1.2935} &(0.433,0.266) & \multirow{4}{*}{0.022} & \multirow{4}{*}{0.304} & \multirow{4}{*}{0.511} & \multirow{2}{*}{0} & \multirow{2}{*}{0} & \multirow{2}{*}{0} \\
\cline{3-3}
& & (0.567,0.734) & & & & & & \\
\cline{3-3}
\cline{7-9}
 & & (0.433,0.638) & & & &\multirow{2}{*}{1} &\multirow{2}{*}{0} &\multirow{2}{*}{0} \\
\cline{3-3}
 & & (0.567,0.362) & & & & & & \\
\hline
\multirow{4}{*}{$(T^2, TST^2U, P_{13}, P_{13})$} & \multirow{4}{*}{0.5023} &(0.933,0.266) & \multirow{4}{*}{0.022} & \multirow{4}{*}{0.304} & \multirow{4}{*}{0.489} & \multirow{2}{*}{1} & \multirow{2}{*}{0} & \multirow{2}{*}{0} \\
\cline{3-3}
 & & (0.067,0.734) & & & & & & \\
\cline{3-3}
\cline{7-9}
 & & (0.933,0.638) & & & &\multirow{2}{*}{0} &\multirow{2}{*}{0} &\multirow{2}{*}{0} \\
\cline{3-3}
 & & (0.067,0.362) & & & & & & \\
\hline
\multirow{2}{*}{$(U, 1, P_{12}, P_{13})$} & \multirow{2}{*}{17.3286} &(0.071,0.167) & \multirow{2}{*}{0.024} & \multirow{2}{*}{0.344} & \multirow{2}{*}{0.512} & \multirow{2}{*}{0} & \multirow{2}{*}{0} & \multirow{2}{*}{0} \\
\cline{3-3}
 & & (0.929,0.667) & & & & & & \\
\hline
\multirow{2}{*}{$(U, 1, P_{13}, P_{13})$} & \multirow{2}{*}{16.4425} &(0.571,0.167) & \multirow{2}{*}{0.024} & \multirow{2}{*}{0.344} & \multirow{2}{*}{0.488} & \multirow{2}{*}{1} & \multirow{2}{*}{0} & \multirow{2}{*}{0} \\
\cline{3-3}
 & & (0.429,0.667) & & & & & & \\
\hline
\multirow{2}{*}{$(T^2, TST^2U, P_{13}, P_{12})$} & \multirow{2}{*}{10.0552} &(0.276,0.165) & \multirow{2}{*}{0.022} & \multirow{2}{*}{0.297} & \multirow{2}{*}{0.614} & \multirow{2}{*}{0} & \multirow{2}{*}{0} & \multirow{2}{*}{0} \\
\cline{3-3}
 & & (0.724,0.835) & & & & & & \\
\hline\hline
\end{tabular}
\caption{\label{tab:Uac_chi2_NH}
The results of the $\chi^{2}$ analysis for the independent and viable cases with $(G_{l}, G_{\nu})=(Z_{2}^{ST^2SU}, Z_{2}^{S})$ under the
assumption of NH neutrino mass spectrum. $\chi^2_{\mathrm{min}}$ is the global minimum of $\chi^2$ at the best fitting values $(\theta_{l}, \theta_{\nu})_{\mathrm{bf}}$ for $\theta_{l}$ and $\theta_{\nu}$. We give the values of the mixing angles $\sin^2\theta_{13}$, $\sin^2\theta_{12}$, $\sin^2\theta_{23}$ and the $CP$ violating phases $\delta_{CP}$, $\alpha_{21}$ and $\alpha_{31}$ for $(\theta_{l}, \theta_{\nu})=(\theta_{l}, \theta_{\nu})_{\mathrm{bf}}$.}
\end{center}
\end{table}

\begin{table}[hptb!]
\footnotesize
\begin{center}
\begin{tabular}{|c|c|c|c|c|c|c|c|c|}
\hline\hline
\multirow{2}{*}{$(X_{l}, X_{\nu}, P_{l}, P_{\nu})$} & \multirow{2}{*}{$\chi^{2}_{\mathrm{min}}$}  & \multirow{2}{*}{$(\theta_{l},\theta _{\nu})_{\mathrm{bf}}/\pi$} & \multirow{2}{*}{$\sin^{2}\theta_{13}$} & \multirow{2}{*}{$\sin^{2}\theta_{12}$} & \multirow{2}{*}{$\sin^{2}\theta_{23}$} & \multirow{2}{*}{$\delta_{CP}/\pi$} & $\alpha_{21}/\pi$ & $\alpha_{31}/\pi$ \\
& & & & & & &(mod 1) &(mod 1)\\
\hline
\multirow{2}{*}{$(T^2, 1, P_{12}, P_{13})$} & \multirow{2}{*}{17.6652} & (0.571,0.417) & \multirow{2}{*}{0.024} & \multirow{2}{*}{0.344} & \multirow{2}{*}{0.512} & \multirow{2}{*}{0}  & \multirow{2}{*}{0} & \multirow{2}{*}{0.5} \\
\cline{3-3}
 & & (0.429,0.917) & & & & & & \\
\hline
\multirow{2}{*}{$(T^2, 1, P_{13}, P_{13})$} & \multirow{2}{*}{20.5458} &(0.071,0.417) & \multirow{2}{*}{0.024} & \multirow{2}{*}{0.344} & \multirow{2}{*}{0.488} & \multirow{2}{*}{1} & \multirow{2}{*}{0} & \multirow{2}{*}{0.5} \\
\cline{3-3}
 & & (0.929,0.917) & & & & & & \\
\hline
\multirow{4}{*}{$(T^2, SU, P_{12}, P_{13})$} & \multirow{4}{*}{3.3575} &(0.433,0.567) & \multirow{4}{*}{0.022} & \multirow{4}{*}{0.304} & \multirow{4}{*}{0.511} & \multirow{2}{*}{0.542} & \multirow{2}{*}{0.209} & \multirow{2}{*}{0.147} \\
\cline{3-3}
 & & (0.567,0.933) & & & & & & \\
\cline{3-3}
\cline{7-9}
 & & (0.567,0.567) & & & &\multirow{2}{*}{1.458} &\multirow{2}{*}{0.791} &\multirow{2}{*}{0.853} \\
\cline{3-3}
& & (0.433,0.933) & & & & & & \\
\hline
\multirow{4}{*}{$(T^2, SU, P_{13}, P_{13})$} & \multirow{4}{*}{5.9412} &(0.933,0.567) & \multirow{4}{*}{0.022} & \multirow{4}{*}{0.304} & \multirow{4}{*}{0.489} &\multirow{2}{*}{1.542} &\multirow{2}{*}{0.208} &\multirow{2}{*}{0.146} \\
\cline{3-3}
 & & (0.067,0.933) & & & & & & \\
\cline{3-3}
\cline{7-9}
 & & (0.067,0.567) & & & &\multirow{2}{*}{0.458} &\multirow{2}{*}{0.792} &\multirow{2}{*}{0.854} \\
\cline{3-3}
 & & (0.933,0.933) & & & & & & \\
\hline
\multirow{4}{*}{$(T^2, TST^2U, P_{12}, P_{13})$} & \multirow{4}{*}{3.3575} &(0.433,0.266) & \multirow{4}{*}{0.022} & \multirow{4}{*}{0.304} & \multirow{4}{*}{0.511} & \multirow{2}{*}{0} & \multirow{2}{*}{0} & \multirow{2}{*}{0} \\
\cline{3-3}
 & & (0.567,0.734) & & & & & & \\
\cline{3-3}
\cline{7-9}
 & & (0.433,0.638) & & & &\multirow{2}{*}{1} &\multirow{2}{*}{0} &\multirow{2}{*}{0} \\
\cline{3-3}
 & & (0.567,0.362) & & & & & & \\
\hline
\multirow{4}{*}{$(T^2, TST^2U, P_{13}, P_{13})$} & \multirow{4}{*}{5.9412} & (0.933,0.266) & \multirow{4}{*}{0.022} & \multirow{4}{*}{0.304} & \multirow{4}{*}{0.489} & \multirow{2}{*}{1} & \multirow{2}{*}{0} & \multirow{2}{*}{0} \\
\cline{3-3}
 & & (0.067,0.734) & & & & & & \\
\cline{3-3}
\cline{7-9}
 & & (0.933,0.638) & & & &\multirow{2}{*}{0} &\multirow{2}{*}{0} &\multirow{2}{*}{0} \\
\cline{3-3}
 & & (0.067,0.362) & & & & & & \\
\hline
\multirow{2}{*}{$(U, 1, P_{12}, P_{13})$} & \multirow{2}{*}{17.6652} &(0.071,0.167) & \multirow{2}{*}{0.024} & \multirow{2}{*}{0.344} & \multirow{2}{*}{0.512} & \multirow{2}{*}{0} & \multirow{2}{*}{0} & \multirow{2}{*}{0} \\
\cline{3-3}
 & & (0.929,0.667) & & & & & & \\
\hline
\multirow{2}{*}{$(U, 1, P_{13}, P_{13})$} & \multirow{2}{*}{20.5458} &(0.571,0.167) & \multirow{2}{*}{0.024} & \multirow{2}{*}{0.344} & \multirow{2}{*}{0.488} & \multirow{2}{*}{1} & \multirow{2}{*}{0} & \multirow{2}{*}{0} \\
\cline{3-3}
 & & (0.429,0.667) & & & & & & \\
\hline
\multirow{2}{*}{$(T^2, TST^2U, P_{13}, P_{12})$} & \multirow{2}{*}{2.2779} &(0.276,0.165) & \multirow{2}{*}{0.022} & \multirow{2}{*}{0.297} & \multirow{2}{*}{0.614} & \multirow{2}{*}{0} & \multirow{2}{*}{0} & \multirow{2}{*}{0} \\
\cline{3-3}
 & & (0.724,0.835) & & & & & & \\
\hline\hline
\end{tabular}
\caption{\label{tab:Uac_chi2_IH}
The results of the $\chi^{2}$ analysis for the independent and viable cases with $(G_{l}, G_{\nu})=(Z_{2}^{ST^2SU}, Z_{2}^{S})$ under the
assumption of IH neutrino mass spectrum. $\chi^2_{\mathrm{min}}$ is the global minimum of $\chi^2$ at the best fitting values $(\theta_{l}, \theta_{\nu})_{\mathrm{bf}}$ for $\theta_{l}$ and $\theta_{\nu}$. We give the values of the mixing angles $\sin^2\theta_{13}$, $\sin^2\theta_{12}$, $\sin^2\theta_{23}$ and the $CP$ violating phases $\delta_{CP}$, $\alpha_{21}$ and $\alpha_{31}$ for $(\theta_{l}, \theta_{\nu})=(\theta_{l}, \theta_{\nu})_{\mathrm{bf}}$.}
\end{center}
\end{table}

\clearpage

\section{\label{sec:quark_Z2xCP}Quark flavor mixing from residual symmetry $Z_2\times CP$ in up and down quark sectors }

The Lagrangian for the quark masses and the charged current interactions reads as
\begin{equation}
\label{eq:quk01}
\mathcal{L}=-\overline{U}_{R}m_{U}U_{L}-\overline{D}_{R}m_{D}D_{L}+\frac{g}{\sqrt{2}}\overline{U}_{L}\gamma^{\mu}D_{L}W_{\mu}^{+}+h.c.\,,
\end{equation}
where $U_R=(u_R, c_R, t_R)^{T}$, $U_L=(u_L, c_L, t_L)^{T}$, $D_R=(d_R, s_R, b_R)^{T}$ and $D_L=(d_L, s_L, b_L)^{T}$ denote the three left-handed and right-handed up type quark and down type quark fields respectively. It is well-known that the mass matrices $m_{U}$ and $m_{D}$ can be diagonalized by bi-unitary transformations,
\begin{equation}
V_{u}^{\dagger}m_{U}U_{u}=\text{diag}(m_{u},m_{c},m_{t})\equiv\hat{m}_{U},\quad V_{d}^{\dagger}m_{D}U_{d}=\text{diag}(m_{d},m_{s},m_{b})\equiv\hat{m}_{D}\,.
\end{equation}
The CKM matrix is given by
\begin{equation}
\label{eq:quk06}U_{CKM}=U_{u}^{\dagger}U_{d}\,.
\end{equation}
In this section, we assume that the parent flavor and CP symmetry is broken down to $Z^{g_{u}}_2\times X_{u}$ and $Z^{g_{d}}_2\times X_{d}$ in the up and down quark sectors respectively, where $g_u$ and $g_d$ denote the generators of the $Z_2$ residual flavor symmetry groups with $g^2_u=g^2_d=1$. Similar to the lepton sector, we assign the three generations of left-handed quarks to a three-dimensional representation $\mathbf{3}$. The mass matrices $m_{U}$ and $m_{D}$ respect the residual symmetries $Z^{g_{u}}_2\times X_{u}$ and $Z^{g_{d}}_2\times X_{d}$ respectively, and they should fulfill
\begin{eqnarray}
\nonumber&&\rho^{\dagger}_{\mathbf{3}}(g_{u})m^{\dagger}_{U}m_{U}\rho_{\mathbf{3}}(g_{u})=m^{\dagger}_{U}m_{U},\quad X^{\dagger}_{u}m^{\dagger}_{U}m_{U}X_{u}=(m^{\dagger}_{U}m_{U})^{*}\,,\\
&&\rho^{\dagger}_{\mathbf{3}}(g_{d})m^{\dagger}_{D}m_{D}\rho_{\mathbf{3}}(g_{d})=m^{\dagger}_{D}m_{D},\quad X^{\dagger}_{d}m^{\dagger}_{D}m_{D}X_{d}=(m^{\dagger}_{D}m_{D})^{*}\,.
\end{eqnarray}
Following the procedures presented in section~\ref{subsec:framework_Z2xCP}, the constraints on the unitary transformations $U_u$ and $U_d$ from the postulated residual symmetries can be straightforwardly extracted. A critical step is the Takagi factorization of the residual $CP$ transformations $X_u$ and $X_{d}$ which have the following properties
\begin{eqnarray}
\label{eq:quk32} X_{u}&=&\Sigma_{u}\Sigma_{u}^{T},\qquad \Sigma_{u}^{\dagger}\rho_{\mathbf{3}}(g_{u})\Sigma_{u}=\text{diag}(1,-1, -1),\\
\label{eq:quk33}
X_{d}&=&\Sigma_{d}\Sigma_{d}^{T},\qquad \Sigma_{d}^{\dagger}\rho_{\mathbf{3}}(g_{d})\Sigma_{d}=\text{diag}(1,-1, -1)\,.
\end{eqnarray}
Then the remnant symmetries enforce the CKM mixing matrix is of the form
\begin{equation}
\label{eq:quk37} U_{CKM}=Q_{u}P_{u}S_{23}(\theta_{u})\Sigma_{u}^{\dagger}\Sigma_{d}S_{23}(\theta_{d})P_{d}Q_{d}\,,
\end{equation}
where $Q_u$ and $Q_d$ are generic diagonal matrices of phases, they can be removed by utilizing the rephasing freedom of the up and down quarks, and $P_u$ and $P_d$ are permutation matrices. Similar to the master formula of the lepton flavor mixing in Eq.~\eqref{eq:z2lep57}, the CKM mixing matrix is determined up to possible permutations of rows and columns, and it depends on two free parameters $\theta_{u}$ and $\theta_{d}$ which can take values between $0$ and $\pi$.

In the same fashion as section~\ref{subsec:criterion_Z2xCP}, we can find the condition under which the CKM matrices predicted by two distinct residual symmetries are equivalent. We generically denote the combination $U_q\equiv \Sigma^{\dagger}_u\Sigma_d$ for any two postulated residual symmetries as
\begin{equation}
\label{eq:quk58}
U_q=\left(\begin{array}{ccc}
a_{1} &~ a_{2} &~a_{3}\\
a_{4} &~a_{5} &~a_{6}\\
a_{7} &~a_{8} &~a_{9} \end{array}
\right),\quad
U^{\prime}_q=\left( \begin{array}{ccc}
b_{1} &~b_{2} &~b_{3}\\
b_{4} &~b_{5} &~b_{6}\\
b_{7} &~b_{8} &~b_{9} \end{array}
\right)\,,
\end{equation}
where $a_1$ and $b_1$ are fixed by remnant symmetries up to an overall phase. The corresponding CKM mixing matrices cannot be effectively the same one if $|a_1|\neq|b_1|$. In the following, we shall focus on the case of $a_1=b_1=0$. The results for the most general case $|a_1|=|b_1|\neq0, 1$ are summarized in appendix~\ref{sec:equivalence_quark_n0_n1_app}. After some straightforward algebra the conditions of equivalence can be described as follows.

\begin{itemize}[labelindent=-0.6em, leftmargin=1.0em]

\item{$b_{2}^{2}+b_{3}^{2}\neq0$,~ $b_{4}^{2}+b_{7}^{2}\neq 0$}

The assumed remnant symmetries would lead to the same quark mixing pattern if the following equalities are satisfied,
\begin{eqnarray}
\nonumber
&&|a^2_{2}+a^2_{3}|=|b^2_{2}+b^2_{3}|,\quad (a_{2}b_{2}+a_{3}b_{3})(a^{*}_{2}b^{*}_{3}-a^{*}_{3}b^{*}_{2})\in\mathbb{R}\,, \\
\nonumber&&|a_{4}^2+a_{7}^2|=|b_{4}^2+b_{7}^2|,\quad (a_{4}b_{4}+a_{7}b_{7})(a^{*}_{4}b^{*}_{7}-a^{*}_{7}b^{*}_{4})\in \mathbb{R} \,,\\
\label{eq:quk62}&&~t_{i}T_{j}-t_{j}T_{i}=0,\quad |t_i|=|T_i|,\quad i,j=5, 6, 8, 9\,,
\end{eqnarray}
where
\begin{eqnarray}
\nonumber
t_{5}&=&(xb_{5}+yb_{6})z+(xb_{8}+yb_{9})w,~~~t_{6}=(xb_{6}-yb_{5})z+(xb_{9}-yb_{8})w,\\
\label{eq:quk63}
t_{8}&=&(xb_{8}+yb_{9})z-(xb_{5}+yb_{6})w,~~~t_{9}=(xb_{9}-yb_{8})z-(xb_{6}-yb_{5})w\,,
\end{eqnarray}
and
\begin{eqnarray}
  \nonumber
T_{5}&=&(b_{2}^{2}+b_{3}^{2})(b_{4}^{2}+b_{7}^{2})a_{5},~~~T_{6}=(b_{2}^{2}+b_{3}^{2})(b_{4}^{2}+b_{7}^{2})a_{6},\\
  \label{eq:quk64}
T_{8}&=&(b_{2}^{2}+b_{3}^{2})(b_{4}^{2}+b_{7}^{2})a_{8},~~~T_{9}=(b_{2}^{2}+b_{3}^{2})(b_{4}^{2}+b_{7}^{2})a_{9}\,,
\end{eqnarray}
with
\begin{equation}
\label{eq:quk65}
x=a_{2}b_{2}+a_{3}b_{3},~~y=a_{2}b_{3}-a_{3}b_{2},~~z=a_{4}b_{4}+a_{7}b_{7},~~w=a_{4}b_{7}-a_{7}b_{4}\,.
\end{equation}

\item{$b_{2}^2+b_{3}^2=0$,~ $b_{4}^2+b_{7}^2\neq 0$}

The necessary and sufficient conditions for the equivalence of these two CKM matrices are found to be
\begin{eqnarray}
\nonumber&&|a_{4}^{2}+a_{7}^{2}|=|b_{4}^{2}+b_{7}^{2}|,\quad (a_{4}b_{4}+a_{7}b_{7})(a^{*}_{4}b^{*}_{7}-a^{*}_{7}b^{*}_{4})\in\mathbb{R}\,,\\
\label{eq:quk68}&&a_2b_2+a_3b_3=0,\quad t_{5}T_{8}-t_{8}T_{5}=0,\quad |t_5|=|T_5|,\quad |t_8|=|T_8|\,,
\end{eqnarray}
with
\begin{eqnarray}
\nonumber
t_{5}&=&(zb_5+wb_8)a_2^{*},\quad t_{8}=(zb_8-wb_5)a_2^{*}\,,\\
\label{eq:quk70}
T_{5}&=&(b_4^{2}+b_7^{2})b_2^{*}a_5,\quad T_{8}=(b_4^{2}+b_7^{2})b_2^{*}a_8\,.
\end{eqnarray}

\item{$b_{2}^2+b_{3}^2\neq 0$,~ $b_{4}^2+b_{7}^2=0$}

In this case, the equivalence condition is given by
\begin{eqnarray}
\nonumber&&|a_{2}^2+a_{3}^2|=|b_{2}^2+b_{3}^2|,\quad (a_{2}b_{2}+a_{3}b_{3})(a^{*}_{2}b^{*}_{3}-a^{*}_{3}b^{*}_{2})\in\mathbb{R}\,,\\
\label{eq:quk74}&&a_4b_4+a_7b_7=0,\quad t_{5}T_{6}-t_{6}T_{5}=0,\quad |t_5|=|T_5|,\quad |t_6|=|T_6|\,,
\end{eqnarray}
where
\begin{eqnarray}
\nonumber
t_{5}&=&(xb_5+yb_6)a_4^{*},\quad t_{6}=(xb_6-yb_5)a_4^{*},\\
\label{eq:quk75}
T_{5}&=&(b_2^{2}+b_3^{2})b_4^{*}a_5,\quad T_{6}=(b_2^{2}+b_3^{2})b_4^{*}a_6\,.
\end{eqnarray}

\item{$b_{2}^2+b_{3}^2=0,~ b_{4}^2+b_{7}^2=0$}

After the freedom to redefine the free parameters $\theta_{u}$ and $\theta_{d}$ is taken into account, the same quark mixing pattern would be obtained if the parameters $a_i$ and $b_i$ are subject to the following constraints,
\begin{equation}
a_2b_2+a_3b_3=0,\quad a_4b_4+a_7b_7=0\,.
\end{equation}
Notice that if the conditions of any of the above four cases are satisfied under the substitutions
\begin{eqnarray}
\nonumber &&a_{1}\rightarrow a_{1},\quad a_{2}\rightarrow a_{2},\quad a_{3}\rightarrow s_{2}a_{2},\\
\nonumber &&a_{4}\rightarrow a_{4},\quad a_{5}\rightarrow a_{5},\quad a_{6}\rightarrow s_{2}a_{6},\\
\label{eq:quk79} &&a_{7}\rightarrow s_{1}a_{7},\quad a_{8}\rightarrow s_{1}a_{8},\quad a_{9}\rightarrow s_{1}s_{2}a_{9}\,,
\end{eqnarray}
with $s_{1,2}=\pm1$, the assumed remnant symmetries would give rise to the same quark mixing.
\end{itemize}

\begin{figure}[hptb!]
\centering
\includegraphics[width=0.5\textwidth]{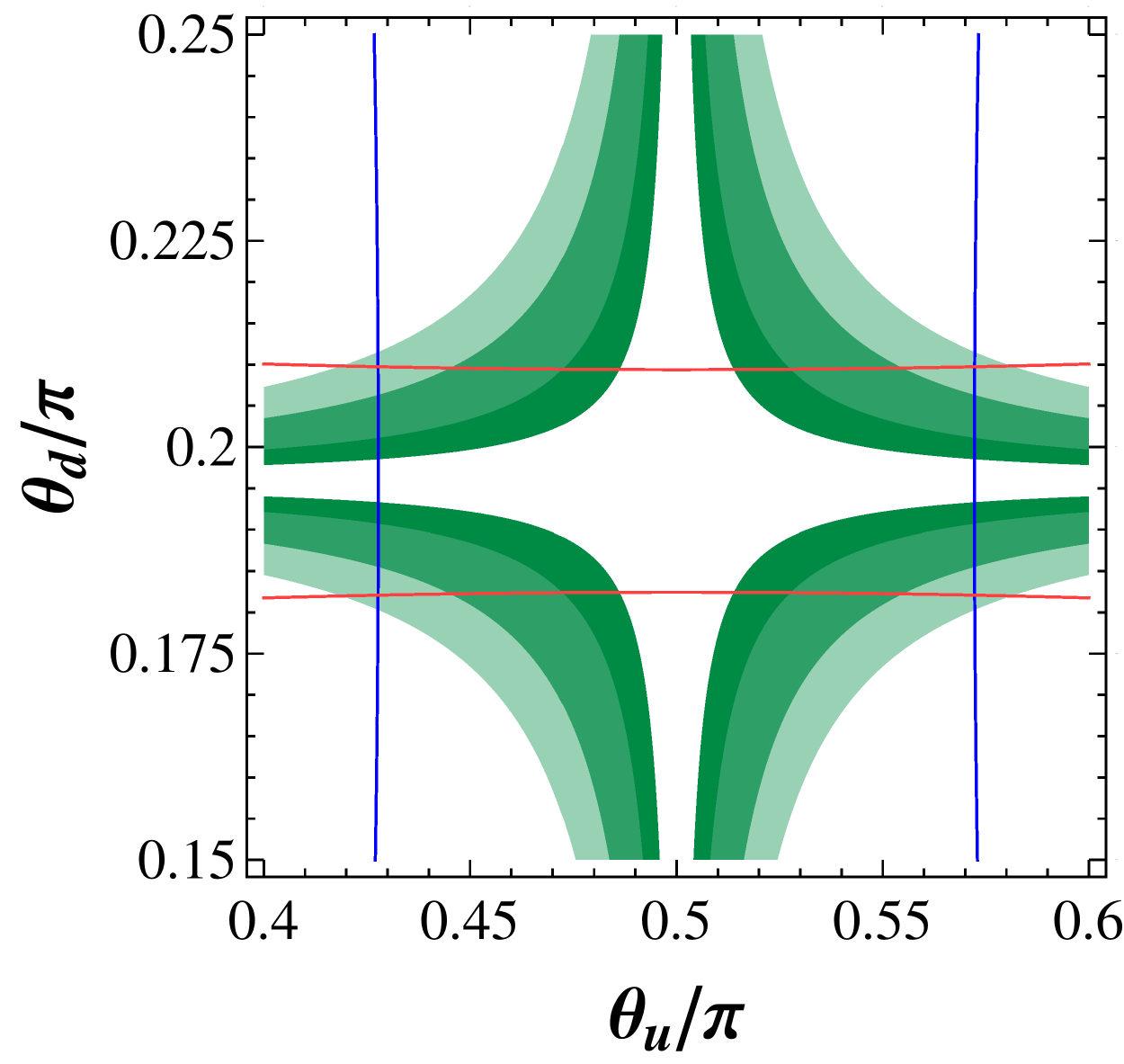}
\caption{\label{fig:contour_quark} The contour plot of $\sin\theta^{q}_{13}$, $\sin\theta^{q}_{12}$ and $\sin\theta^{q}_{23}$
in the $\theta_{d}-\theta_{u}$ plane. The blue and red lines denote the central values of $\sin\theta^{q}_{12}$ and $\sin\theta^{q}_{23}$ respectively.
The different shading areas from dark green to light green represent three interesting regions of $\sin\theta^{q}_{13}$ such as $0.5(\sin\theta^{q}_{13})_{\mathrm{bf}}\rightarrow (\sin\theta^{q}_{13})_{\mathrm{bf}}$, $(\sin\theta^{q}_{13})_{\mathrm{bf}}\rightarrow 2(\sin\theta^{q}_{13})_{\mathrm{bf}}$ and $2(\sin\theta^{q}_{13})_{\mathrm{bf}}\rightarrow 3(\sin\theta^{q}_{13})_{\mathrm{bf}}$, where we use $(\sin\theta^{q}_{13})_{\mathrm{bf}}=0.00368$.}
\end{figure}

So far the CKM mixing matrix has been measured quite accurately. The present global fit result for the magnitude of each CKM matrix element is~\cite{Bona:2006ah}
\begin{equation}
\label{eq:CKM_expe}|U_{\text{CKM}}|=\left(\begin{array}{ccc}
0.97431\pm0.00015 ~ &~   0.22512\pm0.00067  ~&~  0.00365\pm0.00012  \\
0.22497\pm0.00067  ~& ~  0.97344\pm0.00015  ~&~  0.04255\pm0.00069  \\
0.00869\pm0.00014  ~ & ~ 0.04156\pm0.00056   ~& ~ 0.999097\pm0.000024
\end{array}
\right)\,.
\end{equation}
The full fit values of three quark mixing angles read as~\cite{Bona:2006ah}
\begin{equation}
\label{eq:sin_angles_quark_best}\sin\theta^{q}_{12}=0.22497\pm0.00069,~ \sin\theta^{q}_{23}=0.04229\pm0.00057,~
\sin\theta^{q}_{13}=0.00368\pm0.00010\,.
\end{equation}
Now let us concentrate on the $S_4$ flavor symmetry group as an illustrative example. Considering all the possible residual subgroup $Z_2\times CP$ arising from the original $S_4$ and $CP$ symmetry, we find the fixed element can be $0$, $1/2$, $1/\sqrt{2}$ or $1$. According to experimental data shown in Eq.~\eqref{eq:CKM_expe}, vanishing (13) or (31) element of the CKM matrix is a good leading order approximation, since the (13) and (31) entries are very small and this tiny discrepancy could be easily resolved in an explicit model with small corrections. All the three quark mixing angles except $\theta^{q}_{13}$ can be accommodated very well for the representative remnant symmetries $G_{u}=Z^{ST^2SU}_2\times T^2$ and $G_{d}=Z^{T^2U}_2\times T^2$ in the up and down quark sectors respectively. The corresponding Takagi factorization matrices $\Sigma_u$ and $\Sigma_d$ are determined to be
\begin{equation}
\Sigma_u=\frac{1}{\sqrt{6}}\left(\begin{array}{ccc}
2   &~ 0   & ~ -\sqrt{2}\\
e^{i\pi/3}   &~ -\sqrt{3}e^{i\pi/3}   &~  \sqrt{2}e^{i\pi/3}\\
e^{-i\pi/3}   &~ \sqrt{3}e^{-i\pi/3}   &~  \sqrt{2}e^{-i\pi/3}
\end{array}
\right),\quad \Sigma_d=\frac{1}{\sqrt{2}}\left(\begin{array}{ccc}
 0  & ~0  &~ \sqrt{2} \\
 -e^{i\pi/3} &~  e^{i\pi/3}    &~  0 \\
 e^{-i\pi/3} &~  e^{-i\pi/3}   &~  0
\end{array}
\right)\,.
\end{equation}
For the permutation matrices $P_{u}=1$ and $P_{d}=P_{13}$, we find $U_{CKM}(1,3)=0$ and the CKM matrix takes the form
\begin{small}
\begin{equation}
\hskip-0.12in U_{CKM}=\frac{1}{\sqrt{3}}\left(
\begin{array}{ccc}
\sqrt{2}\cos\theta_d+\sin\theta_d &~ \cos\theta_d-\sqrt{2}\sin\theta_d &~ 0 \\
\left(\sqrt{2}\sin\theta_d-\cos\theta_d\right) \sin\theta_u &~ \left(\sqrt{2} \cos\theta_d+\sin\theta_d\right)\sin\theta_u &~ \sqrt{3}\cos\theta_u\\
\left(\sqrt{2}\sin\theta_d-\cos\theta_d\right)\cos\theta_u &~ \left(\sqrt{2} \cos\theta_d+\sin\theta_d\right)\cos\theta_u &~ -\sqrt{3} \sin\theta_u \\
\end{array}
\right)\,,
\end{equation}
\end{small}
from which we can extract the quark mixing angles as
\begin{equation}
\sin^{2}\theta^{q}_{13}=0,\quad \sin^{2}\theta^{q}_{12}=\frac{1}{3}(\cos\theta_{d}-\sqrt{2}\sin\theta_{d})^{2},\quad \sin^{2}\theta^{q}_{23}=\cos^{2}\theta_{u}\,.
\end{equation}
The best fitting values of $\theta^{q}_{12}$ and $\theta^{q}_{23}$ in Eq.~\eqref{eq:sin_angles_quark_best} can be obtained for
\begin{equation}
(\theta_{u},\theta_{d})=(0.513\pi,0.124\pi),~~ (0.513\pi,0.268\pi),~~ (0.487\pi,0.124\pi),~~(0.487\pi,0.268\pi)\,.
\end{equation}
We expect that the small mixing angle $\theta^{q}_{13}$ as well as the $CP$ violation phase can be generated by higher order contributions in a concrete model. For the values $P_{u}=P_{13}$ and $P_{d}=1$, we have $U_{CKM}(3,1)=0$. The CKM mixing matrix is given by
\begin{small}
\begin{equation}
U_{CKM}=\frac{1}{\sqrt{3}}\left(
\begin{array}{ccc}
 -\sqrt{3} \sin\theta_u &~    \left(\sqrt{2} \cos\theta_d+\sin\theta_d\right)\cos\theta_u &~ \left(\sqrt{2} \sin \theta_d-\cos\theta_d\right)\cos\theta_u \\
 \sqrt{3} \cos\theta_u &~ \left(\sqrt{2} \cos\theta_d+\sin\theta_d\right) \sin\theta_u
   &~ \left(\sqrt{2} \sin\theta_d-\cos\theta_d\right) \sin\theta_u \\
 0 &~ \cos\theta_d-\sqrt{2} \sin\theta_d &~
   \sqrt{2} \cos\theta_d+\sin\theta_d
\end{array}
\right)\,.
\end{equation}
\end{small}
The mixing angles read
\begin{eqnarray}
  \nonumber
\sin^{2}\theta^{q}_{13}&=&\frac{1}{3} \cos ^2\theta _u \left(\cos \theta _d-\sqrt{2} \sin \theta _d\right){}^2,\\
\nonumber
\sin^{2}\theta^{q}_{12}&=&\frac{4 \cos ^2\theta _u \left(\sin \theta _d+\sqrt{2} \cos \theta _d\right){}^2}{9-3\cos2\theta _u+2\cos^2\theta_u \left(2\sqrt{2}\sin2\theta_d+\cos2\theta_d\right)},\\
\label{eq:quk96}
\sin^{2}\theta^{q}_{23}&=&\frac{4 \sin ^2\theta_u \left(\cos\theta_d-\sqrt{2}\sin \theta_d\right){}^2}{9-3\cos 2 \theta _u+2 \cos ^2\theta _u \left(2 \sqrt{2} \sin 2 \theta _d+\cos 2 \theta _d\right)}\,.
\end{eqnarray}
In this case, the central values of $\theta^{q}_{12}$ and $\theta^{q}_{23}$ can be obtained for
\begin{equation}
(\theta_{u},\theta_{d})=(0.428\pi, 0.182\pi),~~(0.428\pi,0.210\pi),~~(0.572\pi,0.182\pi),~~(0.572\pi, 0.210\pi)\,.
\end{equation}
We display the contour plot of $\sin\theta^{q}_{13}$ $\sin\theta^{q}_{12}$ and $\sin\theta^{q}_{23}$
in the plane $\theta_{d}$ versus $\theta_{u}$ in figure~\ref{fig:contour_quark}.
If the best fit values of $\theta^{q}_{12}$ and $\theta^{q}_{23}$ are reproduced, we see that $\sin\theta^q_{13}$ would be approximately three times as large as its measured value. However, accordance with the experimental data could be easily achieved in a concrete model after subleading corrections are taken into account.

\section{\label{sec:single_res_CP} Lepton flavor mixing from single residual CP transformation in the neutrino sector}

In this section we shall instead consider the scenario in which the residual symmetry of the charged lepton sector is an abelian subgroup and the neutrino mass matrix $m_{\nu}$ is invariant under a single residual $CP$ transformation $X_{\nu}$. In order to avoid degenerate neutrino masses, $X_{\nu}$ should be a symmetric unitary matrix with $X_{\nu}=X^{T}_{\nu}$ and $X_{\nu}X^{*}_{\nu}=1$~\cite{Chen:2014wxa,Chen:2015nha}. As a result, $m_{\nu}$ is invariant under the action of $X_{\nu}$,
\begin{equation}
X_{\nu}^{T}m_{\nu}X_{\nu}=m_{\nu}^{*}\,.
\end{equation}
Without reconstructing the neutrino mass matrix, from this equation we can derive that the unitary transformation $U_{\nu}$ which is the a diagonalization matrix of $m_{\nu}$
with $U_{\nu}^{T}m_{\nu}U_{\nu}=\text{diag}(m_{1},m_{2},m_{3})$, is subject to the following constraint~\cite{Chen:2014wxa,Chen:2015nha},
\begin{equation}
U^{\dagger}_{\nu}X_{\nu}U^{*}_{\nu}=\text{diag}(\pm1, \pm1, \pm1)\equiv Q^2_{\nu}\,,
\end{equation}
where $Q_{\nu}$ is a diagonal matrix with non-vanishing entries $\pm1$ and $\pm i$ to make the light neutrino masses positive definite, and it can be parameterized as Eq~\eqref{eq:Qnu_para}. Performing Takagi factorization $X_{\nu}=\Sigma_{\nu}\Sigma^T_{\nu}$ where
$\Sigma_{\nu}$ is unitary, we obtain
\begin{equation}
\left(\Sigma^T_{\nu}U^{*}_{\nu}Q_{\nu}\right)^{T}\left(\Sigma^T_{\nu}U^{*}_{\nu}Q_{\nu}\right)=1\,.
\end{equation}
Therefore $\Sigma^T_{\nu}U^{*}_{\nu}Q_{\nu}$ is a real orthogonal matrix
\begin{equation}
\Sigma^T_{\nu}U^{*}_{\nu}Q_{\nu}=O_{3\times3}\,,
\end{equation}
where $O_{3\times 3}$ can be parameterized as
\begin{eqnarray}
\nonumber O_{3\times3}= \left( \begin{array}{ccc} 1 & 0 & 0 \\ 0& \cos\theta_{1} & \sin\theta_{1}\\ 0 &  -\sin\theta_{1} & \cos\theta_{1} \end{array} \right) \left( \begin{array}{ccc} \cos\theta_{2} & 0 & \sin\theta_{2} \\ 0 & 1& 0 \\ -\sin\theta_{2} & 0 & \cos\theta_{2} \end{array} \right) \left( \begin{array}{ccc} \cos\theta_{3} & \sin\theta_{3} & 0 \\ -\sin\theta_{3} & \cos\theta_{3} & 0 \\ 0 & 0 & 1 \end{array} \right)\,,
\end{eqnarray}
where the fundamental interval of the real parameters $\theta_{1, 2, 3}$ is $[0, \pi)$. Thus the neutrino mixing matrix is determined to be of the form~\cite{Chen:2015siy,Chen:2016ica}
\begin{equation}
U_{\nu}=\Sigma_{\nu} O_{3\times3}Q_{\nu}\,.
\end{equation}
The flavor symmetry is assumed to be broken to an abelian subgroup $G_{l}$ in the charged lepton sector, and the generator of $G_{l}$ is denoted as $g_{l}$~\footnote{Here we assume $G_{l}$ is generated by a single generator, and the generalization to the case in which $G_l$ has several generators is straightforward.}. The charged lepton mass matrix $m_l$ would fulfill
\begin{equation}
\rho^{\dagger}_{\mathbf{3}}(g_{l})m^{\dagger}_{l}m_{l}\rho_{\mathbf{3}}(g_{l})=m^{\dagger}_{l}m_{l}\,.
\end{equation}
Thus we find that the unitary transformation $U_{l}$ which diagonalizes $m^{\dagger}_{l}m_{l}$ is constrained to satisfy
\begin{equation}
U^{\dagger}_{l}\rho_{\mathbf{3}}(g_{l})U_{l}=\rho^{diag}_{\mathbf{3}}(g_{l})\,,
\end{equation}
where $\rho^{diag}_{\mathbf{3}}(g_{l})$ is a diagonal phase matrix. That is to say, the charged lepton mixing matrix $U_{l}$ can be obtained by diagonalizing the representation matrix of the generator $g_{l}$ without resorting to the mass matrix. Here we assume that the residual symmetry $G_{l}$ can distinguish among the three charged leptons, consequently $U_{l}$ is uniquely  determined up to permutations and phases of its column vectors. As a result, the PMNS mixing matrix is found to be of the form
\begin{equation}
\label{eq:PMNS_master_single}U_{PMNS}=Q_{l}P_{l}U_{l}^{\dagger}\Sigma_{\nu} O_{3\times3}Q_{\nu}\,,
\end{equation}
where $P_{l}$ is an arbitrary three dimensional permutation matrix, $Q_{l}$ is a diagonal unitary matrix and it can be absorbed into the charged lepton fields. If two pairs of residual subgroups $\{G_l, X_\nu \}$ and $\{G'_l, X'_\nu\}$ are related by a similarity transformation $\Omega$,
\begin{equation}
\label{eq:lep45}
\rho_{\mathbf{3}}(g_l')=\Omega \rho_{\mathbf{3}}(g_l)\Omega^\dagger,~~~X_\nu'=\Omega X_\nu\Omega^T\,,
\end{equation}
both pairs would lead to the same result for $U_{PMNS}$. The reason is because that the Takagi factorization of $X'_{\nu}$ is $\Omega\Sigma_{\nu}$ and $\rho_{\mathbf{3}}(g_l')$ is diagonalized by $\Omega U_{l}$.

\subsection{\label{subsec:equivalence_single}Condition for the equivalence of two mixing patterns }

Let us assume two different residual symmetries $\{G_{l}, X_{\nu}\}$ and $\{G'_{l}, X'_{\nu}\}$, accordingly the PMNS matrices are predicted to be
\begin{eqnarray}
\label{eq:lep48_v2}
U_{PMNS}&=&Q_{l}P_{l}U_{l}^{\dagger}\Sigma_{\nu} O_{3\times3}Q_{\nu}\,,\\
\label{eq:lep49_v2}
U_{PMNS}'&=&Q_{l}'P'_{l}U_{l}'^{\dagger}\Sigma'_{\nu} O_{3\times3}'Q_{\nu}'\,.
\end{eqnarray}
For any given value of the real orthogonal matrix $O_{3\times 3}$, if one can always find a corresponding orthogonal matrix $O'_{3\times 3}$ as well as $Q_{l}'$, $P'_{l}$ and $Q_{\nu}'$,
such that the equality
\begin{equation}
\label{eq:PMNS_eq_single}U_{PMNS}=U'_{PMNS}
\end{equation}
is fulfilled, then these two mixing patterns would be essentially the same. From Eq.~\eqref{eq:PMNS_eq_single} we can obtain the condition
\begin{equation}
\label{eq:sim_con_single}UO_{3\times3}=Q_{L}P_{L}U'O_{3\times3}'Q_{N}\,,
\end{equation}
with
\begin{equation}
U\equiv U_{l}^{\dagger}\Sigma_{\nu},~~U'\equiv U_{l}'^{\dagger}\Sigma'_{\nu}\,,~~ P_{L}\equiv P^{T}_{l}P'_{l},~~Q_{L}\equiv P^{T}_{l}Q_{l}^{\dagger}Q_{l}'P_{l}\,,~~ Q_{N}\equiv Q_{\nu}'Q_{\nu}^{\dagger}\,.
\end{equation}
Both sides of Eq.~\eqref{eq:sim_con_single} multiply with their transpose, we have
\begin{equation}
\label{eq:lep54_v2}
UU^{T}=Q_{L}P_{L}U'O_{3\times3}'Q_{N}^{2}O_{3\times3}'^{T}U'^{T}P^{T}_{L}Q_{L}\,.
\end{equation}
Notice that $Q_{N}$ is a diagonal matrix with entries $\pm1$ and $\pm i$, and Eq.~\eqref{eq:lep54_v2} is satisfied for a generic orthogonal matrix  $O'_{3\times3}$. This requires $Q_{N}^{2}=\pm\text{diag}(1, 1, 1)$, and $Q_{N}^{2}$ can be set to be an identity matrix by choosing suitable values of $Q_{\nu}$ and $Q_{\nu}'$. Thus the condition for the equivalence of the two mixing patterns in this scenario simplifies into
\begin{equation}
\label{eq:lep55_v2}
UU^{T}=Q_{L}P_{L}U'U'^{T}P^{T}_{L}Q_{L}\,.
\end{equation}
Inversely, if we can find a permutation matrix $P_{L}$ and a phase matrix $Q_{L}$ such that Eq.~\eqref{eq:lep55_v2} is fulfilled, the postulated residual symmetries would lead to the same lepton mixing pattern.

\subsection{Examples in $S_4$ and CP }

In this section, we shall analyze the lepton mixing patterns which arise from the breaking of the flavor group $S_4$ and $CP$ symmetry to an abelian subgroup $G_{l}$ in the charged lepton sector and to a residual $CP$ $X_{\nu}$ in the neutrino sector. We shall consider all possibilities for $G_{l}$, i.e., $G_{l}=Z_3, Z_4, K_4$ and all possible residual $CP$ transformation $X_{\nu}$ which should be a unitary symmetric matrix,
\begin{equation}
\label{eq:lep65}
X_{\nu}= \{1,S,T,T^2,STS,ST^2S,U,SU,TST^2U,T^2STU\}\,,
\end{equation}
where we do not distinguish between the abstract elements of the $S_4$ group and their representation matrices in $\mathbf{3}$ for simplicity of notation. In fact it is not necessary to study the mixing patterns comprehensively for all possible combinations of $G_{l}$ and $X_{\nu}$. By applying the general equivalence criterion in Eq.~\eqref{eq:lep55_v2}, we find there are only five independent cases with $\left(G_{l},X_{\nu}\right)=\left(Z_{3}^{T}, \mathbf{1}\right)$, $\left(Z_{3}^{T}, S\right)$, $\left(Z_{3}^{T}, U\right)$, $\left(Z_{3}^{T}, SU\right)$ and $\big(K_{4}^{(S,U)},T\big)$. In the following, we take into account all possible row permutations of the mixing matrix in each case, the predictions for lepton mixing angles and $CP$ violation phases as well as neutrinoless double decay would be investigated.

\begin{itemize}[labelindent=-0.7em, leftmargin=1.2em]

\item{$G_{l}=Z_{3}^{T},X_{\nu}=1$}

In this case, the unitary matrices $U_{l}$ and $\Sigma_{\nu}$ are given by,
\begin{equation}
U_{l}=\left(\begin{array}{ccc} 1~  & ~ 0~  & ~ 0~ \\ 0  &  1  &  0 \\0  &  0  &  1 \end{array}\right),\qquad \Sigma_{\nu}=\left( \begin{array}{ccc} 1~  & ~ 0~  & ~ 0~ \\ 0  &  1  &  0 \\0  &  0  &  1 \end{array} \right)\,.
\end{equation}
Moreover, we find that the six row permutations of the mixing matrix lead to the same mixing pattern. Consequently we shall choose $P_l=1$ without loss of generality, and thus the PMNS matrix is of the form,
\begin{small}
\begin{eqnarray}
\nonumber&&\hskip-0.1in U_{PMNS}=P_{l}U_{l}^{\dagger}\Sigma O_{3\times3}Q_{\nu}\\
\nonumber&&\hskip-0.3in\qquad=\left( \begin{array}{ccc} \cos\theta_{2}\cos\theta_{3} & \cos\theta_{2}\sin\theta_{3} & \sin\theta_{2} \\ -\cos\theta_{3}\sin\theta_{1}\sin\theta_{2}-\cos\theta_{1}\sin\theta_{3} & \cos\theta_{1}\cos\theta_{3}-\sin\theta_{1}\sin\theta{2}\sin\theta{3} & \cos\theta_{2}\sin\theta_{1}\\ -\cos\theta_{1}\cos\theta_{3}\sin\theta_{2}+\sin\theta_{1}\sin\theta_{3} & -\cos\theta_{3}\sin\theta_{1}-\cos\theta_{1}\sin\theta_{2}\sin\theta_{3} & \cos\theta_{1}\cos\theta_{2} \end{array} \right)Q_{\nu}\,,
\end{eqnarray}
\end{small}
where the unphysical phase matrix $Q_{l}$ on the far left is omitted. The mixing angles and $CP$ violation phases can be read off as
\begin{equation}
\begin{split}
&\sin^{2}\theta_{13}=\sin^{2}\theta_{2},~~ \sin^{2}\theta_{12}=\sin^{2}\theta_{3},~~ \sin^{2}\theta_{23}=\sin^{2}\theta_{1},\\
&\sin\delta_{CP}=\sin\alpha_{21}=\sin\alpha_{31}=0\,.
\end{split}
\end{equation}
We see that all the three $CP$ phases are predicted to be trivial, the measured values of the lepton mixing angles can be reproduced for certain values of the parameters $\theta_{1, 2, 3}$.

\item{$G_{l}=Z_{3}^{T},X_{\nu}=S$}

This case differs from the previous one in the value of the residual $CP$ transformation $X_{\nu}$, and we have
\begin{equation}
U_{l}=\left( \begin{array}{ccc}
~1~ & ~0~ & ~0~\\
0   &  1  &  0 \\
0   &  0  &  1 \\
\end{array} \right),\qquad
\Sigma_{\nu}=\frac{1}{\sqrt{6}}\left( \begin{array}{ccc} ~0~ & ~2i~ & ~\sqrt{2}~  \\
~\sqrt{3}\,i~ & ~-i~ & ~\sqrt{2}~ \\
~-\sqrt{3}\,i~ & ~-i~ & ~\sqrt{2}~ \end{array} \right)\,.
\end{equation}
The six row permutations of the PMNS matrix are related through shifts in the free parameters $\theta_{1, 2, 3}$. We take $P_{l}=1$ and then the lepton mixing angles can be extracted as follows
\begin{eqnarray}
\nonumber
 \sin^{2}\theta_{13}&=&\frac{1}{6} (3-\cos 2 \theta _1) \cos ^2\theta _2\,,\\
\nonumber
\sin^{2}\theta_{12}&=&\sin ^2\theta _3+\frac{2 (\cos 2 \theta _1+3) \cos 2 \theta _3-2 \sin 2 \theta _1 \sin \theta _2 \sin 2 \theta _3}{\cos 2 \theta _1+(\cos 2 \theta _1-3) \cos 2 \theta _2+9}\,,\\
\nonumber
\sin^{2}\theta_{23}&=&\frac{1}{2}-\frac{2 \sqrt{3} \sin \theta _1 \sin 2 \theta _2}{\cos 2 \theta _1+(\cos 2 \theta _1-3) \cos 2 \theta _2+9}
\end{eqnarray}
and the CP-odd weak basis invariants are given by
\begin{eqnarray}
\nonumber
J_{CP}&=&\frac{1}{96\sqrt{6}}\Big[-20 \sin \theta _1 \sin 2 \theta _3 \cos \theta _2 \cos ^2\theta _1+4 \left(\cos 3 \theta _1-5 \cos \theta _1\right)\sin 2 \theta _2 \cos 2 \theta _3\\
\nonumber
&~&+\left(\sin 3 \theta _1-15 \sin \theta _1\right) \sin 2 \theta _3 \cos3\theta_2\Big]\,,\\
\nonumber
I_{1}&=&\frac{(-1)^{k_{1}}}{36 \sqrt{2}}\Big[8 \sin 2 \theta _1 \sin ^2\theta _2 \cos 2 \theta _3+\big(\left(7 \cos 2 \theta _1+3\right)\sin \theta _2 -\left(\cos 2 \theta _1-3\right)\sin3 \theta _2 \big)\sin 2 \theta _3 \Big]\,,\\
I_{2}&=&\frac{(-1)^{k_{2}}}{9} \Big[\sqrt{2} \big( \left(\cos 2 \theta _1-3\right)\sin \theta _2 \cos \theta _3-\sin 2 \theta _1 \sin \theta _3 \big)\sin \theta _3 \cos ^2\theta _2\Big]\,.
\end{eqnarray}
We perform a numerical analysis by treating the free parameters $\theta_{1,2,3}$ as random numbers in the range of $[0,\pi]$. The three mixing angles $\theta_{12}$, $\theta_{13}$ and $\theta_{23}$ as well as $CP$ violating phases $\delta_{CP}$, $\alpha_{21}$ and $\alpha_{31}$ are calculated for each random point, and only points which agree with the global fit data at 3$\sigma$ level with global fit data are retained. We plot the correlations among the mixing angles and $CP$ phases in figure~\ref{fig:Z3TS_single}. We see that any value of Dirac $CP$ phase $\delta_{CP}$ in the interval $[0, 2\pi]$ can be achieved. However, the Majorana phases are strongly constrained, and they values lie in the ranges $\alpha_{21}$(mod $\pi$)$\in[0, 0.13\pi]\cup[0.87\pi, \pi]$ and $\alpha_{31}$(mod $\pi$)$\in[0, 0.25\pi]\cup[0.75\pi, \pi]$.

\begin{figure}[t!]
\centering
\includegraphics[height=0.2\textwidth]{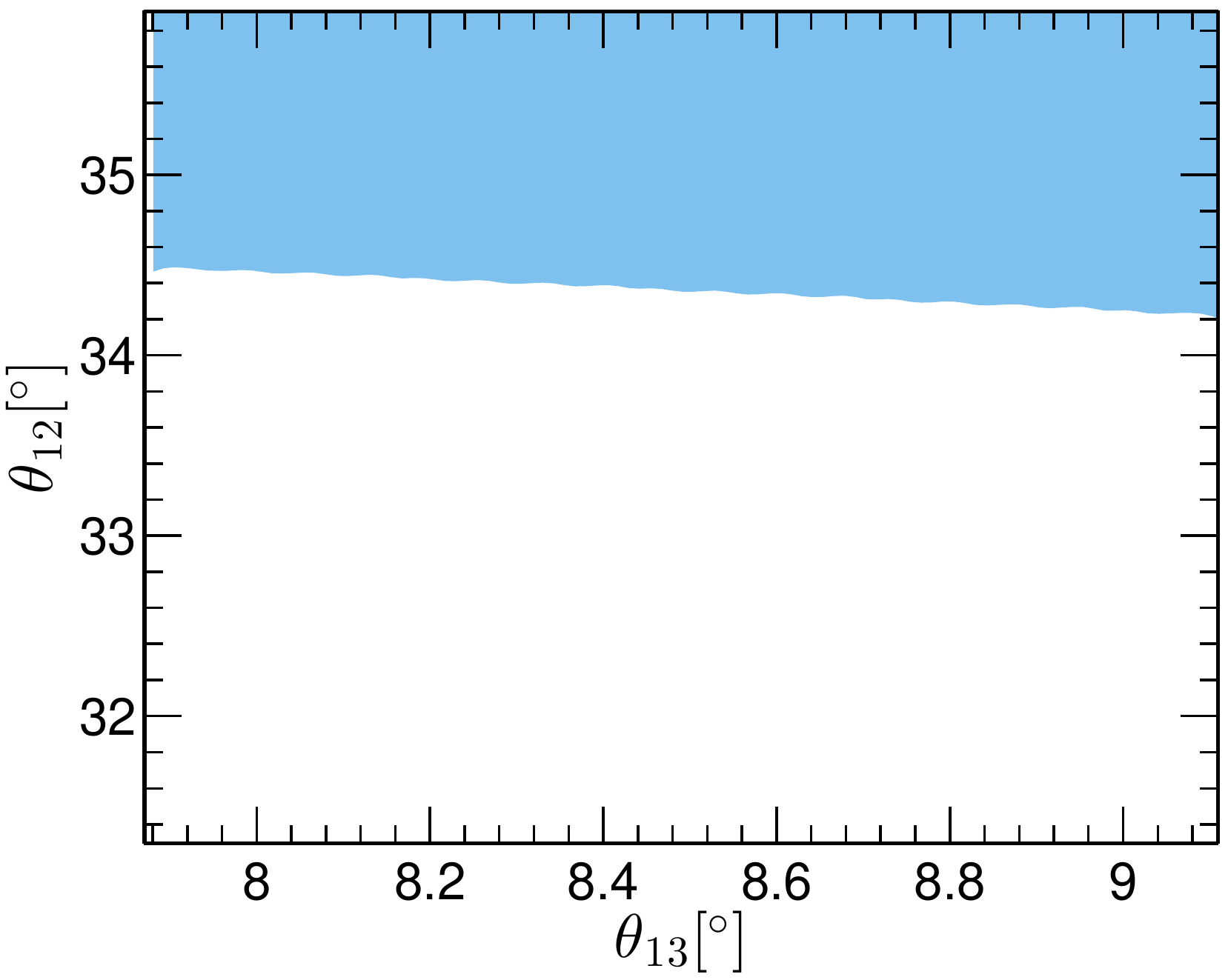}
\includegraphics[height=0.2\textwidth]{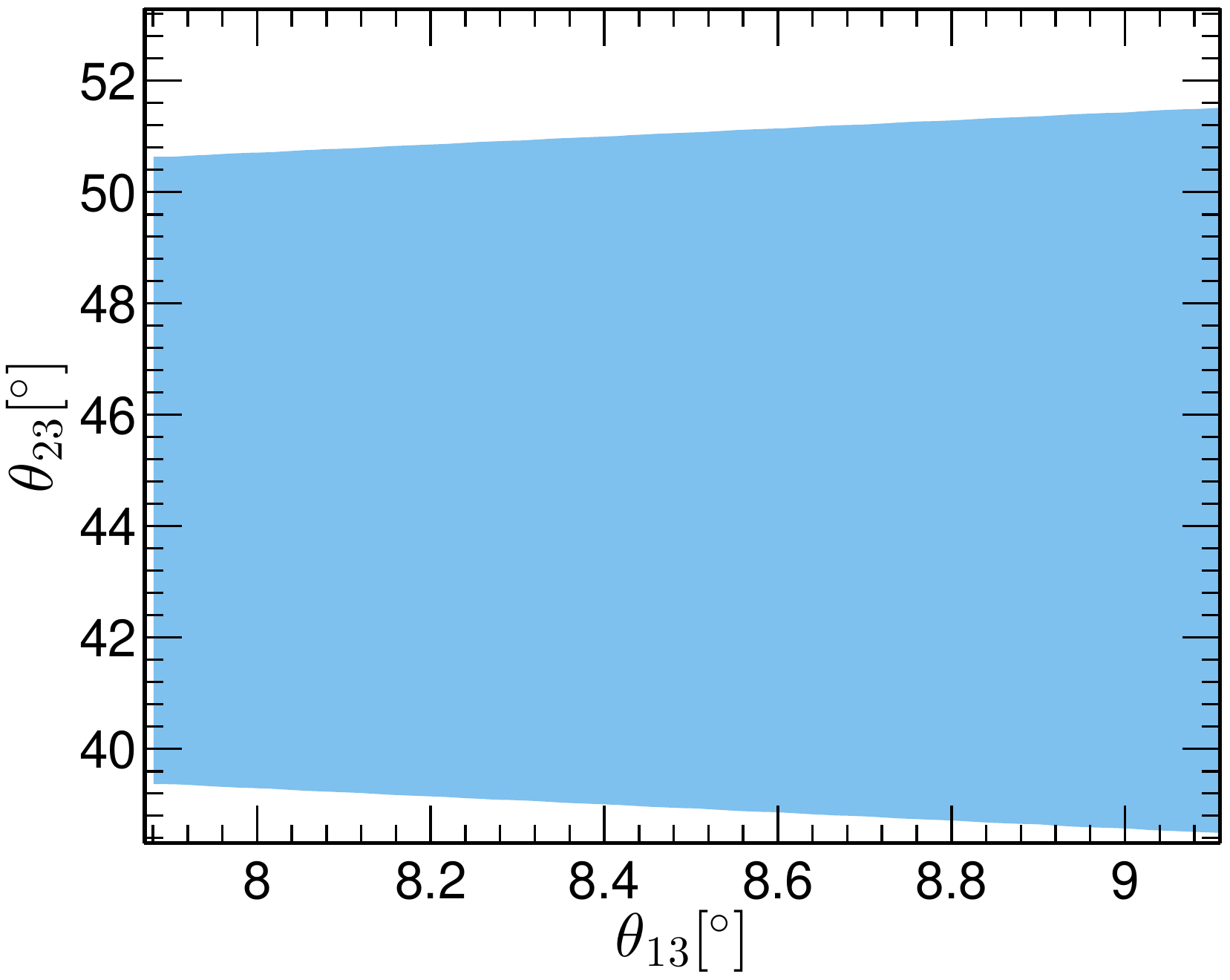}
\includegraphics[height=0.2\textwidth]{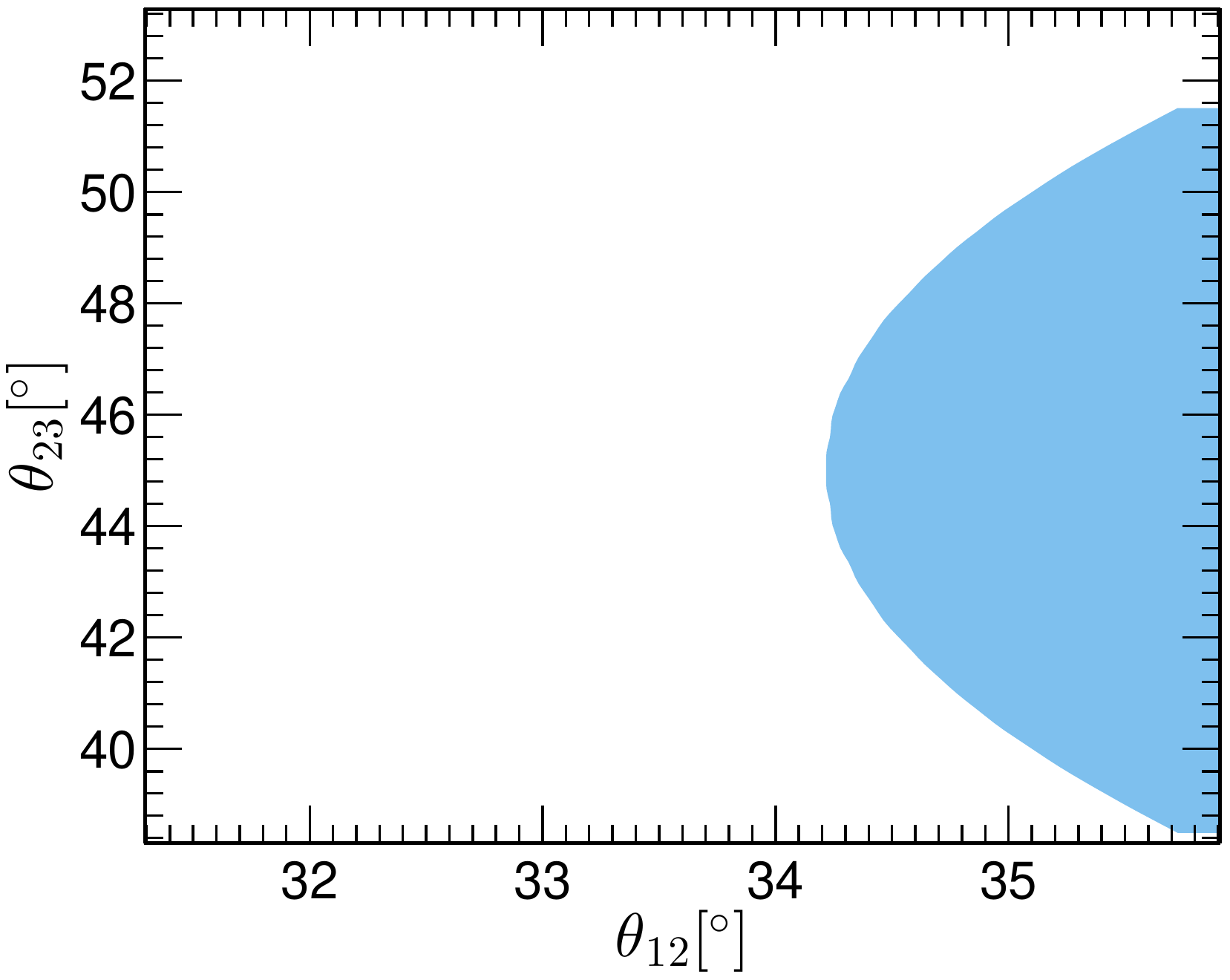}
\includegraphics[height=0.2\textwidth]{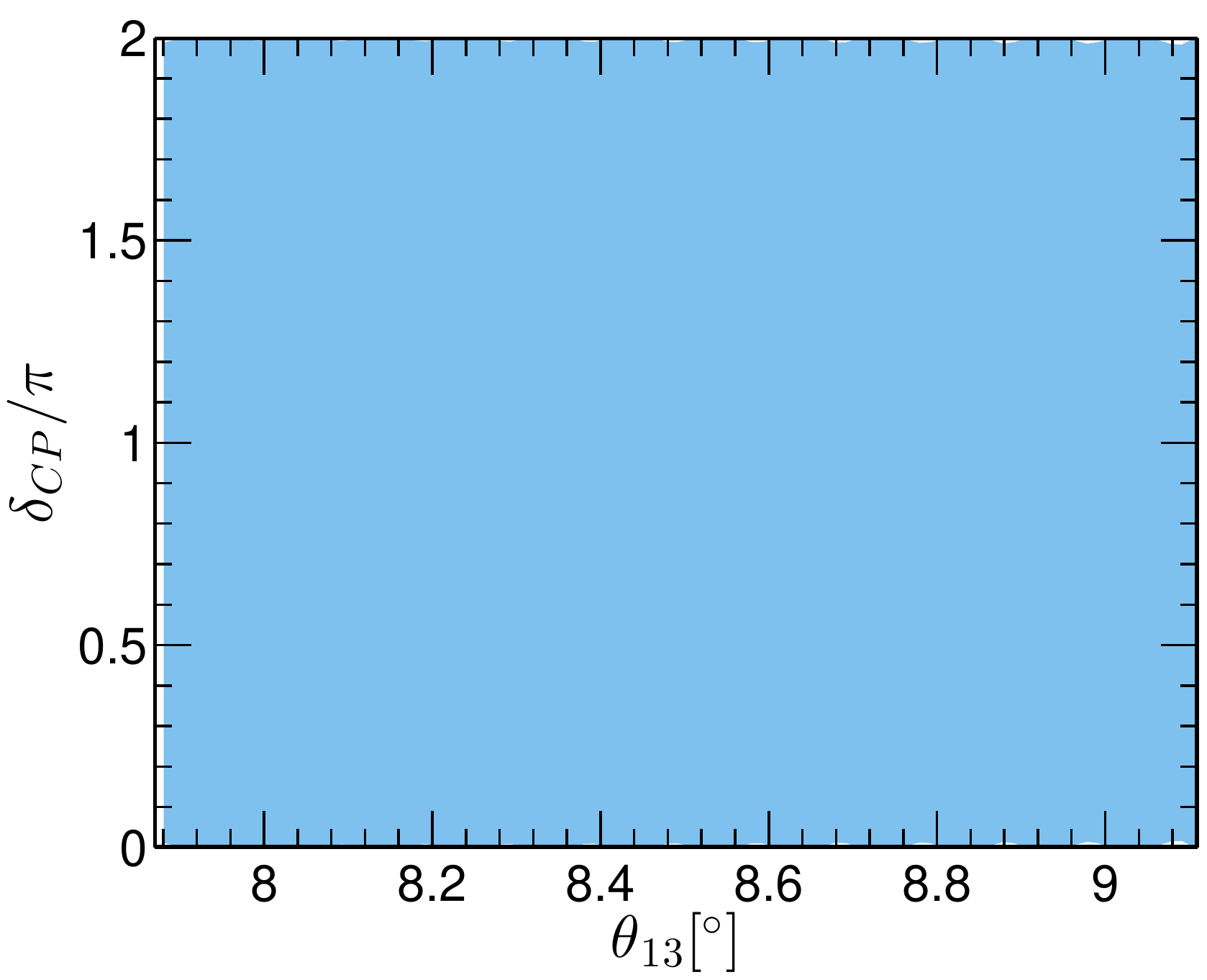}
\includegraphics[height=0.2\textwidth]{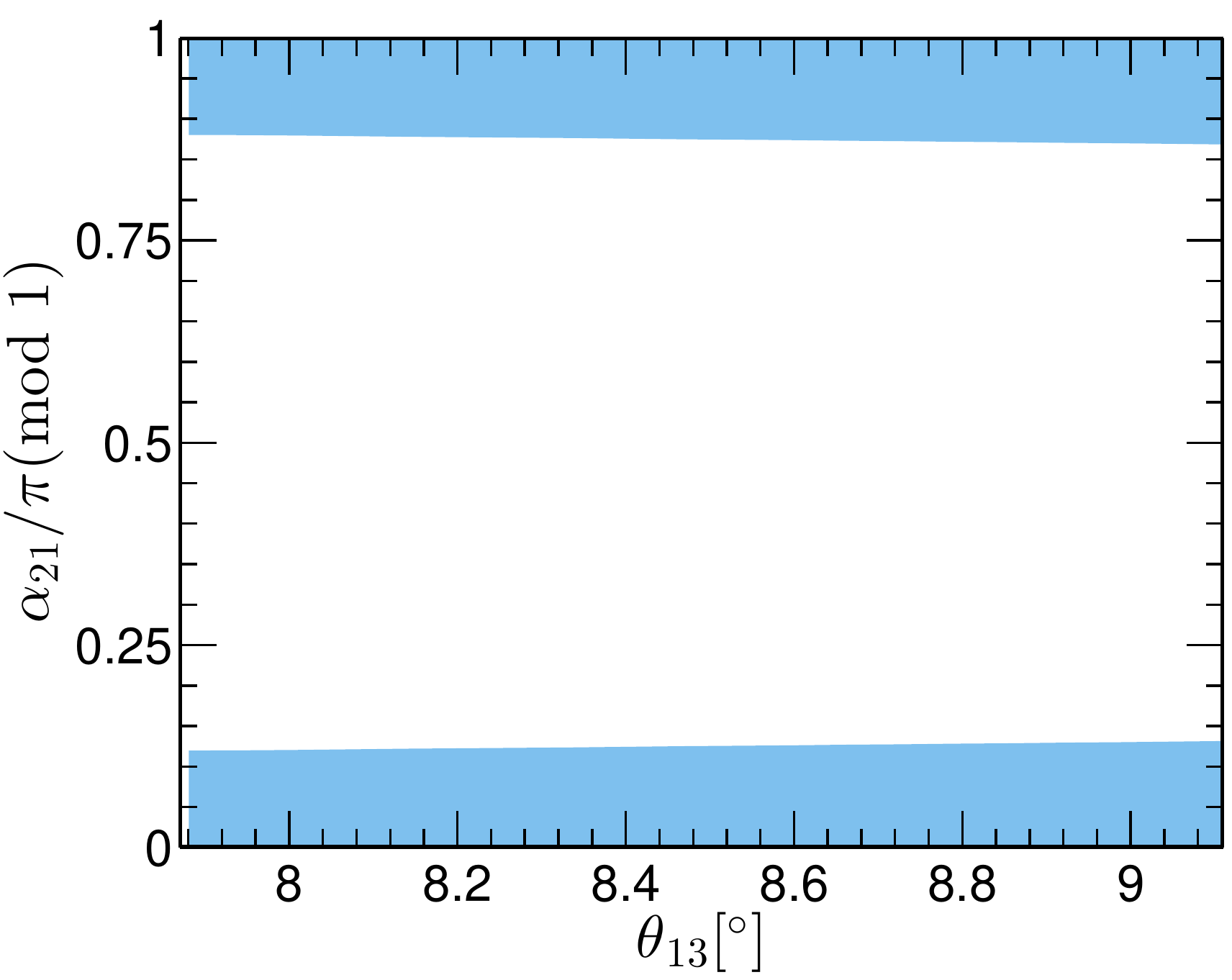}
\includegraphics[height=0.2\textwidth]{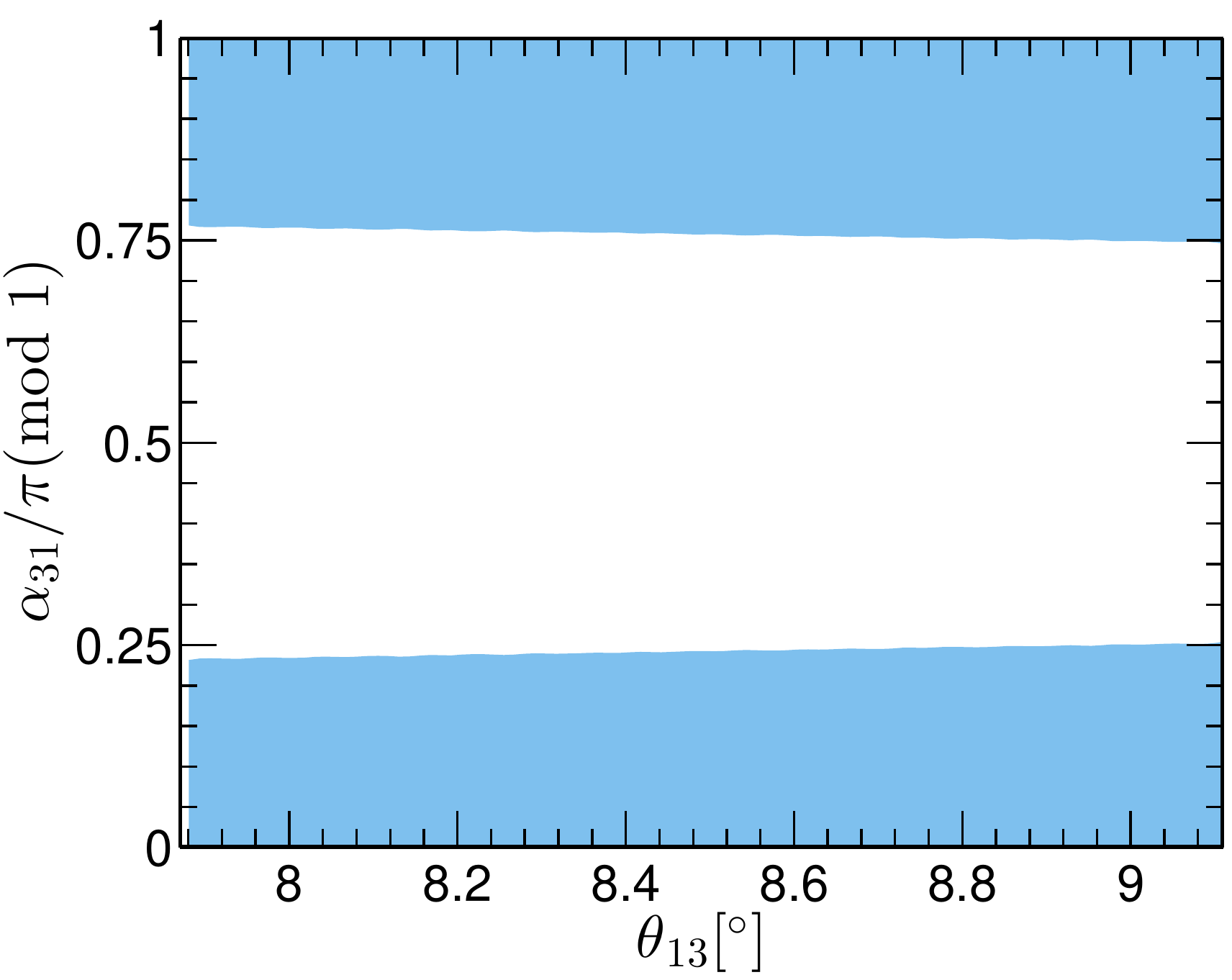}
\includegraphics[height=0.2\textwidth]{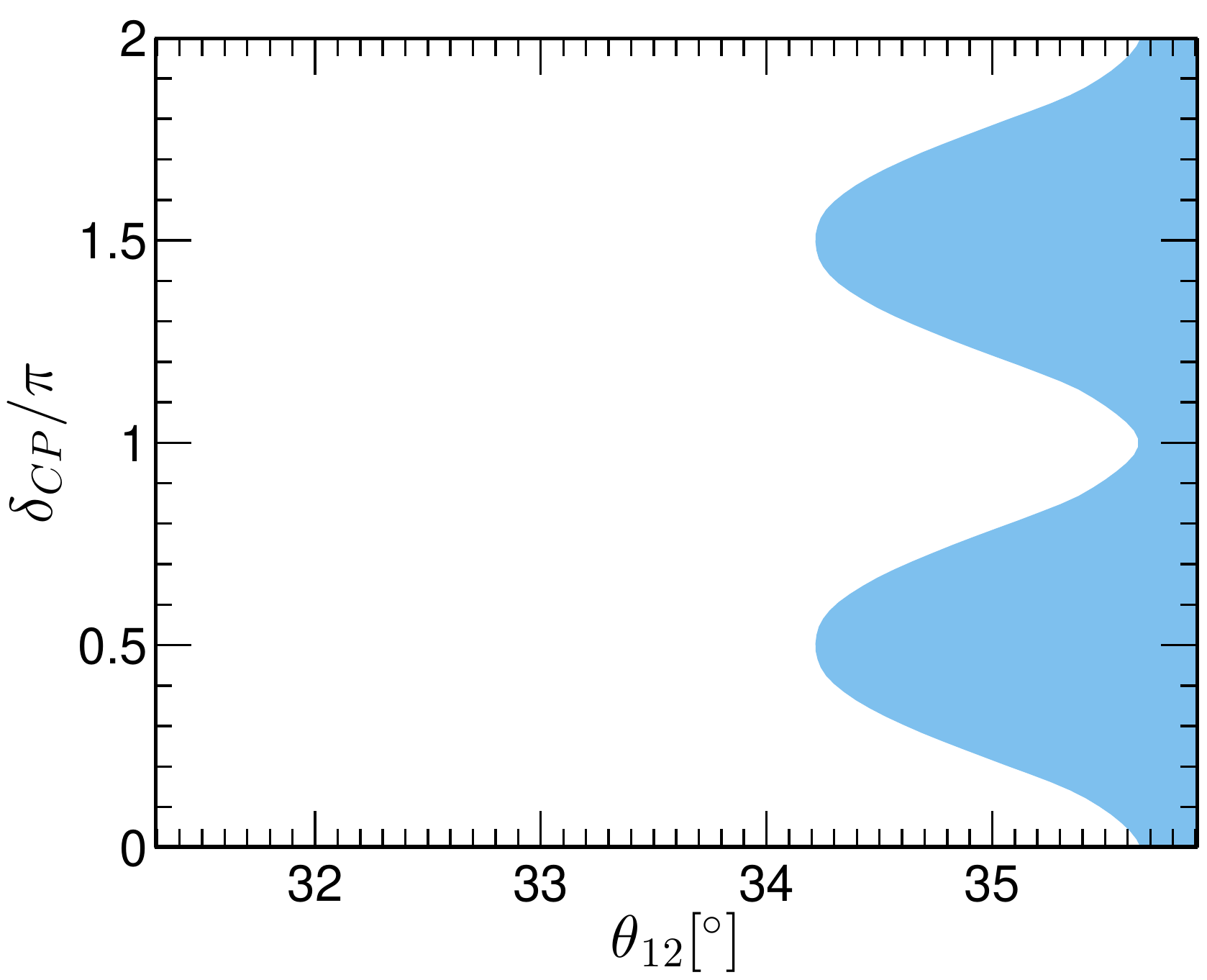}
\includegraphics[height=0.2\textwidth]{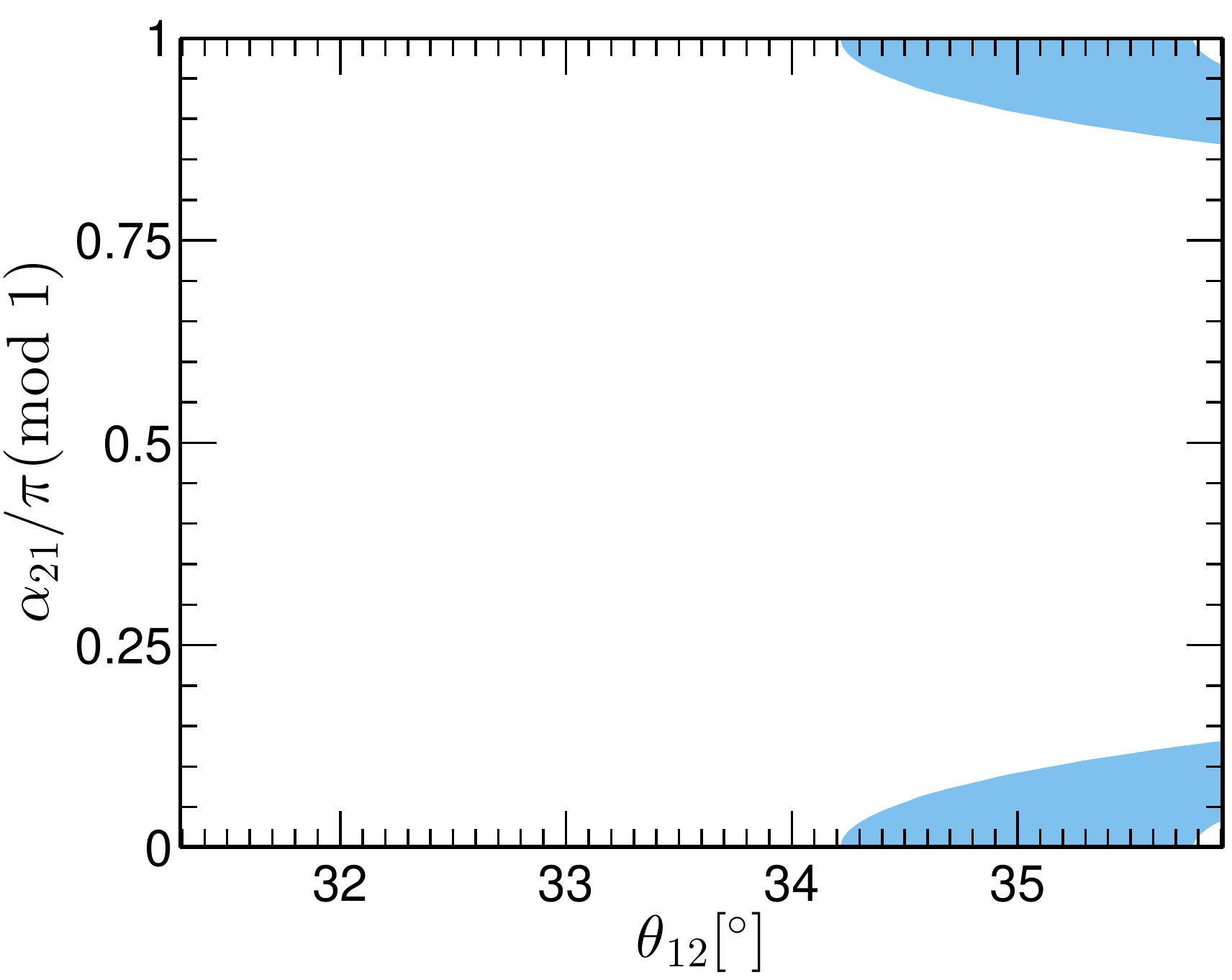}
\includegraphics[height=0.2\textwidth]{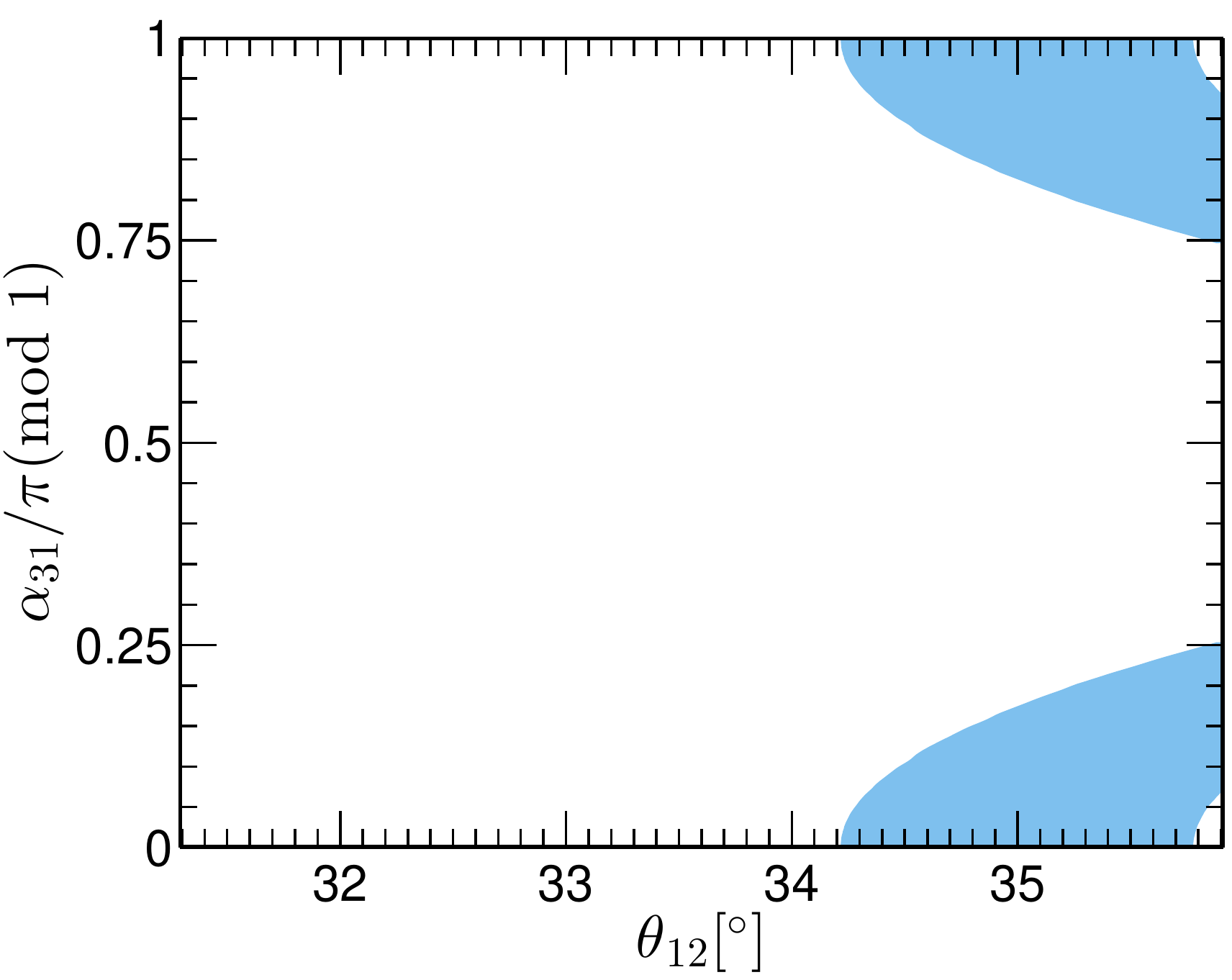}
\includegraphics[height=0.2\textwidth]{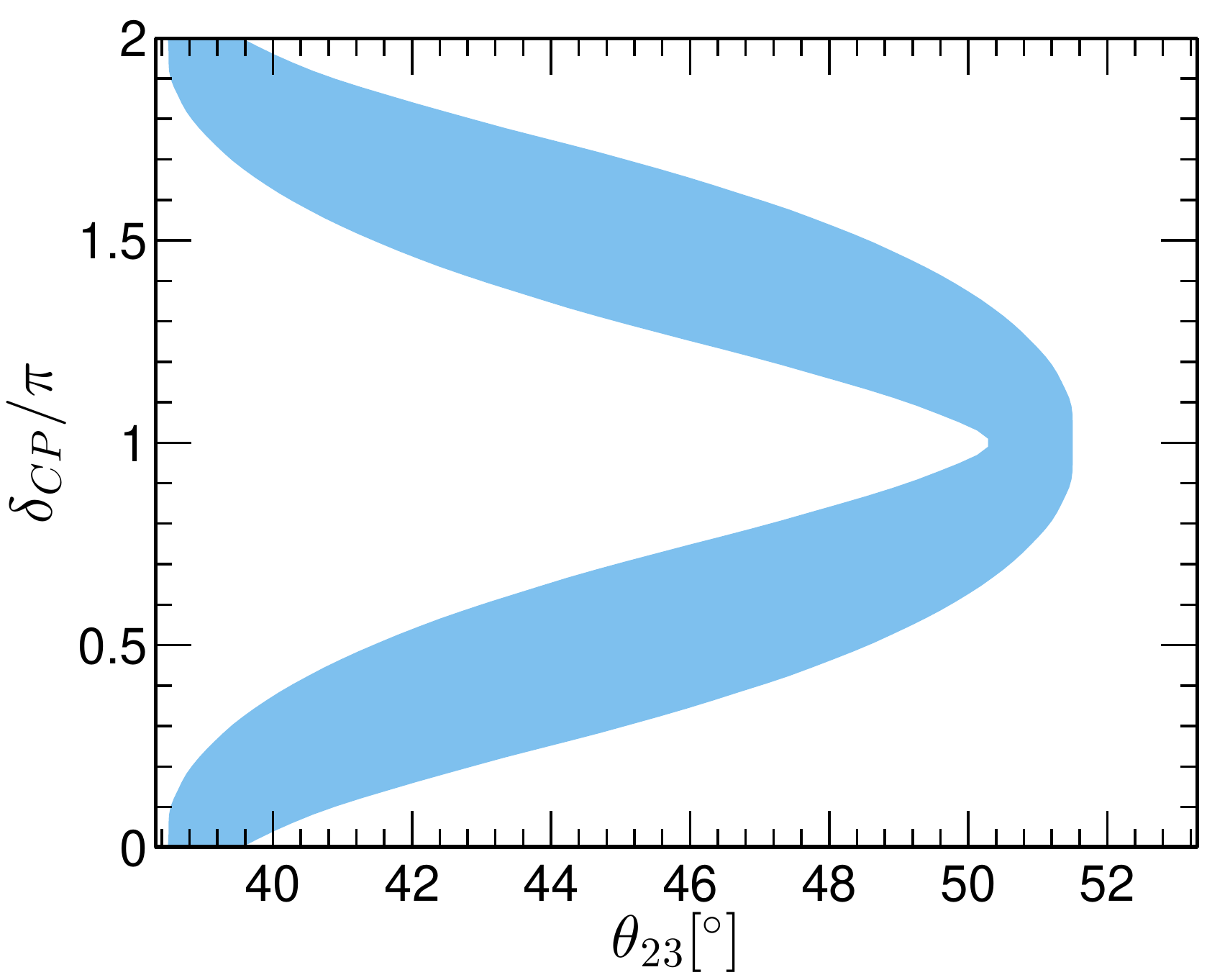}
\includegraphics[height=0.2\textwidth]{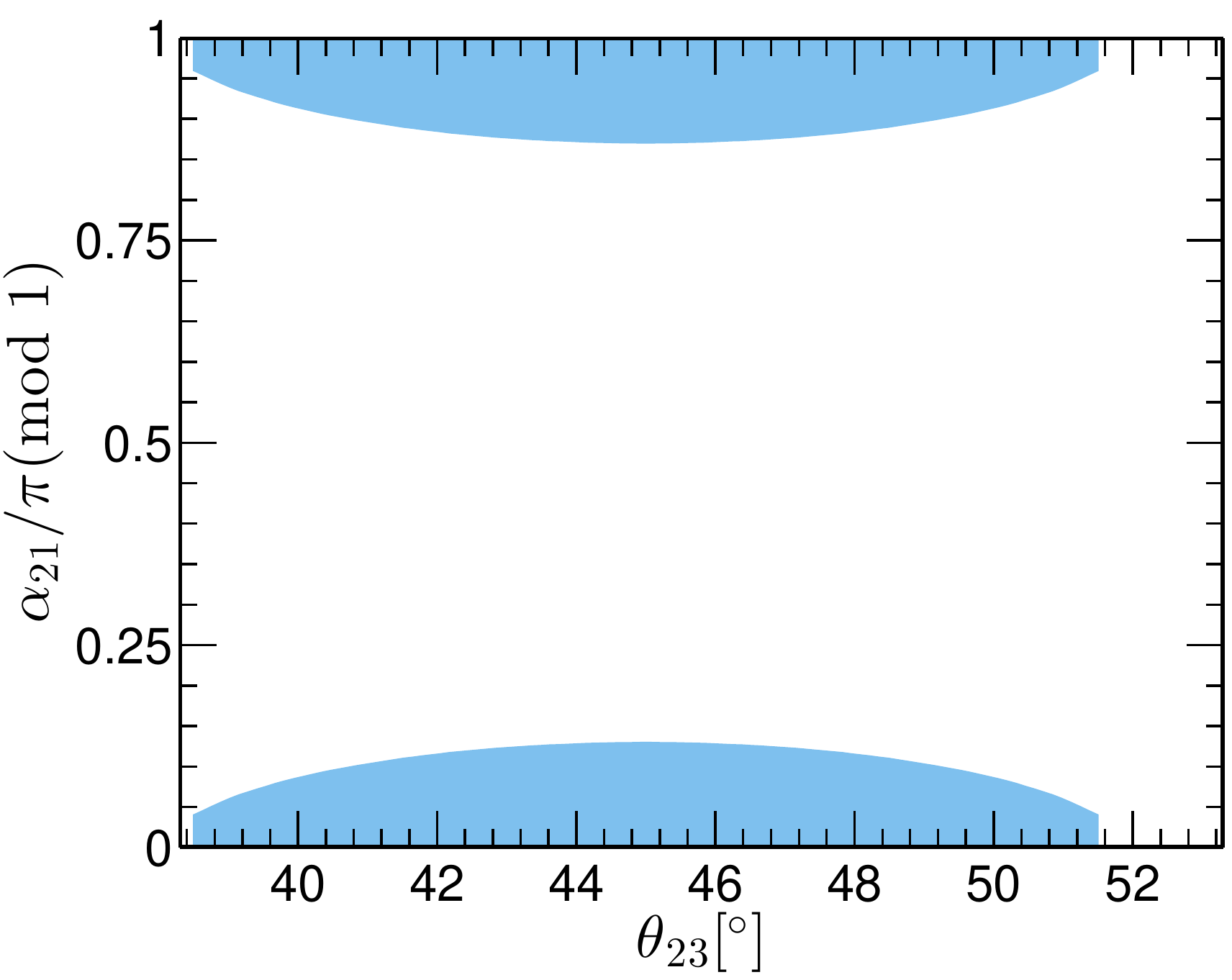}
\includegraphics[height=0.2\textwidth]{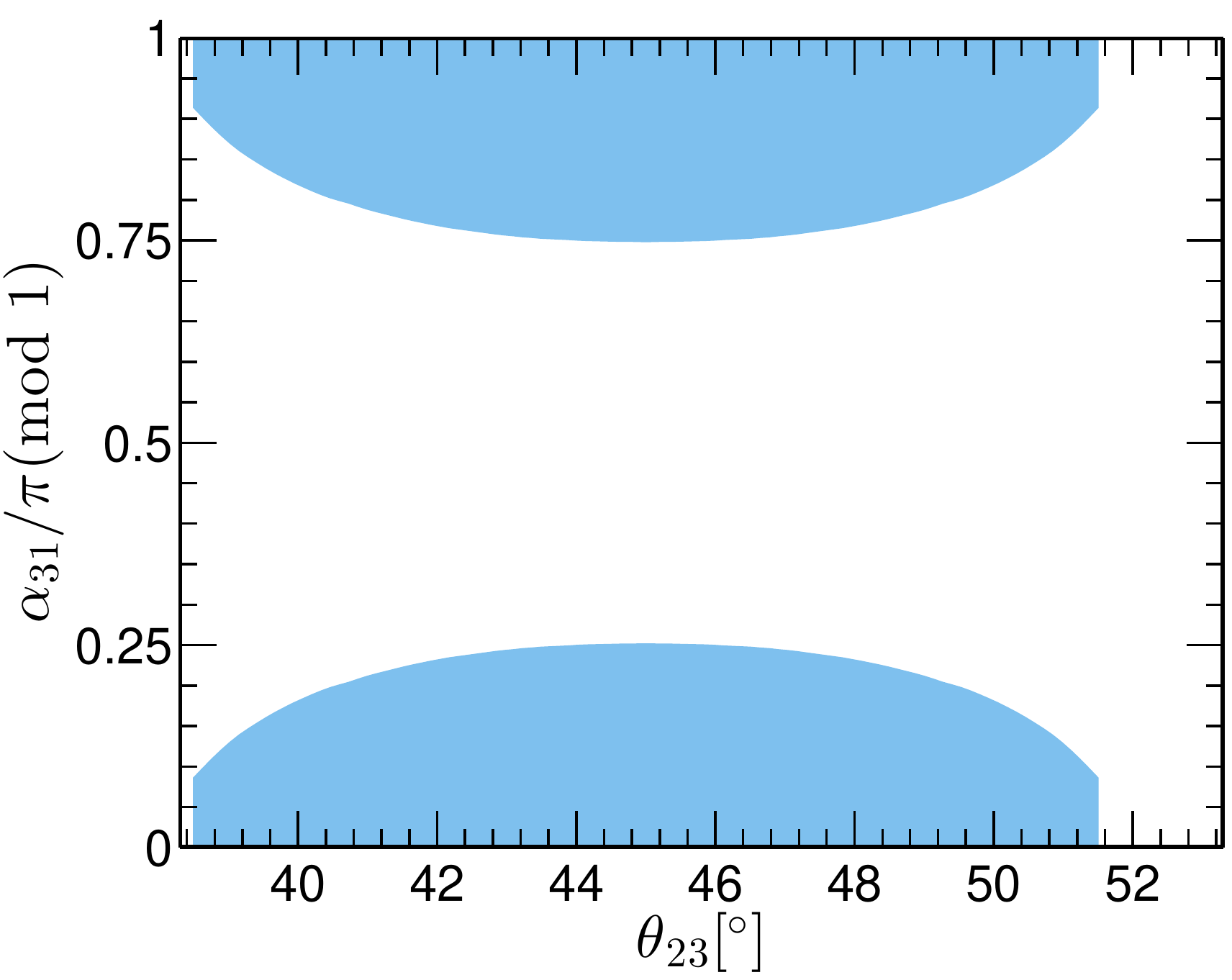}
\includegraphics[height=0.2\textwidth]{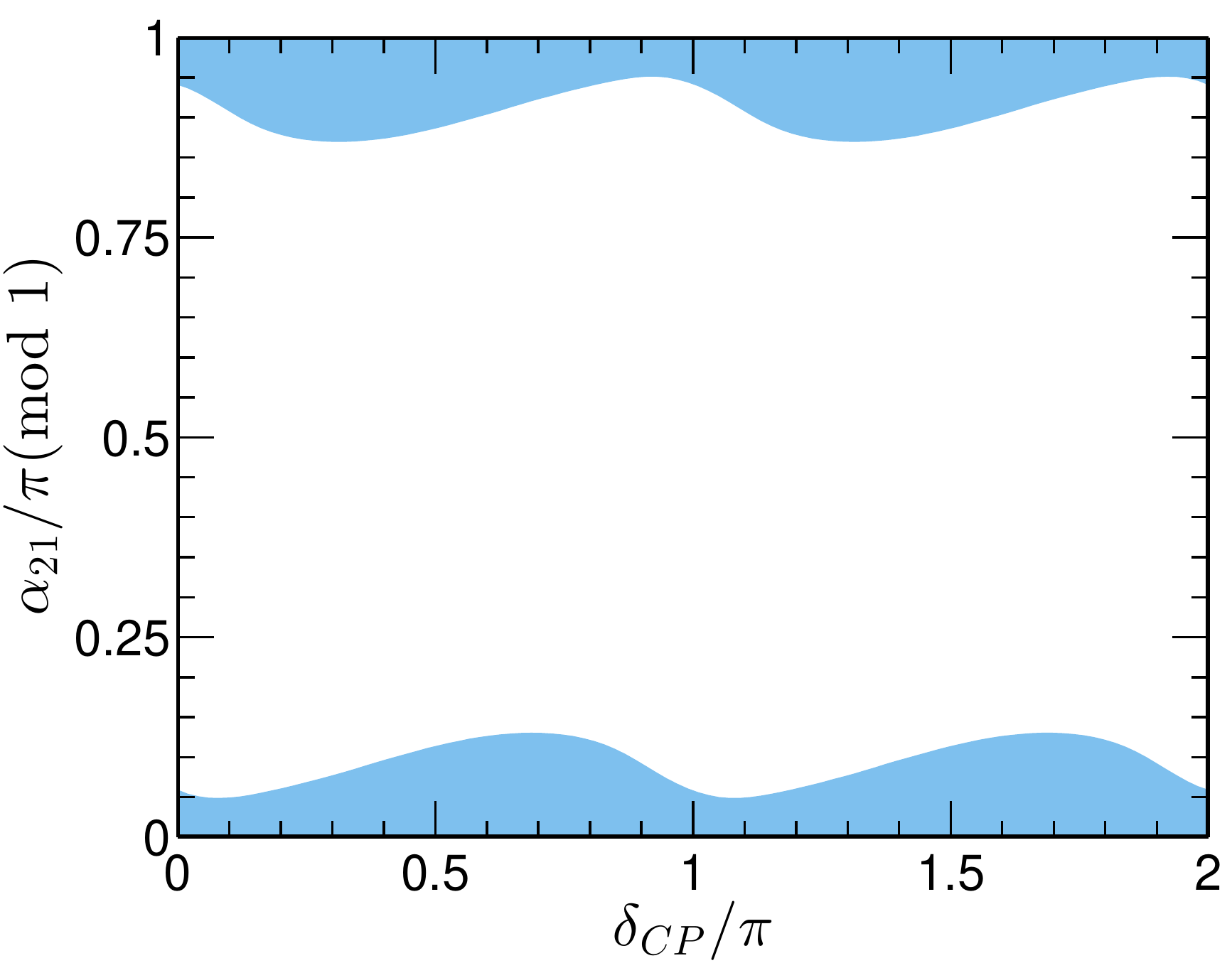}
\includegraphics[height=0.2\textwidth]{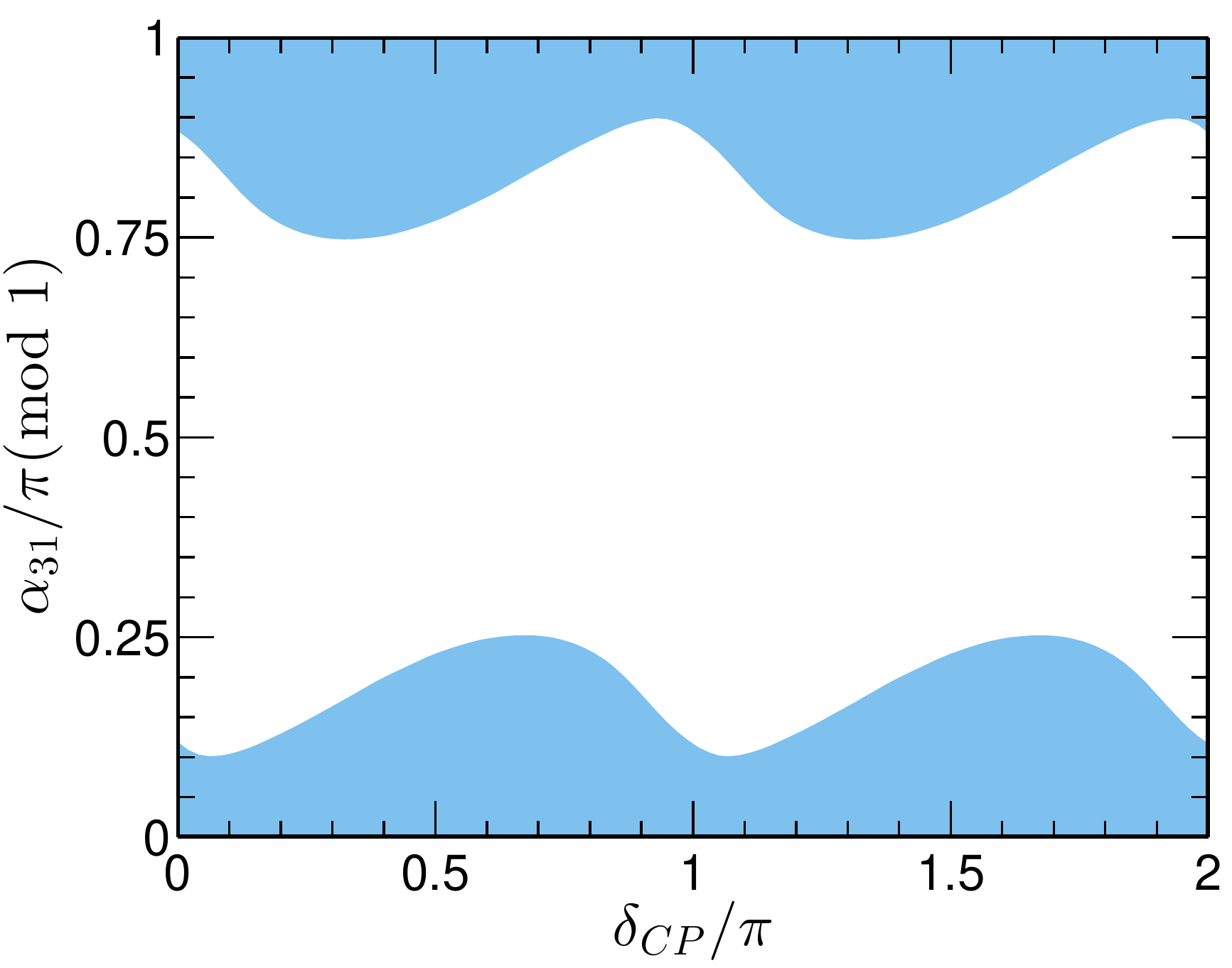}
\includegraphics[height=0.2\textwidth]{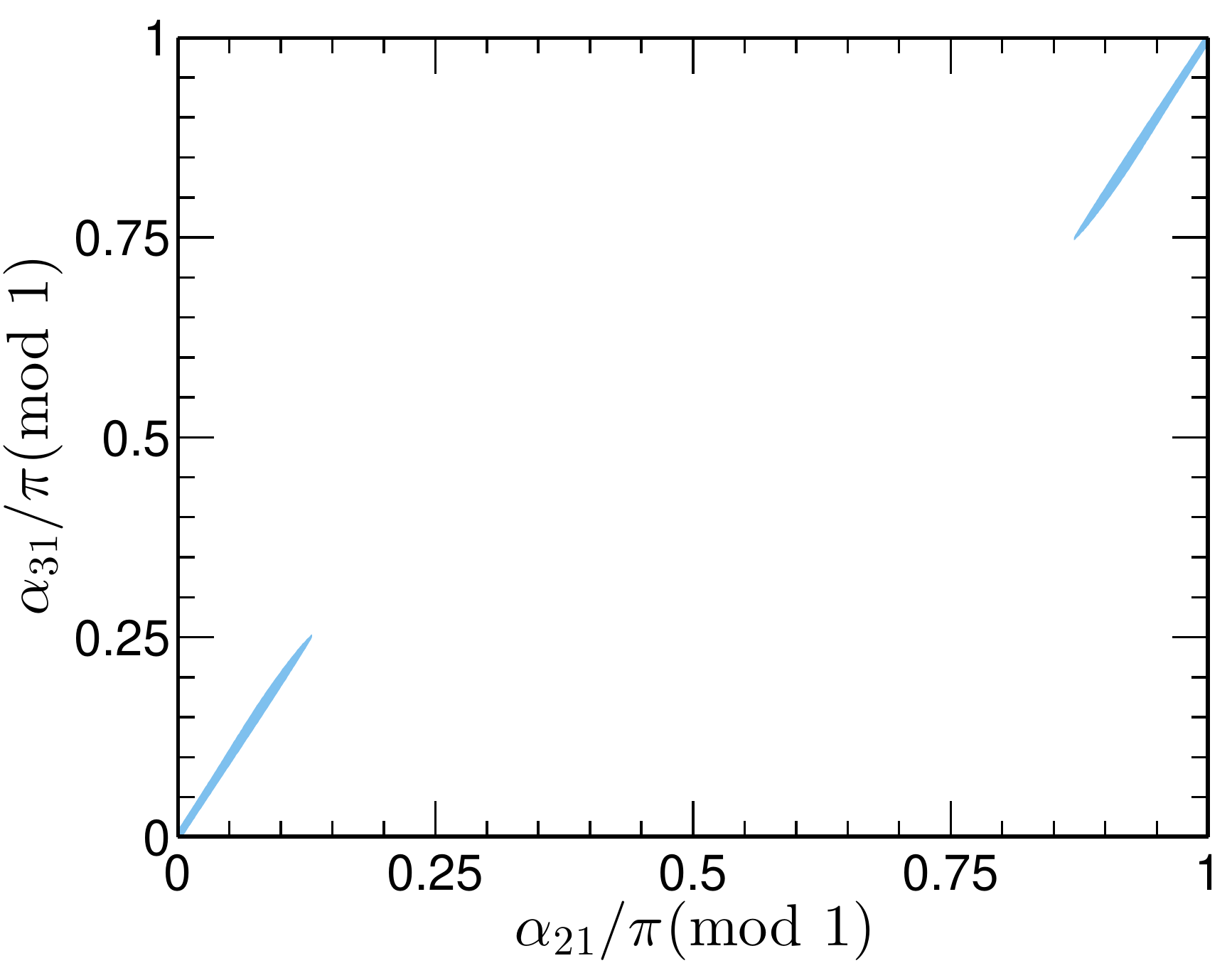}
\caption{\label{fig:Z3TS_single}Correlations between different mixing parameters in the case of $(G_{l},X_{\nu})=(Z_{3}^{T},S)$, where the three lepton mixing angles are required to be compatible with the experimental data at $3\sigma$ level~\cite{Gonzalez-Garcia:2014bfa}. }
\end{figure}

\item{$G_{l}=Z_{3}^{T},X_{\nu}=U$}

This case is exactly the $\mu-\tau$ reflection symmetry in the charged lepton diagonal basis. One can straightforwardly read out $U_{l}$ and $\Sigma_{\nu}$ as follows,
\begin{equation}
U_{l}=\left(
\begin{array}{ccc}
 ~1~  &  ~0~  &  ~0~ \\
 0  &  1  &  0 \\
 0  &  0  &  1 \end{array} \right),\qquad
\Sigma_{\nu}=\frac{1}{\sqrt{2}}\left(
\begin{array}{ccc}
 ~0~ & ~\sqrt{2}\,i~ & ~0~  \\
i & 0 & -1 \\
i & 0 & 1
\end{array} \right)\,.
\end{equation}
Out of the six possible row permutations only $P_{l}=1$ and $P_{l}=P_{23}$ lead to a pattern compatible
with data. The PMNS matrices arising form $P_{l}=1$ or $P_{l}=P_{23}$ are equivalent. The others give rise to either $\tan\theta_{13}=\sin\theta_{23}$ or $\tan\theta_{13}=\cos\theta_{23}$  which does not allow both $\theta_{13}$ and $\theta_{23}$ to be fitted well simultaneously. For the case of $P_{l}=1$, the lepton mixing angles and the $CP$ violation phases are found to be of the form
\begin{eqnarray}
\nonumber&&\sin^{2}\theta_{13}=\sin^2\theta_1\cos^2\theta_2,\qquad \sin^{2}\theta_{23}=\frac{1}{2},\\ \nonumber&&\sin^{2}\theta_{12}=\sin^2\theta_3+\frac{4(\cos\theta_1\cos2\theta_3-\sin\theta_1 \sin\theta_2\sin2\theta_3)\cos\theta_1}{\cos2\theta_1-2\sin^2\theta_1\cos2\theta_2+3},\\
&&|\sin\delta_{CP}|=1,\qquad \sin\alpha_{21}=\sin\alpha_{31}=0\,.
\end{eqnarray}
Hence both the atmospheric mixing angle $\theta_{23}$ and the Dirac phase $\delta_{CP}$ are predicted to be maximal while the solar as well as reactor mixing angles are not constrained. There are evidences showing that the Dirac $CP$ violating phase $\delta_{CP}$ is close to $-\pi/2$ (or $3\pi/2$)~\cite{T2K_delta_CP,NovA_delta_CP}. If these data are further confirmed in near future, this mixing pattern would be an excellent leading order approximation.

\item{$G_{l}=Z_{3}^{T},X_{\nu}=SU$}

We can read out $U_{l}$ and $\Sigma_{\nu}$ as
\begin{equation}
U_{l}=\left(
\begin{array}{ccc}
 ~1~ & ~0~  &  ~0~ \\
 0  &  1  &  0 \\
 0  &  0  &  1 \end{array} \right),~~~~~~
\Sigma_{\nu}=\frac{1}{\sqrt{30}}\left(
\begin{array}{ccc}
\sqrt{6}\,i ~&~ 2 i ~&~ -2\sqrt{5} \\
 0 ~&~ 5i ~&~ \sqrt{5} \\
 2\sqrt{6}\,i ~&~ -i ~&~ \sqrt{5}
\end{array}
\right)\,.
\end{equation}
For the six possible permutations of rows, only the mixing patterns with $P_{l}=1$ and $P_{l}=P_{23}$ can accommodate the experimental data of the mixing angles for certain values of the parameters $\theta_{1, 2, 3}$. The PMNS matrices arising from $P_{l}=1$ and $P_{l}=P_{23}$ are essentially the same if redefinition of $\theta_{1,2,3}$ and relabeling of $k_{1, 2}$ are taken into account. Using the actual form of the PMNS matrix given in Eq.~\eqref{eq:PMNS_master_single}, we find
\begin{eqnarray}
\nonumber
\sin^{2}\theta_{13}&=&\frac{1}{15} \Big[(\sqrt{3} \sin \theta _2+\sqrt{2} \sin \theta _1 \cos \theta _2){}^2+10 \cos ^2\theta _1 \cos ^2\theta _2\Big]\,,\\
\nonumber
\sin^{2}\theta_{12}&=&\sin ^2\theta _3+\frac{4 (2 \cos 2 \theta _1-3) \cos 2 \theta _3-2 (4 \sin 2 \theta _1 \sin \theta _2+\sqrt{6} \cos \theta _1 \cos \theta _2)\sin 2 \theta _3 }{2 \sqrt{6} \sin \theta _1 \sin 2 \theta _2+8 \cos 2 \theta _1 \cos ^2\theta _2+3 \cos 2 \theta _2-21}\,,\\
\sin^{2}\theta_{23}&=&\frac{5 (2 \cos 2 \theta _1-3) \cos ^2\theta _2}{2 \sqrt{6} \sin \theta _1 \sin 2 \theta _2+8 \cos 2 \theta _1 \cos ^2\theta _2+3 \cos 2 \theta _2-21}
\end{eqnarray}
and
\begin{eqnarray}
\nonumber
J_{CP}&=&\frac{1}{144 \sqrt{5}}\Big[\big(\sqrt{6} \sin 3 \theta _1 (\cos 3 \theta _2-5 \cos \theta _2)-2 \sqrt{6} \sin \theta _1 (\cos \theta _2+3 \cos 3 \theta _2)\\
\nonumber&~&+36\sin\theta_2\cos^2\theta_2\big)\sin2\theta_3+4\sqrt{6}(\cos3\theta_1-2\cos\theta_1)\sin2\theta_2\cos2\theta_3\Big]\,,\\
\nonumber
I_{1}&=&\frac{(-1)^{k_{1}}}{90 \sqrt{5}}\Big[\big[\sqrt{6} \big((\cos 3 \theta _2-5 \cos \theta _2)\sin 3 \theta _1 +10 \sin \theta _1 \cos ^3\theta _2\big)\\
\nonumber
&~&+(25\sin\theta_2-7\sin3\theta_2)\cos2\theta_1\big]\sin2\theta_3+\big((10-22\cos2\theta_2)\sin2\theta_1 \\
\nonumber&~&+4\sqrt{6}\sin2\theta_2\cos3\theta_1\big)\cos2\theta_3 \Big]\,,\\
\nonumber
I_{2}&=&\frac{(-1)^{k_{2}}}{90 \sqrt{5}}\Big[\big(5 \sqrt{6} \sin \theta _1 \sin \theta _2 \sin 2 \theta _2-\sqrt{6} \sin 3 \theta _1 (\cos 3 \theta _2-5 \cos \theta _2)\\
\nonumber
&~&-(5\sin\theta_2-7\sin3\theta_2)\cos2\theta_1\big)\sin2\theta_3-10\sin2\theta_1\cos^2\theta_2+(17\cos2\theta_2+5)\sin2\theta_1\cos2\theta_3\\
\nonumber
&~&-\sqrt{6}\big((8\cos2\theta_1+1)\cos2\theta_3+5\big)\sin2\theta_2\cos\theta _1\Big]\,.
\end{eqnarray}
The numerical results for the correlations among different mixing parameters are shown in figure~\ref{fig:Z3TSU_single}. We notice that both Majorana phases $\alpha_{21}$ and $\alpha_{31}$ are determined to be around $0$ and $\pi$, the solar mixing angle $\theta_{12}$ near its $3\sigma$ upper limit $\theta_{12}\sim35^{\circ}$ is preferred, and atmospheric mixing angle $\theta_{23}$ and Dirac phase $\delta_{CP}$ are correlated. The forthcoming reactor and long baseline neutrino experiments which are expected to make precise measurement of $\theta_{12}$, $\theta_{23}$ and $\delta_{CP}$, have the potential to exclude this mixing pattern.

\begin{figure}[hptb!]
\centering
\includegraphics[height=0.2\textwidth]{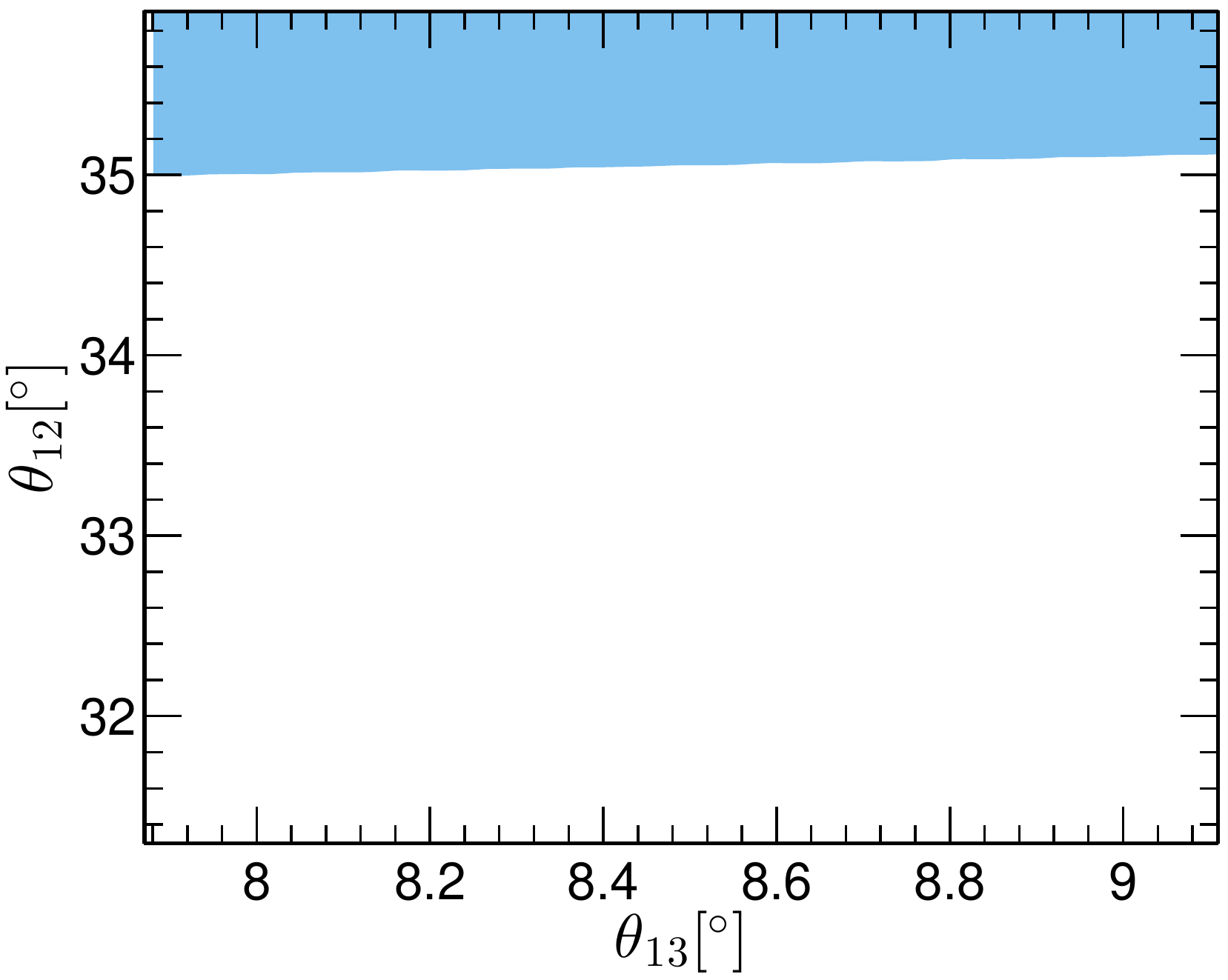}
\includegraphics[height=0.2\textwidth]{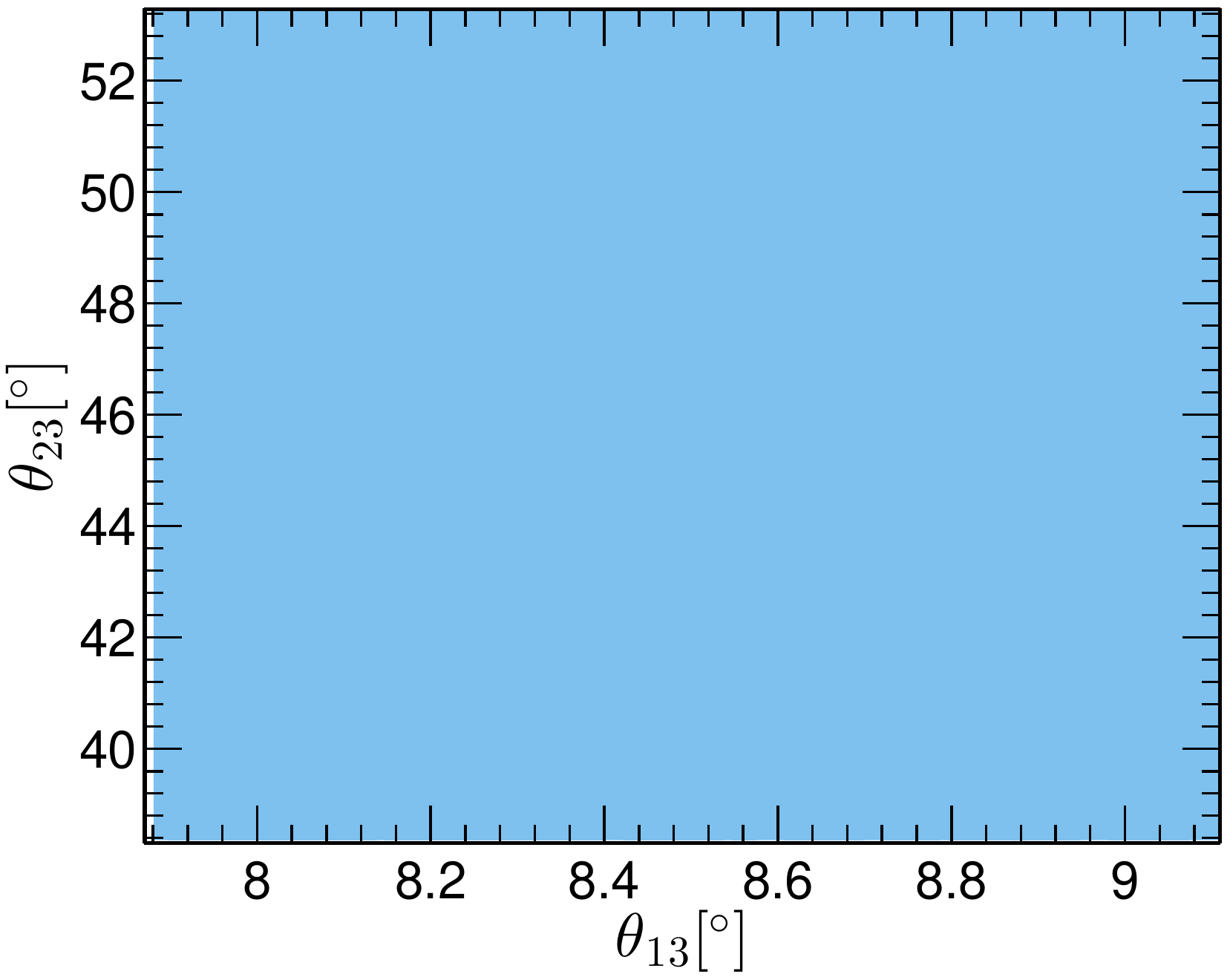}
\includegraphics[height=0.2\textwidth]{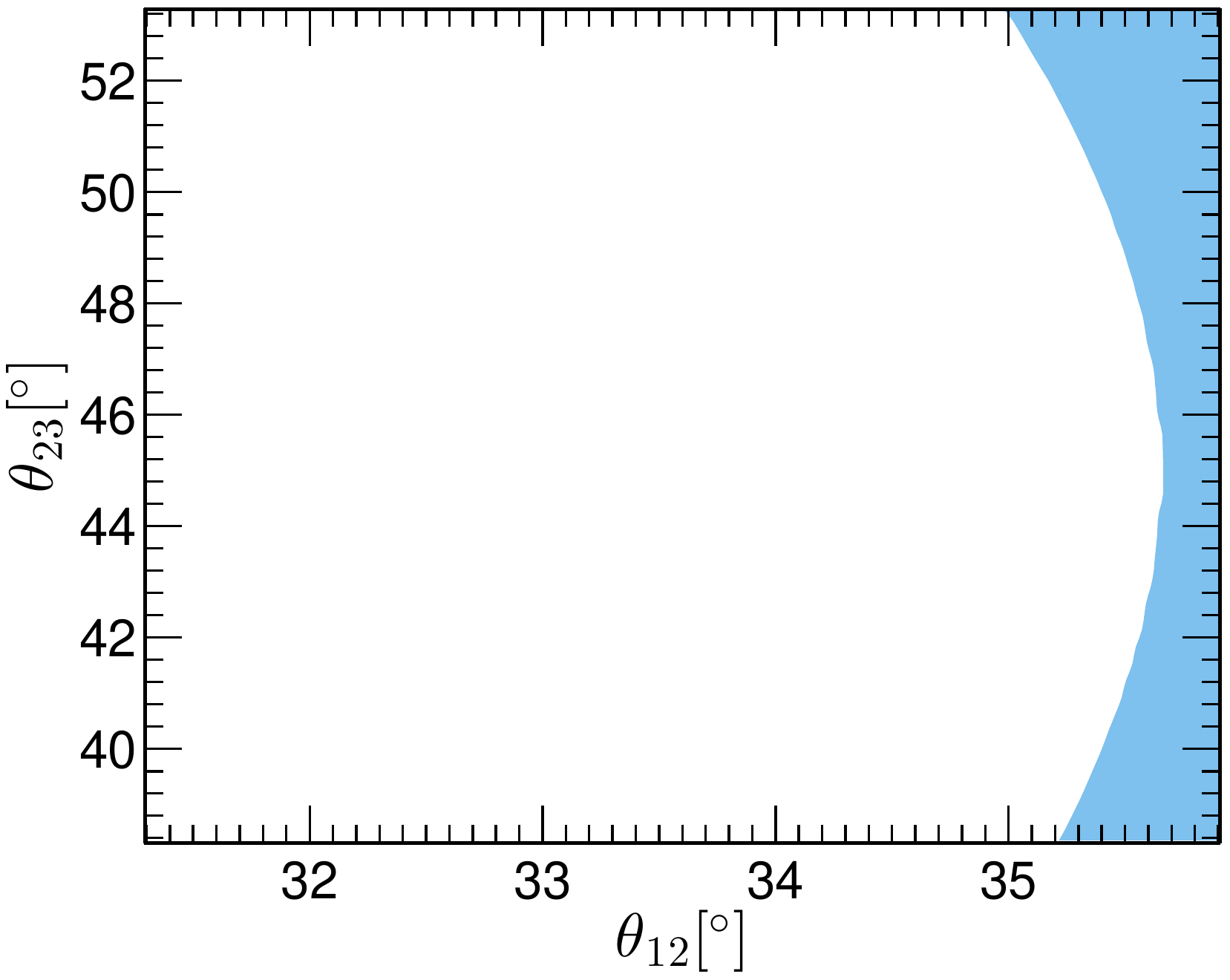}
\includegraphics[height=0.2\textwidth]{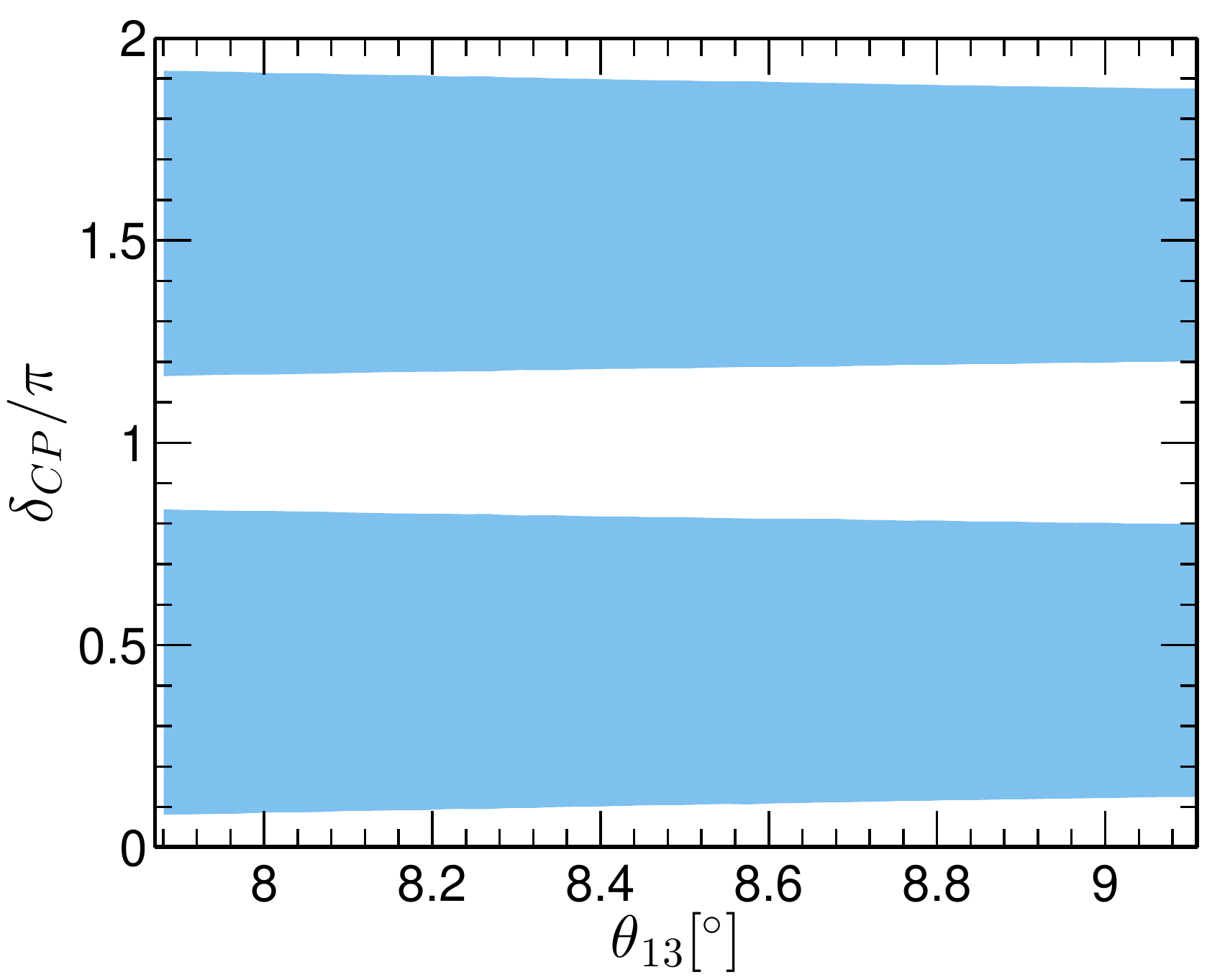}
\includegraphics[height=0.2\textwidth]{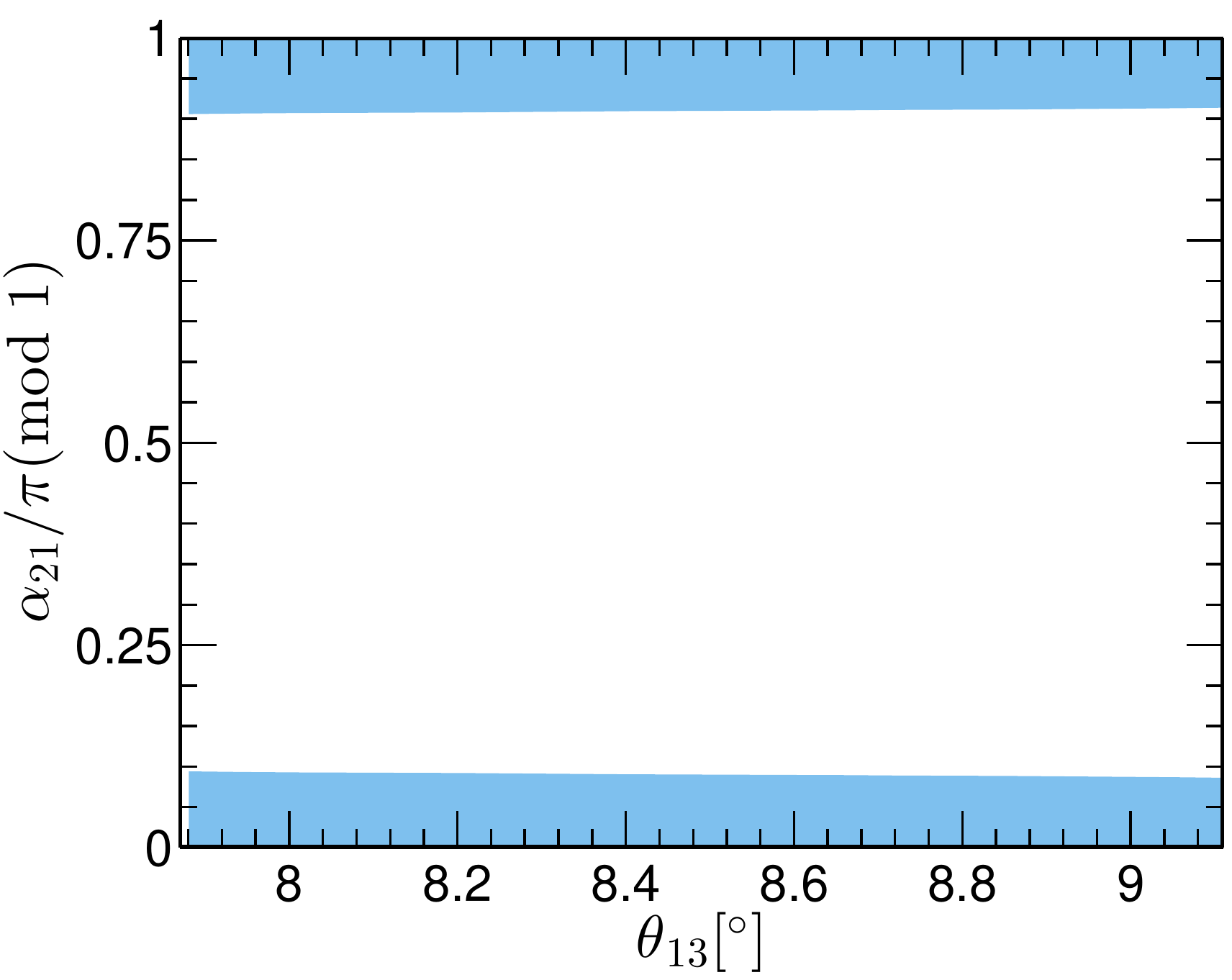}
\includegraphics[height=0.2\textwidth]{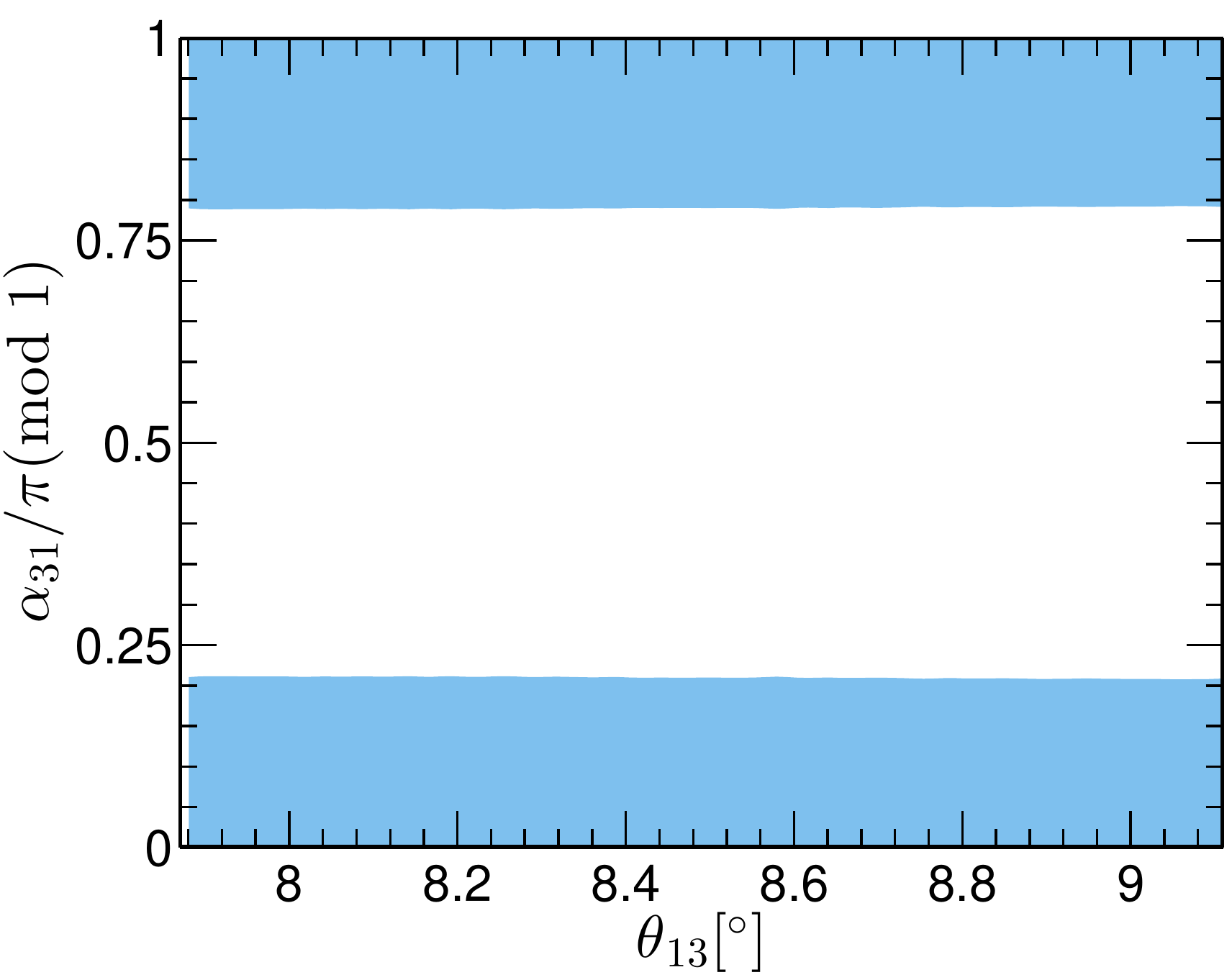}
\includegraphics[height=0.2\textwidth]{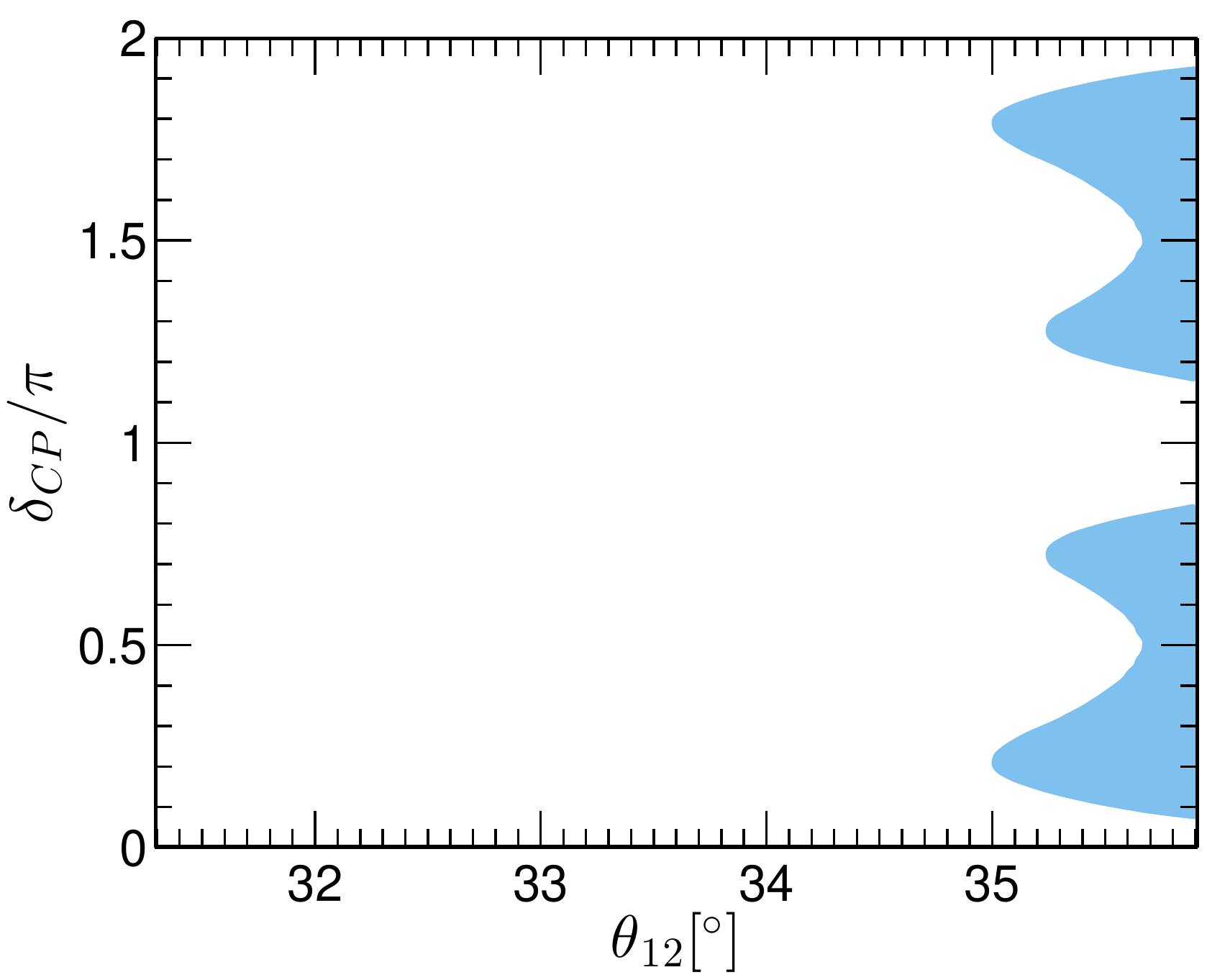}
\includegraphics[height=0.2\textwidth]{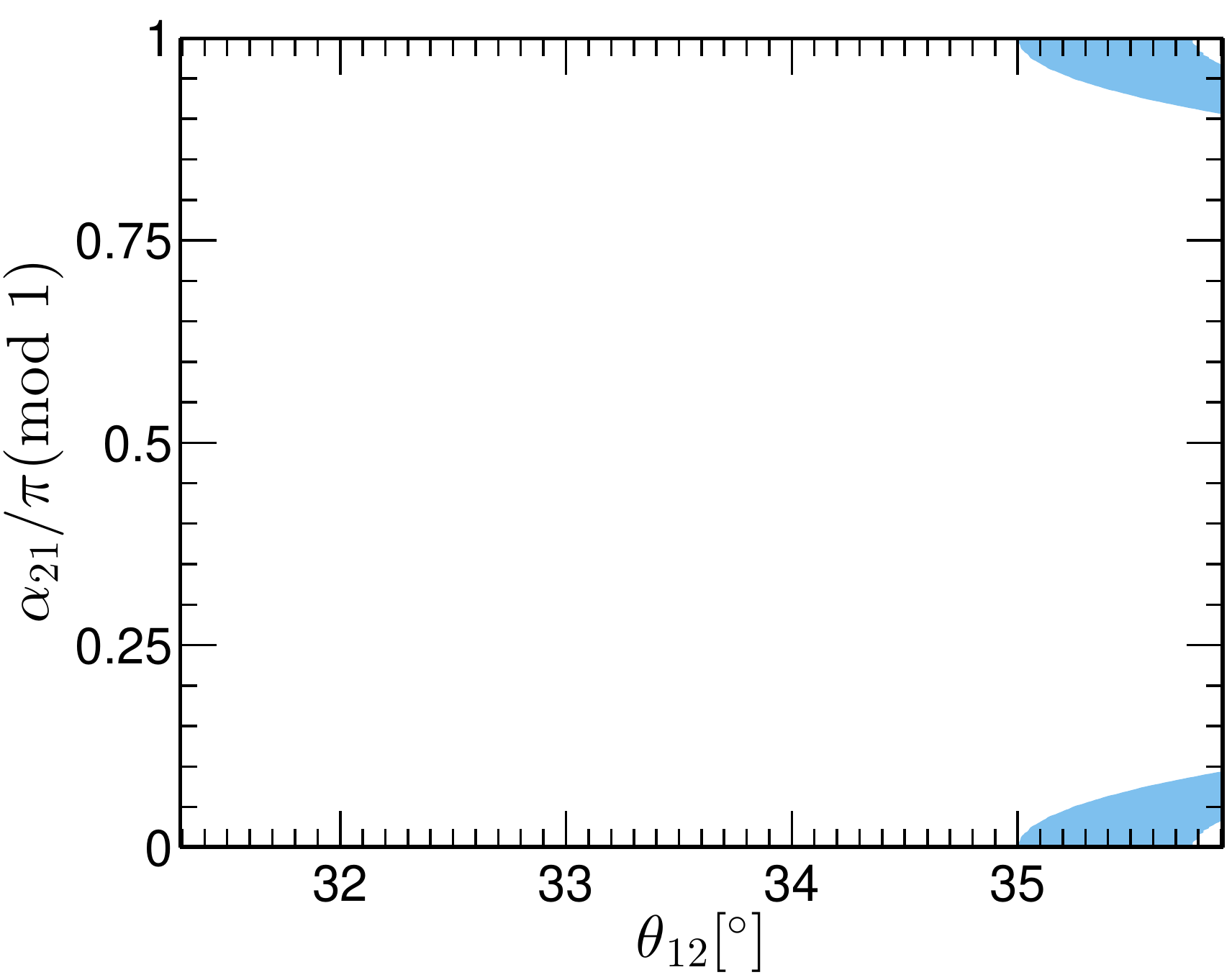}
\includegraphics[height=0.2\textwidth]{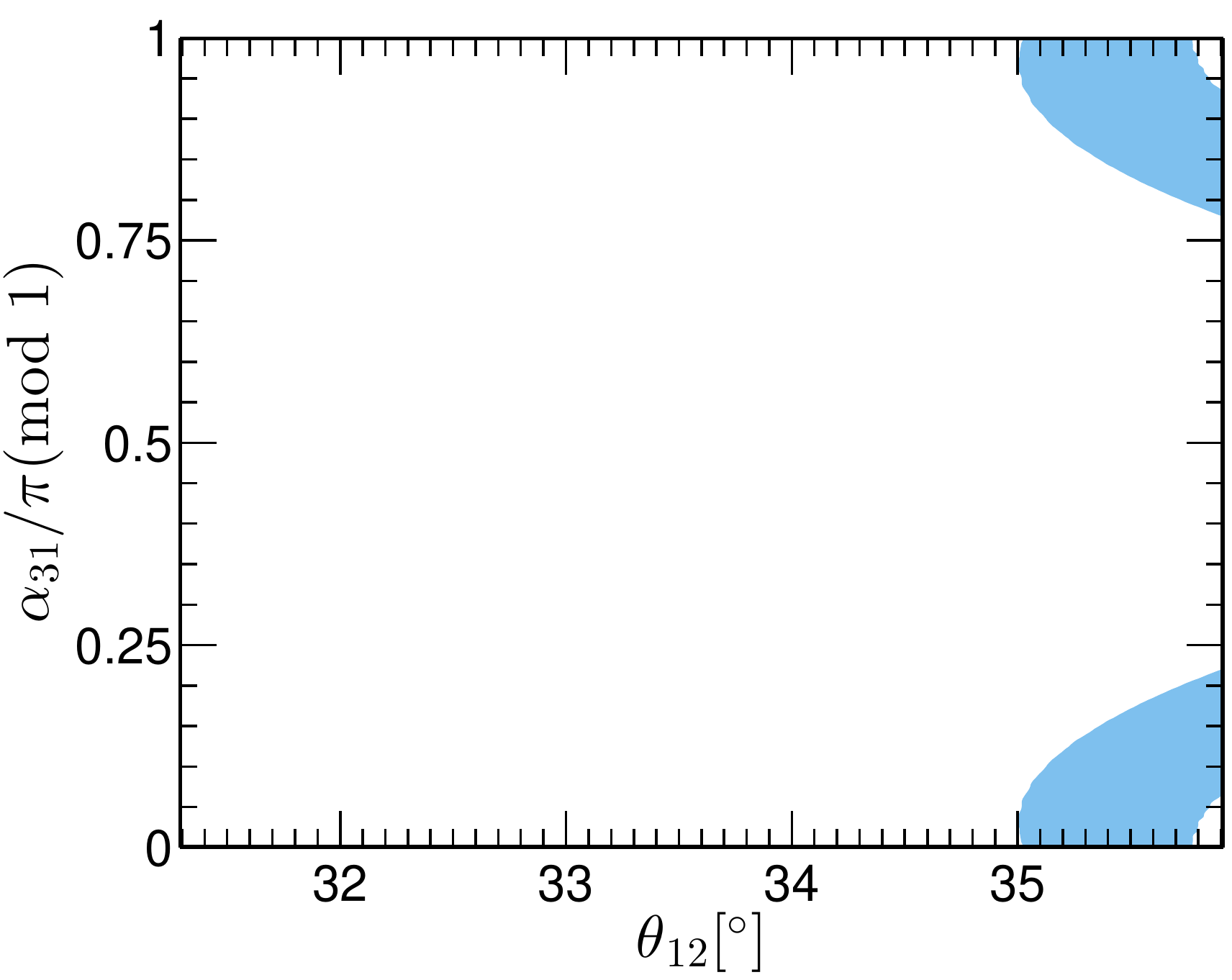}
\includegraphics[height=0.2\textwidth]{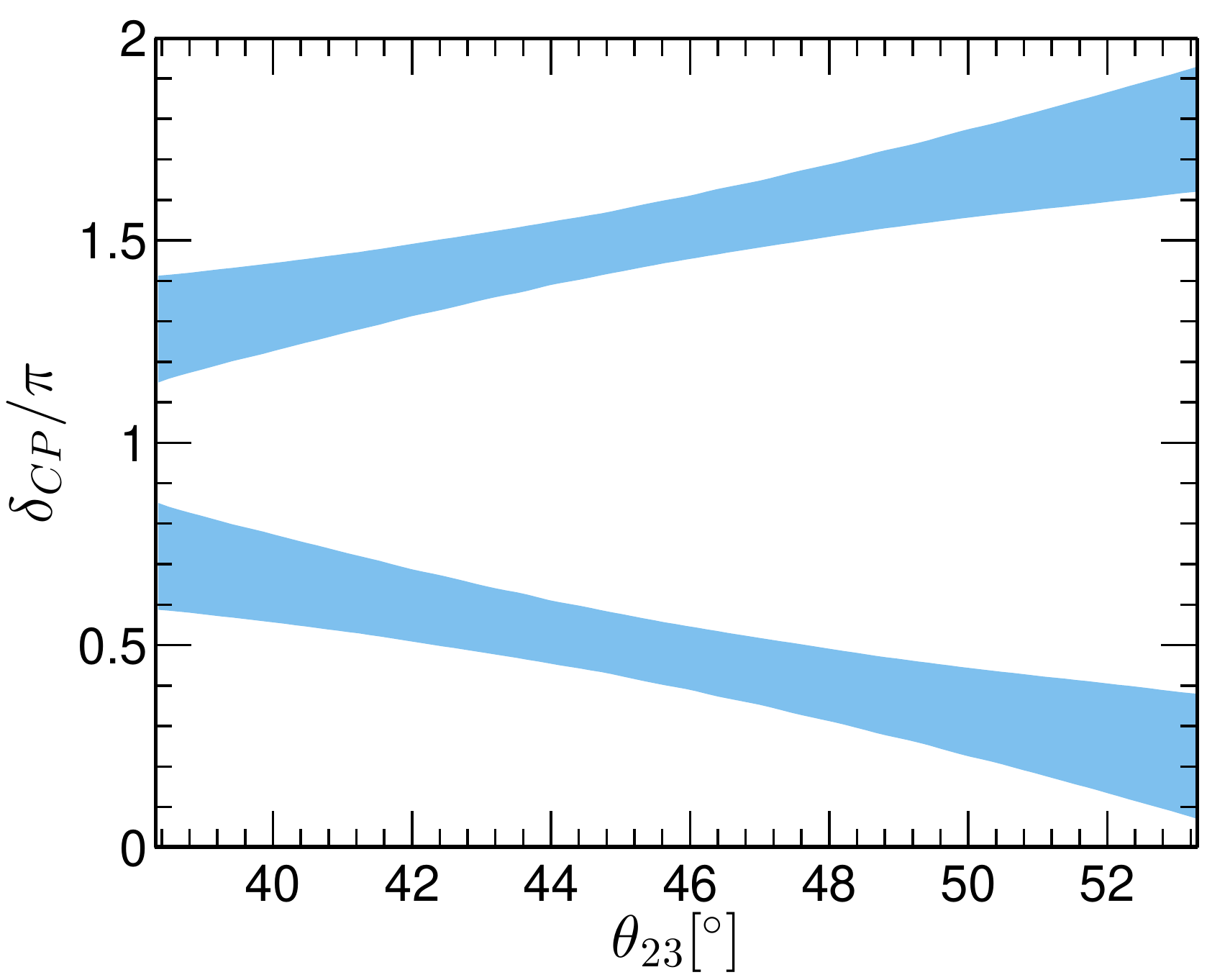}
\includegraphics[height=0.2\textwidth]{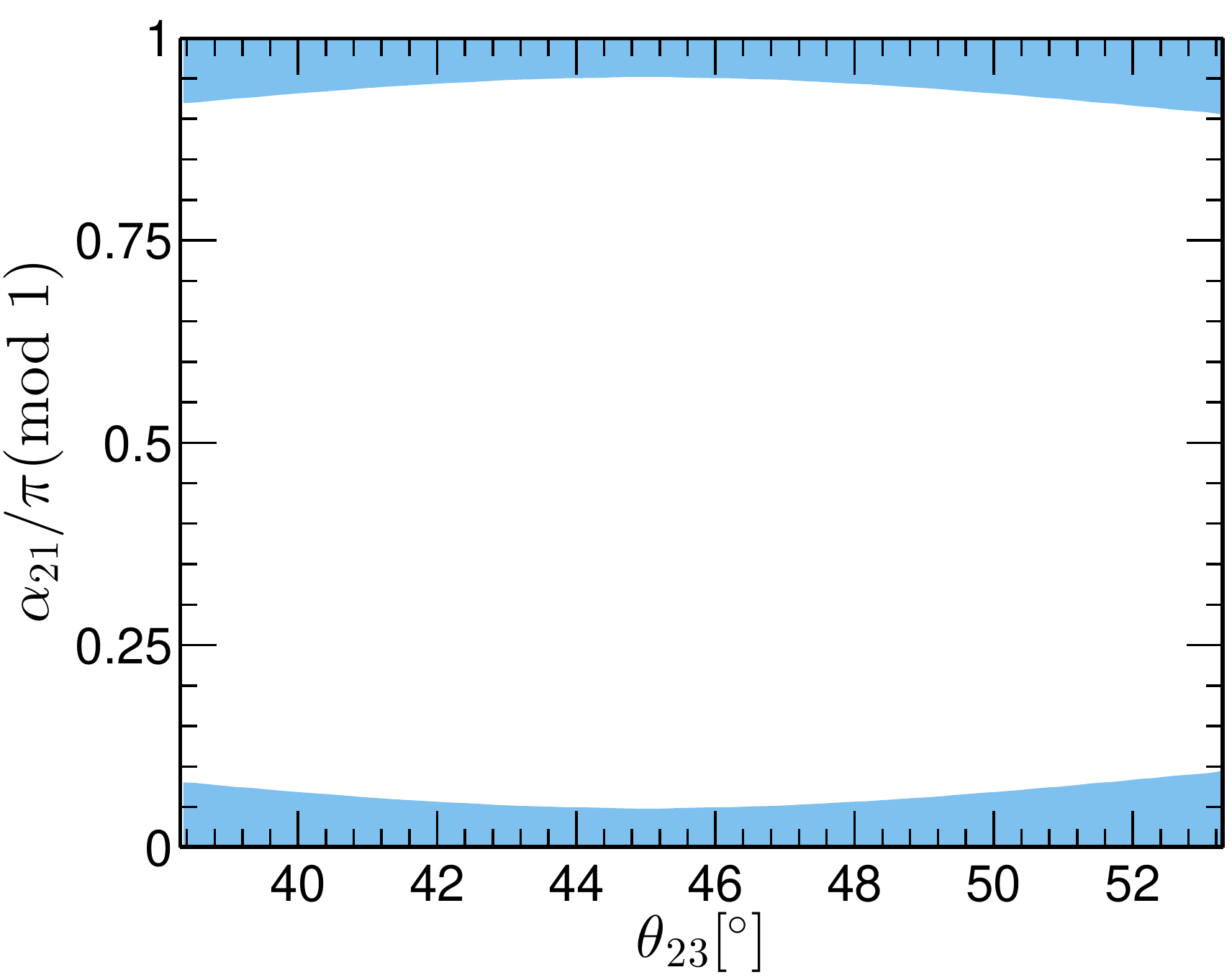}
\includegraphics[height=0.2\textwidth]{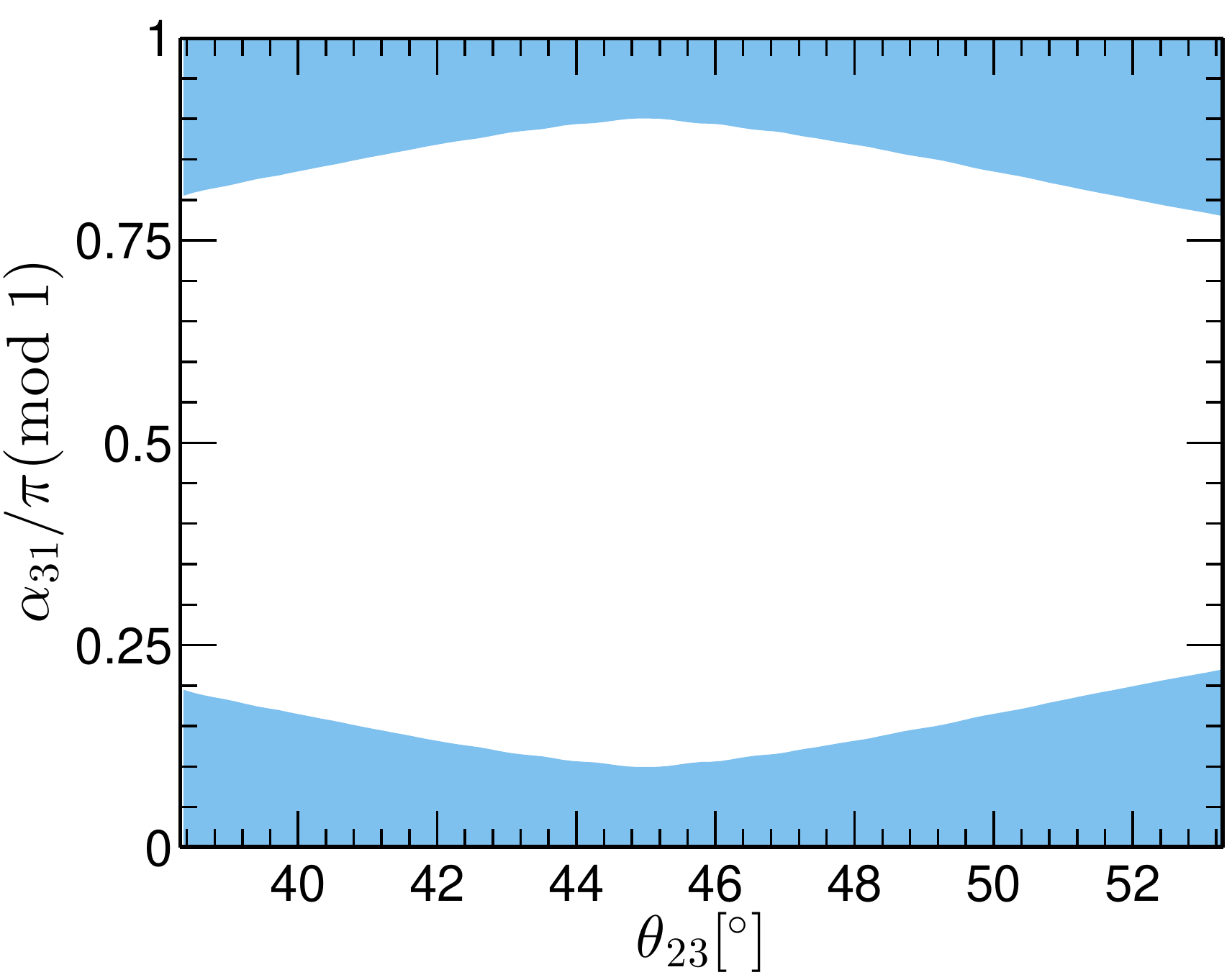}
\includegraphics[height=0.2\textwidth]{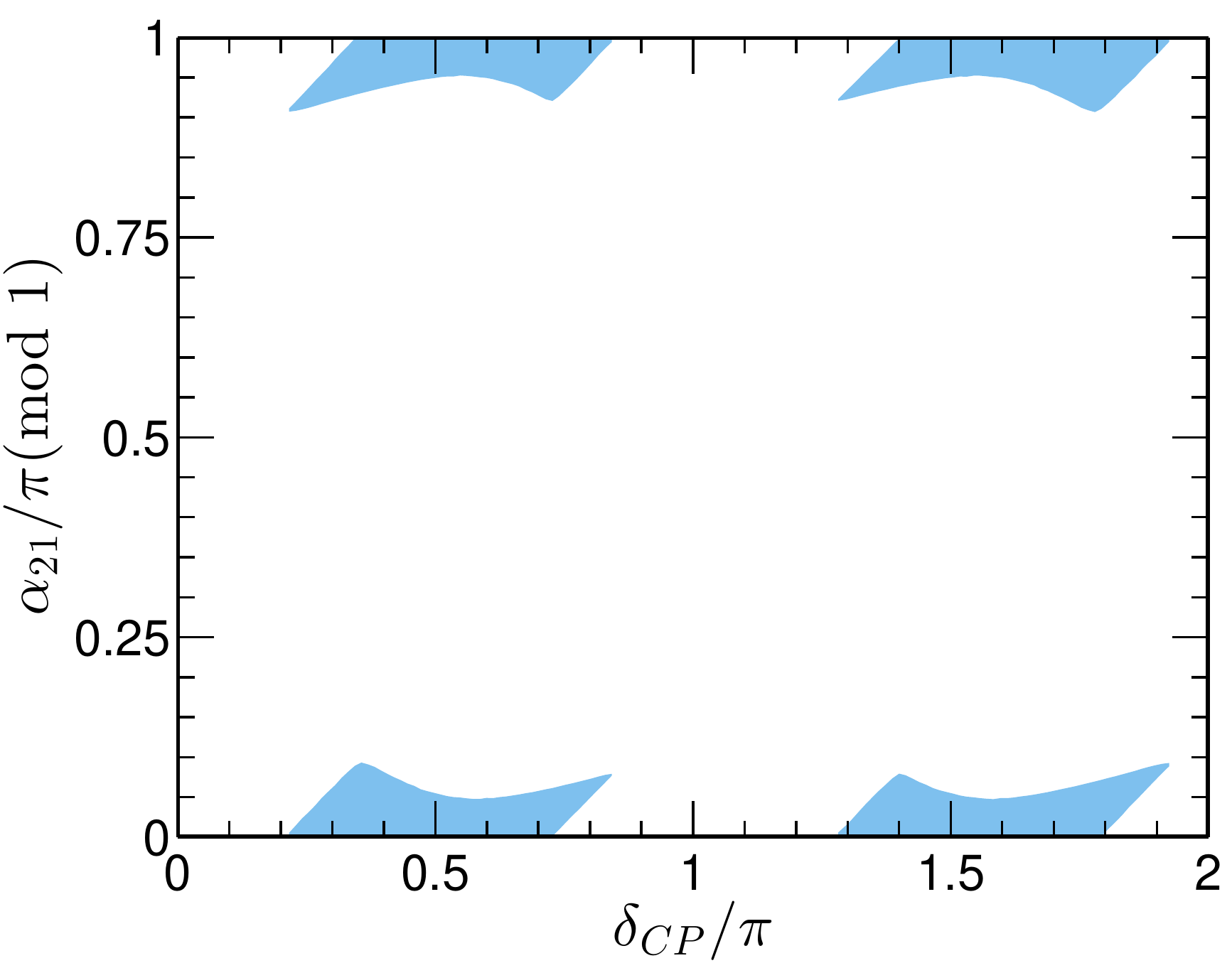}
\includegraphics[height=0.2\textwidth]{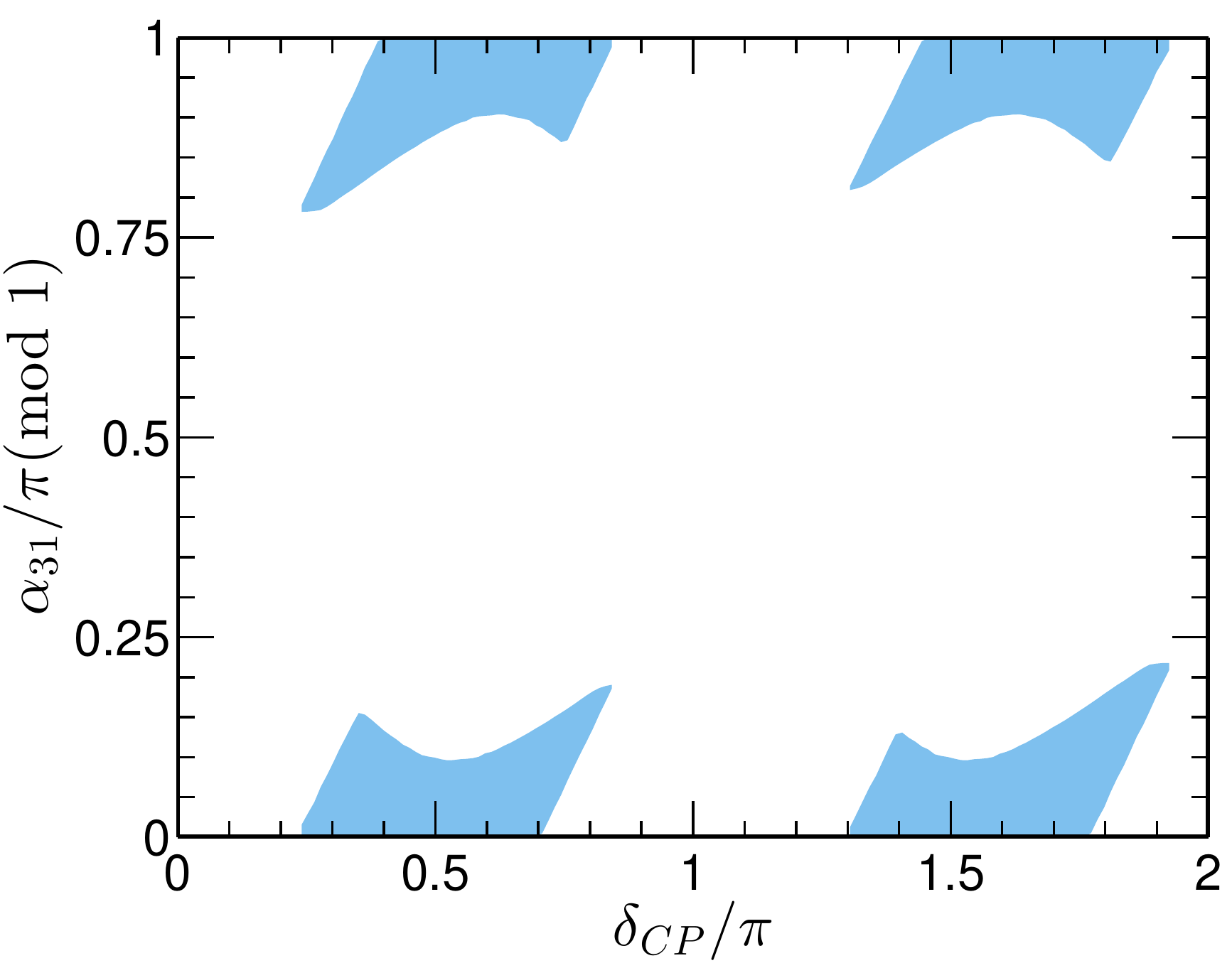}
\includegraphics[height=0.2\textwidth]{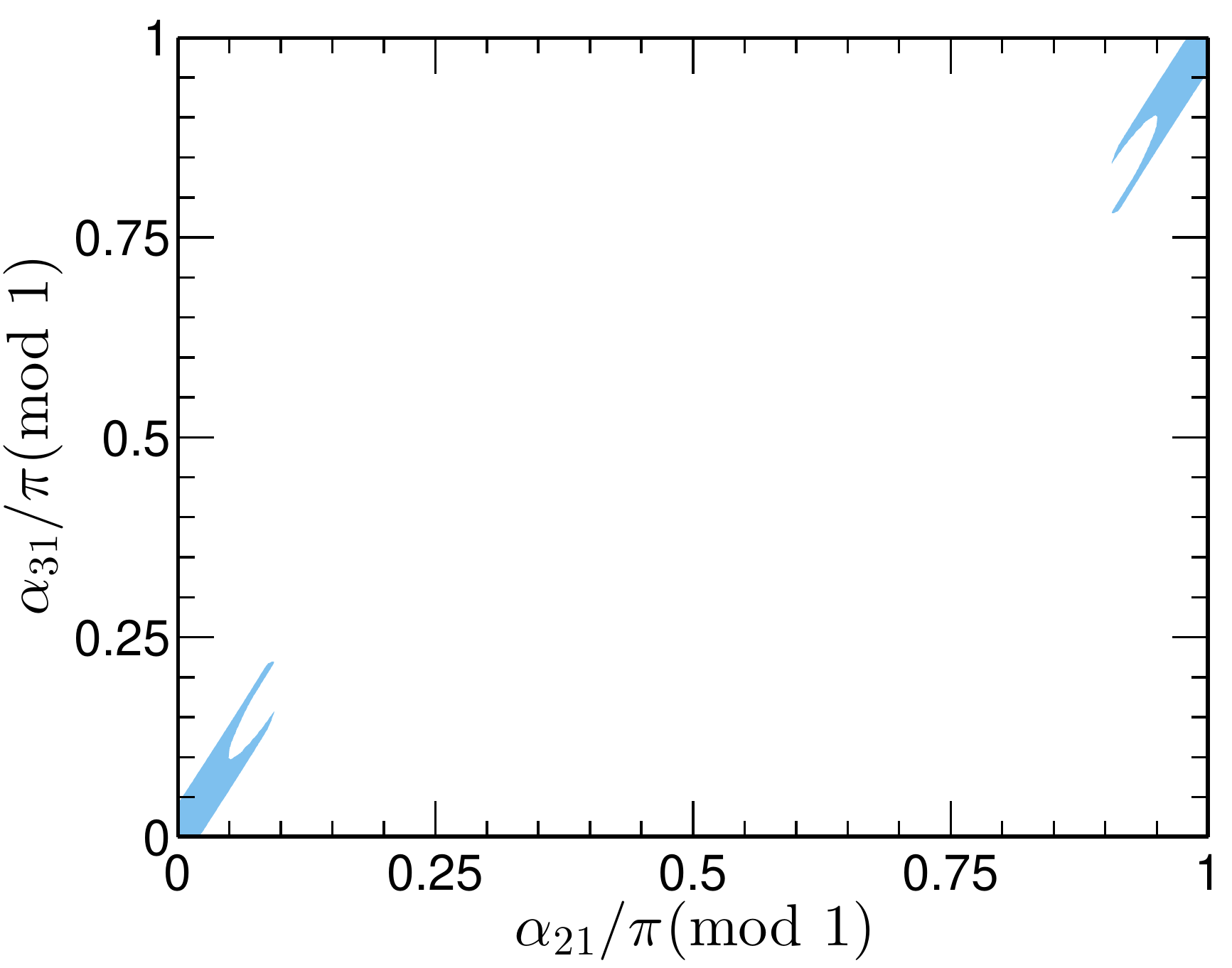}
\caption{\label{fig:Z3TSU_single}Correlations between different mixing parameters in the case of $(G_{l},X_{\nu})=(Z_{3}^{T},SU)$, where the three lepton mixing angles are required to be compatible with the experimental data at $3\sigma$ level~\cite{Gonzalez-Garcia:2014bfa}.}
\end{figure}

\item{$G_{l}=K_{4}^{(S,U)},X_{\nu}=T$}

The unitary transformations $U_{l}$ and $\Sigma_{\nu}$ are fixed to be
\begin{equation}
U_{l}=\frac{1}{\sqrt{6}}\left(
\begin{array}{ccc}
 2 ~&~ \sqrt{2} ~&~ 0 \\
 -1 ~&~ \sqrt{2} ~&~ \sqrt{3} \\
 -1 ~&~ \sqrt{2} ~&~ -\sqrt{3}
\end{array}
\right),\qquad \Sigma_{\nu}=\left(
\begin{array}{ccc}
 ~1~ & ~0~ & ~0~ \\
 0 & e^{-i\frac{\pi}{3}} & 0  \\
 0  & 0 & e^{i\frac{\pi}{3}}
\end{array}
\right)\,.
\end{equation}
The agreement with experimental data on lepton mixing angles can only be achieved for $P_{l}=1$, $P_{l}=P_{13}$, $P_{l}=P_{23}$ and $P_{l}=P_{23}P_{13}$. The two permutations $P_{l}=1$ and $P_{l}=P_{23}$ lead to equivalent PMNS mixing matrices as $P_{l}=P_{13}$ and $P_{l}=P_{23}P_{13}$ respectively. In the case of $P_{l}=P_{23}P_{13}$, we can read out the mixing angles and $CP$ invariants as follows,
\begin{eqnarray}
\nonumber\sin^{2}\theta_{13}&=&\frac{1}{4} (\sin 2 \theta _1+2) \cos ^2\theta_2\,,\\
\nonumber
\sin^{2}\theta_{12}&=&\sin ^2\theta _3+\frac{(\sin 2 \theta _1-2) \cos 2 \theta _3+\sin \theta _2 \sin 2 \theta _3 \cos 2 \theta _1}{\cos 2 \theta _2+\sin 2 \theta _1 \cos ^2\theta _2-3}\,,\\
\nonumber
\sin^{2}\theta_{23}&=&\frac{2 \big(\sqrt{2} \sin (\theta _1+\frac{\pi}{4}) \sin 2 \theta _2-\sin ^2\theta _2\big)}{3 (\cos 2 \theta _2+\sin 2 \theta _1 \cos ^2\theta _2-3)}+\frac{1}{3}\,,\\
\nonumber
J_{CP}&=&\frac{1}{128 \sqrt{3}}\Big[4 \sin \theta _2 \sin 2 \theta _3+4 \sin 3 \theta _2 \sin 2 \theta _3-8 \sqrt{2} (\sin 2 \theta _1+2) \sin 2 \theta _2 \cos (\theta _1+\frac{\pi }{4}) \cos2\theta_3\\
\nonumber
&~&
-2 \sqrt{2} (\sin 2 \theta _1+4) \sin (\theta _1+\frac{\pi }{4}) \sin 2 \theta _3 \cos 3 \theta _2\\
\nonumber
&~&
-\sqrt{2}\big(3 \sin (\theta _1+\frac{\pi }{4})+5 \cos (3 \theta _1+\frac{\pi }{4})\big) \sin 2 \theta _3 \cos \theta _2\Big]\,,\\
\nonumber
I_{1}&=&\frac{(-1)^{k_{1}}}{32} \sqrt{3} \Big[\big[\big(4 \cos ^2\theta _2+\sin 2 \theta _1 (\cos 2 \theta _2-3)\big)\sin 2 \theta _3+4 \sin \theta _2 \cos 2 \theta _1 \cos 2 \theta _3\big]\sin \theta _2\Big]\,,\\
I_{2}&=&\frac{(-1)^{k_{2}+1}}{8} \sqrt{3} \Big[\big((\sin 2 \theta _1+2) \sin \theta _2 \cos \theta _3+\sin \theta _3 \cos 2 \theta _1\big)\sin \theta _3 \cos ^2\theta _2\Big]\,.
\end{eqnarray}
For another independent permutation $P_{l}=P_{13}$, the atmospheric angle changes from $\theta_{23}$ to $\pi/2-\theta_{23}$, the Dirac phase turns out to be $\pi+\delta_{CP}$, and the expressions of the other mixing parameters are not changed. The numerical results for $P_{l}=P_{13}$ and $P_{l}=P_{23}P_{13}$ are plotted in figure~\ref{fig:K4SUT_single}. There are no preferred values of $\delta_{CP}$ within the viable parameter space. The atmospheric mixing angle $\theta_{23}$ is non-maximal, and it lies in the interval $[38.3^{\circ}, 40.5^{\circ}]\cup[49.5^{\circ}, 51.7^{\circ}]$.

\begin{figure}[hptb!]
\centering
\includegraphics[height=0.2\textwidth]{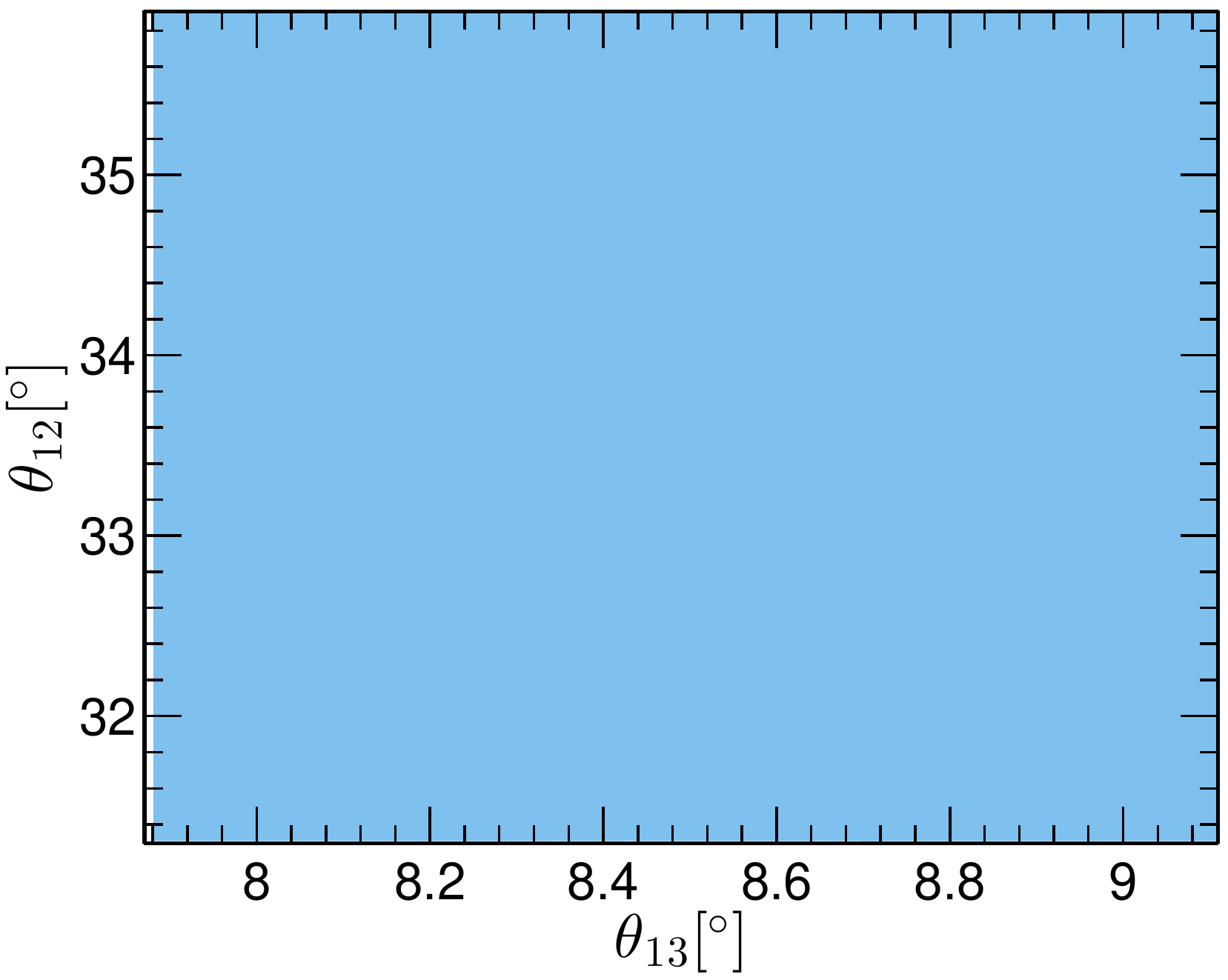}
\includegraphics[height=0.2\textwidth]{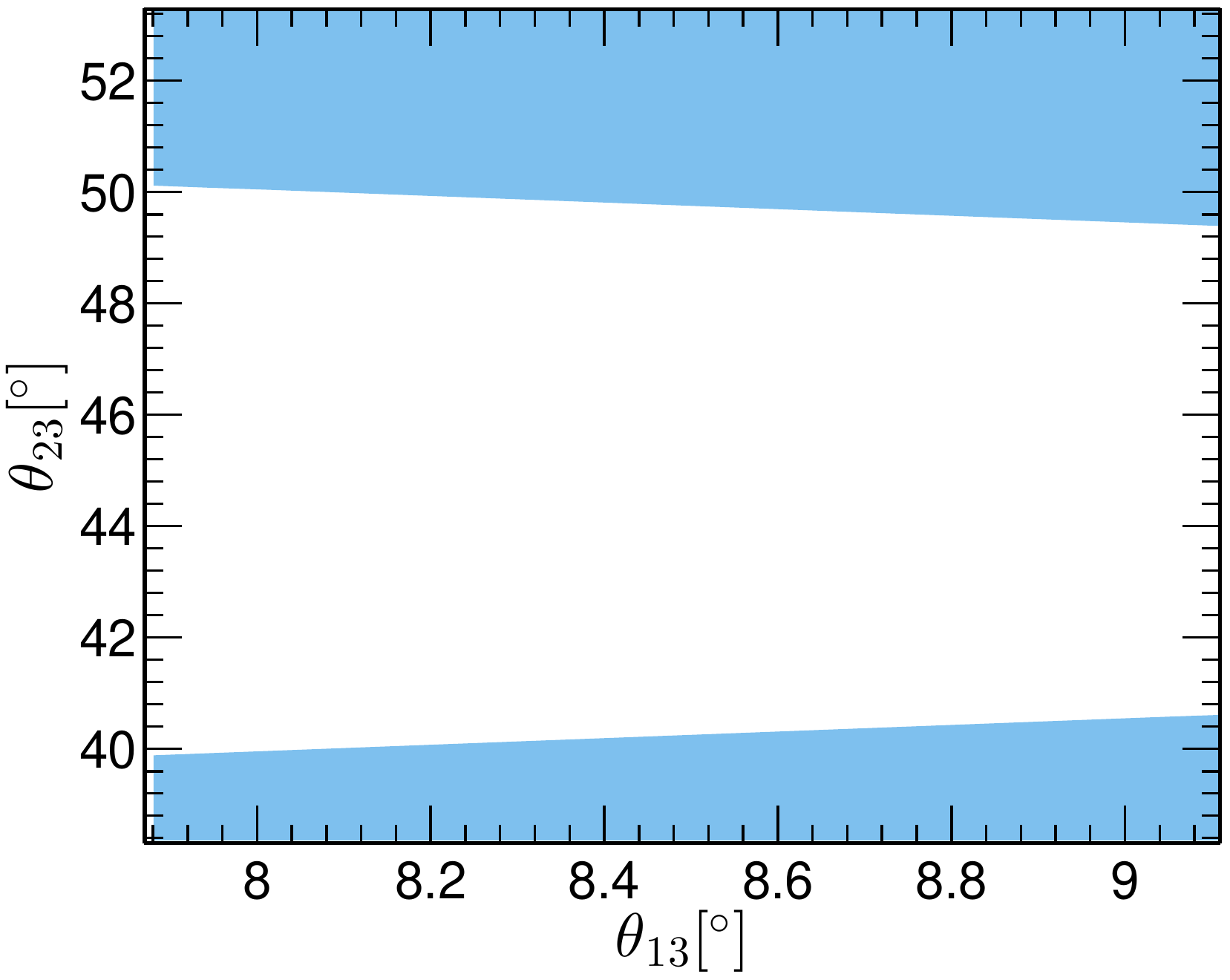}
\includegraphics[height=0.2\textwidth]{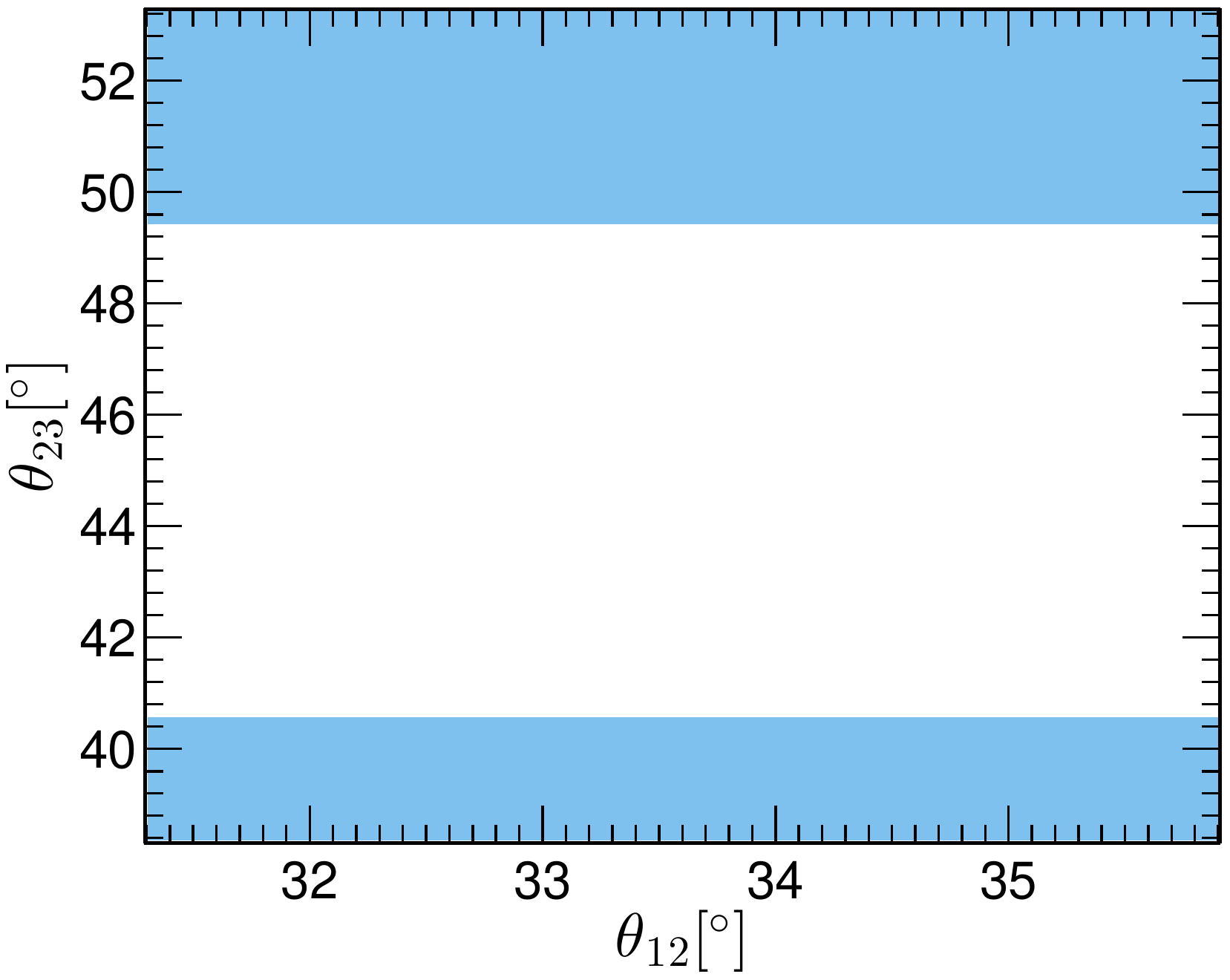}
\includegraphics[height=0.2\textwidth]{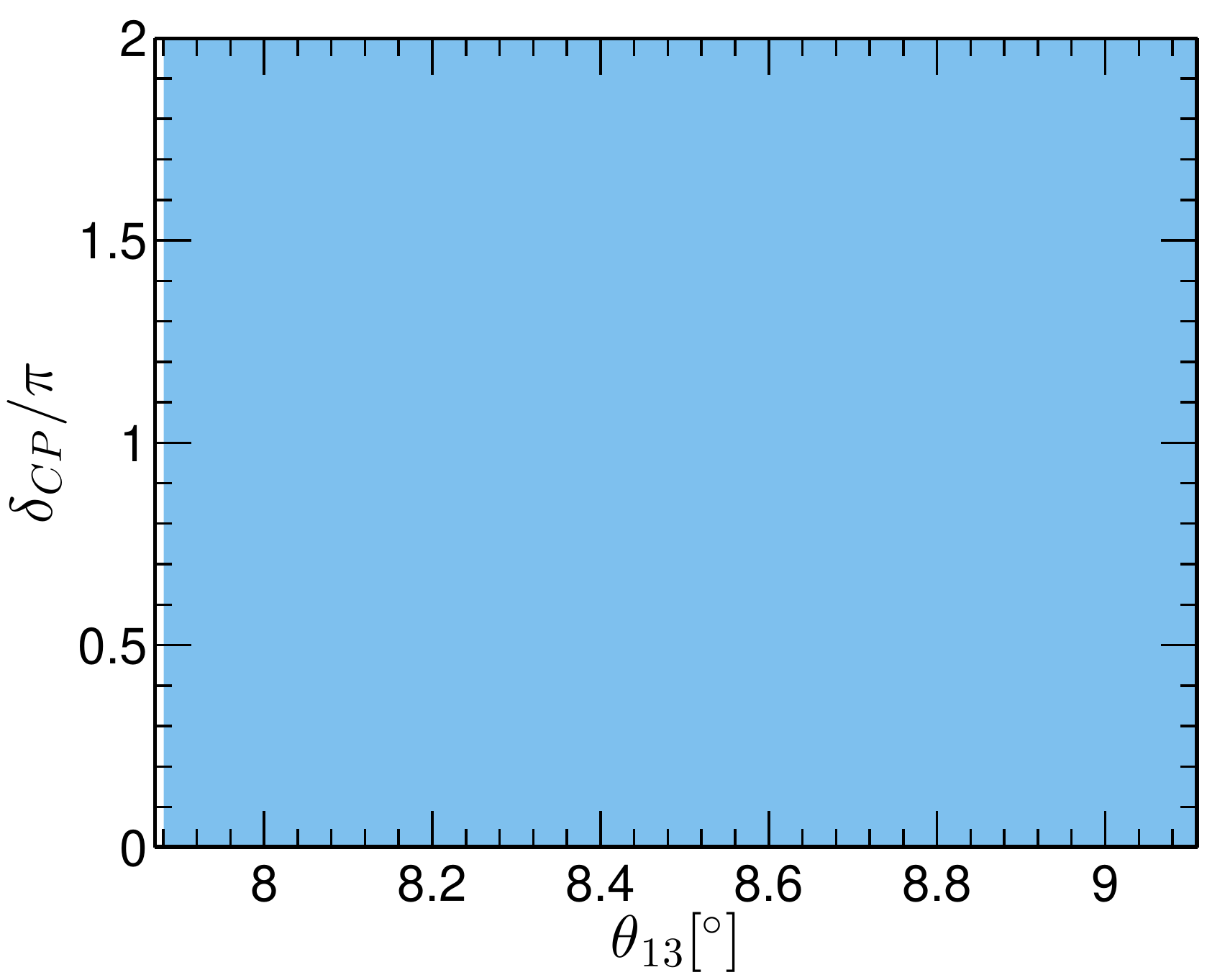}
\includegraphics[height=0.2\textwidth]{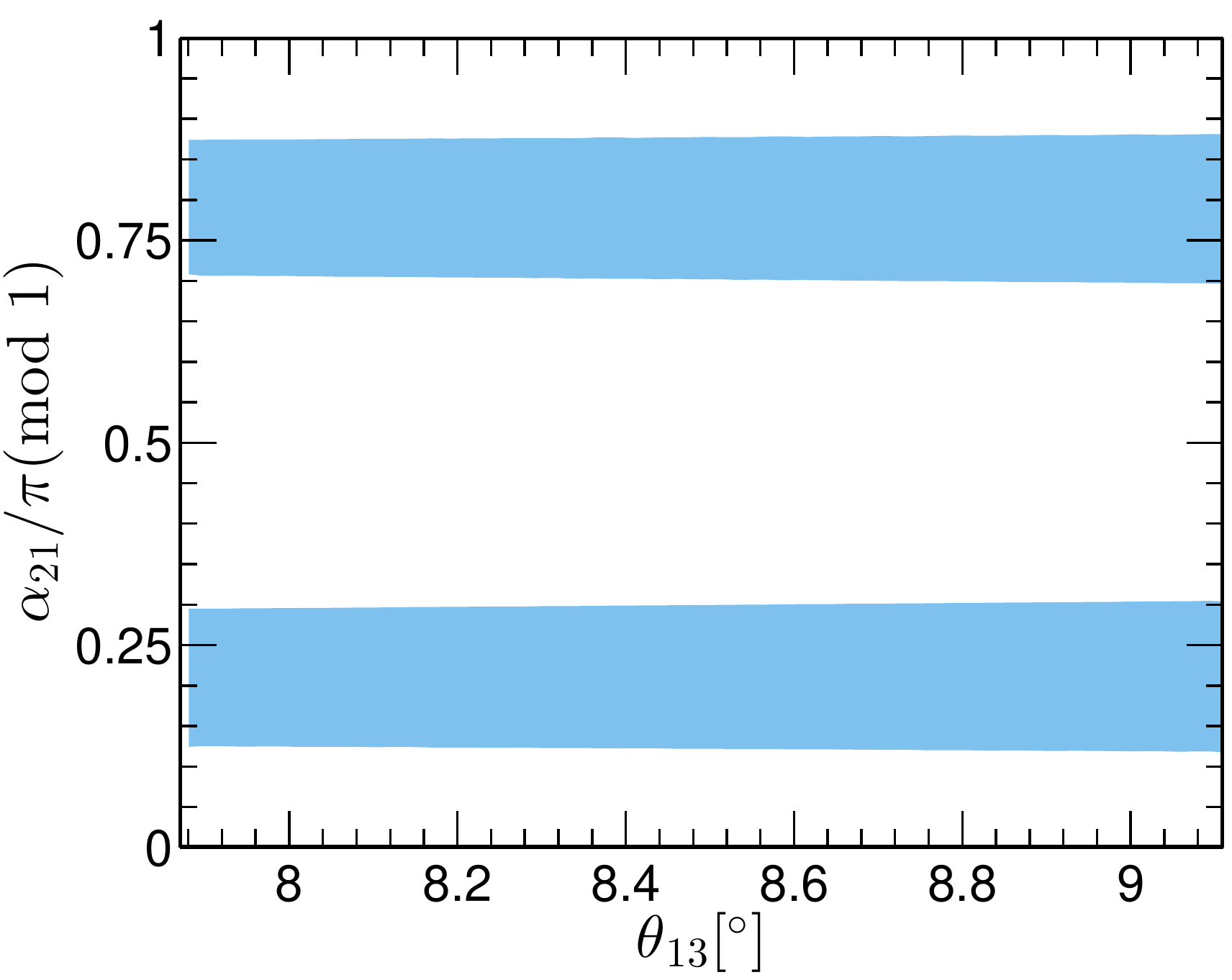}
\includegraphics[height=0.2\textwidth]{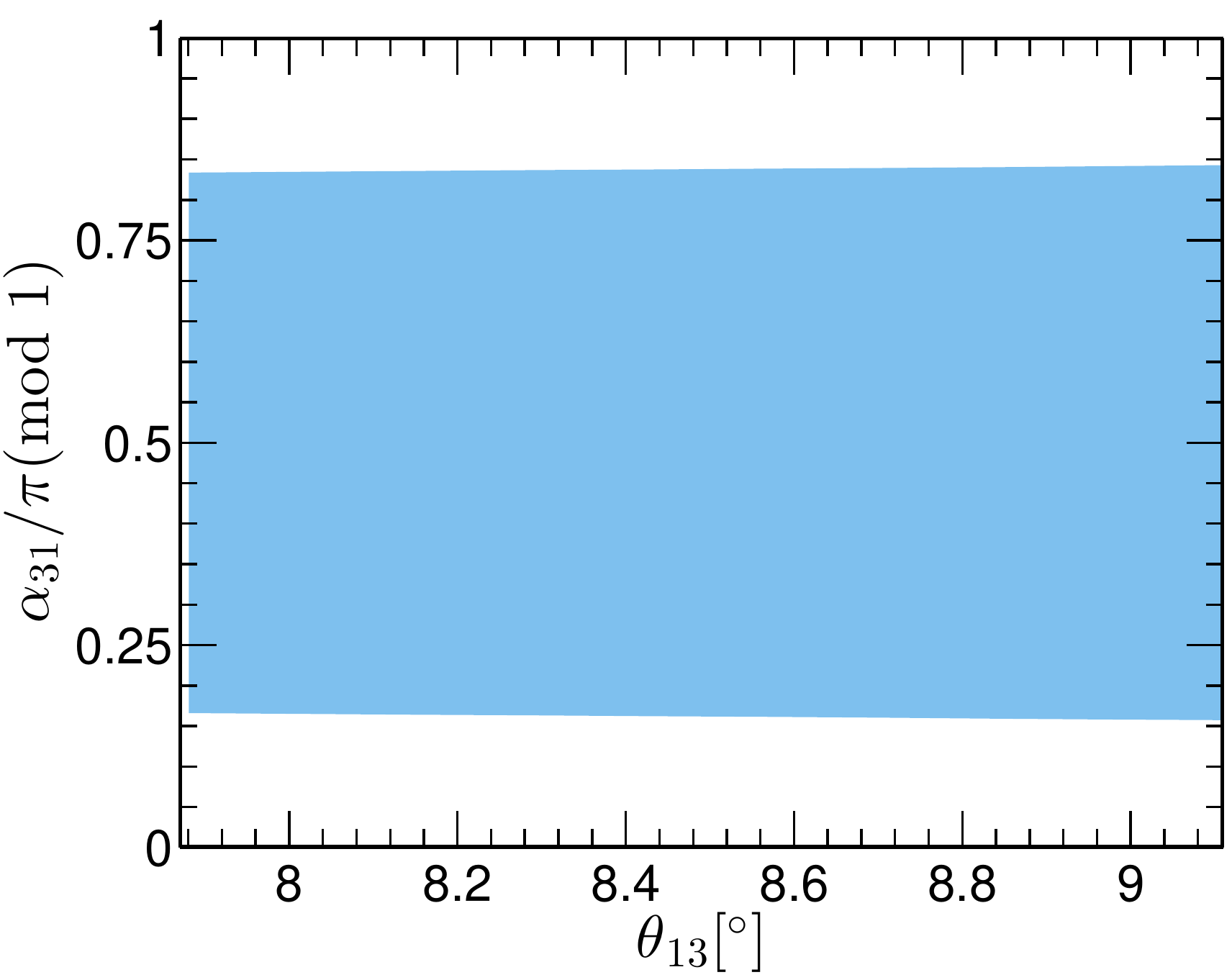}
\includegraphics[height=0.2\textwidth]{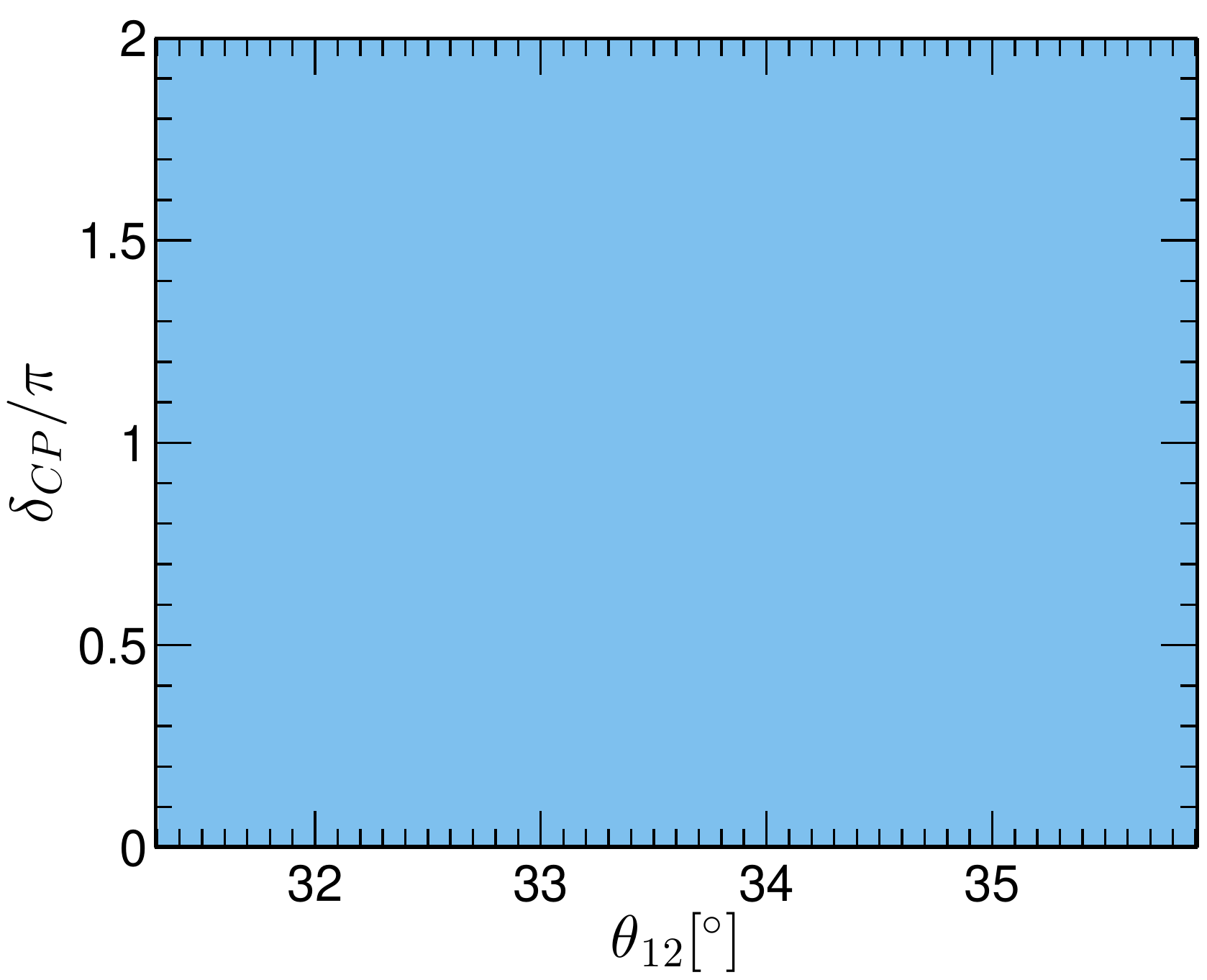}
\includegraphics[height=0.2\textwidth]{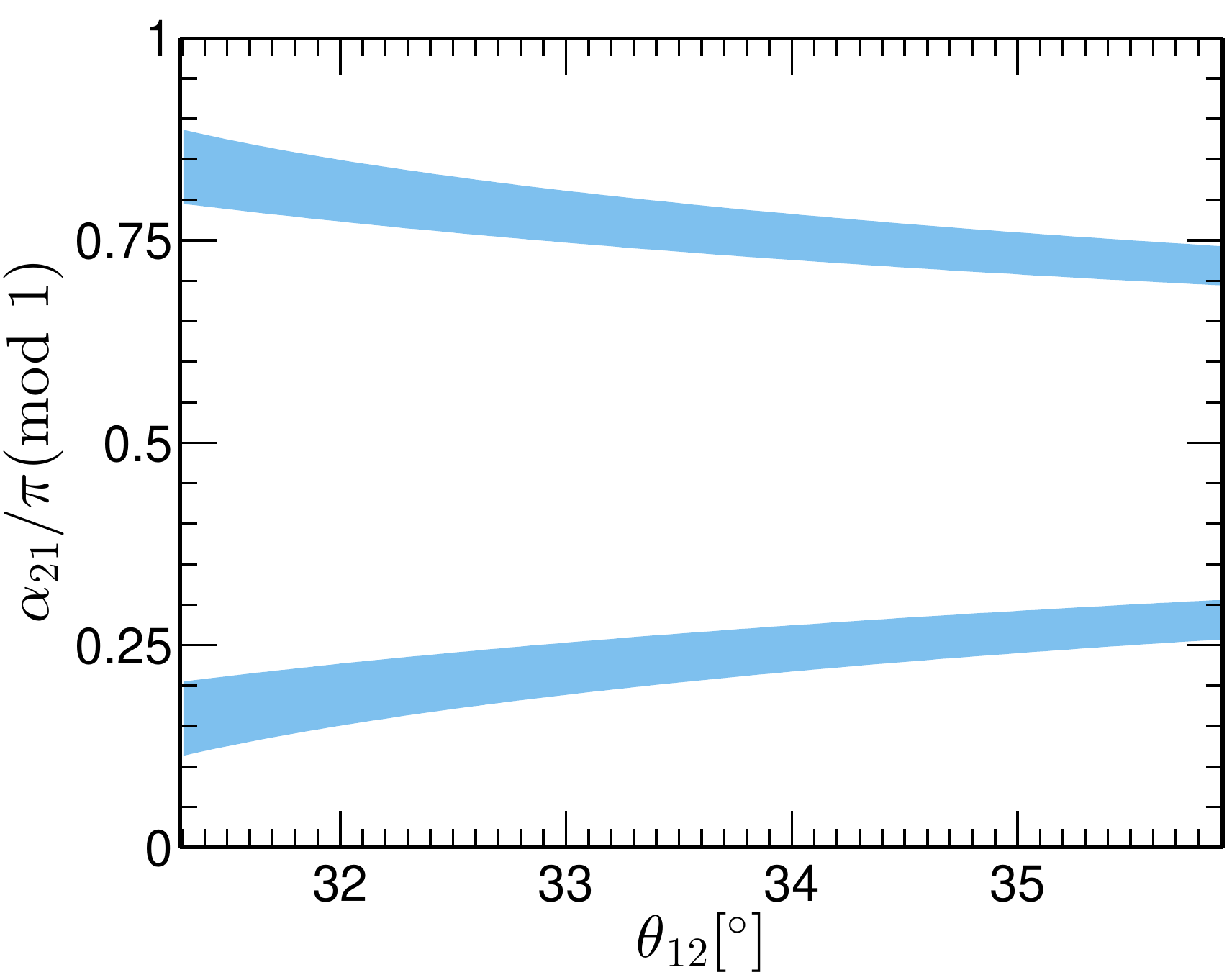}
\includegraphics[height=0.2\textwidth]{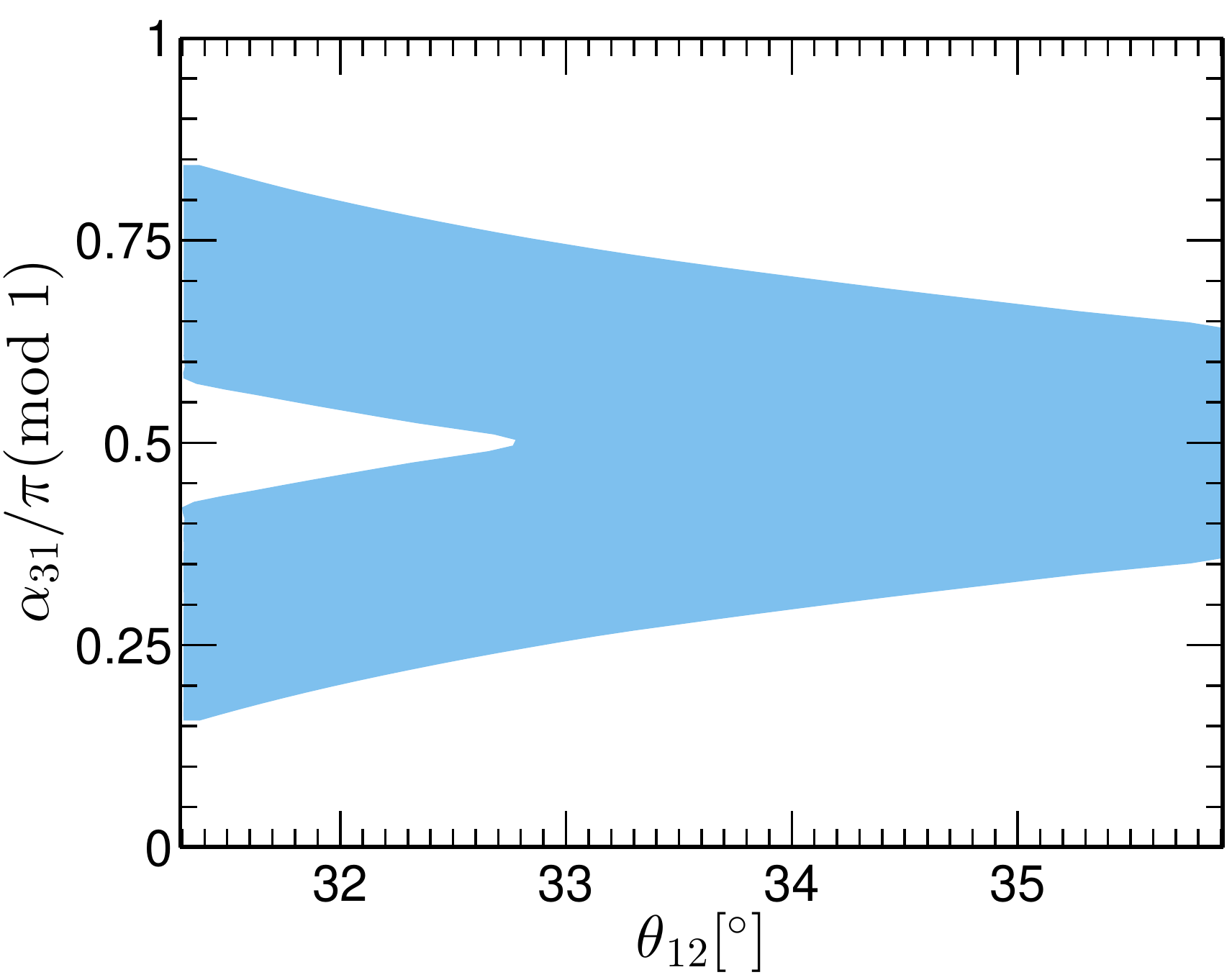}
\includegraphics[height=0.2\textwidth]{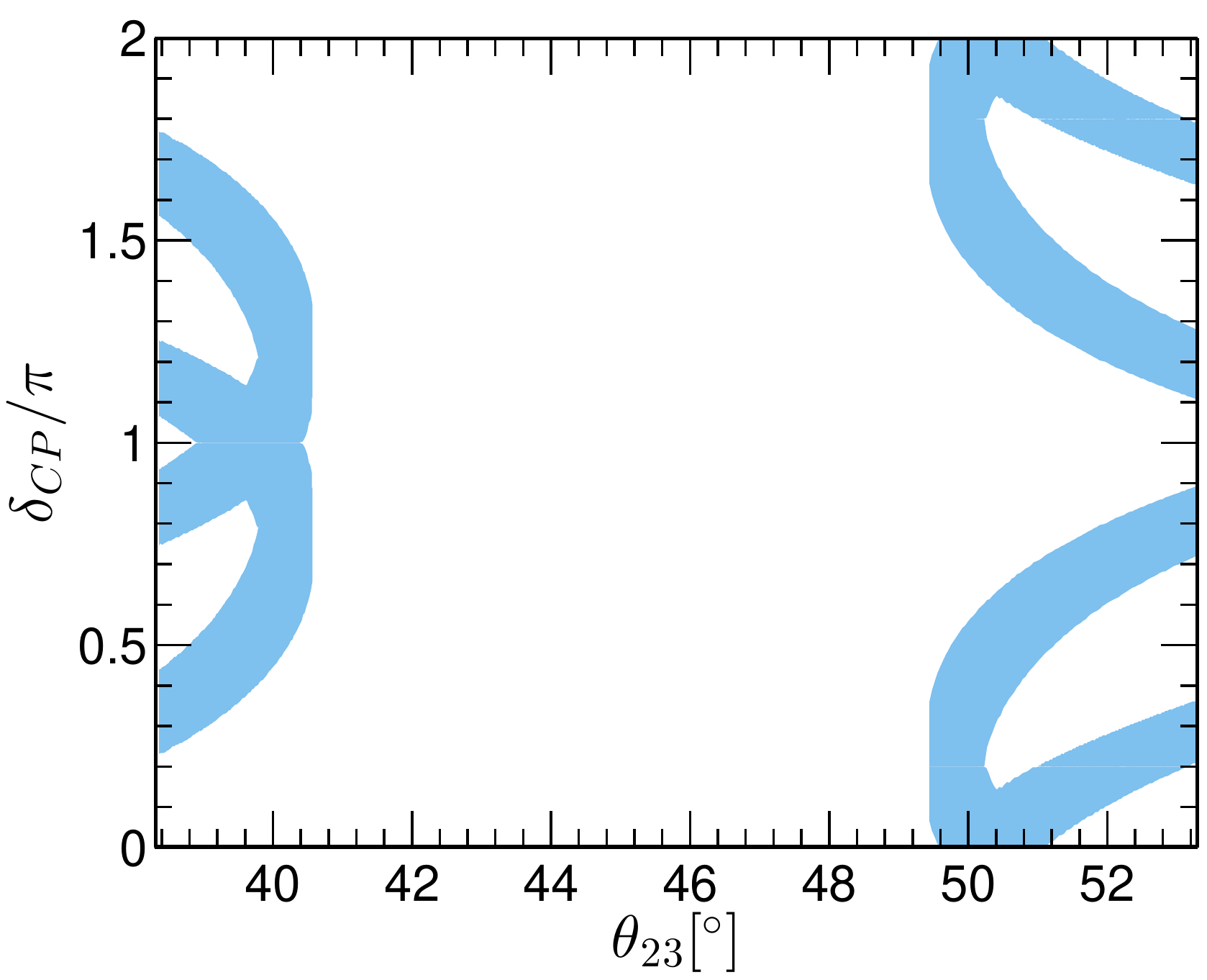}
\includegraphics[height=0.2\textwidth]{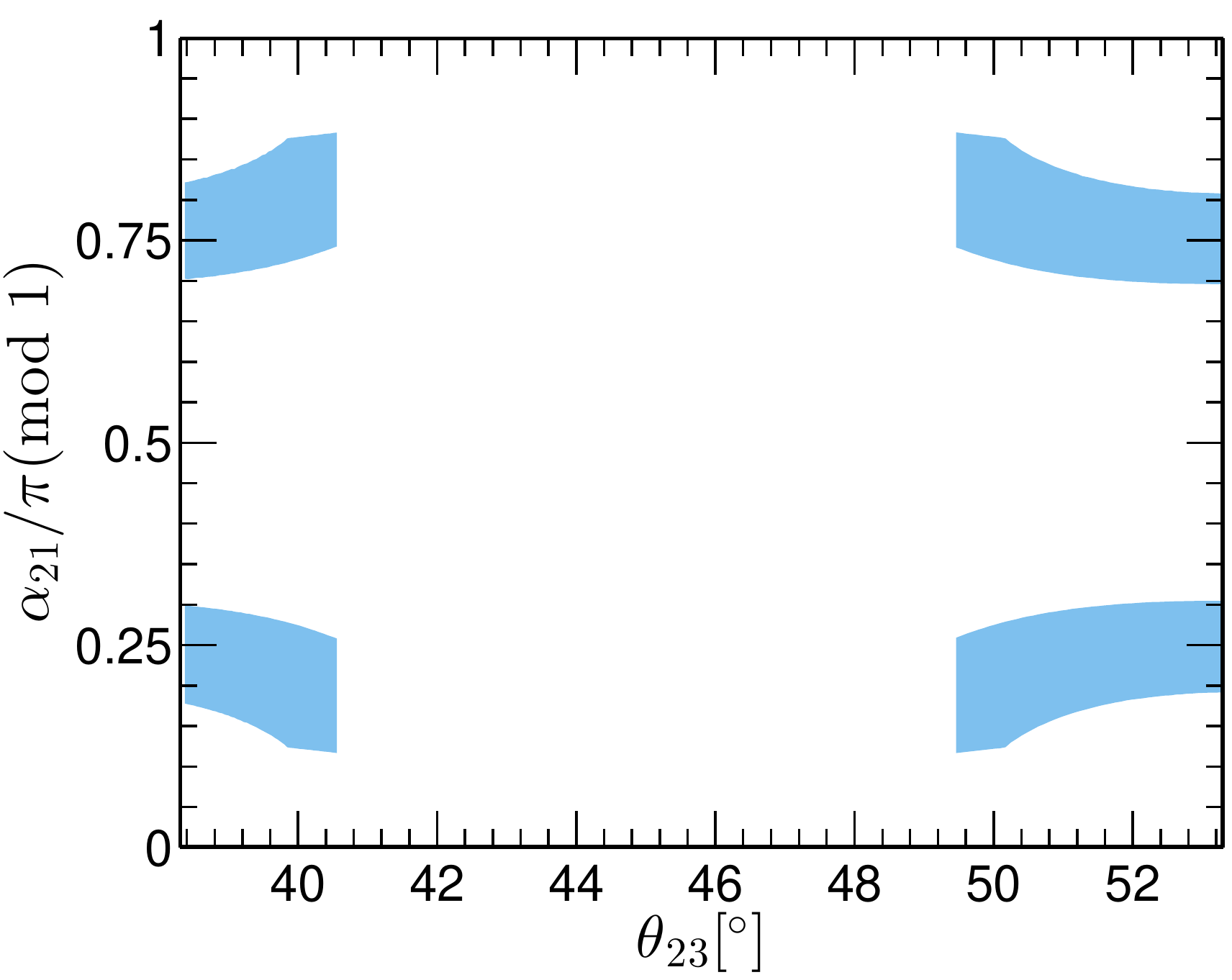}
\includegraphics[height=0.2\textwidth]{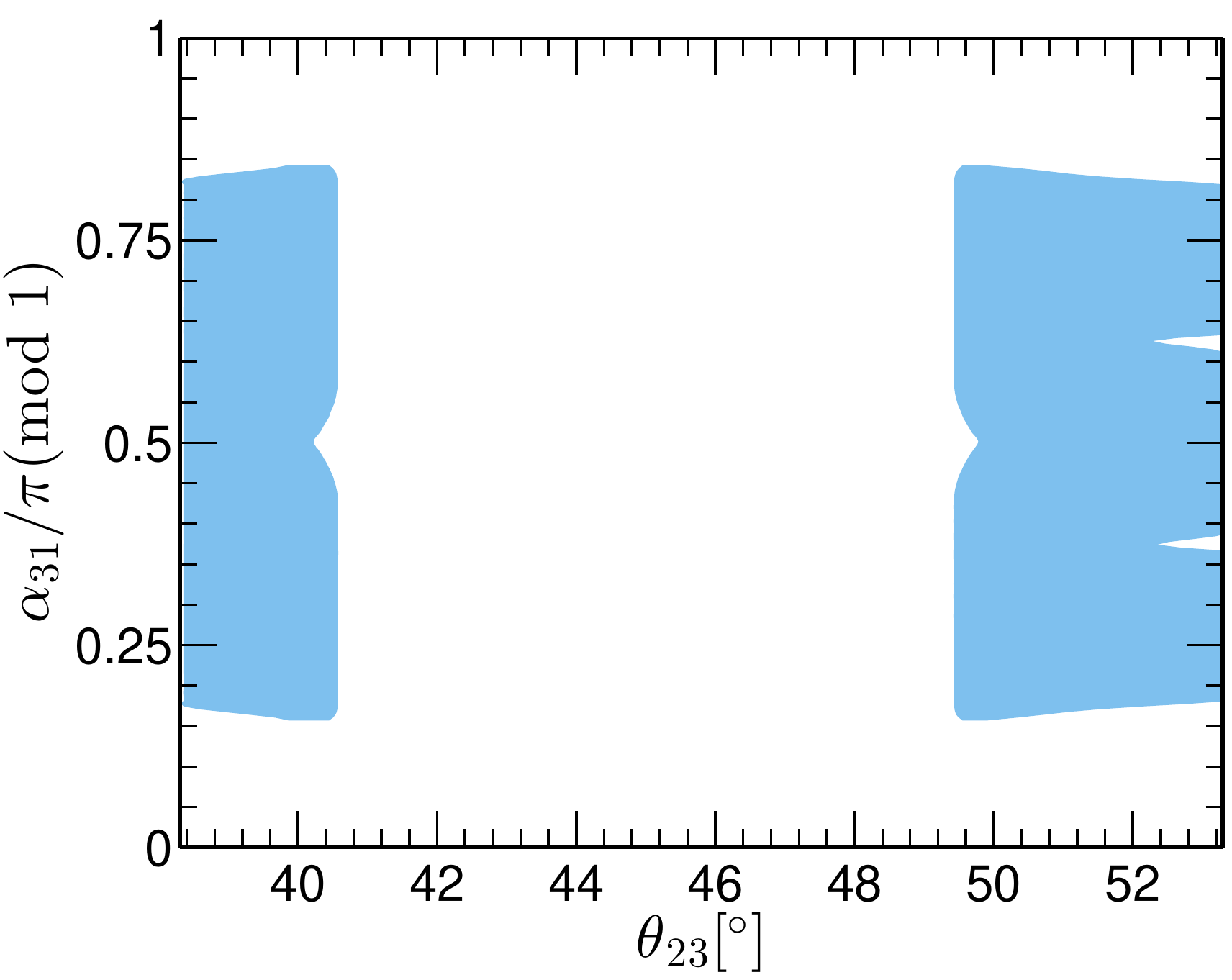}
\includegraphics[height=0.2\textwidth]{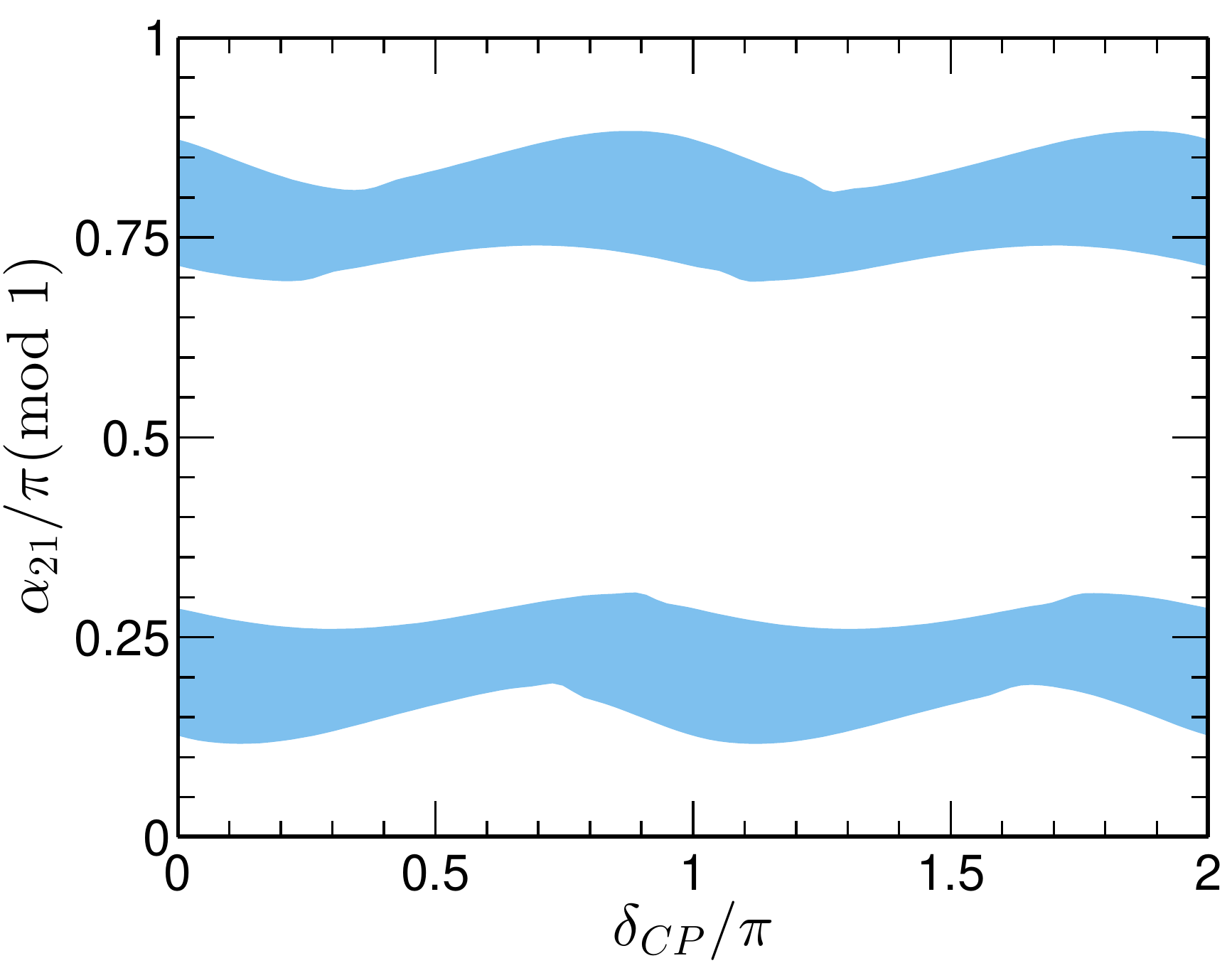}
\includegraphics[height=0.2\textwidth]{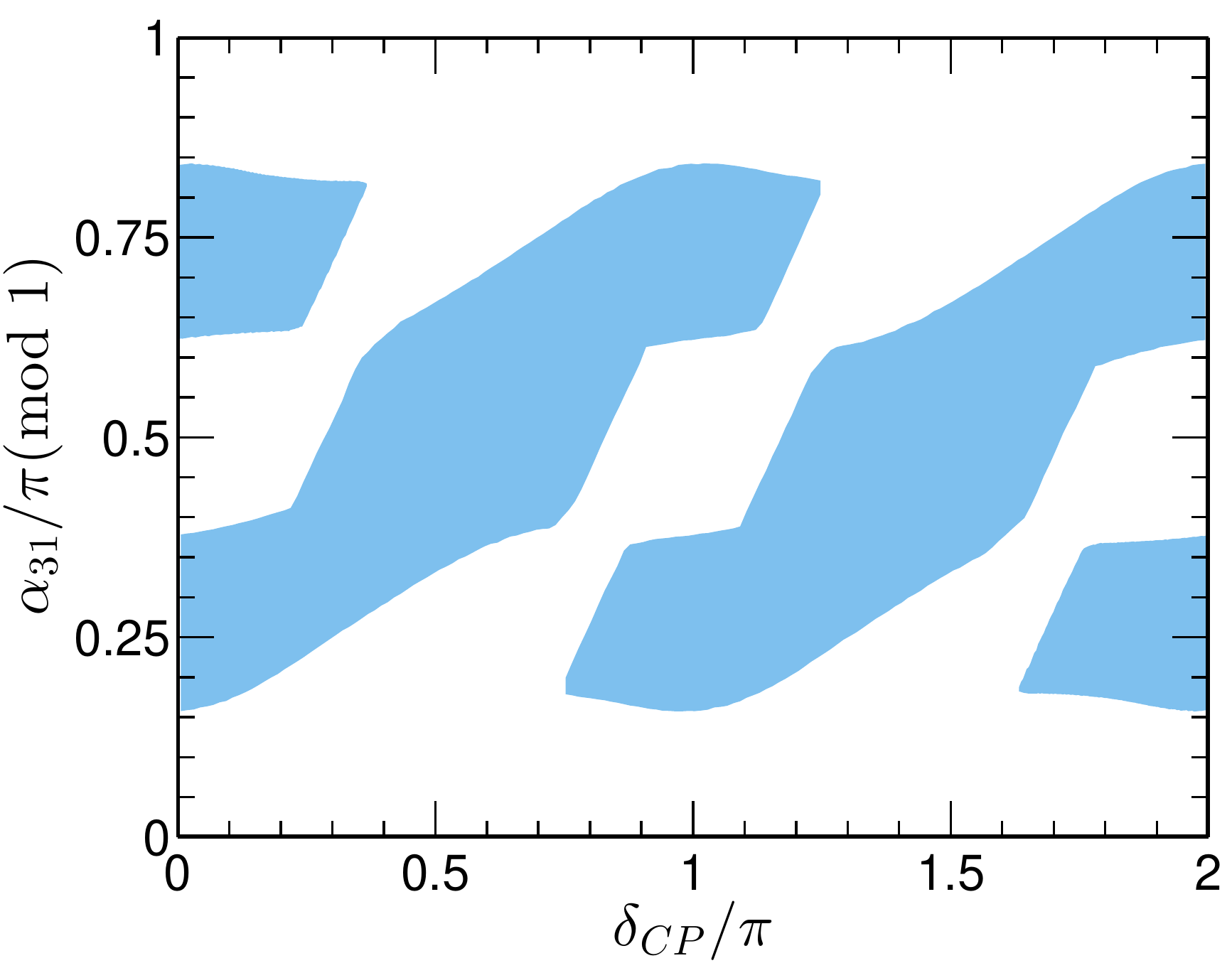}
\includegraphics[height=0.2\textwidth]{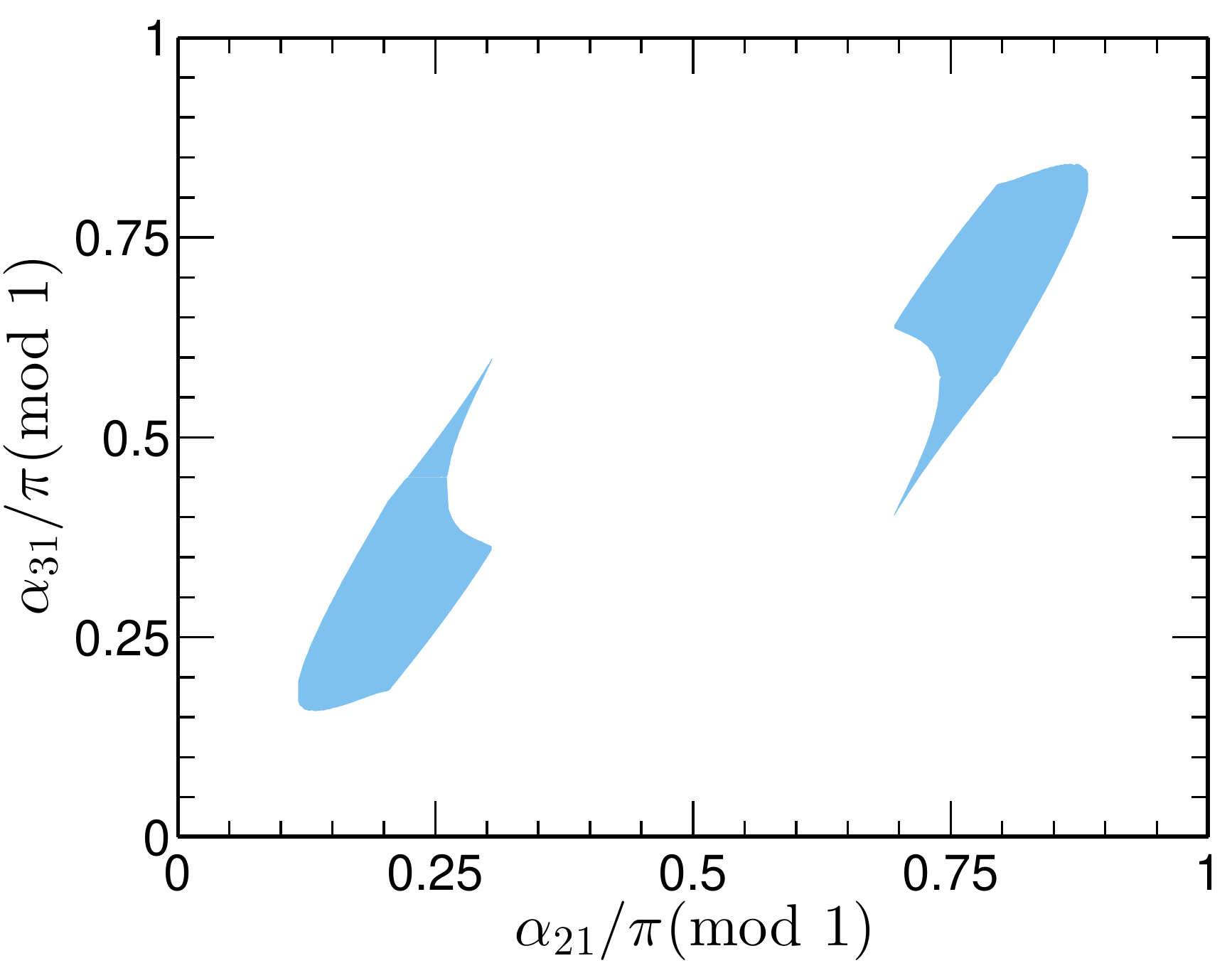}
\caption{\label{fig:K4SUT_single}Correlations between different mixing parameters in the case of $(G_{l},X_{\nu})=(K_{4}^{(S,U)},T)$, where the three lepton mixing angles are required to be compatible with the experimental data at $3\sigma$ level~\cite{Gonzalez-Garcia:2014bfa}. }
\end{figure}

\end{itemize}

\begin{figure}[hptb!]
\centering
\begin{tabular}{ >{\centering\arraybackslash} m{6.5cm} >{\centering\arraybackslash} m{6.5cm} }
\includegraphics[width=0.4\textwidth]{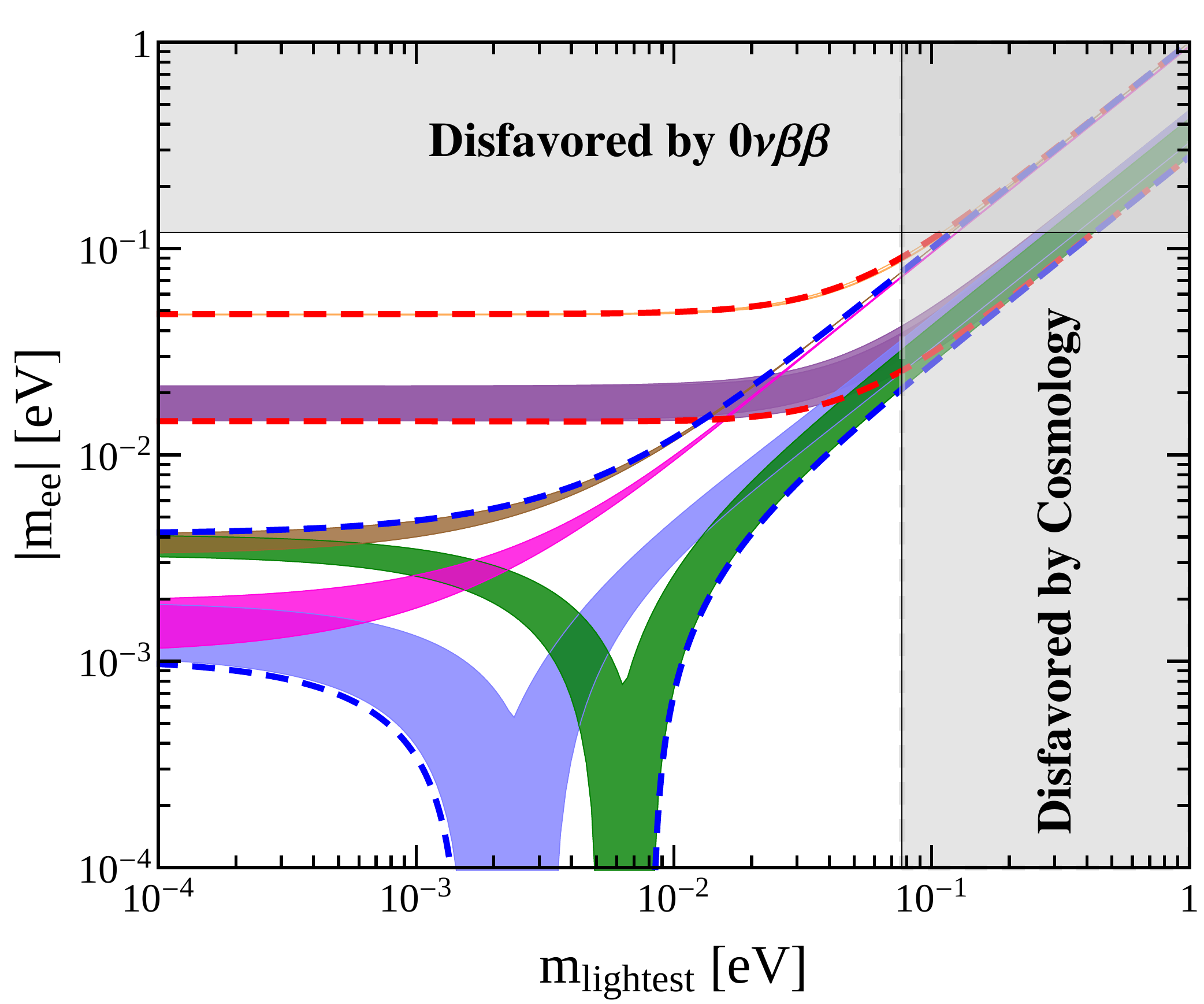}&
\includegraphics[width=0.4\textwidth]{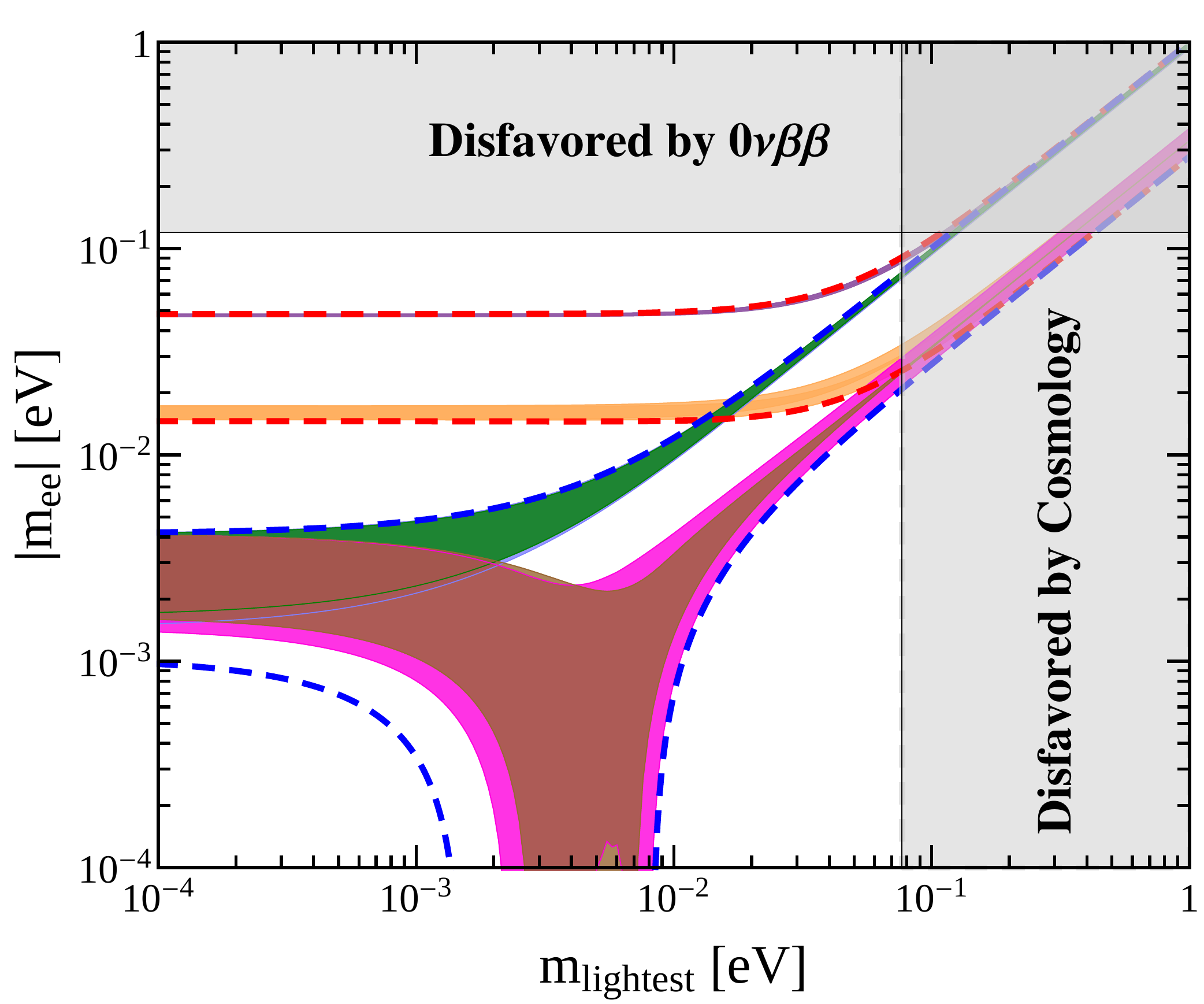}\\
\includegraphics[width=0.4\textwidth]{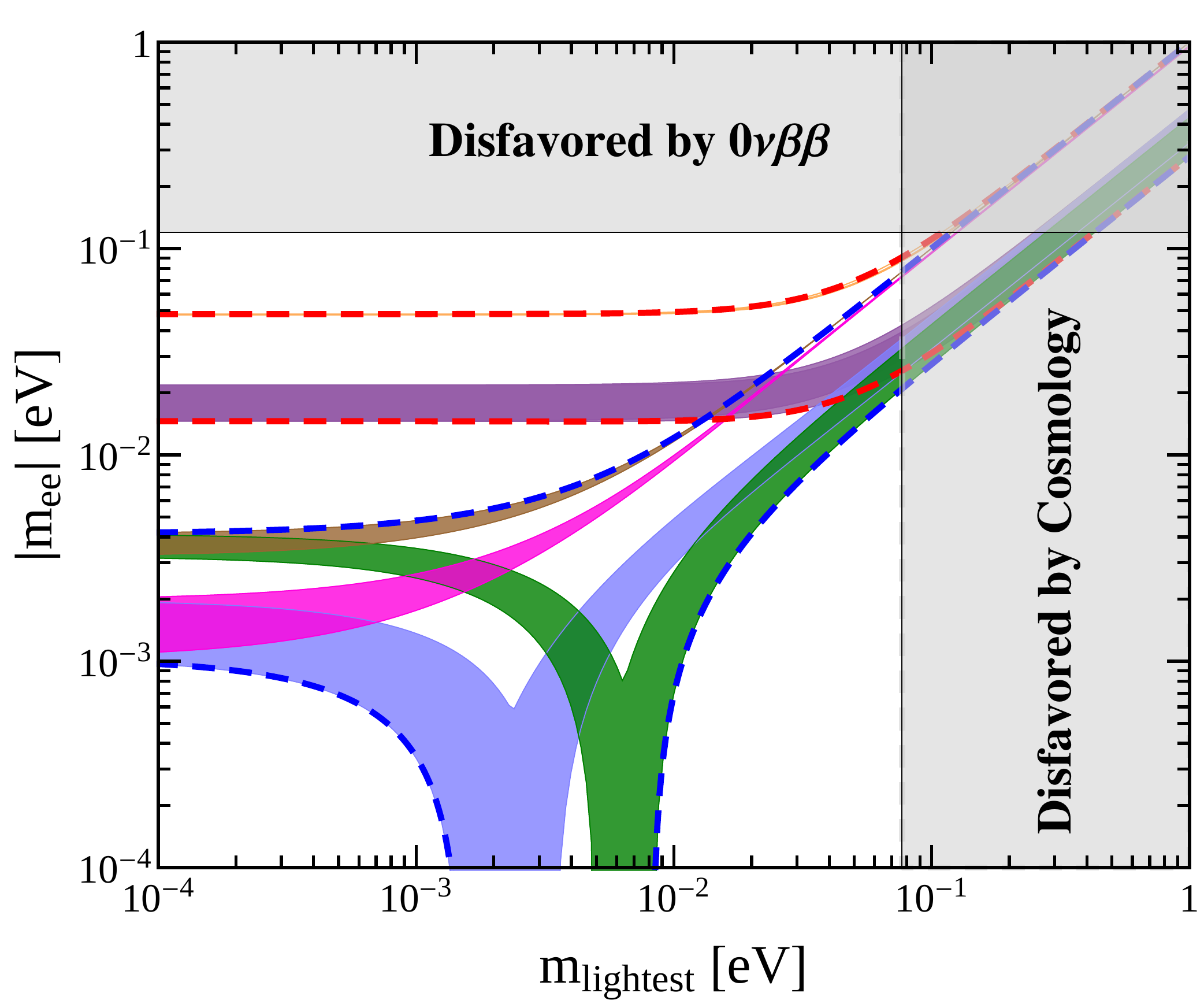}&
\includegraphics[width=0.4\textwidth]{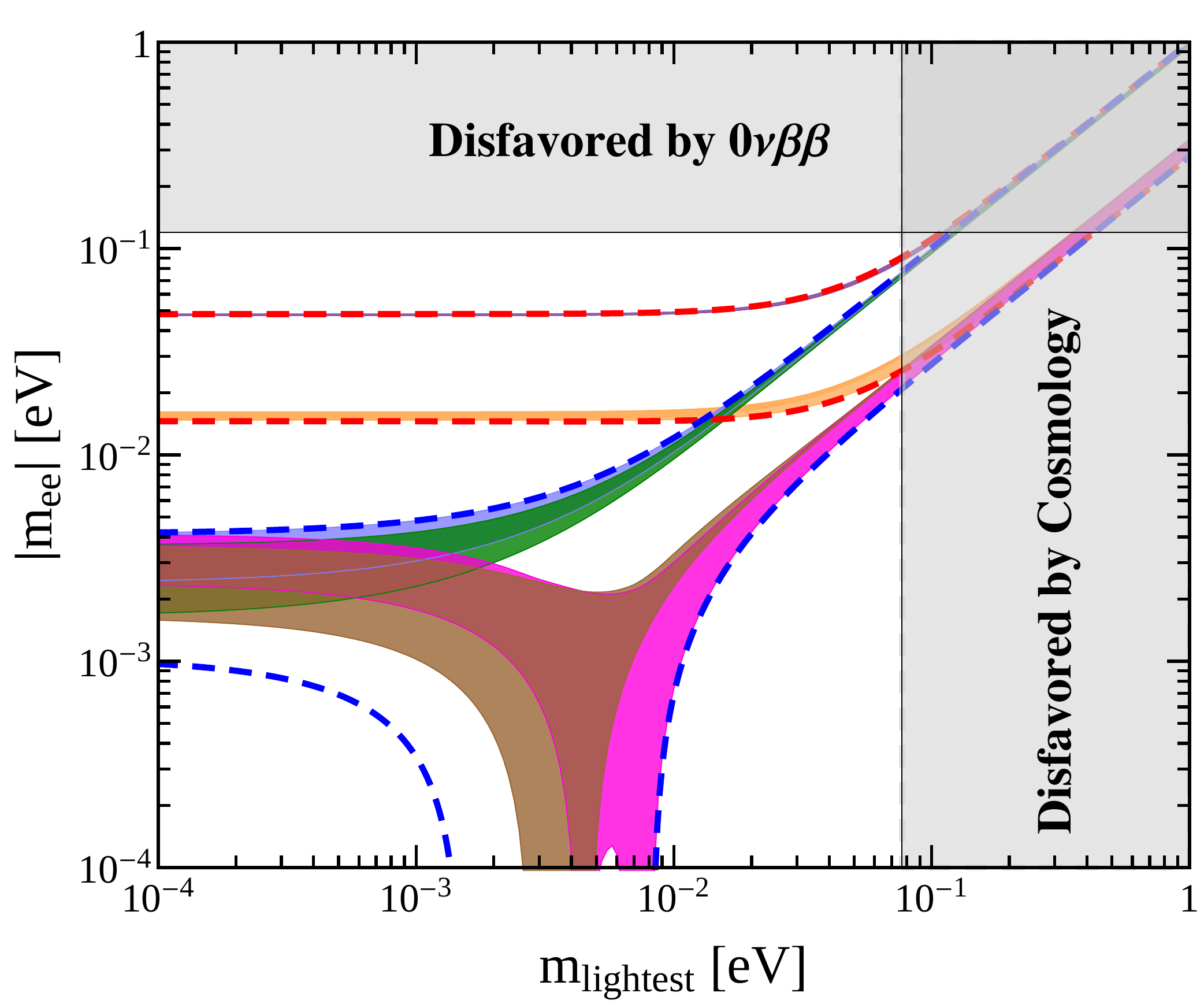}\\
\includegraphics[width=0.4\textwidth]{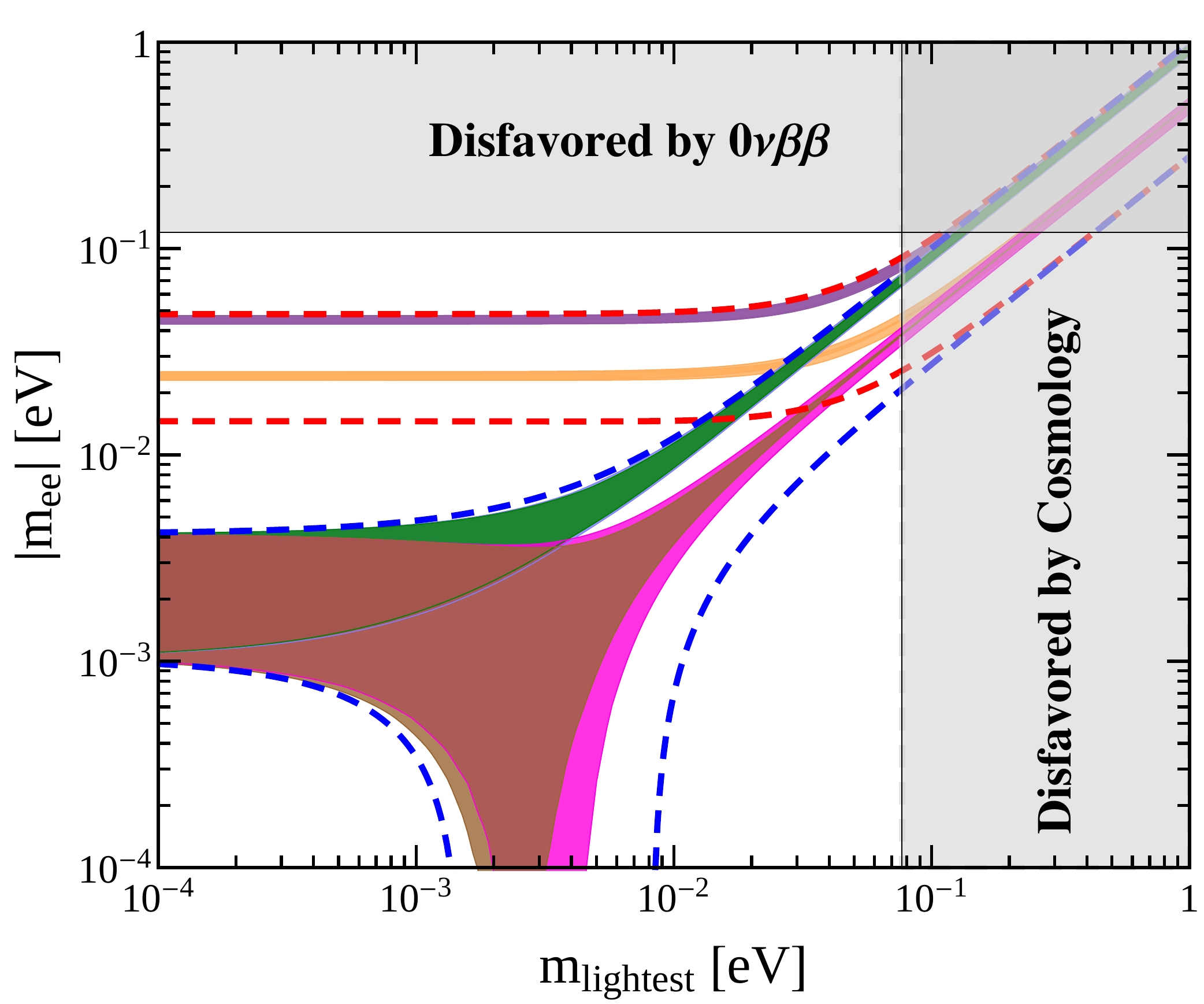}&~~~~~\quad
\includegraphics[width=0.32\textwidth]{note_v4.pdf}
\end{tabular}
\caption{\label{fig:0nubb_single_CP}The allowed regions of the effective Majorana mass $|m_{ee}|$ with respect to the lightest neutrino mass. The red (blue) dashed lines indicate the most general allowed regions for IH(NH) neutrino mass spectrum obtained by varying the mixing parameters over their 3$\sigma$ ranges~\cite{Gonzalez-Garcia:2014bfa}. The top row
corresponds to the residual symmetry $(G_{l}, X_{\nu})=(Z^{T}_3, 1)$ on the left and $(G_{l}, X_{\nu})=(Z^{T}_3, S)$ on the right, the middle row is for $(G_{l}, X_{\nu})=(Z^{T}_3, U)$ and $(G_{l}, X_{\nu})=(Z^{T}_3, SU)$, and the bottom row for $(G_{l}, X_{\nu})=(K^{(S, U)}_4, T)$.  The present most stringent upper limits $|m_{ee}|<0.120$ eV from EXO-200~\cite{Auger:2012ar, Albert:2014awa} and KamLAND-ZEN~\cite{Gando:2012zm} is shown by horizontal grey band. The vertical grey exclusion band is the current limit on the lightest neutrino masses from the cosmological data $\sum m_i<0.230$ eV at $95\%$ confidence level obtained by the Planck collaboration~\cite{Ade:2013zuv}. }
\end{figure}

Moreover, we explore the phenomenological predictions for neutrinoless double beta ($0\nu\beta\beta$) decay in each case. The effective mass $|m_{ee}|$ as a function of the lightest neutrino mass is plotted in figure~\ref{fig:0nubb_single_CP}. We find that $|m_{ee}|$ is around 0.015 eV, 0.024 eV or 0.048 eV for IH spectrum while $|m_{ee}|$ depends on the neutrino masses and it is strongly suppressed to be small than $10^{-4}$ eV for certain values of the lightest neutrino mass in case of NH.

\section{\label{sec:conclusion}Summary and conclusions}

In recent years, discrete flavor symmetry in combination with $CP$ symmetry has been pursued to describe the experimental data on lepton mixing in particularly to predict the $CP$ violating phases. Generally it is assumed that the original flavor and CP symmetry are broken down to an abelian subgroup and $Z_2\times CP$ in the charged lepton and neutrino sectors respectively. In this work we have considered other possible choices for the residual symmetry. In the first scenario, the residual subgroups preserved by the neutrino and charged lepton mass matrices are of the structure $Z_2\times CP$. The lepton mixing matrix is found to depend on two free parameters $\theta_{l}$ and $\theta_{\nu}$ which vary between $0$ and $\pi$, and generally one element is fixed to be certain constant by the residual symmetry. The procedure to extract the PMNS mixing matrix is presented. Moreover, we derive the criterion to determine whether two distinct remnant subgroups lead to the same mixing pattern if the freedom of redefining $\theta_{l}$ and $\theta_{\nu}$ is taken into account. In order to show concrete examples and find new interesting mixing patterns, we have performed a comprehensive analysis for the popular $S_4$ flavor symmetry group. All possible residual groups $Z_2\times CP$ have been considered, and we find eighteen phenomenologically viable cases which can accommodate the experimentally measured values of the mixing angles for particular values of $\theta_{l}$ and $\theta_{\nu}$, as shown in tables~\ref{tab:Uab}-\ref{tab:Uac_chi2_IH}. This scheme is quite predictive since the allowed regions of $\theta_{l}$ and $\theta_{\nu}$ are strongly constrained in order to accommodate the experimentally measured values of the mixing angles. In light of the recent experimental results of $\delta_{CP}\sim3\pi/2$ from T2K and NO$\nu$A~\cite{T2K_delta_CP,NovA_delta_CP}, the cases with $(G_{l}, G_{\nu}, X_{l}, X_{\nu}, P_{l}, P_{\nu})=(Z_{2}^{ST^2SU}, Z_{2}^{TU}, T^2, T, P_{12}, P_{12})$, $(Z_{2}^{ST^2SU}, Z_{2}^{S}, T^2, SU, P_{12}, P_{13})$, $(Z_{2}^{ST^2SU}, Z_{2}^{S}, T^2, SU, P_{12}, P_{13})$ are slightly preferred because they predict the Dirac phase  could be $1.569\pi$, $1.458\pi$ and $1.542\pi$ respectively. In all the eighteen cases, the effective Majorana mass $|m_{ee}|$ are determined to be around 0.015 eV, 0.028 eV or 0.048 eV for IH which are within the sensitivity of the near future $0\nu\beta\beta$ decay experiments

Discrete flavor symmetry has also been employed to explain the quark flavor mixing described by the well-known CKM matrix as well. Extensive scan of finite groups shows that only the Cabbibo mixing in the quark sector can be reproduced at leading order without resorting to special model dependent corrections~\cite{Holthausen:2013vba,Yao:2015dwa}, regardless of whether the three left-handed quark fields are assigned to an irreducible triplet or doublet plus singlet. In the approach with flavor and CP symmetry, if the remnant symmetries preserved by the down and up quark mass matrices are chosen to be an abelian subgroup and $Z_2\times CP$, the correct size of the quark mixing angles and CP phase still can not be obtained. In this work we propose the scheme with the residual symmetry $Z_2\times CP$ in both the up and down quark sectors. The expression for the CKM matrix and the equivalence condition are derived. From the $S_4$ flavor group along with a CP symmetry, we find an interesting leading order quark mixing pattern in which the experimentally preferred values of the quark mixing angles $\theta^{q}_{12}$ and $\theta^{q}_{23}$ can be accommodated while $\theta^{q}_{13}$ is a bit large. It could be brought into agreement with the experimental data in a concrete model with small subleading corrections. We comment that large flavor groups can accommodate well the precisely measured CKM mixing matrix without corrections in this approach~\cite{work_future}.

Furthermore we consider another type of residual symmetry. The postulated flavor and CP symmetry is broken to an abelian subgroup contained in the flavor group in the charged lepton sector and to a single remnant CP transformation in the neutrino sector. The lepton mixing angles and CP violation phases are determined in terms of three free parameters $\theta_{1,2,3}$ in the interval $\left[0, \pi\right)$. In general this scenario is less predictive than the previous one, each mixing parameter can vary in a relatively  wide range. As an example, we find that the flavor group $S_4$ combined with CP symmetry gives rise to five independent mixing patterns which can describe the experimental data on lepton mixing angles. The correlation between different mixing parameters and the predictions for the neutrinoless double beta decay are studied. Given the above rich results from the $S_4$ group, we expect that many other new mixing patterns compatible with experimental data could be obtained in our proposal for other choice of the flavor symmetry group such as $A_5$ and $\Delta(6n^2)$.

In the present work, we propose alternative schemes to understand the puzzle of quark and lepton flavor mixings from flavor and CP symmetry. The implications of our proposal for the flavor mixing are completely determined by the assumed residual symmetries and are independent of the underlying theory, they are just a consequence of group theory. It is interesting to construct explicit models to dynamically achieve the breaking patterns of flavor and CP symmetry. The required size of $\theta_{l}$ and $\theta_{\nu}$ (or $\theta_{1,2,3}$) as well as the charged lepton mass hierarchy should be obtained in such models.

\section*{Acknowledgements}

This work is supported by the National Natural Science Foundation of China under Grant Nos. 11275188, 11179007 and 11522546.

\appendix

\section{\label{sec:S4_group_app}Group theory of $S_4$}

$S_4$ is the permutation group of four distinct objects, and geometrically it is the symmetry group of a regular octahedron.
$S_4$ can be defined by three generators $S$, $T$ and $U$ which satisfy~\cite{Ding:2013hpa,Li:2013jya,Li:2014eia}
\begin{equation}
S^2=T^3=U^2=(ST)^3=(SU)^2=(TU)^2=(STU)^4=1\,.
\end{equation}
The 24 elements of the group belong to five conjugacy classes
\begin{eqnarray}\nonumber
1 {C}_1&=& ~\{1\}\ ,  \\ \nonumber
3 {C}_2&=&~\{S, TST^2, T^2ST\} \, \\
6{C}_2^{\prime}&=& ~\{U, TU, SU, T^2U, STSU,ST^2SU \} \, \\ \nonumber
8 {C}_3&=&~ \{T,ST,TS, STS, T^2, ST^2, T^2S, ST^2S\} \, \\ \nonumber
6C_4&=&~\{STU, TSU, T^2SU, ST^2U, TST^2U, T^2STU\}\,,
\end{eqnarray}
where $kC_n$ designates a conjugacy class of $k$ elements whose order is $n$. The group structure of $S_4$ has been studied in detail in Ref.~\cite{Ding:2009iy}. The residual flavor symmetry group can only be abelian group in order to avoid degenerate mass spectrum. The abelian subgroups of $S_4$ are given as follows,

\begin{itemize}[labelindent=-0.8em, leftmargin=1.2em]

\item{$Z_2$ subgroups}
\begin{equation}
\label{eq:Z2-subgroups}
\begin{array}{lll}
Z_2^{ST^{2}SU}=\{1,ST^{2}SU\},& ~~~ Z_2^{TU}=\{1,TU\},& ~~~ Z_2^{STSU}=\{1,STSU\},\\
Z_2^{T^2U}=\{1,T^2U\},&~~~ Z_2^{U}=\{1,U\}, &~~~ Z_2^{SU}=\{1,SU\}, \\
Z_2^{S}=\{1,S\},&~~~ Z_2^{T^2ST}=\{1,T^2ST\}, &~~~ Z_2^{TST^{2}}=\{1,TST^{2}\}\,,
\end{array}
\end{equation}
where the superscripts denote the generators of the subgroups. The first six $Z_{2}$ subgroups are related to each other by group conjugation, and the last three subgroups are conjugate to each other as well.
\item{$Z_3$ subgroups}
\begin{equation}
\label{eq:Z3-subgroups}
\begin{array}{ll}
Z_3^{ST}=\{1,ST,T^{2}S\},&\qquad Z_3^{T}=\{1,T,T^{2}\},\\
Z_3^{STS}=\{1,STS,ST^2S\},&\qquad Z_3^{TS}=\{1,TS,ST^{2}\}\,.
\end{array}
\end{equation}
All the above $Z_3$ subgroups are conjugate among each other.
\item{$Z_4$ subgroups}
\begin{eqnarray}\label{eq:Z4-subgroups}
\nonumber&&Z_4^{TST^{2}U}=\{1,TST^{2}U,S,T^{2}STU\},\quad Z_4^{ST^2U}=\{1,ST^2U,TST^{2},T^2SU\},\\
&&Z_4^{TSU}=\{1,TSU,T^2ST,STU\}\,,
\end{eqnarray}
which are related with each other under group conjugation.
\item{$K_4$ subgroups}
\begin{equation}\label{eq:K4-subgroups}
\begin{array}{l}
K^{(S, TST^{2})}_4\equiv Z_2^{S}\times Z_2^{TST^{2}}=\{1,S,TST^{2},T^{2}ST\},\\
K^{(S,U)}_4\equiv Z_2^{S}\times Z_2^{U}=\{1,S,U,SU\}, \\
K^{(TST^{2}, T^{2}U)}_4\equiv Z_2^{TST^{2}}\times Z_2^{T^{2}U}\equiv \{1,TST^{2},T^2U,ST^{2}SU\}, \\
K^{(T^{2}ST, TU)}_4\equiv Z_2^{T^{2}ST}\times Z_2^{TU}=\{1,T^{2}ST,TU,STSU\}\,,
\end{array}
\end{equation}
where $K^{(S, TST^{2})}_4$ is a normal subgroup of $S_4$, and the remaining three $K_4$ subgroups are conjugate to each other.
\end{itemize}
The group $S_4$ has five irreducible representations: two singlets $\mathbf{1}$ and $\mathbf{1}'$, one doublet $\mathbf{2}$, and two triplets $\mathbf{3}$ and $\mathbf{3}'$. The representation matrices for the generators $S$, $T$ and $U$ in each of the irreducible representations are summarized in table~\ref{tab:representation}. Notice that the representations $\mathbf{3}$ and $\mathbf{3}'$ differ in the overall sign of the generator $U$. As has been shown in previous work~\cite{Ding:2013hpa,Li:2013jya}, the generalized CP transformation compatible with the $S_4$ flavor symmetry is of the
same form as the flavor group transformation in our working basis.

\begin{table}[t!]
\begin{center}
\begin{tabular}{|c|c|c|c|}\hline\hline
 ~~  &  $S$  &   $T$    &  $U$  \\ \hline
~~~${\bf 1}$, ${\bf 1^\prime}$ ~~~ & 1   &  1  & $\pm1$  \\ \hline
   &   &    &    \\ [-0.16in]
${\bf 2}$ &  $\left( \begin{array}{cc}
    1&0 \\
    0&1
    \end{array} \right) $
    & $\left( \begin{array}{cc}
    \omega&0 \\
    0&\omega^2
    \end{array} \right) $
    & $\left( \begin{array}{cc}
    0&1 \\
    1&0
    \end{array} \right)$\\ [0.12in]\hline
   &   &    &    \\ [-0.16in]
${\bf 3}$, ${\bf 3^\prime}$ & $\frac{1}{3} \left(\begin{array}{ccc}
    -1& 2  & 2  \\
    2  & -1  & 2 \\
    2 & 2 & -1
    \end{array}\right)$
    & $\left( \begin{array}{ccc}
    1 & 0 & 0 \\
    0 & \omega^{2} & 0 \\
    0 & 0 & \omega
    \end{array}\right) $
    & $\mp\left( \begin{array}{ccc}
    1 & 0 & 0 \\
    0 & 0 & 1 \\
    0 & 1 & 0
    \end{array}\right)$
\\[0.22in] \hline\hline
\end{tabular}
\caption{\label{tab:representation}The representation matrices of the generators $S$, $T$ and $U$ in different irreducible representations of $S_4$, where $\omega=e^{2\pi i/3}$.}
\end{center}
\end{table}

\section{\label{sec:equivalence_quark_n0_n1_app}Equivalence conditions for two CKM matrices with $|a_{1}|=|b_{1}|\neq 0,1$ }

Following the methods in section~\ref{subsec:criterion_Z2xCP} and section~\ref{sec:quark_Z2xCP}, we can find out the criterion to determine whether two distinct residual symmetries of the structure $Z_2\times CP$ in both the up and down type quark sectors lead to the same CKM matrix for the general case with $|a_{1}|=|b_{1}|\neq 0,1$, if possible shifts of the free parameters $\theta_{u}$ and $\theta_d$ are considered. The expression for the combination $U_q\equiv \Sigma^{\dagger}_u\Sigma_d$ is written as Eq.~\eqref{eq:quk58}. One can always set $a_1$ and $b_1$ to be real and positive by redefining the quark fields. We shall report the results in the following.

\begin{itemize}[labelindent=-0.6em, leftmargin=1.0em]
\item{$b_{2}^{2}+b_{3}^{2}\neq 0$,~ $b_{4}^{2}+b_{7}^{2}\neq 0$}

In this case, the conditions under which essentially the same quark mixing is obtained, are given by
\begin{eqnarray}
\nonumber
&&|a_{2}^2+a_{3}^2|=|b_{2}^2+b_{3}^2|,\qquad (a_{2}b_{2}+a_{3}b_{3})(a^{*}_{2}b^{*}_{3}-a^{*}_{3}b^{*}_{2})\in \mathbb{R},\\
\nonumber&&|a_{4}^2+a_{7}^2|=|b_{4}^2+b_{7}^2|,\qquad (a_{4}b_{4}+a_{7}b_{7})(a^{*}_{4}b^{*}_{7}-a^{*}_{7}b^{*}_{4})\in \mathbb{R}\,,\\
\nonumber&&a_{5}=\frac{(xb_{5}+yb_{6})z+(xb_{8}+yb_{9})w}{(b_{2}^{2}+b_{3}^{2})(b_{4}^{2}+b_{7}^{2})},~~a_{6}=\frac{(xb_{6}-yb_{5})z+(xb_{9}-yb_{8})w}{(b_{2}^{2}+b_{3}^{2})(b_{4}^{2}+b_{7}^{2})}\,,\\
\label{eq:z2lep86}&&
a_{8}=\frac{(xb_{8}+yb_{9})z-(xb_{5}+yb_{6})w}{(b_{2}^{2}+a_{3}^{2})(b_{4}^{2}+b_{7}^{2})},~~a_{9}=\frac{(xb_{9}-yb_{8})z-(xb_{6}-yb_{5})w}{(b_{2}^{2}+b_{3}^{2})(b_{4}^{2}+b_{7}^{2})}\,,
\end{eqnarray}
with
\begin{eqnarray}
\label{eq:z2lep87}
x\equiv a_{2}b_{2}+a_{3}b_{3},~~y\equiv a_{2}b_{3}-a_{3}b_{2},~~z\equiv a_{4}b_{4}+a_{7}b_{7},~~w\equiv a_{4}b_{7}-a_{7}b_{4}\,.
\end{eqnarray}

\item{$b_{2}^2+b_{3}^2=0$, ~$b_{4}^2+b_{7}^2\neq 0$}

The equivalent conditions are found to be
\begin{eqnarray}
\nonumber&&a_2b_2+a_3b_3=0,\quad  |a_{4}^{2}+a_{7}^{2}|=|b_{4}^{2}+b_{7}^{2}|,\quad (a_{4}b_{4}+a_{7}b_{7})(a^{*}_{4}b^{*}_{7}-a^{*}_{7}b^{*}_{4})\in \mathbb{R}\,,\\
&&t_{i}T_{j}-t_{j}T_{i}=0,\quad t_i/T_i\in\mathbb{R},\quad\text{with}\quad i,j=5, 6, 8, 9\,,
\end{eqnarray}
where
\begin{eqnarray}
\nonumber
&&t_{5}=va_{5}b_2-(zb_{5}+wb_{8})a_2,~~t_{6}=-va_{6}b_2+(zb_{6}+wb_{9})a_2,\\
\nonumber
&&t_{8}=va_{8}b_2-(zb_{8}-wb_{5})a_2,~~t_{9}=-va_{9}b_2+(zb_{9}-wb_{6})a_2\,,\\
\nonumber
&&T_{5}=-iva_{5}b_2-i(zb_{6}+wb_{9})a_3,~~T_{6}=iva_{6}b_2-i(zb_{5}+wb_{8})a_3,\\
\label{eq:z2lep98}
&&T_{8}=-iva_{8}b_2-i(zb_{9}-wb_{6})a_3,~~T_{9}=iva_{9}b_2-i(zb_{8}-wb_{5})a_3\,,
\end{eqnarray}
with
\begin{equation}
\label{eq:z2lep99}
v\equiv b_{4}^2+b_{7}^2\,.
\end{equation}

\item{$b_{2}^2+b_{3}^2\neq 0$,~ $b_{4}^2+b_{7}^2=0$}

The resulting CKM matrices would be related through redefinition of the parameters $\theta_{u}$ and $\theta_{d}$ if the following constraints are fulfilled,
\begin{eqnarray}
\nonumber&& a_4b_4+a_7b_7=0,\quad |a_{2}^2+a_{3}^2|=|b_{2}^2+b_{3}^2|,\quad (a_{2}b_{2}+a_{3}b_{3})(a^{*}_{2}b^{*}_{3}-a^{*}_{3}b^{*}_{2})\in\mathbb{R}\,,\\
&&t'_{i}T'_{j}-t'_{j}T'_{i}=0,\quad t'_i/T'_i\in\mathbb{R},\quad\text{with}\quad i,j=5, 6, 8, 9\,,
\end{eqnarray}
where
\begin{eqnarray}
\nonumber
&&t'_{5}=ua_{5}b_4-(xb_{5}+yb_{6})a_4,~~t'_{6}=ua_{6}b_4-(xb_{6}-yb_{5})a_4,\\
\nonumber
&&t'_{8}=-ua_{8}b_4+(xb_{8}+yb_{9})a_4,~~t'_{9}=-ua_{9}b_4+(xb_{9}-yb_{8})a_4\,,\\
\nonumber
&&T'_{5}=-iua_{5}b_4-i(xb_{8}+yb_{9})a_7,~~T'_{6}=-iua_{6}b_4-i(xb_{9}-yb_{8})a_7,\\
\label{eq:z2lep98}
&&T'_{8}=iua_{8}b_4-i(xb_{5}+yb_{6})a_7,~~T'_{9}=iua_{9}b_4-i(xb_{6}-yb_{5})a_7\,,
\end{eqnarray}
with
\begin{equation}
u\equiv b^2_2+b^2_3\,.
\end{equation}

\item{$b_{2}^2+b_{3}^2=0$,~ $b_{4}^2+b_{7}^2=0$}

The postulated residual symmetries would give rise to the same quark mixing pattern if the following conditions are satisfied,
\begin{equation}
a_2b_2+a_3b_3=0,\quad a_4b_4+a_7b_7=0,\quad a^2_2a_4(b_2b_5-b_3b_6)=b^2_2b_4(a_2a_5-a_3a_6)\,.
\end{equation}
Note that the above results are valid up to the transformations in Eq.~\eqref{eq:quk79}.

\end{itemize}

\end{document}